Opinion

# Mitochondrial Genomes of Domestic Animals Need Scrutiny

Ni-Ni Shi[1,2,‡], Long Fan[3,‡], Yong-Gang Yao[2,4], Min-Sheng Peng[1,2,*], Ya-Ping Zhang[1,2,5,*]

[1] State Key Laboratory of Genetic Resources and Evolution, and Yunnan Laboratory of Molecular Biology of Domestic Animals, Kunming Institute of Zoology, Chinese Academy of Sciences, Kunming, Yunnan 650223, China

[2] Kunming College of Life Science, University of Chinese Academy of Sciences, Kunming, Yunnan 650204, China

[3] School of Life Sciences, The Chinese University of Hong Kong, Shatin, Hong Kong, China

[4] Key Laboratory of Animal Models and Human Disease Mechanisms, Kunming Institute of Zoology, Chinese Academy of Sciences, Kunming, Yunnan 650223, China

[5] Laboratory for Conservation and Utilization of Bio-Resources & Key Laboratory for Microbial Resources of the Ministry of Education, Yunnan University, Kunming, Yunnan 650091, China

[‡] These authors contributed equally to this work.

* Correspondence:

Dr. Min-Sheng Peng

State Key Laboratory of Genetic Resources and Evolution, and Yunnan Laboratory of Molecular Biology of Domestic Animals, Kunming Institute of Zoology, Chinese Academy of Sciences, Kunming, China

Tel: +86 871-6518 9265

E-mail: pengminsheng@mail.kiz.ac.cn

Or

Prof. Ya-Ping Zhang

State Key Laboratory of Genetic Resources and Evolution, and Yunnan Laboratory of Molecular Biology of Domestic Animals, Kunming Institute of Zoology, Chinese Academy of Sciences, Kunming, China

Tel: +86 871-6519 9030

Fax: +86 871-6519 5430

E-mail: zhangyp@mail.kiz.ac.cn

**Abstract**

More than 1000 complete or near-complete mitochondrial DNA (mtDNA) sequences have been deposited in GenBank for eight common domestic animals (i.e. cattle, dog, goat, horse, pig, sheep, yak and chicken) and their close wild ancestors or relatives. Nevertheless, few efforts have been performed to evaluate the sequence data quality, which heavily impact the original conclusion. Herein, we conducted a phylogenetic survey of these complete or near-complete mtDNA sequences based on mtDNA haplogroup trees for the eight animals. We show that, errors due to artificial recombination, surplus of mutations, and phantom mutations, do exist in 14.5% (194/1342) of mtDNA sequences and shall be treated with wide caution. We propose some caveats for mtDNA studies of domestic animals in the future.

## Text

Mitochondrial DNA (mtDNA) is one of the most widely used markers in exploring genetic diversity and tracing evolutionary history for human and domestic and wild animals. Since 2006, we witnessed the change from partial mtDNA sequence (e.g. control region) to complete mitochondrial genome in abundant studies of various domestic animals (Wang *et al.* 2014). Hitherto more than 1000 complete or near-complete mtDNA sequences have been deposited in GenBank for eight common domestic animals (i.e. cattle, dog, goat, horse, pig, sheep, yak and chicken) and their close wild ancestors or relatives (Table 1). However, most researchers took those reported mtDNA genomes of domestic animals in their data-mining and comparative analyses without any scrutiny for data quality. There are warnings as some sequences showing to contain sequencing errors (Achilli *et al.* 2012; Miao *et al.* 2013) and nuclear mtDNAs (NUMTs) contaminations (Hassanin *et al.* 2010), similar to a condition for human mtDNA studies (Yao *et al.* 2008; Yao *et al.* 2009).

Most of released mtDNA genome sequences of domestic animals were generated through several PCR and Sanger sequencing reactions (e.g. Pang *et al.* 2009; Yu *et al.* 2013). These practices are labor-intensive and prone to have errors, as has been well demonstrated in human mtDNA data generated through similar experimental protocols (Bandelt *et al.* 2006). It triggers us to ask: do the similar errors also occur in the mitochondrial genome data of domestic animals? Herein, we summarized three major kinds of errors (Table 2), and screened these errors in 1342 complete or near-complete mtDNA sequences from eight domestic animals and their close wild ancestors or relatives (Table 1).

We adopted the phylogenetic strategy developed in human mtDNA analyses (Yao et al. 2009) based on the high-resolution genealogy (i.e. mtDNA haplogroup tree). In brief, mtDNA haplogroup tree can be constructed with the parsimony-like methods such as networks (Bandelt et al. 1999). And the haplogroups can be defined, each with diagnostic mutational motif - a string of characteristic mutations shared exclusively by the members belonging to certain haplogroup that define the internal branch directing to the haplogroup (internal node) in the tree (van Oven & Kayser 2009). When comparing with the defined reference sequence (Bandelt et al. 2014), variants in each sequence can be scored. According the haplogroup tree, the variants can be mapped on the internal branch as diagnostic mutations or on the terminal branch (tip). As a result, sequences with anomalous variants showing conflicts with their known phylogenetic status can be identified and shall be treated with caution. This phylogenetic method has been proved to be powerful and sensitive for human mtDNA data quality assessment (Bandelt *et al.* 2006; Kong *et al.* 2008; Salas *et al.* 2005a; Yao *et al.* 2009).

Since 2007, the mtDNA haplogroup trees for pig (Wu *et al.* 2007), cattle (Achilli *et al.* 2009; Achilli *et al.* 2008; Bonfiglio *et al.* 2010; Bonfiglio *et al.* 2012), horse (Achilli *et al.* 2012), chicken (Miao *et al.* 2013), and sheep (Lancioni *et al.* 2013) have been constructed. In terms of the available trees, we followed the proposed caveats (Bandelt *et al.* 2006; Kong *et al.* 2008; Yao *et al.* 2008; Yao *et al.* 2009) to investigate the data quality. We discovered that potential errors occurred in cattle, horse, pig, and chicken (Table 3; Table S1). For instance, in five pig mitochondrial genomes published recently (Yu *et al.* 2013), two sequences (GenBank accession numbers KC250273 and KC469586) were problematic (Figure 1a). The mis-added variants C9553T-C9605T in KC250273 (which belongs to haplogroup D1e) are the diagnostic motif of haplogroup E1a. Similarly, motifs T2374C (i.e. @2374)-T2613C, A5672G-C5678T-T5708C-T5753C, and C9156T-C9157T in KC469586 (which belongs to haplogroup D1a1a) characterize haplogroup E. The missed and mis-added mutations most likely represent as "recombinants" of separate segments from different samples of haplogroups E1a and E that are common in European pigs. The errors of artificial recombination are likely due to sample mix-up or contamination with European pig breed(s).

We followed the same approach to check mitochondrial genome data from other domestic animals. We first reconstructed mtDNA haplogroup trees for dog, goat, and yak (Table S2-S4) based on the parsimony-like method as described elsewhere (Miao *et al.* 2013; Wu *et al.* 2007). The hierarchical haplogroup nomenclature systems were updated from previous studies (Doro *et al.* 2014; Pang *et al.* 2009; Wang *et al.* 2010), with referring to the rules in human mtDNA phylogeny (van Oven & Kayser 2009). Not unexpectedly, several potential errors are detected in the mtDNA sequences of these animals (Table 3; Table S1). For instance, a dog sequence (EU789672) from our previous study (Pang *et al.* 2009) was suspected to be corroded by artificial recombination. The diagnostic variants G2232A and T2227C-T2837C-G3296A, which define haplogroups A1’2’3 and A3, respectively, are missing (Figure 1b). It was likely due to sample mix-up with replacing the fragment around nucleotide position 2200 – 3300 of EU789672 with other sample of non-haplogroup A1’2’3. We re-amplified this mtDNA and sequenced the entire mitochondrial genome. Indeed, the suspected artificial recombination is confirmed (Figure 1b). The corrected sequence has been updated in GenBank as KM113774.

In addition to artificial recombination, we show that, phantom mutation and poor sequencing quality do exist in mtDNA data of domestic animals (Table 3; Table S1). When using the flawed sequences, it is prone to get erroneous claims. In some cases, it is expected that, the surplus of mutations can “increase” length of branch in the phylogeny and “accelerate” molecular evolution (e.g. detection of positive selection and violation of molecular clock). Indeed, there’re several cases reported in human mtDNA studies (e.g. Liu *et al.* 2012; Salas *et al.* 2005b). Moreover, phantom mutations most likely due to technological pitfalls in next-generation sequencing platforms (especially for homopolymers) are detected (Table S5). Similarly, data qualities of some human mtDNA sequences generated via next-generation sequencing platforms were far from being satisfied (Bandelt & Salas 2012).

Along with the progress of sequencing techniques, accumulation of mitochondrial genome sequences from domestic animals is accelerating. The common practice of generating and analyzing mtDNA data should be carried out with the necessary caution. Analyses based on traditional phylogenetic software, at least in the related literatures (e.g. Pang *et al.* 2009; Yu *et al.* 2013), are inefficient to discern those errors. Thus, we suggest researchers to follow some caveats from human mtDNA studies during experiments and data analyses (Bandelt *et al.* 2006; Kong *et al.* 2008; Salas *et al.* 2005a; Yao *et al.* 2008; Yao *et al.* 2009). Phylogenetic analyses based on mtDNA haplogroup trees are recommended, although it may be difficult for a non-expert or a novice in mtDNA analysis. Developing bioinformatic tools to make the related analyses convenient and efficient should be encouraged in the future. Furthermore, we suggest that these flawed sequences identified in this study (Table S1) should be never used in future data-mining analyses, unless a correct one was updated. We welcome the researchers of the original study take our warnings and double check these sequences tagged with "flawed", just as what we did for the dog sequence EU789672.

**Acknowledgments**

This work was supported by the 973 program (2013CB835200, 2013CB835204). The Youth Innovation Promotion Association, Chinese Academy of Sciences (CAS) provided support to M-S.P. M-S.P. is a member of Youth Innovation Promotion Association, CAS.

## Table Legends

Table 1. Summary of (near-)complete mtDNA genome sequences analyzed in this study.

Table 2. Three kinds of common errors occurring in mtDNA data analyzed in this study.

Table 3. mtDNA sequences with potential errors in eight domestic animals.

## Figure Legends

Figure 1. The flawed mitochondrial genome sequences in pigs (a) and dogs (b).

The nucleotide positions in the sequences from pigs and dogs were scored relative to the reference sequences EF545567 and EU789787, respectively. Transitions are shown on the branches and transversions are further annotated by adding suffixes. Deletions and insertions are indicated by 'd' and '+', respectively. Mutations toward a base identical-by-state to the reference sequences are indicated with the prefix @. Suspected mutations due to errors are in red; and their corresponding phylogenetic status in haplogroup trees are underlined.

## Supporting Information

Table S1. Potential errors identified in eight domestic animals.

Table S2. mtDNA haplogroup tree of dog and gray wolf.

Table S3. mtDNA haplogroup tree of goat.

Table S4. mtDNA haplogroup tree of yak.

Table S5. Potential errors detected in next-generation sequencing data.

Table 1 Summary of (near-)complete mtDNA genome sequences analyzed in this work.

| **Animals** | **Related Species** | **Released by GenBank*** | **Reference sequence** | **Haplogroup nomenclature** |
|---|---|---|---|---|
| Cattle and Aurochs | *Bos taurus* | 266 | V00654 | I, P, Q, R, and T |
| | *B. indicus* | 10 | | |
| | *B. primigenius* | 2 | | |
| Yak and wild yak | *B. grunniens* | 79 | GQ464259# | A – E |
| Dog and gray wolf | *Canis lupus* | 447 | EU789787# | A – F |
| Horse and Przewalski's horse | *Equus caballus* | 254 | JN398377 | A – R |
| | *E. przewalskii* | 9 | | |
| Chicken and Red Junglefowl | *Gallus gallus* | 66 | AP003321 | A – I and X – Z |
| Pig and wild boar | *Sus scrofa* | 127 | EF545567 | A, D, and E |
| Sheep | *Ovis aries* | 47 | AF010406 | A – E |
| Goat | *Capra hircus* | 35 | GU068049 | A – C |

Note: *Access time: June 1st, 2014. #Reference sequences are defined in this work.

Table 2. Three kinds of common errors occurring in mtDNA data analyzed in this study.

| **Errors** | **Common Phenotypes** | **Major Causes** |
|---|---|---|
| Artificial recombination | Missing diagnostic mutations; mis-added diagnostic mutations of different haplogroups | Sample mix-up; contamination |
| Surplus of mutations | Excessive unusual mutations, especially transversions, indels and heteroplasmic mutations | Sequencing errors; NUMTs contamination |
| Phantom mutation | Mutations are laboratory-specific and occur repeatedly on different haplogroup backgrounds | Low quality of sequencing; technical pitfalls of NGS |

Table 3. mtDNA genome sequences with potential errors in eight domestic animals.

| **Animals** | **Released by GenBank** | **Artificial recombination** | **Surplus of mutations** | **Phantom mutations** |
|---|---|---|---|---|
| Cattle | 278 | 1 | 1 | 7+34* |
| Chicken | 66 | | 1 | |
| Dog | 447 | 11 | 5 | |
| Goat | 35 | | 4 | |
| Horse | 263 | 20 | 2 | 1+71* |
| Pig | 127 | 7 | 19 | 15 |
| Sheep | 47 | | 2 | |
| Yak | 79 | 7 | 1 | |

Note: *detected in next-generation sequencing data (Table S5).

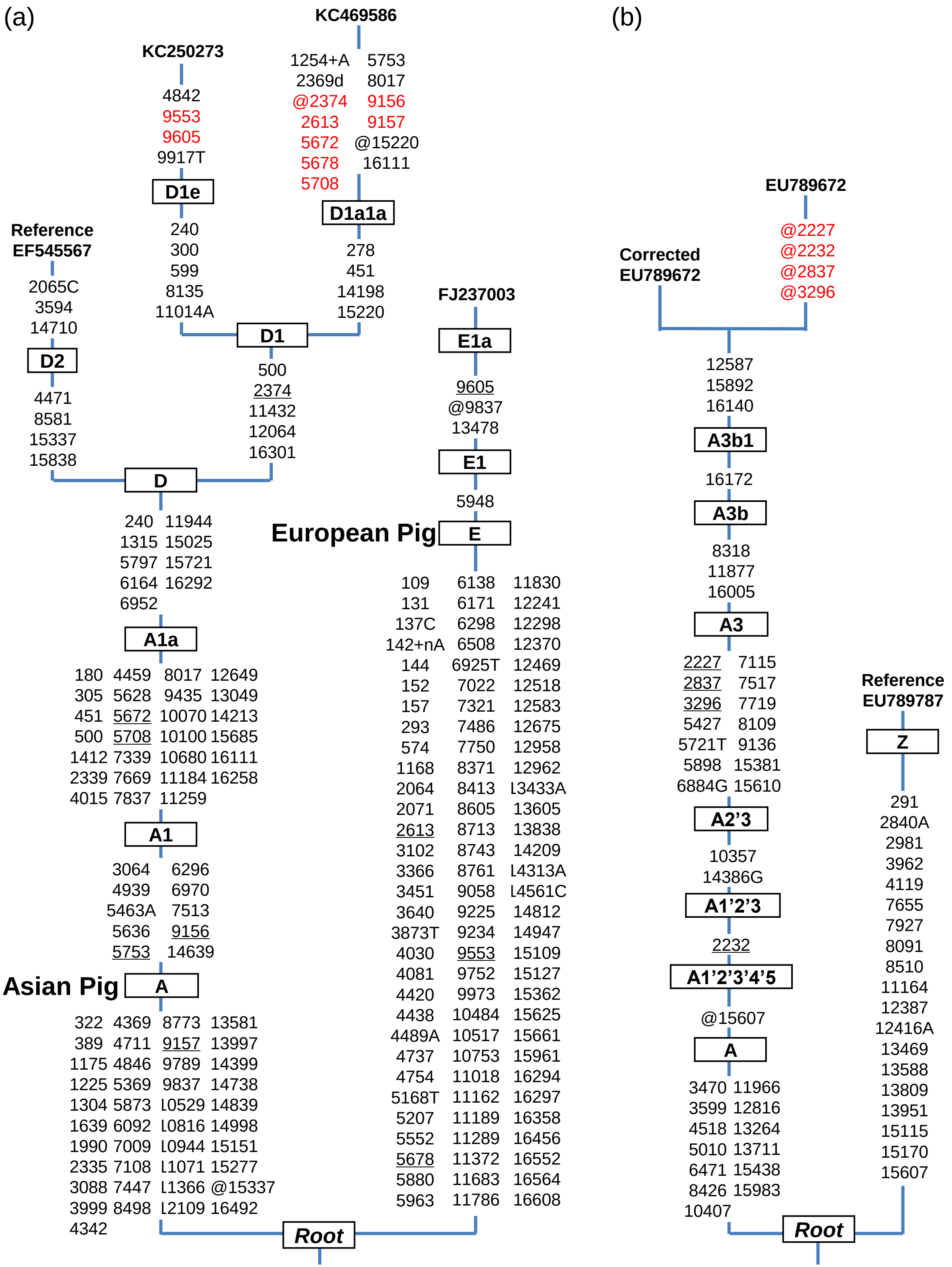

**Supporting Information**

Table S1. Potential errors identified in eight domestic animals.

| **Sequence** | **Potential Error** | **Haplogroup Most Matched Sequence** | **Variants** | **Missing Diagnostic Mutations** | **Private** | **Sources** |
|---|---|---|---|---|---|---|
| AB074962(Cattle and Aurochs) | Phantom mutations: 105d, and missing mutations 587+C, 2536A, 9682C. | T4 | 105d, 169, 6178, 12158, 13310C, 15510, 16042, 16093, 16302 | T4:587+C, 2536A, 9682C | T4:105d, 6178 | Mannen,H., Morimoto,M. and Tsuji,S. Unpublished |
| AB074963(Cattle and Aurochs) | Phantom mutations: 105d, and missing mutations 587+C, 2536A, 9682C. | T4a | 105d, 169, 11174, 12158, 13310C, 14565, 15510, 16042, 16093, 16302 | T4a:587+C, 2536A, 9682C | T4a:105d, 14565 | Mannen,H., Morimoto,M. and Tsuji,S. Unpublished |
| AB074964(Cattle and Aurochs) | Phantom mutations: 105d, and missing mutations 587+C, 2536A, 9682C. | T4 | 105d, 169, 12158, 13310C, 15510, 16093, 16302 | T4:587+C, 2536A, 9682C, 16042 | T4:105d | Mannen,H., Morimoto,M. and Tsuji,S. Unpublished |
| AB074965(Cattle and Aurochs) | Phantom mutations: 105d, and missing | T3k1 | 105d, 169, 4301, 6424, 10889, 12908, 13310C, 14140, 16055, 16112, 16122 | T3k1:587+C, 2536A, 9682C, 15910 | T3k1:105d, 4301, 6424, 14140, 16112, 16122 | Mannen,H., Morimoto,M. and Tsuji,S. Unpublished |
| | mutations 587+C, 2536A, 9682C. | | | | | |
| AB074966(Cattle and Aurochs) | Phantom mutations: 105d, and missing mutations 587+C, 2536A, 9682C. | T3k, T3q, T3r_V00654 | 105d, 169, 221+C, 8400, 12908, 13310C, 14122, 16055, 16058, 16088 | T3k:587+C, 2536A, 9682C; T3q:587+C, 2536A, 9682C | T3k:105d, 221+C, 8400, 14122, 16055, 16058, 16088; T3q:105d, 221+C, 8400, 12908, 14122, 16055, 16088; T3r_V00654:105d, 169, 221+C, 8400, 12908, 13310C, 14122, 16055, 16058, 16088 | Mannen,H., Morimoto,M. and Tsuji,S. Unpublished |
| AB074967(Cattle and Aurochs) | Phantom mutations: 105d, and missing mutations 587+C, 2536A, 9682C. | T3n, T3r_V00654 | 105d, 169, 2232, 13310C, 16119, 16302 | T3n:587+C, 2536A, 9682C | T3n:105d, 2232, 16302; T3r_V00654:105d, 169, 2232, 13310C, 16119, 16302 | Mannen,H., Morimoto,M. and Tsuji,S. Unpublished |
| AB074968(Cattle and Aurochs) | Phantom mutations: 105d, and missing mutations 587+C, 2536A, 9682C. | T3n, T3r_V00654 | 105d, 169, 2232, 13310C, 16104, 16119 | T3n:587+C, 2536A, 9682C | T3n:105d, 2232, 16104; T3r_V00654:105d, 169, 2232, 13310C, 16104, 16119 | Mannen,H., Morimoto,M. and Tsuji,S. Unpublished |
| DQ124403 (Cattle and Aurochs) | Artificial recombination: missing mutations 518,737, 3931, 3976, 3985, 4000, 4328, 4442, 4937, 7304, 7330, | I | 8, 106, 166, 173, 206, 221+C, 232, 233, 248, 249, 296, 300, 587+C, 649, 761, 816, 1158, 1474, 1492, 1677, 1824, 1860A, 1869, 2016, 2099, 2536A, 2558, 2575, 2953, 2977, 2979, 2988, 2989, 2990, 3051, 3071, 3136, 3145, 3241, 3325, 3335, 3379, 3439, 3535, 3550, 3600, 3793, 3829, 3874, 4562, 4730, 4733, 4769, 4871, 5285, 5501, 5531, 5614, 5705, 5743, 5783, 5890, 5917, 5998, 6115T, 6235, 6340, 6367, 6379, 6436, 6460, 6727, 6772, 6881, 6922, 8210, 8308, 8370C, 8466, 8494, | I:518, 737, 3931, 3976, 3985, 4000, 4328, 4442, 4937, 7304, 7330, 7356, 7358, 7361, 7514, 7830, 7851, 8045, 8168, | I:173, 221+C, 649, 2575, 5705, 9302, 9480, 12158, 12582, 13371, 16042, 16093 | Shin,H.D. and Kim,L.H. Unpublished |
| | 7356, 7358, 7361, 7514, 7830, 7851, 8045, 8168, 8188, 8285, 9068, 12135, 12178, 12234, 12377, 15579, 15593, 15818, 15951, 15953G, 15959, 15994, 16022, 16049, 16057, 16058, 16074, 16082, 16084, 16102, 16109, 16113, 16116, 16117, 16119, 16121, 16122, 16130, 16137, 16138, 16143d, 16147, 16196, 16200+A, 16229, 16247, 16248, 16300, 16301 of haplogroup I; mis-added mutations 12158,16042,16093 from haplogroup | | 8503, 8571, 8749, 8984, 9005, 9038, 9245, 9302, 9480, 9581, 9602, 9682C, 9767, 9891, 9930C, 9978, 10039, 10066, 10137, 10153, 10268, 10322, 10331, 10445, 10590, 10621, 10691G, 10849, 11000, 11035, 11068, 11134, 11200, 11266, 11329, 11407, 11419, 11803, 11842, 12158, 12433, 12468, 12469, 12513, 12582, 12622, 12672, 12675, 12684, 12750, 12801, 12900Y, 12923T, 12924, 13005, 13056A, 13098, 13104, 13275, 13310C, 13371, 13380, 13433, 13437, 13554, 13564, 13584, 13677, 13689, 13692, 13882, 13908A, 13909, 14036, 14066, 14120, 14129, 14138, 14255, 14315, 14411, 14416, 14503, 14606, 14825, 14858, 14897, 15105, 15134, 15146, 15287, 15308, 15326, 15563, 15605, 15617, 15627, 15629, 15741, 15751, 16042, 16093 | 8188, 8285, 9068, 12135, 12178, 12234, 12377, 15579, 15593, 15818, 15951, 15953G, 15959, 15994, 16022, 16049, 16057, 16058, 16074, 16082, 16102, 16109, 16113, 16116, 16117, 16119, 16121, 16122, 16130, 16137, 16138, 16143d, 16147, 16196, 16229, 16247, 16248, 16300, 16301 | | |
| | T3a1. | | | | | |
| GU256940 (Cattle and Aurochs) | Surplus of mutations: mis-added mutations 3821, 3879, 3883, 3940, 3955, 3979, 3982, 3988, 4018A, 4021, 7054, 7069, 7081, 7114, 7132, 7135, 7145 of sequences of Yak and wild yak. | I1 | 8, 106, 166, 173, 206, 232, 233, 248, 249, 296, 300, 518, 587+C, 722, 737, 761, 816, 1158, 1474, 1492, 1495, 1600d, 1677, 1824, 1860A, 1869, 2016, 2099, 2536A, 2558, 2575, 2634, 2953, 2977, 2979, 2988, 2989, 2990, 3071, 3136, 3145, 3241, 3325, 3335, 3379, 3439, 3535, 3550, 3600, 3793, 3821, 3829, 3874, 3879, 3883, 3931, 3940, 3955, 3979, 3982, 3985, 3988, 4000, 4018A, 4021, 4328, 4442, 4562, 4730, 4733, 4769, 4871, 4937, 4988A, 5285, 5501, 5531, 5614, 5783, 5890, 5917, 5998, 6115T, 6235, 6340, 6367, 6379, 6436, 6460, 6727, 6772, 6881, 7054, 7069, 7081, 7114, 7132, 7135, 7145, 7304, 7330, 7356, 7358, 7361, 7514, 7830, 7851, 8045, 8168, 8188, 8194, 8210, 8285, 8308, 8370C, 8466, 8494, 8503, 8514, 8571, 8646, 8749, 8984, 9005, 9038, 9068, 9245, 9302, 9480, 9581, 9602, 9682C, 9767, 9891, 9912, 9930C, 9978, 10039, 10066, 10071, 10137, 10153, 10268, 10322, 10331, 10445, 10590, 10621, 10691G, 10849, 11000, 11035, 11068, 11134, 11200, 11266, 11329, 11407, 11419, 11803, 11842, 12135, 12178, 12234, 12377, 12433, 12468, 12469, 12513, 12672, 12675, 12684, 12750, 12801, 12900, 12923T, 12924, 13005, 13056A, 13098, 13101A, 13104, 13275, 13310C, 13371, 13380, 13433, 13437, 13554, 13564, 13584, 13628, 13677, 13689, 13692, 13882, 13908A, 13909, 14027, 14036, 14066, 14120, 14129, 14138, 14255, 14315, 14411, 14416, 14503, 14897, 15105, 15134, 15146, 15287, 15308, 15326, | I1:3051, 3976, 5743, 6922, 12622, 14606, 14825, 14858 | I1:1495, 1600d, 3821, 3879, 3883, 3940, 3955, 3979, 3982, 3988, 4018A, 4021, 7054, 7069, 7081, 7114, 7132, 7135, 7145, 8646, 9912, 14027, 16084, 16200+A | Wang,Z., Shen,X., Liu,B., Su,J., Yonezawa,T., Yu,Y., Guo,S., Ho,S.Y.W., Vila,C., Hasegawa,M. and Liu,J. Unpublished |
| | | | 15563, 15579, 15593, 15605, 15617, 15627, 15629, 15741, 15751, 15818, 15951, 15953G, 15959, 15994, 16022, 16049, 16057, 16058, 16074, 16082, 16084, 16102, 16109, 16113, 16116, 16117, 16119, 16121, 16122, 16130, 16137, 16138, 16143d, 16147, 16196, 16200+A, 16229, 16247, 16248, 16300, 16301 | | | |
| X52392(Chicken) | Surplus of mutations | A | 50-52d, 167, 210, 212, 225, 246, 315, 686, 707d, 859d, 1215, 1699C, 2071, 2209, 2678, 3841-3842d, 3952+G, 5106, 5832, 5934C, 7098d, 7105-7106d, 7539C, 9806, 10270, 10447, 10958d, 12104, 12689, 13884, 14393+C, 14406d, 14853, 15445, 15693T, 15875, 16131 | A:12961, 16263, 16371 | A:50-52d, 210, 686, 707d, 859d, 2678, 3841-3842d, 3952+G, 5934C, 7098d, 7105-7106d, 7539C, 10958d, 14393+C, 14406d, 14853, 15445, 16131 | Desjardins,P. and Morais,R.J. Mol. Biol. 212 (4), 599-634 (1990) |
| AY729880(Dog and gray wolf) | Surplus of mutations. | A1a2'3'4'5'6, A1a3, A1a3a1 | 291, 517, 520, 520+TA, 521, 543+C, 640+A, 896A, 1169+G, 1183+A, 1449, 1881G, 1882A, 1883, 1886, 1887, 1888C, 1889C, 1890C, 1893T, 1894, 1895, 1897, 1898, 1976+C, 2232, 2840A, 2963, 2981, 3122A, 3197, 3407, 3470, 3599, 3962, 4119, 4518, 4647, 5287-5288d, 5368, 5445, 6066, 6471, 7007+A, 7015, 7655, 7927, 8091, 8183T, 8222, 8324, 8426, 8510, 8704, 8765G, 9915-9916d, 10168, 10407, 10920, 10995, 11164, 11703, 11712, 11761G, 11771, 11966, 11967T, 12027, 12387, 12416A, 12678, 12687T, 12784G, 12791, 12816, 13266, 13469, 13588, 13597, 13621, 13663, 13780, 13809, 13951, 15115, 15170, 15217, 15438, 15514+AGGTAAACCCTTCTCCCCTCCCC, 15607, 15983, 16005, 16065+T,16140 | A1a234'5'6:5010, 8282, 13264, 13711, 16140; A1a3:5010, 8282, 8537, 13264, 13711, 16140; A1a3a1:5010, 6630, 8282, 13264, 13711, 16140 | A1a234'5'6:517, 520, 520+TA, 521, 543+C, 640+A, 896A, 1169+G, 1183+A, 1449, 1881G, 1882A, 1883, 1886, 1887, 1888C, 1889C, 1890C, 1893T, 1894, 1895, 1897, 1898, 1976+C, 3122A, 4647, 5287-5288d, 5368, 5445, 6066, 7007+A, 7015, 8183T, 8222, 9915-9916d, 10168, 11703, 11712, 11761G, 11771, 11967T, 12027, 12678, 12687T, 12784G, 13266, 13597, 13621, 13663, 13780, 15514+AGGTAAACCCTTCTCCCCTCCCC, 16065+T<br>A1a3:517, 520, 520+TA, 521, 543+C, 640+A, 896A, 1169+G, 1183+A, 1449, 1881G, 1882A, 1883, 1886, 1887, 1888C, 1889C, 1890C, 1893T, 1894, 1895, 1897, 1898, 1976+C, 3122A, 4647, 5287-5288d, 5368, 5445, 6066, 7007+A, 7015, 8183T, 8222, 9915-9916d, 11703, 11712, 11761G, 11771, 11967T, 12027, 12678, 12687T, 12784G, 13266, 13597, 13621, 13663, 13780, 15514+AGGTAAACCCTTCTCCCCTCCCC,16065+T<br>A1a3a1:517, 520, 520+TA, 521, 543+C, 640+A, 896A, 1169+G, 1183+A, 1449, 1881G, 1882A, 1883, 1886, 1887, 1888C, 1889C, 1890C, 1893T, 1894, 1895, 1897, 1898, 1976+C, 3122A, 4647, 5287-5288d, 5368, 5445, 6066, 7007+A, 7015, 8183T, 8222, 9915-9916d, 11703, 11712, 11761G, 11771, 11967T, 12027, 12678, 12687T, 12784G, 13266, 13597, 13621, 13663, 13780, 15514+AGGTAAACCCTTCTCCCCTCCCC,16065+T | Zhu,S., Xu,Q. and Chang,H. Unpublished |
| EU789758(Dog and gray wolf) | Artificial recombination: missing mutation 6258 from haplogroup B1; | B1c | 16, 291, 381A, 445, 1454, 1756, 2185, 2813, 2840A, 2981, 3962, 4119, 4205, 4278, 4391, 4647, 6111, 6471, 6765, 7655, 7927, 8091, 8222C, 8226, 8510, 8570, 8737, 8761, 8818, 8878, 8992, 9220, 9826, 10260, 10443, 10536T, 10545, 11164, 11253, 11405, 11951, 12387, 12416A, 12668, 13469, 13588, 13597, 13621, | 6258 | B1: 6471 15918d | Pang JF, Kluetsch C, Zou XJ, Zhang AB, Luo LY, Angleby H, |
| | mis-added mutations 6471 of haplogroup A1'2'3'4'5'6. | | 13663, 13780, 13809, 13951, 14674, 14933, 15115, 15170, 15575, 15592, 15607, 15612, 15619, 15623, 15632, 15780, 15795, 15892, 15918d, 15935, 16140, 16141 | | | Ardalan A, Ekström C, Sköllermo A, Lundeberg J, Matsumura S, Leitner T, Zhang YP, Savolainen P. Mol. Biol. Evol. 26 (12), 2849-2864 (2009) |
| JF342814( Dog and gray wolf) | Artificial recombination: missing mutations 11951, 15575, 15592, 15612, 15619, 15623, 15632, 15780, 15795,15892, 15935 from haplogroup B1a4; mis-added mutations 11966 of haplogroup A1'2'3'4'5'6. | B1a4 | 16, 291, 381A, 445, 1454, 1756, 2185, 2813, 2840A, 2891N, 2892N, 2893N, 2894N, 2895N, 2896N, 2897N, 2898N, 2899N, 2900N, 2901N, 2981, 3029C, 3962, 4119, 4205, 4278, 4391, 4647, 6258, 6765, 7015, 7655, 7927, 8091, 8102, 8222C, 8226, 8510, 8570, 8737, 8761, 8818, 8878, 8992, 9183N, 9184N, 9185N, 9186N, 9220, 9286N, 9287N, 9288N, 9289N, 9290N, 9291N, 9292N, 9293N, 9294N, 9826, 10260, 10443, 10536T, 10545, 10560, 10617A, 11164, 11179, 11253, 11405, 11966, 12387, 12416A, 12668, 13429, 13469, 13588, 13597, 13621, 13663, 13780, 13781, 13809, 13951, 14674, 14933, 15115, 15170, 15514+AGGTAAACCCTTCTCCCCTCCCC, 15607, 15983, 16005,16140 | B1a4:11951, 15575, 15592, 15612, 15619, 15623, 15632, 15780, 15795, 15892, 15935, 16141 | B1a4:2891N, 2892N, 2893N, 2894N, 2895N, 2896N, 2897N, 2898N, 2899N, 2900N, 2901N, 9183N, 9184N, 9185N, 9186N, 9286N, 9287N, 9288N, 9289N, 9290N, 9291N, 9292N, 9293N, 9294N, 10617A, 11966, 13781, 15514+AGGTAAACCCTTCTCCCCTCCCC, 15983, 16005 | Imes DL, Wictum EJ, Allard MW, Sacks BN. Forensic Sci Int Genet.6 (5), 630-639 (2012) |
| JF342821((Dog and gray wolf) | Artificial recombination: missing mutations 4235, 4504 of haplogroup C. | C1b_EU408293 | 16, 291, 381A, 557, 1046, 1204, 1454, 1709, 1748, 1756, 2840A, 2981, 3962, 4119, 4518, 5625, 6258, 7059, 7113, 7187A, 7364, 7655, 7927, 8091, 8161, 8222C, 8226, 8510, 8761, 8878, 8992, 9223, 9709, 9866+A, 10536T, 10779, 10788, 10866, 11103, 11164, 11253, 11325, 11403, 11405, 11551T, 11575C, 11598C, 11842, 12125, 12275, 12333, 12387, 12416A, 12639, 13115, 13469, 13588, 13621, 13663, 13780, 13809, 13951, 14611, 14650, 15115, 15170, 15188, 15511, 15514+AGGTAAACCCTTCTTCCCTCCCC, 15591, 15601, 15607, 15619, 15630, 15780, 15892, 15918d, 15935,16174 | C1b_EU408293:4235, 4504, 14695 | C1b_EU408293:4518, 8161, 11551T, 11598C, 15514+AGGTAAACCCTTCTTCCCTCCCC, 15918d | Imes DL, Wictum EJ, Allard MW, Sacks BN. Forensic Sci Int Genet.6 (5), 630-639 (2012) |
| JF342824(Dog and gray wolf) | Artificial recombination: missing mutations 1873, 2855 of haplogroup D1; mis-added mutations 2232 from haplogroup A'1'2'3. | D1a1 | 162, 291, 658, 1454, 1689, 2232, 2280N, 2281N, 2656, 2840A, 2981, 3035, 3452, 3466, 3629, 3938, 3941, 3951, 3962, 4119, 4467, 4485, 4518, 4592, 4596, 5520, 5938, 5974N, 5975N, 5976N, 6054, 6093, 6258, 6768, 6861, 6882, 7619N, 7620N, 7621N, 7622N, 7623N, 7624N, 7655, 7927, 8091, 8109, 8226, 8391, 8510, 8671, 8761, 8783, 8854, 8971, 8992, 9286N, 9287N, 9288N, 9289N, 9290N, 9291N, 9292N, 9293N, 9294N, 9295N, 9836, 9887, 10063, 10162, 10198, 10314, 10349, 10536T, 10683, 10728, 11164, 11253, 11405, 11803, 11900, 11987, 12387, 12416A, 12668, 12821, 12971, 13469, 13588, 13597, 13663, 13780, 13809, 13951, 14611, 14650, 15115, 15170, 15290, 15514+AGGTAAACCCTTCTCCCCTCCCC, 15605, 15607, 15612, 15614, 15616, 15619, 15623, 15780, 15795, 15828, 15892, 15935, 15939,16139 | D1a1:1873, 2855, 14507, 14809 | D1a1:2232, 2280N, 2281N, 4518, 5974N, 5975N, 5976N, 7619N, 7620N, 7621N, 7622N, 7623N, 7624N, 9286N, 9287N, 9288N, 9289N, 9290N, 9291N, 9292N, 9293N, 9294N, 9295N, 12971, 14650, 15514+AGGTAAACCCTTCTCCCCTCCCC | Imes DL, Wictum EJ, Allard MW, Sacks BN. Forensic Sci Int Genet.6 (5), 630-639 (2012) |
| JF342826(Dog and gray wolf) | Surplus of mutations: sequencing errors 2982G, 2983A. | B1a | 16, 291, 381A, 445, 1454, 1756, 2185, 2813, 2840A, 2981, 2982G, 2983A, 2984T, 2985N, 2986N, 2987N, 2988N, 2989N, 2990N, 2991N, 3029C, 3962, 4119, 4205, 4278, 4391, 4518, 4647, 6258, 6765, 6916+C, 6928+T, 7015, 7655, 7927, 8091, 8102, 8222C, 8226, 8510, 8570, 8737, 8761, 8818, 8878, 8992, 9220, 9826, 10260, 10443, 10536T, 10545, 11164, 11179, 11253, 11405, 11951, 12387, 12416A, 12668, 13469, 13588, 13597, 13621, 13663, 13780, 13809, 13951, 14674, 14933, 15115, 15170, 15514+AGGTAAACCCTTCTTCCCTCCCC, 15575, 15592, 15607, 15612, 15619, 15632, 15780, 15795, 15892, 15935,16140,16141 | B1a:15623 | B1a:2982G,2983A,2985N, 2986N, 2987N, 2988N, 2989N, 2990N, 2991N, 15514+AGGTAAACCCTTCTTCCCTCCCC | Imes DL, Wictum EJ, Allard MW, Sacks BN. Forensic Sci Int Genet.6 (5), 630-639 (2012) |
| JF342827(Dog and gray wolf) | Artificial recombination: missing mutations 6471 of haplogroup A1'2'3'4'5'6; mis-added mutations 6765, 7015 from haplogroup B1a. | A1a3b | 291, 2232, 2840A, 2963, 2981, 3197, 3407, 3470, 3599, 3962, 4119, 4518, 5010, 6765, 7015, 7024N, 7025N, 7026N, 7027N, 7028N, 7029N, 7030N, 7031N, 7032N, 7033N, 7034N, 7035N, 7036N, 7037N, 7038N, 7039N, 7040N, 7041N, 7042N, 7043N, 7044N, 7045N, 7046N, 7047N, 7048N, 7049N, 7050N, 7051N, 7052N, 7053N, 7054N, 7055N, 7056N, 7057N, 7058N, 7059N, 7060N, 7061N, 7062N, 7063N, 7064N, 7065N, 7066N, 7067N, 7068N, 7069N, 7070N, 7071N, 7072N, 7073N, 7074N, 7075N, 7076N, 7077N, 7078N, 7079N, 7080N, 7081N, 7082N, 7083N, 7084N, 7085N, 7086N, 7087N, 7088N, 7089N, 7090N, 7091N, 7092N, 7093N, 7094N, 7095N, 7096N, 7097N, 7098N, 7099N, 7100N, 7101N, 7102N, 7103N, 7104N, 7105N, 7106N, 7107N, 7108N, 7109N, 7110N, 7111N, 7112N, 7113N, 7114N, 7115N, 7116N, 7117N, 7118N, 7119N, 7120N, 7121N, 7122N, 7123N, 7124N, 7125N, 7126N, 7127N, 7128N, 7129N, 7130N, 7131N, 7132N, 7133N, 7134N, 7247A, 7619N, 7620N, 7621N, 7655, 7927, 8091, 8282, 8324, 8426, 8510, 8537, 8704, 8765G, 9286N, 9287N, 9288N, 9289N, | A1a3b:6471 | A1a3b:6765, 7015, 7024N, 7025N, 7026N, 7027N, 7028N, 7029N, 7030N, 7031N, 7032N, 7033N, 7034N, 7035N, 7036N, 7037N, 7038N, 7039N, 7040N, 7041N, 7042N, 7043N, 7044N, 7045N, 7046N, 7047N, 7048N, 7049N, 7050N, 7051N, 7052N, 7053N, 7054N, 7055N, 7056N, 7057N, 7058N, 7059N, 7060N, 7061N, 7062N, 7063N, 7064N, 7065N, 7066N, 7067N, 7068N, 7069N, 7070N, 7071N, 7072N, 7073N, 7074N, 7075N, 7076N, 7077N, 7078N, 7079N, 7080N, 7081N, 7082N, 7083N, 7084N, 7085N, 7086N, 7087N, 7088N, 7089N, 7090N, 7091N, 7092N, 7093N, 7094N, 7095N, 7096N, 7097N, 7098N, 7099N, 7100N, 7101N, 7102N, 7103N, 7104N, 7105N, 7106N, 7107N, 7108N, 7109N, 7110N, 7111N, 7112N, 7113N, 7114N, 7115N, 7116N, 7117N, 7118N, 7119N, 7120N, 7121N, 7122N, 7123N, 7124N, 7125N, 7126N, 7127N, 7128N, 7129N, 7130N, 7131N, 7132N, 7133N, 7134N, 7247A, 7619N, 7620N, 7621N, 9286N, 9287N, 9288N, | Imes DL, Wictum EJ, Allard MW, Sacks BN. Forensic Sci Int Genet.6 (5), 630-639 (2012) |
| | | | 9290N, 9291N, 9292N, 9293N, 9294N, 9295N, 9296N, 9297N, 10168, 10407, 10920, 10995, 11164, 11966, 12387, 12416A, 12791, 12816, 13264, 13469, 13588, 13711, 13809, 13951, 14477, 14731N, 14732N, 14733N, 14734N, 14735N, 14736N, 15115, 15170, 15217, 15438, 15514+AGGTAAACCCTTCTCCCCTCCCC, 15607, 15983, 16005,16140 | | 9289N, 9290N, 9291N, 9292N, 9293N, 9294N, 9295N, 9296N, 9297N, 14731N, 14732N, 14733N, 14734N, 14735N, 14736N, 15514+AGGTAAACCCTTCTCCCCTCCCC | |
| JF342841(Dog and gray wolf) | Artificial recombination: missing mutations 6402, 6471, 6555 of haplogroup A1a1c; mis-added mutations 6765, 7015 from haplogroup B1a. | A1c1a1a_EU408263 | 291, 2232, 2646, 2840A, 2981, 3407, 3470, 3599, 3962, 4119, 4304, 4518, 4941, 5010, 6467, 6765, 7015, 7594, 7655, 7927, 8091, 8324, 8426, 8510, 8594+A, 8704, 8765G, 10407, 10614T, 10920, 11164, 11966, 12203, 12387, 12416A, 12791, 12816, 13264, 13469, 13588, 13711, 13809, 13951, 14731N, 14732N, 14733N, 14734N, 14735N, 14980, 15115, 15170, 15438, 15514+AGGTAAACCCTTCTCCCCTCCCC, 15600, 15935, 15983,16132+2C,16140,16141 | A1c1a1a_EU408263:1351, 6402, 6471, 6555 | A1c1a1a_EU408263:6467, 6765, 7015, 8594+A, 14731N, 14732N, 14733N, 14734N, 14735N, 15514+AGGTAAACCCTTCTCCCCTCCCC, | Imes DL, Wictum EJ, Allard MW, Sacks BN. Forensic Sci Int Genet.6 (5), 630-639 (2012) |
| JF342853(Dog and gray wolf) | Artificial recombination: missing | C2d | 16, 291, 381A, 557, 1204, 1454, 1709, 1748, 1756, 2981, 3495, 3962, 4119, 4235, 4504, 5625, 6258, 6712A, 6741, 6864, 7059, 7655, 7927, 8091, 8282, | C2d:2840A, 8222C, 8226, 11575C, 15511 | C2d:8282, 8324, 8426, 14887, 15514+AGGTAAACCCTTCTTCCCTCCCC, 15918d,16173C | Imes DL, Wictum EJ, Allard |
| | mutations 8222C, 8226 of haplogroup C; mis-added mutations 8324, 8426 from haplogroup A1. | | 8324, 8426, 8510, 8761, 8878, 8992, 9223, 9709, 9866+A, 10536T, 10779, 10788, 11164, 11250, 11253, 11325, 11403, 11405, 11962, 12125, 12275, 12333, 12349A, 12387, 12416A, 12639, 13469, 13588, 13621, 13663, 13780, 13794, 13809, 13951, 14611, 14650, 14887, 15115, 15170, 15188, 15375, 15514+AGGTAAACCCTTCTTCCCTCCCC, 15591, 15607, 15619, 15630, 15690, 15780, 15892, 15918d, 15935,16173C | | | MW, Sacks BN. Forensic Sci Int Genet.6 (5), 630-639 (2012) |
| JF342868(Dog and gray wolf) | Artificial recombination: missing mutations 9709, 9866+A, 10536T of haplogroup C1'2. | C2 | 16, 291, 381A, 557, 1204, 1454, 1709, 1748, 1756, 2840A, 2981, 3495, 3962, 4119, 4235, 4504, 5625, 6258, 6712A, 6741, 6864, 7059, 7655, 7927, 8091, 8222C, 8226, 8510, 8761, 8878, 8992, 9223, 10779, 10788, 11164, 11250, 11253, 11325, 11403, 11405, 11575C, 11669N, 11670N, 11671N, 11672N, 11673N, 11674N, 11675N, 11962, 12125, 12275, 12333, 12349A, 12387, 12416A, 12639, 13469, 13588, 13621, 13663, 13780, 13794, 13809, 13951, 14611, 14650, 14695, 15115, 15170, 15188, 15375, 15511, 15514+AGGTAAACCCTTCTTCCCTCCCC, 15533, 15591, 15607, 15619, 15630, 15690, 15780, 15892, 15918d, 15935 | C2:9709, 9866+A, 10536T | C2:11669N, 11670N, 11671N, 11672N, 11673N, 11674N, 11675N, 15514+AGGTAAACCCTTCTTCCCTCCCC, 15533, 15918d | Imes DL, Wictum EJ, Allard MW, Sacks BN. Forensic Sci Int Genet.6 (5), 630-639 (2012) |
| JF342869(Dog and gray wolf) | Artificial recombination: missing mutations 10260, 10443, 10545 of haplogroup B1. | B1a2 | 16, 291, 381A, 445, 1454, 1756, 2185, 2813, 2840A, 2981, 3029C, 3962, 4119, 4205, 4278, 4391, 4647, 6258, 6765, 7015, 7655, 7927, 8091, 8102, 8222C, 8226, 8510, 8570, 8737, 8761, 8818, 8878, 8992, 9220, 9826, 10536T, 11164, 11179, 11253, 11405, 11951, 12387, 12416A, 12668, 13469, 13588, 13597, 13621, 13663, 13780, 13809, 13951, 14546, 14674, 14933, 15115, 15170, 15514+AGGTAAACCCTTCTTCCCTCCCC, 15575, 15592, 15607, 15612, 15619, 15623, 15632, 15780, 15795, 15892, 15935 | B1a2:10260, 10443, 10545, 16140, 16141 | B1a2:15514+AGGTAAACCCTTCTTCCCTCCCC | Imes DL, Wictum EJ, Allard MW, Sacks BN. Forensic Sci Int Genet.6 (5), 630-639 (2012) |
| JF342893(Dog and gray wolf) | Surplus of mutations: sequencing error:13056-13059d 13063A 13064 13065. | A1a2_JF342893 | 44, 291, 713, 2232, 2840A, 2895N, 2896N, 2897N, 2898N, 2899N, 2963, 2981, 3197, 3407, 3470, 3599, 3962, 4119, 4518, 5010, 5232, 6471, 7655, 7927, 8091, 8282, 8324, 8426, 8510, 8704, 8765G, 9287N, 9288N, 9289N, 9290N, 9291N, 9583, 10407, 10920, 10995, 11164, 11966, 12387, 12416A, 12791, 12816, 13056-13059d, 13063A, 13064, 13065, 13264, 13469, 13588, 13711, 13765, 13809, 13951, 14082, 15115, 15170, 15217, 15438, 15514+AGGTAAACCCTTCTCCCCTCCCC, 15607, 15983, 16005,16140 | | A1a2_JF342893:2895N, 2896N, 2897N, 2898N, 2899N, 9287N, 9288N, 9289N, 9290N, 9291N, 15514+AGGTAAACCCTTCTCCCCTCCCC | Imes DL, Wictum EJ, Allard MW, Sacks BN. Forensic Sci Int Genet.6 (5), 630-639 (2012) |
| JF342894(Dog and gray wolf) | Surplus of mutations. | A1a3a | 291, 2232, 2279N, 2280N, 2281N, 2282N, 2283N, 2284N, 2840A, 2895N, 2896N, 2897N, 2898N, 2899N, 2963, 2981, 3197, 3407, 3470, 3599, 3962, 4119, 4518, 5010, 6471, 6630, 7655, 7927, 8091, 8282, 8324, 8426, 8510, 8537, 8704, 8765G, 8992, 10168, 10407, 10920, 10995, 11164, 11476C, 11483C, 11504d, 11512, 11533d, 11548, 11550-11551d, 11560-11561d, 11574d, 11585d, 11596C, 11597-11599d, 11624d, 11636d, 11966, 12271Y, 12273+TC, 12387, 12416A, 12791, 12816, 13105G, 13264, 13469, 13588, 13711, 13809, 13951, 15115, 15170, 15217, 15438, 15514+AGGTAAACCCTTCTCCCCTCCCC, 15607, 15983, 16005,16140 | | A1a3a:2279N, 2280N, 2281N, 2282N, 2283N, 2284N, 2895N, 2896N, 2897N, 2898N, 2899N, 8992, 11476C, 11483C, 11504d, 11512, 11533d, 11548, 11550-11551d, 11560-11561d, 11574d, 11585d, 11596C, 11597-11599d, 11624d, 11636d, 12271, 12273+TC, 13105G, 15514+AGGTAAACCCTTCTCCCCTCCCC | Imes DL, Wictum EJ, Allard MW, Sacks BN. Forensic Sci Int Genet.6 (5), 630-639 (2012) |
| JF342901(Dog and gray wolf) | Artificial recombination: mis-added mutations 16, 381A, 445 from haplogroup B1. | A1c1a1a | 16, 291, 381A, 399, 445, 1351, 2232, 2840A, 2981, 3407, 3470, 3599, 3962, 4119, 4304, 4518, 4941, 5010, 6402, 6471, 6555, 7594, 7655, 7927, 8091, 8324, 8426, 8510, 8704, 8765G, 10407, 10614T, 10920, 10995, 11164, 11966, 12203, 12387, 12416A, 12791, 12816, 13264, 13469, 13588, 13711, 13809, 13951, 14980, 15115, 15170, 15438, 15514+AGGTAAACCCTTCTCCCCTCCCC, 15600, 15935, 15983, 16132+2C,16140,16141 | | A1c1a1a:16, 381A, 399, 445, 10995, 15514+AGGTAAACCCTTCTCCCCTCCCC | Imes DL, Wictum EJ, Allard MW, Sacks BN. Forensic Sci Int Genet.6 (5), 630-639 (2012) |
| EU408251 (Dog and gray wolf) | Artificial recombination: missing mutations 9709, 9866+A, 10779, 10788 of haplogroup C; mis-added mutations 10407, 10614T, 10920 from haplogroup A1. | C2 | 16, 291, 381A, 557, 1204, 1454, 1709, 1748, 1756, 2840A, 2981, 3495, 3599R, 3962, 4119, 4235Y, 4504, 5625, 6258, 6712A, 6741, 6864, 7059, 7655, 7834, 7927, 8091, 8222C, 8226, 8510, 8761, 8878, 8992, 9223Y, 10407, 10536W, 10614T, 10920R, 11164, 11250, 11253, 11325, 11403, 11405, 11575C, 11962, 12125, 12275, 12333, 12349A, 12387, 12416A, 12639, 13469, 13588, 13621, 13663, 13780, 13794, 13809, 13951, 14611, 14650, 14695, 15115, 15170, 15188, 15375, 15511, 15514+AGGTAAACCCTTCTTCCCTCCCC, 15591Y, 15607, 15619, 15630, 15690, 15780, 15892, 15918d, 15935 | C2:9709, 9866+A, 10779, 10788 | C2:3599, 7834, 10407, 10614T, 10920, 15514+AGGTAAACCCTTCTTCCCTCCCC, 15918d | Webb,K.M. and Allard,M.W. J. Forensic Sci. 54 (2), 275-288 (2009) |
| DQ480503 (Dog and gray wolf) | Surplus of mutations. | Q1_KF661039 | 291, 381A, 767, 1187+T, 1454, 1685, 1756, 2196, 2840A, 2945, 2981, 3314, 3476, 3533, 3962, 4119, 4497, 4748, 4904, 4947, 5278, 6258, 6948, 7109, 7232, 7463, 7526, 7655, 7927, 8091, 8391, 8510, 8566, 8761, 8878, 8992, 9079, 9805, 10278, 10321, 10536T, 11164, 11253, 11405, 11477, 12010, 12387, 12407A, 12416A, 12689, 12704, 12740, 12743, 12746, 12750, 12752, 12794, 12795, 12806, 12821, 12833, 12842, 12845, 12857, 12860, 12863, 12878, 12893, 12896, 12899G, 12903, 12926, 12929, 12935, 12962, 13046, 13070A, 13118, 13166, 13175, 13178, 13244, 13246, 13265A, 13331, 13430, 13469, 13588, 13617, 13621, 13663, 13780, 13809, 13951, 14022, 14440, 14611, 15115, 15170, 15378, 15514+AGGTAAACCCTTCTTCCCTCCCC, 15601, 15780, 15794, 15797, 15800, 15892, 15918d, 15935 | | Q1_KF661039:1187+T, 7232, 10321, 12689, 12704, 12740, 12743, 12746, 12750, 12752, 12794, 12795, 12806, 12821, 12833, 12842, 12845, 12857, 12860, 12863, 12878, 12893, 12896, 12899G, 12903, 12926, 12929, 12935, 12962, 13046, 13070A, 13118, 13175, 13178, 13244, 13246, 15514+AGGTAAACCCTTCTTCCCTCCCC, 15918d, 15935 | Bjornerfeldt, S., Webster,M.T. and Vila,C. Genome Res. 16 (8), 990-994 (2006) |
| GU229278 (Goat) | Surplus of mutations. | A2a1a | 177G, 179, 180-181d, 1395+A, 2140, 2141, 2208, 4654, 4942, 5127, 5710, 6523, 7840, 8828, 8834, 8858, 8861, 8864, 9311, 10028, 11062, 11128, 12252, 12257, 12266, 12277, 12278, 12281C, 12287, 12290, 12293, 12299, 12300, 12314, 12318, 12335, 12338, 12341, 12362, 12368C, 12377, 12380, 12383, 12386, 12392C, 12393, 12396, 12397, 12398A, 12399, 12404, 12408, 12410, 12419, 12422, 12428, 12446, 12452, 12470T, 12476, 12484, 12530, 12531, 12533, 12536, 12537, 12539, 12557, 12579, 12590, 12593, 12594, 12599, 12608, 12623, 12641, 12671, 12677, 12689, 12716, 12717, 12719, 12725, 12737, 12743, 12761, 12762, 12763, 12779, 12793, 12800, 12801, 12812, 12824, 12836, 12851, 12878, 12881, 12884, 12899, 12927, 12932, 12950, 12953, 12984, 12990, 12992, 12998, 13001, 13007, 13008, 13016, 13019, 13020, 13022T, 13028, 13031, 13034, 13049, 13067, 13070, 13072C, 13073, 13091, 13114C, 13119, 13124, 13130, 13133, 13137, 13146A, 13151, 13163, 13187A, 13214, 13226, 13244, 13250, 13256, 13258, 13259, 13260, 13261G, 13262, 13271, 13274, 13275A, 13280, 13286, 13290, 13293, 13304C, 13316, 13319, 13323, 13343, 13357, 13358G, 13371, 13382, 13400, 13425, 13427, 13430, 13445, 13450T, 13457, 13477, 13483, 13496, 13502, 13503, 13520, 13523, 13532, 13535, 13548, 13551, 13582, 13597, 13639, 13675, 13680A, 13728, 13732, 13741, 13753, 13756, 13759, 13768, 13792, 13800, 13810, 13813, 13843, 13846, 13867, 15530, 15616, 15636, 15645, 15920, 15976, 16000, 16028, 16066, 16435, 16504+C | A2a1a:15811 | A2a1a:177G, 179, 180-181d, 1395+A, 2140, 2141, 2208, 4654, 5127, 5710, 7840, 8828, 8834, 8858, 8861, 8864, 9311, 12252, 12257, 12266, 12277, 12278, 12281C, 12287, 12290, 12293, 12299, 12300, 12314, 12318, 12335, 12338, 12341, 12362, 12368C, 12377, 12383, 12386, 12392C, 12393, 12396, 12397, 12398A, 12399, 12404, 12408, 12410, 12419, 12422, 12428, 12446, 12452, 12470T, 12476, 12484, 12530, 12531, 12533, 12536, 12537, 12539, 12557, 12579, 12590, 12593, 12594, 12599, 12608, 12623, 12641, 12671, 12677, 12689, 12716, 12717, 12719, 12725, 12737, 12743, 12761, 12762, 12763, 12779, 12793, 12800, 12801, 12812, 12824, 12836, 12851, 12878, 12881, 12884, 12899, 12927, 12932, 12950, 12953, 12984, 12990, 12992, 12998, 13001, 13007, 13008, 13016, 13019, 13020, 13022T, 13028, 13031, 13049, 13067, 13070, 13072C, 13073, 13091, 13114C, 13119, 13124, 13130, 13133, 13137, 13146A, 13151, 13163, 13187A, 13214, 13226, 13244, 13250, 13256, 13258, 13259, 13260, 13261G, 13262, 13271, 13274, 13275A, 13280, 13286, 13290, 13293, 13304C, 13316, 13319, 13323, 13343, 13357, 13358G, 13371, 13382, 13400, 13425, 13427, 13430, 13445, 13450T, 13457, 13477, 13483, 13496, 13502, 13503, 13520, 13523, 13532, 13535, 13548, 13551, 13582, 13597, 13639, 13675, 13680A, 13728, 13732, 13741, 13753, 13756, 13759, 13768, 13792, 13800, 13810, 13813, 13843, 13846, 13867, 15530, 15616, 15636, 15645, 15920, 15976, 16504+C | Wu,Y.P. and Ma,Y.H. Direct Submission |
| GU229279 (Goat) | Surplus of mutations. | B_KF952601 | 177G, 179, 180-181d, 205, 1395+A, 1850, 2092, 2208, 2417, 2850, 3195, 3469, 4033, 4802, 4942, 4964, 4988d, 5602, 6097, 6523, 7053, 7301, 7343, 7661, 7840, 7886, 8056, 8079, 8481, 8636, 8731, 8960, 9011, 9441, 9673, 9827, 9920C, 10028, 10480, 10639, 10687, 10814, 11062, 11089, 11128, 11173, 11365, 11691, 11702, 11872, 12252, 12257, 12266, 12277, 12278, 12281C, 12287, 12290, 12293, 12299, 12300, 12314, 12318, 12335, 12338, 12341, 12362, 12368C, 12377, 12380, 12383, 12386, 12392C, 12393, 12396, 12397, 12398A, 12399, 12404, 12408, 12410, 12419, 12422, 12428, 12446, 12452, 12470T, 12476, 12484, 12530, 12531, 12533, 12536, 12537, 12539, 12557, 12579, 12590, 12593, 12594, 12599, 12608, 12623, 12641, 12671, 12677, 12689, 12716, 12717, 12719, 12725, 12737, 12743, 12761, 12762, 12763, 12779, 12793, 12800, 12801, 12812, 12824, 12836, 12851, 12878, 12881, 12884, 12899, 12927, 12932, 12950, 12953, 12984, 12990, 12992, 12998, 13001, 13007, 13008, 13016, 13019, 13020, 13022T, 13028, 13031, 13034, 13049, 13067, 13070, 13072C, 13073, 13091, 13114C, 13119, 13124, 13130, 13133, 13137, 13146A, 13151, 13163, 13187A, 13214, 13226, 13244, 13250, 13256, 13258, 13259, 13260, 13261G, 13262, 13271, 13274, 13275A, 13280, 13286, 13290, 13293, 13304C, 13316, 13319, 13323, 13343, 13357, 13358G, 13371, 13382, 13400, 13425, 13427, 13430, 13445, 13450T, 13457, 13477, 13478C, 13483, 13496, 13502, 13503, 13520, 13523, 13532, 13535, 13548, 13551, 13582, 13583C, 13597, 13675, 13680A, 13728, 13732, 13741, 13753, 13756, 13759, 13800, 13810, 13813, 13843, 13846, 13867, 14746, 14873, 15004, 15220, 15448A, 15449, 15450, 15451, 15543, 15616, 15631, 15636, 15637, | B_KF952601:1120+C, 8819, 12584, 13100A, 13178, 13681, 13762, 14728, 15977d | B_KF952601:177G, 179, 180-181d, 1395+A, 2092, 2417, 2850, 4802, 4964, 4988d, 9441, 9920C, 11089, 12252, 12257, 12266, 12277, 12278, 12281C, 12287, 12290, 12293, 12299, 12300, 12314, 12318, 12335, 12338, 12341, 12362, 12368C, 12377, 12386, 12392C, 12393, 12396, 12397, 12398A, 12399, 12404, 12408, 12410, 12419, 12422, 12428, 12446, 12452, 12470T, 12476, 12484, 12530, 12531, 12533, 12536, 12537, 12539, 12557, 12579, 12590, 12593, 12594, 12599, 12608, 12623, 12641, 12671, 12677, 12689, 12716, 12717, 12719, 12725, 12737, 12743, 12761, 12762, 12763, 12779, 12793, 12800, 12801, 12812, 12824, 12836, 12851, 12878, 12881, 12884, 12899, 12927, 12932, 12950, 12953, 12984, 12990, 12992, 12998, 13001, 13007, 13008, 13016, 13019, 13020, 13022T, 13028, 13031, 13049, 13067, 13070, 13072C, 13073, 13091, 13114C, 13119, 13124, 13130, 13133, 13137, 13146A, 13151, 13163, 13187A, 13214, 13226, 13244, 13250, 13256, 13258, 13259, 13260, 13261G, 13262, 13271, 13274, 13275A, 13290, 13293, 13304C, 13316, 13319, 13323, 13343, 13357, 13358G, 13371, 13382, 13400, 13425, 13427, 13430, 13445, 13450T, 13457, 13477, 13478C, 13483, 13496, 13502, 13503, 13520, 13523, 13532, 13535, 13548, 13551, 13582, 13583C, 13597, 13675, 13680A, 13728, 13732, 13753, 13759, 13800, 13810, 13813, 13843, 13846, 13867, 15448A, 15449, 15450, 15451, 15974, 15976, 15977, 16000, 16028, 16040, 16485 | Wu,Y.P. and Ma,Y.H. Direct Submission |
| | | | 15728, 15807, 15833, 15843, 15864, 15893, 15913, 15920, 15930, 15947, 15950, 15965, 15967, 15969, 15973, 15974, 15975, 15976, 15977, 15981, 15982, 15983, 16000, 16006, 16010, 16011, 16019, 16027, 16028, 16037, 16040, 16045, 16066, 16083, 16232, 16281, 16458, 16477, 16485, 16514 | | | |
| GU229280 (Goat) | Surplus of mutations. | C | 165, 177G, 179, 180-181d, 213, 1192, 1369, 1395+A, 1850, 2092, 2208, 2517, 2886C, 2945, 3003, 3132, 3195, 3279, 3294, 3469, 3600, 3771, 4570, 4627, 4708, 4714, 4891, 4942, 5386, 5542, 5602, 5713, 5776, 6094, 6325, 6391, 6415, 6493, 6502, 6514, 6520, 6523, 6529, 6577, 6637, 6706, 6736, 6808, 7053, 7109, 7112, 7280, 7313, 7319, 7346, 7376, 7406A, 7421, 7427, 7658, 7840, 7886, 8056, 8079, 8105, 8265, 8301C, 8487, 8624, 8852, 8960, 9062, 9112, 9206, 9347, 9548, 9600, 9827, 9995, 10028, 10303, 10357A, 10450, 10546, 10564, 10630T, 10639, 10951, 11062, 11128, 11257, 11386, 11569, 11771, 11957, 11969, 12146, 12155, | C:944, 1026, 1114, 1118-1119d, 6031, 6538, 7118, 7918, 8114, 10107, 11074, 12626, 12638, 12791, 13157, 13292, 15771, 15893, 15930, 15981 | C:177G, 179, 180-181d, 1395+A, 2092, 2517, 2945, 3279, 5386, 5542, 6493, 6502, 6514, 6520, 6529, 6577, 6637, 6706, 6736, 6808, 7053, 7109, 7112, 7280, 7313, 7319, 7346, 7376, 7406A, 7421, 7427, 7886, 8105, 8624, 10028, 10630T, 11386, 11771, 12155, 12192, 12252, 12257, 12266, 12277, 12278, 12281C, 12287, 12290, 12293, 12299, 12300, 12318, 12335, 12338, 12341, 12368C, 12377, 12392C, 12393, 12396, 12397, 12398A, 12399, 12404, 12408, 12410, 12419, 12422, 12428, 12446, 12452, 12470T, 12476, 12484, 12530, 12531, 12533, 12536, 12537, 12557, 12579, 12590, 12593, 12594, | Wu,Y.P. and Ma,Y.H. Direct Submission |
| | | | 12192, 12252, 12257, 12266, 12277, 12278, 12281C, 12287, 12290, 12293, 12299, 12300, 12314, 12318, 12335, 12338, 12341, 12362, 12368C, 12377, 12380, 12383, 12386, 12392C, 12393, 12396, 12397, 12398A, 12399, 12404, 12408, 12410, 12419, 12422, 12428, 12446, 12452, 12470T, 12476, 12484, 12530, 12531, 12533, 12536, 12537, 12539, 12557, 12579, 12590, 12593, 12594, 12599, 12608, 12623, 12641, 12671, 12677, 12689, 12716, 12717, 12719, 12725, 12737, 12743, 12761, 12762, 12763, 12779, 12793, 12800, 12801, 12812, 12824, 12836, 12851, 12878, 12881, 12884, 12899, 12927, 12932, 12950, 12953, 12984, 12990, 12992, 12998, 13001, 13007, 13008, 13016, 13019, 13020, 13022T, 13028, 13031, 13034, 13049, 13067, 13070, 13072C, 13073, 13091, 13114C, 13119, 13124, 13130, 13133, 13137, 13146A, 13151, 13163, 13187A, 13214, 13226, 13244, 13250, 13256, 13258, 13259, 13260, 13261G, 13262, 13271, 13274, 13275A, 13280, 13286, 13290, 13293, 13304C, 13316, 13319, 13323, 13343, 13357, 13358G, 13371, 13382, 13400, 13425, 13427, 13430, 13445, 13450T, 13457, 13477, 13483, 13496, 13502, 13503, 13520, 13523, 13532, 13535, 13548, 13551, 13582, 13583C, 13597, 13639, 13675, 13680A, 13728, 13732, 13741, 13753, 13756, 13759, 13768, 13792, 13800, 13810, 13813, 13843, 13846, 13867, 14044, 14163, 14290, 14386, 14518, 14527, 14746, 15004, 15053, 15265, 15515, 15543, 15554, 15616, 15631, 15636, 15637, 15662, 15675, 15689, 15699, 15728, 15744, 15763, 15797, 15807, 15822, 15842, 15843, 15862, 15864, 15866, 15870, 15909, 15920, 15950, 15969, 15973, 15974, 15975, 15976, 15983, 15984, 15992, 15994, 16000, 16003, 16006, 16010, 16011, 16019, 16022, 16026, 16027, | | 12599, 12608, 12623, 12641, 12671, 12677, 12689, 12716, 12717, 12719, 12725, 12737, 12743, 12761, 12762, 12763, 12779, 12793, 12800, 12801, 12812, 12824, 12836, 12851, 12878, 12881, 12884, 12899, 12927, 12932, 12950, 12953, 12984, 12990, 12992, 12998, 13007, 13008, 13016, 13019, 13020, 13022T, 13028, 13031, 13049, 13067, 13070, 13072C, 13073, 13091, 13114C, 13119, 13124, 13130, 13133, 13137, 13146A, 13151, 13163, 13187A, 13214, 13226, 13244, 13250, 13256, 13258, 13259, 13260, 13261G, 13262, 13271, 13274, 13275A, 13280, 13286, 13290, 13293, 13304C, 13316, 13319, 13323, 13343, 13357, 13358G, 13371, 13382, 13400, 13425, 13427, 13430, 13445, 13450T, 13457, 13477, 13483, 13496, 13503, 13520, 13523, 13532, 13535, 13548, 13551, 13582, 13583C, 13597, 13639, 13675, 13680A, 13728, 13732, 13753, 13759, 13768, 13792, 13800, 13810, 13813, 13843, 13846, 13867, 14163, 15265, 15515, 15842, 15976, 15992, 16011, 16028, 16054, 16430, 16494 | |
| | | | 16028, 16038, 16040, 16043, 16045, 16054, 16066, 16084C, 16150, 16151, 16232, 16325, 16363, 16376, 16430, 16440, 16448, 16460, 16477, 16494, 16495, 16514 | | | |
| GU229281 (Goat) | Surplus of mutations. | A2a1a | 177G, 179, 180-181d, 294, 1120A, 1395+A, 2119, 2639, 3195, 3740, 4277, 4942, 5085, 5950, 6334G, 6523, 6575, 7091, 7109, 7190, 7464, 7840, 8056, 8475, 8897, 8969, 10002, 10028, 11062, 11128, 11173, 11198, 11308, 11323, 11377, 11470, 11488, 11492, 11497, 11503, 11521, 11809, 12002, 12006, 12035, 12047, 12068, 12098, 12104, 12107, 12137, 12146, 12155, 12161, 12188A, 12209, 12221, 12252, 12257, 12266, 12277, 12278, 12281C, 12287, 12290, 12293, 12299, 12300, 12314, 12318, 12335, 12338, 12341, 12362, 12368C, 12377, 12380, 12383, 12386, 12392C, 12393, 12396, 12397, 12398A, 12399, 12404, 12408, 12410, 12419, 12422, 12428, 12446, 12452, 12470T, 12476, 12484, 12530, 12531, 12533, 12536, 12537, 12539, 12557, 12579, 12590, 12593, 12594, 12599, 12608, 12623, 12641, 12671, 12677, 12689, 12716, 12717, 12719, 12725, 12743, 12761, 12762, 12763, 12779, 12793, 12800, 12801, 12812, 12824, 12836, 12840, 12851, 12878, 12881, 12884, 12899, 12927, 12932, 12950, 12953, 12984, 12990, 12992, 12998, 13001, 13007, 13008, 13016, 13019, 13020, 13022T, 13028, 13031, 13034, 13049, 13067, 13070, 13072C, 13073, | A2a1a:15811 | A2a1a:177G, 179, 180-181d, 294, 1120A, 1395+A, 2119, 2639, 3195, 3740, 4277, 5085, 5950, 6334G, 6575, 7091, 7109, 7190, 7464, 7840, 8056, 8475, 8897, 8969, 10002, 11173, 11198, 11308, 11323, 11377, 11470, 11488, 11492, 11497, 11503, 11521, 11809, 12002, 12006, 12035, 12047, 12068, 12098, 12104, 12107, 12137, 12146, 12155, 12161, 12188A, 12209, 12221, 12252, 12257, 12266, 12277, 12278, 12281C, 12287, 12290, 12293, 12299, 12300, 12314, 12318, 12335, 12338, 12341, 12362, 12368C, 12377, 12383, 12386, 12392C, 12393, 12396, 12397, 12398A, 12399, 12404, 12408, 12410, 12419, 12422, 12428, 12446, 12452, 12470T, 12476, 12484, 12530, 12531, 12533, 12536, 12537, 12539, 12557, 12579, 12590, 12593, 12594, 12599, 12608, 12623, 12641, 12671, 12677, 12689, 12716, 12717, 12719, 12725, 12743, 12761, 12762, 12763, 12779, 12793, 12800, 12801, 12812, 12824, 12836, 12840, 12851, 12878, 12881, 12884, 12899, 12927, 12932, 12950, 12953, 12984, 12990, 12992, 12998, 13001, 13007, 13008, 13016, 13019, 13020, 13022T, 13028, 13031, 13049, 13067, 13070, | Wu,Y.P. and Ma,Y.H. Direct Submission |
| | | | 13091, 13114C, 13119, 13124, 13130, 13133, 13137, 13146A, 13151, 13163, 13187A, 13214, 13226, 13250, 13256, 13258, 13259, 13260, 13261G, 13262, 13271, 13274, 13275A, 13280, 13286, 13290, 13293, 13304C, 13316, 13319, 13323, 13343, 13357, 13358G, 13371, 13382, 13400, 13425, 13427, 13431, 13439, 13445, 13450T, 13457, 13477, 13483, 13502, 13503, 13508, 13514, 13523, 13526, 13535, 13548, 13551, 13582, 13597, 13639, 13675, 13680A, 13728, 13732, 13741, 13753, 13756, 13759, 13768, 13792, 13800, 13810, 13813, 13843, 13846, 13867, 13894, 14317, 14443, 14500, 14746, 15238, 15289, 15543, 15616, 15631, 15636, 15771, 15807, 15810, 15818, 15873, 15911, 15913, 15920, 15930, 15969, 15973, 15978, 15983, 15992, 16000, 16006, 16019, 16027, 16028, 16038, 16048, 16062, 16066, 16435, 16460, 16470, 16472, 16477, 16504d, 16514, 16588 | | 13072C, 13073, 13091, 13114C, 13119, 13124, 13130, 13133, 13137, 13146A, 13151, 13163, 13187A, 13214, 13226, 13250, 13256, 13258, 13259, 13260, 13261G, 13262, 13271, 13274, 13275A, 13280, 13286, 13290, 13293, 13304C, 13316, 13319, 13323, 13343, 13357, 13358G, 13371, 13382, 13400, 13425, 13427, 13431, 13439, 13445, 13450T, 13457, 13477, 13483, 13502, 13503, 13508, 13514, 13523, 13526, 13535, 13548, 13551, 13582, 13597, 13639, 13675, 13680A, 13728, 13732, 13741, 13753, 13756, 13759, 13768, 13792, 13800, 13810, 13813, 13843, 13846, 13867, 13894, 14317, 14443, 14500, 14746, 15238, 15289, 15543, 15616, 15631, 15636, 15771, 15807, 15810, 15818, 15873, 15911, 15913, 15920, 15930, 15969, 15973, 15978, 15983, 15992, 16006, 16019, 16027, 16038, 16048, 16062, 16460, 16470, 16472, 16477, 16504d, 16514, 16588 | |
| AY584828(Horse and Przewalski's horse) | Surplus of mutations. | ABC, C, C2 | 358d, 1095, 1096A, 1098, 1099, 1146, 1147, 1224C, 1253, 1267, 1335, 1382, 1451, 1493, 1526, 1546, 1851, 2104, 2157, 2159, 2172, 2188, 2227+T, 2339A, 2803A, 3205, 3800, 4062, 4583, 4884, 4993, 5239+A, 5277+A, 5500, 5546, 5637G, 6076, 6784, 6850, 6908, 6912, 7001, 7008A, 7722A, 9239, 9248C, 9664, 9734T, 10123C, 10124G, 10427, 10441T, 10895, 11046, 11129, 11240, 11543, 12921, 12995, 13079, 13086G, 13411, 13555A, 13685, 13709, 13742G, 13996T, 14252, 14656, 15274, 15385+T, 15404, 15406, 15492, 15535, 15569C, 15594, 15599, 15615T, 15659, 15660C, 15661, 15662, 15664, 15665, 15700G, 15708G, 15717, 15768, 15776T, 15804, 15824, 15853, 15854, 15855, 15867, 15868, 16034C, 16089, 16268-16331d, 16368, 16388A, 16393A, 16396C, 16404, 16405A, 16406C, 16429T, 16430G, 16448T | ABC:158, 356, 1387d, 2788, 10214, 15647, 15823; C:158, 356, 957, 1387d, 2238, 2788, 4599, 7942, 10214, 10217, 12352, 15647, 15823, 15953, 15971, 16110; C2:158, 267, 356, 957, 1387d, 2238, 2788, 4599, 7942, 10214, 10217, 12352, 15647, 15823, 15953, 15971, 16110 | ABC:358d, 1095, 1096A, 1098, 1099, 1146, 1147, 1224C, 1253, 1267, 1335, 1382, 1451, 1493, 1526, 1546, 1851, 2104, 2157, 2159, 2172, 2188, 2227+T, 2339A, 2803A, 3205, 3800, 4583, 4884, 4993, 5239+A, 5277+A, 5500, 5546, 5637G, 6076, 6850, 6908, 6912, 7008A, 7722A, 9248C, 9664, 9734T, 10123C, 10124G, 10427, 10441T, 10895, 11046, 11129, 12921, 12995, 13079, 13086G, 13411, 13555A, 13685, 13709, 13742G, 13996T, 14252, 14656, 15274, 15385+T, 15404, 15406, 15535, 15569C, 15594, 15615T, 15659, 15660C, 15661, 15662, 15664, 15665, 15700G, 15708G, 15768, 15776T, 15804, 15824, 15853, 15854, 15855, 15868, 16034C, 16089, 16268-16331d, 16388A, 16393A, 16396C, 16404, 16405A, 16406C, 16429T, 16430G, 16448T; C:358d, 1095, 1096A, 1098, 1099, 1146, 1147, 1224C, 1253, 1267, 1335, 1382, 1451, 1493, 1526, 1546, 1851, 2104, 2157, 2159, 2172, 2188, 2227+T, 2339A, 2803A, 3205, 4583, 5239+A, 5277+A, 5546, 5637G, 6850, 6908, 6912, 7008A, 7722A, 9248C, 9734T, 10123C, 10124G, 10427, 10441T, 10895, 12921, 12995, 13086G, 13411, 13555A, 13685, 13709, 13742G, 13996T, 14252, 14656, 15274, 15385+T, 15404, 15406, 15535, 15569C, 15594, 15615T, 15659, 15660C, 15661, 15662, 15664, 15665, 15700G, 15708G, 15768, 15776T, 15804, 15824, 15853, 15854, 15855, 15868, 16034C, 16089, 16268-16331d, 16388A, 16393A, 16396C, 16404, 16405A, 16406C, 16429T, 16430G, 16448T; C2:358d, 1095, 1096A, 1098, 1099, 1146, 1147, 1224C, 1253, 1267, 1335, 1382, 1451, 1493, 1526, 1546, 1851, 2104, 2157, 2159, 2172, 2188, 2227+T, 2339A, 2803A, 3205, 4583, 5239+A, | Han,S.-H., Kang,M.-C., Oh,J.-H., Song,J.-H., Oh,Y.-S., Kim,J.-H.,Jung,Y.-H. and Oh,M.-Y. Unpublished |
| | | | | | 5277+A, 5546, 5637G, 6850, 6908, 6912, 7008A, 7722A, 9248C, 9734T, 10123C, 10124G, 10427, 10441T, 10895, 12921, 12995, 13086G, 13411, 13555A, 13685, 13709, 13742G, 13996T, 14252, 14656, 15274, 15385+T, 15404, 15406, 15535, 15569C, 15615T, 15659, 15660C, 15661, 15662, 15664, 15665, 15700G, 15708G, 15768, 15776T, 15804, 15824, 15853, 15854, 15855, 15868, 16034C, 16089, 16268-16331d, 16388A, 16393A, 16396C, 16404, 16405A, 16406C, 16429T, 16430G, 16448T | |
| AP013080(Horse and Przewalski's horse) | Artificial recombination: missing mutations 15492,15838 of haplogroup AB; mis-added mutations 222, 382, 387, 416, 5623, 5830, 5881, 5884, | AB, A1a | 158, 222, 356, 382, 387, 416, 5623, 5830, 5881, 5884, 6004, 6784, 10062, 11165, 11210, 11240, 11543, 11552, 16268-16331d, 16368 | AB:4062, 15492, 15647, 15717, 15823 | AB:222, 382, 387, 416, 5623, 5830, 5881, 5884, 6004, 10062, 11165, 11210, 11543, 11552, 16268-16331d; A1a:158, 222, 356, 382, 387, 416, 5623, 5830, 5881, 5884, 6004, 6784, 10062, 11165, 11210, 11240, 11543, 11552, 16268-16331d, 16368 | Wada,K. and Yokohama,M. Unpublished |
| | 6004, 10062, 11165, 11210, 11543, 11552 from haplogroup G. | | | | | |
| AP013082(Horse and Przewalski's horse) | Artificial recombination: missing mutations 13502, 14350, 14626, 14734 of haplogroup G3a; mis-added mutations 14995, 15313 from haplogroup L. | G3a | 158, 222, 356, 382, 387, 416, 1387d, 2788, 2940, 3053, 3576, 4062, 4646, 4669, 4830, 5498, 5830, 5881, 5884, 6004, 6307, 6688, 6784, 7001, 7687, 8005, 8037, 9145, 9239, 9402, 9669, 9741A, 10214, 10376, 10471, 10739, 11165, 11240, 11552, 11842, 12767, 12860, 13049, 13223, 13333, 14995, 15171, 15313, 15492, 15539, 15594, 15599, 15632, 15647, 15663, 15700, 15717, 15867, 16028, 16110, 16127, 16128, 16368 | G3a:13502, 14350, 14626, 14734 | G3a:10739, 14995, 15171, 15313, 16127, 16128 | Wada,K. and Yokohama,M. Unpublished |
| AP013087(Horse and Przewalski's horse) | Artificial recombination: missing mutations 10421, 10613, 12119, 12200, 13520, 14803, 14995, 15313 of haplogroup L3a; mis-added mutations 13502, 13761, 14038, 14554 | L3a | 77, 158, 356, 961, 1375, 1387d, 2788, 2899, 3517, 3942, 4062, 4536, 4646, 4669, 5527, 5815, 5884, 6004, 6307, 6784, 7001, 7516, 7666, 7900, 8005, 8058, 8301, 8319, 8358, 8565, 9239, 9951, 10110, 10214, 10292, 10376, 11240, 11315, 11543, 11693, 11842, 11879, 12767, 12896, 12950, 13049, 13333, 13502, 13761, 14038, 14554, 15491, 15492, 15493, 15531, 15582, 15599, 15600, 15601, 15646, 15717, 15768, 15867, 15868, 15953, 15971, 16065, 16100, 16368, 16404d, 16553, 16560+C, 16626 | L3a:10421, 10613, 12119, 12200, 13520, 14803, 14995, 15313 | L3a:11315, 13502, 13761, 14038, 14554, 16404d, 16560+C | Wada,K. and Yokohama,M. Unpublished |
| | from haplogroup I1. | | | | | |
| AP013100(Horse and Przewalski's horse) | Artificial recombination: missing mutations 16175, 6247 of haplogroup I2a1a; mis-added mutations 5815 from haplogroup L. | I2a1a | 158, 356, 1387d, 1587, 1791, 2614, 2770, 2788, 4062, 4063, 4392, 4646, 4669, 4830, 5061, 5210d, 5815, 5884, 6004, 6307, 6784, 7001, 8005, 8358, 8379, 8792, 9239, 9694, 9948, 10083, 10214, 10238, 10376, 11240, 11543, 11827, 11842, 12404, 12443, 12683, 12767, 13049, 13333, 13502, 13761, 14554, 15492, 15535, 15581, 15582, 15599, 15647, 15706, 15717, 15768, 15823, 15867, 15971, 16077, 16110, 16119, 16127, 16128, 16284-16331d, 16368, 16436 | I2a1a:5210, 6175, 6247 | I2a1a:5210d, 5815, 16127, 16128, 16284-16331d | Wada,K. and Yokohama,M. Unpublished |
| AP013101(Horse and Przewalski's horse) | Artificial recombination: missing mutations 15492, 15823 of haplogroup A1b. | A1a, A1b | 5061, 6784, 9477, 16110, 16292-16331d | A1b:15492, 15823 | A1a:5061, 6784, 9477, 16110, 16292-16331d; A1b:5061, 9477, 16292-16331d | Wada,K. and Yokohama,M. Unpublished |
| EF597512(Horse and Przewalski's horse) | Artificial recombination: missing mutations 1609T, 2339A, 2802, 3070, 3475, 3557, 3800, 4898, 5527, 5827, 6076, 7666, 7900, 8043, 8076, 8150, 8175, 8238, 8280, 8358, 8556T, 8565, 8798, 9332, 10859, 11394, 11492, 11879, 11966, 12029, 12095, 13100, 14803, 14815, 15052A, 15133, 15342, 16108 of haplogroup N2; mis-added mutations 222, 416, 3053, 3576, 4830, 5498, 5669, 5830, 5881, 8037, 11165, 11552, 12860 | N2, N2a | 158, 222, 356, 416, 961, 1387d, 2788, 2940, 3053, 3576, 4062, 4526, 4536, 4605, 4646, 4669, 4830, 4917, 5498, 5669, 5830, 5881, 5884, 6004, 6307, 6457A, 6712, 6784, 7001, 7229C, 7303C, 7432, 8005, 8037, 8383G, 9071, 9086, 9239, 9402, 9540, 9918, 10110, 10173, 10214, 10292, 10376, 10404, 10448, 10460, 10517, 10646, 10893A, 11165, 11240, 11543, 11552, 11842, 12003C, 12024G, 12277G, 12332, 12674, 12767, 12860, 13049, 13333, 13356, 13502, 13567, 13615, 13629, 13720, 13920, 13933, 14270, 14422, 14626, 14671, 15492, 15582, 15598, 15599, 15695G, 15717, 15768, 15803, 15824, 15835, 15866, 15953, 16004, 16065, 16110, 16118, 16127, 16128, 16268-16331d, 16368, 16540A, 16543, 16556, 16626 | N2:1609T, 2339A, 2802, 3070, 3475, 3557, 3800, 4898, 5527, 5827, 6076, 7666, 7900, 8043, 8076, 8150, 8175, 8238, 8280, 8358, 8556T, 8565, 8798, 9332, 10859, 11394, 11492, 11879, 11966, 12029, 12095, 13100, 14803, 14815, 15052A, 15133, 15342, 16108; N2a:1609T, 2339A, 2802, 3070, 3475, 3557, 3800, 4898, 5527, 5827, 6076, 7666, 7900, 8043, 8076, 8150, 8175, 8238, 8280, 8358, 8556T, 8565, 8798, 9332, 10859, 11384, 11394, 11492, 11879, 11966, 12029, | N2:222, 416, 2940, 3053, 3576, 4830, 5498, 5669, 5830, 5881, 6457A, 7229C, 7303C, 8037, 8383G, 10893A, 11165, 11552, 12003C, 12024G, 12277G, 12860, 14270, 15582, 15695G, 16127, 16128, 16268-16331d; N2a:222, 416, 2940, 3053, 3576, 4830, 5498, 5669, 5830, 5881, 6457A, 7229C, 7303C, 8037, 8383G, 10893A, 11165, 11552, 12003C, 12024G, 12277G, 12860, 14270, 15695G, 16127, 16128, 16268-16331d | Xu,S., Luosang,J., Hua,S., He,J., Ciren,A., Wang,W., Tong,X., Liang,Y., Wang,J. and Zheng,X. Genet Genomics 34 (8), 720-729 (2007) |
| | from haplogroup G1. | | | 12095, 13100, 14803, 14815, 15052A, 15133, 15342, 16108 | | |
| EU939445 (Horse and Przewalski's horse) | Surplus of mutations. | H-I | 158, 302, 341, 356, 1258, 1262d, 1293d, 1387d, 1546, 1668, 2788, 3064, 3281, 4062, 4182, 4646, 4669, 4830, 5272, 5884, 6004, 6076, 6103T, 6107C, 6109A, 6110, 6111, 6112A, 6113C, 6121, 6122A, 6123A, 6124T, 6127, 6307, 6565, 6784, 6835, 7243, 7294, 8005, 8071, 8081C, 8174, 8358, 8361, 8556T, 8565, 9209, 9239, 9775, 10083, 10214, 10376, 10731, 10914, 11090, 11101G, 11127G, 11137, 11175, 11200, 11240, 11254A, 11260, 11263G, 11299, 11301G, 11325A, 11342G, 11385G, 11387, 11543, 11640G, 11842, 11989, 12044G, 12372, 12767, 12831T, 13049, 13177, 13232, 13333, 13502, 14395, 14551, 14803, 15297, 15388, 15389, 15395+G, 15452d, 15492, 15582, 15599, 15717, 15768, 15774, 15804, 15824, 15867, 15868, 16068, 16108, 16110, 16127, 16128, 16368 | H-I:7001, 13761, 15647, 15823 | H-I:302, 341, 1258, 1262d, 1293d, 1546, 1668, 3064, 3281, 4182, 5272, 6076, 6103T, 6107C, 6109A, 6110, 6111, 6112A, 6113C, 6121, 6122A, 6123A, 6124T, 6127, 6565, 6835, 7243, 7294, 8071, 8081C, 8174, 8358, 8361, 8556T, 8565, 9209, 9775, 10731, 10914, 11090, 11101G, 11127G, 11137, 11175, 11200, 11254A, 11260, 11263G, 11299, 11301G, 11325A, 11342G, 11385G, 11387, 11640G, 11989, 12044G, 12372, 12831T, 13177, 13232, 14395, 14551, 14803, 15297, 15388, 15389, 15395+G, 15452d, 15774, 15804, 15824, 15868, 16068, 16108, 16127, 16128 | Jiang,Q., Wei,Y., Huang,Y., Jiang,H., Guo,Y., Lan,G. and Liao,D.J. Mol Biol Rep. 38 (1), 593-599 (2007). |
| KC202960( Horse and Przewalski's horse) | Artificial recombination: missing mutations 222, 387, 2940, 15647, 15663, | G2 | 158, 356, 382, 416, 1387d, 2788, 3053, 3576, 4062, 4646, 4669, 4830, 5498, 5830, 5881, 5884, 6004, 6307, 6688, 6784, 7001, 8005, 8037, 8286, 8361, 9195, 9239, 9304, 9402, 9669, 9741A, 10214, 10376, 10471, 11165, 11240, 11543, 11552, 11842, 12161, 12767, 12860, 13049, 13118, 13223, 13333, 13502, 13589, 14350, | G2:222, 387, 2940, 15647, 15663, 16110, | G2:8286, 8361, 9304, 12161, 13118, 13589, 16077 | Qi,A., Davie,A., Wen,L., Qiu,M., Jiang,F., Pacey,T., |
| | 16110 of haplogroup G. | | 14626, 14734, 15492, 15539, 15594, 15599, 15632, 15700, 15717, 15867, 16077, 16368 | | | Twenke,B.,Zhang,Y., Zhou,S. and Liu,B. Unpublished |
| KC202961(Horse and Przewalski's horse) | Artificial recombination: missing mutations 222, 382, 387, 416, 15539, 15663, 16110 of haplogroup G2; mis-added mutations 15773 from haplogroup R. | G2 | 158, 356, 1387d, 2788, 2940, 3053, 3576, 4062, 4646, 4669, 4830, 5498, 5830, 5881, 5884, 6004, 6307, 6688, 6784, 7001, 8005, 8037, 8286, 8361, 9195, 9239, 9402, 9669, 9741A, 10214, 10376, 10471, 11165, 11240, 11543, 11552, 11842, 12161, 12767, 12860, 13049, 13118, 13223, 13333, 13502, 13589, 14350, 14626, 14734, 15492, 15582, 15594, 15599, 15632, 15647, 15700, 15717, 15773, 15867, 16368 | G2:222, 382, 387, 416, 15539, 15663, 16110 | G2:8286, 8361, 12161, 13118, 13589, 15582, 15773 | Qi,A., Davie,A., Wen,L., Qiu,M., Jiang,F., Pacey,T., Twenke,B.,Zhang,Y., Zhou,S. and Liu,B. Unpublished |
| KC202962(Horse and Przewalski's horse) | Artificial recombination: missing mutations 222, 382, 387, 416, 2940, 3053, 15539, 15663,16110 of haplogroup G2; mis-added mutations 287, 3070, 15142 from haplogroup A-Q. | G2 | 158, 287, 356, 575, 961, 1387d, 1696, 2788, 3070, 3576, 4062, 4646, 4669, 4830, 5498, 5830, 5881, 5884, 6004, 6307, 6688, 6784, 7001, 8005, 8037, 8286, 8361, 9195, 9239, 9304, 9402, 9669, 9741A, 10214, 10376, 10471, 11165, 11240, 11543, 11552, 11842, 12161, 12767, 12860, 13049, 13118, 13223, 13333, 13502, 13589, 14350, 14626, 14734, 15142, 15492, 15582, 15594, 15599, 15632, 15647, 15700, 15717, 15867, 16077, 16368 | G2:222, 382, 387, 416, 2940, 3053, 15539, 15663, 16110 | G2:287, 575, 961, 1696, 3070, 8286, 8361, 9304, 12161, 13118, 13589, 15142, 15582, 16077 | Qi,A., Davie,A., Wen,L., Qiu,M., Jiang,F., Pacey,T., Twenke,B.,Zhang,Y., Zhou,S. and Liu,B. Unpublished |
| KC202964(Horse and Przewalski's horse) | Artificial recombination: missing mutations 222, 382, 387, 416, 2940, 3053, 15539, 15663, 16110 of haplogroup G2; mis-added mutations 287, 3070, 15142 from haplogroup A-Q. | G2 | 158, 287, 356, 575, 961, 1387d, 1696, 2788, 3070, 3576, 4062, 4646, 4669, 4830, 5498, 5830, 5881, 5884, 6004, 6307, 6688, 6784, 7001, 8005, 8037, 8286, 8361, 9195, 9239, 9304, 9402, 9669, 9741A, 10214, 10376, 10471, 11165, 11240, 11543, 11552, 11842, 12161, 12767, 12860, 13049, 13118, 13223, 13333, 13502, 13589, 14350, 14626, 14734, 15020, 15142, 15492, 15582, 15594, 15599, 15632, 15647, 15700, 15717, 15768, 15773, 15867, 15953, 16368 | G2:222, 382, 387, 416, 2940, 3053, 15539, 15663, 16110 | G2:287, 575, 961, 1696, 3070, 8286, 8361, 9304, 12161, 13118, 13589, 15020, 15142, 15582, 15768, 15773, 15953 | Qi,A., Davie,A., Wen,L., Qiu,M., Jiang,F., Pacey,T., Twenke,B.,Zhang,Y., Zhou,S. and Liu,B. Unpublished |
| KC202983(Horse and Przewalski's horse) | Artificial recombination: missing mutations 9694, 9948, 10238, 11827, 12443, 12683, 14554 of haplogroup I2; mis-added mutations 10292, 10448, 10646, 10517, 11966 from haplogroup N. | I2 | 158, 356, 1387d, 1587, 1791, 2614, 2770, 2788, 4062, 4063, 4392, 4646, 4669, 4830, 5061, 5210d, 5884, 6004, 6175, 6247, 6307, 6784, 7001, 8005, 8358, 8379, 8792, 9239, 10083, 10173, 10214, 10292, 10376, 10448, 10460, 10517, 11240, 11543, 11842, 11879, 11966, 12404, 12767, 13049, 13333, 13502, 13761, 14626, 15492, 15535, 15581, 15582, 15599, 15647, 15706, 15717, 15768, 15823, 15867, 15971, 16077, 16110, 16119, 16127, 16128, 16368, 16436 | I2:5210, 5217d, 9694, 9948, 10238, 11827, 12443, 12683, 14554 | I2:5210d, 10173, 10292, 10448, 10460, 10517, 11879, 11966, 14626, 15647 | Qi,A., Davie,A., Wen,L., Qiu,M., Jiang,F., Pacey,T., Twenke,B.,Zhang,Y., Zhou,S. and Liu,B. Unpublished |
| KC202991(Horse and Przewalski's horse) | Artificial recombination: missing mutations 12119, 12200, 16065, 16368 of haplogroup L2b. | L2b | 158, 356, 961, 1375, 1387d, 2607, 2788, 2899, 3517, 3942, 4062, 4536, 4646, 4669, 5527, 5815, 5884, 6004, 6307, 6784, 7001, 7516, 7666, 7900, 8005, 8058, 8199, 8301, 8319, 8358, 8403, 8565, 9239, 9951, 10110, 10214, 10292, 10376, 10421, 10613, 11240, 11543, 11693, 11842, 11879, 12767, 12896, 12950, 13049, 13333, 13520, 14803, 14995, 15313, 15491, 15492, 15493, 15531, 15582, 15600, 15646, 15717, 15768, 15867, 15868, 15953, 15971, 16100, 16108, 16268-16331d, 16626 | L2b:12119, 12200, 16065, 16368 | L2b:16108, 16268-16331d | Qi,A., Davie,A., Wen,L., Qiu,M., Jiang,F., Pacey,T., Twenke,B.,Zhang,Y., Zhou,S. and Liu,B. Unpublished |
| KC202997(Horse and Przewalski's horse) | Artificial recombination: missing mutations 10421, 10613, 11693, 11879, 12119, 12200, 12896, 13520, 14803, 15313 of haplogroup L; mis-added mutations 11827, 12443, 12683 from haplogroup N. | L | 158, 356, 961, 1375, 1387d, 1459, 2788, 2899, 3517, 3942, 4062, 4536, 4646, 4669, 5527, 5815, 5884, 6004, 6307, 6784, 7001, 7516, 7666, 7900, 8005, 8058, 8301, 8307, 8319, 8358, 8565, 9239, 9951, 10110, 10214, 10292, 10376, 11240, 11543, 11827, 11842, 12404, 12443, 12683, 12767, 12950, 13049, 13333, 13502, 14659, 14995, 15491, 15492, 15493, 15531, 15582, 15599, 15600, 15601, 15646, 15717, 15768, 15867, 15868, 15953, 15971, 16065, 16100, 16110, 16368, 16404d, 16626 | L:10421, 10613, 11693, 11879, 12119, 12200, 12896, 13520, 14803, 15313 | L:1459, 8307, 11827, 12404, 12443, 12683, 13502, 14659, 15601, 16110, 16404d | Qi,A., Davie,A., Wen,L., Qiu,M., Jiang,F., Pacey,T., Twenke,B.,Zhang,Y., Zhou,S. and Liu,B. Unpublished |
| KC203000(Horse and Przewalski's horse) | Artificial recombination: missing mutations 961, 9951, 10613, 11693, 12119, | L3a | 77, 158, 356, 1375, 1387d, 2788, 2899, 3220, 3517, 3942, 4062, 4536, 4646, 4669, 5527, 5815, 5884, 6004, 6307, 6784, 7001, 7516, 7666, 7900, 8005, 8058, 8301, 8319, 8358, 8565, 9239, 10110, 10214, 10292, 10376, 10421, 11084, 11240, 11543, 11842, 11879, 12404, 12767, 12950, 13049, 13333, 13502, 14803, 14995, | L3a:961, 9951, 10613, 11693, 12119, 12200, 12896, 13520 | L3a:3220, 11084, 12404, 13502, 16108, 16404d, 16632d | Qi,A., Davie,A., Wen,L., Qiu,M., Jiang,F., Pacey,T., |
| | 12200, 12896, 13520 of haplogroup L3a. | | 15313, 15491, 15492, 15493, 15531, 15582, 15599, 15600, 15601, 15646, 15717, 15768, 15867, 15868, 15953, 15971, 16065, 16100, 16108, 16368, 16404d, 16553, 16626, 16632d | | | Twenke,B.,Zhang,Y., Zhou,S. and Liu,B. Unpublished |
| KC203007(Horse and Przewalski's horse) | Artificial recombination: missing mutations 8403, 15313, 15531, 15971, 16100 of haplogroup L2. | L2 | 158, 356, 961, 1375, 1387d, 2607, 2788, 2899, 3517, 3727, 3942, 4062, 4536, 4646, 4669, 5527, 5815, 5884, 6004, 6307, 6784, 7001, 7516, 7666, 7900, 8005, 8058, 8301, 8319, 8358, 8565, 9239, 9951, 10087, 10110, 10214, 10292, 10376, 10421, 10613, 11240, 11543, 11693, 11842, 11879, 12119, 12200, 12767, 12896, 12950, 13049, 13333, 13520, 14803, 14995, 15491, 15492, 15493, 15582, 15600, 15646, 15717, 15768, 15824, 15867, 15868, 15953, 16065, 16108, 16368, 16404d, 16626 | L2:8403, 15313, 15531, 15971, 16100 | L2:3727, 10087, 15824, 16108, 16404d | Qi,A., Davie,A., Wen,L., Qiu,M., Jiang,F., Pacey,T., Twenke,B.,Zhang,Y., Zhou,S. and Liu,B. Unpublished |
| KC203011(Horse and Przewalski's horse) | Artificial recombination: missing mutations 15867, 15868, 15953, 15971, 16065, 16100 of haplogroup L. | L | 158, 356, 961, 1375, 1387d, 1459, 2788, 2899, 3517, 3942, 4062, 4536, 4646, 4669, 5527, 5815, 5884, 6004, 6307, 6784, 7001, 7516, 7666, 7900, 8005, 8058, 8301, 8307, 8319, 8358, 8565, 9239, 9951, 10110, 10214, 10292, 10376, 10421, 10613, 11240, 11543, 11693, 11842, 11879, 12119, 12200, 12314, 12767, 12896, 12950, 13049, 13333, 13520, 14659, 14803, 14995, 15313, 15491, 15492, 15493, 15531, 15582, 15599, 15600, 15601, 15646, 15717, 15768, 16368, 16404d, 16626 | L:15867, 15868, 15953, 15971, 16065, 16100 | L:1459, 8307, 12314, 14659, 15601, 16404d | Qi,A., Davie,A., Wen,L., Qiu,M., Jiang,F., Pacey,T., Twenke,B.,Zhang,Y., Zhou,S. and Liu,B. Unpublished |
| KC203015(Horse and Przewalski's horse) | Artificial recombination: missing mutations 15867, 15868, 15953, 15971, | L | 158, 356, 961, 1375, 1387d, 1459, 2788, 2899, 3517, 3942, 4062, 4536, 4646, 4669, 5527, 5815, 5884, 6004, 6307, 6784, 7001, 7516, 7666, 7900, 8005, 8058, 8301, 8307, 8319, 8358, 8565, 9239, 9951, 10110, 10214, 10292, 10376, 10421, 10613, 11240, 11543, 11693, 11842, 11879, 12119, 12200, 12314, 12767, 12896, | L:15867, 15868, 15953, 15971, 16065, 16100, 16368, 16626 | L:1459, 8307, 12314, 14659, 15601 | Qi,A., Davie,A., Wen,L., Qiu,M., Jiang,F., Pacey,T., |
| | 16065, 16100,16368, 16626 of haplogroup L. | | 12950, 13049, 13333, 13520, 14659, 14803, 14995, 15313, 15491, 15492, 15493, 15531, 15582, 15599, 15600, 15601, 15646, 15717, 15768 | | | Twenke,B.,Zhang,Y., Zhou,S. and Liu,B. Unpublished |
| KC203019(Horse and Przewalski's horse) | Artificial recombination: missing mutations 8319, 8358 of haplogroup L. | L | 158, 356, 961, 1375, 1387d, 1459, 2788, 2899, 3517, 3942, 4062, 4536, 4646, 4669, 5527, 5815, 5884, 6004, 6307, 6784, 7001, 7516, 7666, 7900, 8005, 8058, 8301, 8565, 9239, 9951, 10110, 10214, 10292, 10376, 10421, 10613, 11240, 11543, 11693, 11842, 11879, 12119, 12200, 12767, 12896, 12950, 13049, 13333, 13520, 14335, 14659, 14803, 14995, 15313, 15491, 15492, 15493, 15531, 15582, 15599, 15600, 15646, 15717, 15768, 15867, 15868, 15953, 15971, 16065, 16100, 16133, 16141, 16317, 16368, 16626 | L:8319, 8358 | L:1459, 14335, 14659, 16133, 16141, 16317 | Qi,A., Davie,A., Wen,L., Qiu,M., Jiang,F., Pacey,T., Twenke,B.,Zhang,Y., Zhou,S. and Liu,B. Unpublished |
| KC203022(Horse and Przewalski's horse) | Artificial recombination: missing mutations 1609T, 2339A, 15342, 15656, 15803, 15824, 15866, 15953, 16065, 16118 of haplogroup M. | M | 158, 356, 427, 961, 1387d, 2788, 3070, 3100, 3475, 3800, 4062, 4526, 4536, 4599G, 4605, 4646, 4669, 4898, 5103, 5527, 5827, 5884, 6004, 6076, 6307, 6712, 6784, 7001, 7432, 7666, 7900, 8005, 8043, 8076, 8150, 8175, 8238, 8358, 8556T, 8565, 8798, 9086, 9332, 9540, 10110, 10173, 10214, 10292, 10376, 10448, 10460, 10646, 10859, 11240, 11394, 11492, 11543, 11842, 11848, 11879, 11966, 12029, 12095, 12332, 12767, 13049, 13100, 13333, 13356, 13502, 13615, 13629, 13720, 13920, 13933, 14422, 14626, 14671, 14803, 14815, 15052A, 15133, 15492, 15599, 15614, 15717, 15768, 15823, 15867, 15971, 16077, 16110, 16119, 16127, 16128, 16368, 16540A, 16543, 16556, 16626 | M:1609T, 2339A, 15342, 15656, 15803, 15824, 15866, 15953, 16065, 16118 | M:11848, 15823, 15867, 15971, 16110, 16119 | Qi,A., Davie,A., Wen,L., Qiu,M., Jiang,F., Pacey,T., Twenke,B.,Zhang,Y., Zhou,S. and Liu,B. Unpublished |
| KC203028(Horse and Przewalski's horse) | Artificial recombination: missing mutations 6076, 12029, 12095, 14815, 15052A, 15133, 16368, 16540A, 16543, 16556, 16626 of haplogroup N2a1. | N2a1 | 158, 356, 961, 1387d, 1609T, 2339A, 2788, 2802, 3070, 3475, 3557, 3800, 4062, 4428, 4526, 4536, 4605, 4646, 4669, 4898, 4917, 5527, 5827, 5884, 6004, 6307, 6712, 6784, 7001, 7432, 7666, 7900, 8005, 8043, 8076, 8150, 8175, 8238, 8280, 8358, 8556T, 8565, 8798, 9071, 9086, 9239, 9332, 9402, 9540, 9918, 10110, 10173, 10214, 10292, 10376, 10404, 10448, 10460, 10517, 10646, 10859, 11240, 11384, 11394, 11492, 11543, 11842, 11879, 11966, 12332, 12674, 12767, 13049, 13100, 13333, 13335, 13356, 13502, 13567, 13615, 13629, 13720, 13920, 13933, 14422, 14626, 14671, 14803, 15342, 15492, 15582, 15598, 15599, 15717, 15768, 15803, 15824, 15835, 15866, 15953, 16004, 16065, 16108, 16110, 16118, 16127, 16128 | N2a1:6076, 12029, 12095, 14815, 15052A, 15133, 16368, 16540A, 16543, 16556, 16626 | N2a1:16127, 16128 | Qi,A., Davie,A., Wen,L., Qiu,M., Jiang,F., Pacey,T., Twenke,B.,Zhang,Y., Zhou,S. and Liu,B. Unpublished |
| NC_001640(Horse and Przewalski's horse) | Phantom mutaions: 358d,, 2227+A, 5239+A, 5277+A, 15385+T. | A1a | 358d, 2227+T, 5098T, 5239+A, 5277+A, 10123C, 10124G, 10826G, 10827G, 13685, 13709, 13742G, 13996T, 15385+T | | A1a:358d, 2227+T, 5098T, 5239+A, 5277+A, 10123C, 10124G, 10826G, 10827G, 13685, 13709, 13742G, 13996T, 15385+T | Xu,X. and Arnason,U. Gene 148 (2), 357-362 (1994) |
| AF034253(Pig and wild boar) | Surplus of mutations. | E1a1, E1a1a1 | 109, 124T, 131, 136+C, 144, 152, 157, 180, 240, 293, 305, 322, 389, 451, 500, 574, 691, 703, 705, 833-912d, 1136, 1168, 1175, 1187T, 1204T, 1211T, 1225, 1304, 1315, 1412, 1638d, 1639, 1990, 2057, 2060T, 2064, 2065C, 2071, 2099C, 2122, 2335, 2339, 2388C, 2402, 2483T, 2501+C, 2501T, 2509, 2613, 2737, 2758, 2967, 2989, 3064, 3088, 3102, 3366, 3451, 3594, 3640, 3873T, 3999, 4015, 4030, 4081, 4132G, 4166G, 4191G, 4233G, 4342, 4369, 4420, 4438, 4459, 4471, 4489A, 4711, 4737, 4754, 4846, 4939, 5168T, 5207, 5369, 5463A, 5552, 5628, 5636, 5672, 5678, 5708, 5753, 5797, 5873, 5880, 5948, 5963, 6092, 6138, 6164, 6171, 6296, 6298, 6344, 6508, 6925T, 6952, 6970, 7009, 7022, 7108, 7321, 7339, 7447, 7486, 7513, 7669, 7750, 7837, 8017, 8371, 8413, 8498, 8581, 8605, 8713, 8743, 8761, 8773, 9058, 9156, 9157, 9225, 9234, 9435, 9553, 9605, 9752, 9789, 9973, 10070, 10100, 10484, 10529, 10680, 10753, 10816, 10944, 11018, 11071, 11162, 11184, 11189, 11259, 11289, 11366, 11372, 11683, 11786, 11830, 11944, 12109, 12241, 12298, 12355, 12370, 12469, 12583, 12649, 12675, 12719G, 12958, 12962, 13049, 13051, 13377G, 13433A, 13460, 13478, 13581, 13605, 13838, 13997, 14209, 14213, 14313A, 14399, 14561C, 14639, 14710, 14738, 14812, 14839, 14947, 14998, 15025, 15109, 15127, 15151, 15277, 15362, 15420+AT, 15625, 15661, 15685, 15721, 15838, 15961, 16111, 16258, 16292, 16294, 16297, 16456, 16492, 16552, 16564, 16608, 16681 | E1a1:137C, 142+XA, 10517, 12518, 16358; E1a1a1:137C, 142+XA, 12518, 16358 | E1a1:136+C, 691, 703, 705, 833-912d, 1136, 1187T, 1204T, 1211T, 1638d, 2057, 2060T, 2099C, 2122, 2388C, 2402, 2483T, 2501+C, 2501T, 2509, 2737, 2758, 2967, 2989, 4132G, 4166G, 4191G, 4233G, 6344, 12355, 12719G, 13051, 13377G, 13460, 15420+AT, 16681; E1a1a1:136+C, 180, 691, 703, 705, 833-912d, 1136, 1187T, 1204T, 1211T, 1638d, 2057, 2060T, 2099C, 2122, 2388C, 2402, 2483T, 2501+C, 2501T, 2509, 2737, 2758, 2967, 2989, 4132G, 4166G, 4191G, 4233G, 6344, 12355, 12719G, 13051, 13377G, 13460, 15420+AT, 16681 | Lin,C.S., Sun,Y.L., Liu,C.Y., Yang,P.C., Chang,L.C., Cheng,I.C., Mao,S.J. and Huang,M.C. Gene 236 (1), 107-114 (1999) |
| AY334492(Pig and wild boar) | Surplus of mutations; Phantom mutation 1304. | D1a1 | 66, 278, 436, 451, 500, 567+G, 619, 921+TGCGTACACGTGCGTACACGTGCGTACACGTGCGTACACGTGCGTACACGTGCGTACACG, 1187, 1304, 1315, 1463, 1631T, 2065C, 2336T, 2374, 3259d, 3594, 3642, 4471, 5207, 5290, 5369, 5463A, 5552, 5573, 5597, 5628, 5636, 5672, 5678, 5708, 5753, 5797, 5805, 5873, 5880, 5963, 6092, 6138, 6164, 6171, 6193, 6289, 6296, 6298, 6344, 7258, 7469, 8083A, 8342, 8581, 9409T, 10100, 10264, 10504, 10514G, 11366, 11372, 11513, 11683, 11786, 11830, 11944, 12055, 12109, 12241, 12298, 12370, 12469, 12518, 12583, 12649, 12675, 12879, 12958, 12962, 13037, 13049, 13433A, 13581, 13601, 13605, 13838, 13997, 14044T, 14046, 14099, 14125A, 14209, 14213, 14313A, 14316, 14399, 14561C, 14639, 14710, 14738, 14812, 14839, 15064, 15196, 15220, 15337, 15713T, 15838, 16301, 16681, 16690+C | D1a1:11432, 12064, 14198 | D1a1:66, 436, 567+G, 619, 921+TGCGTACACGTGCGTACACGTGCGTACACGTGCGTACACGTGCGTACACGTGCGTACACG, 1187, 1304, 1315, 1463, 1631T, 2336T, 3259d, 3642, 5207, 5290, 5369, 5463A, 5552, 5573, 5597, 5628, 5636, 5672, 5678, 5708, 5753, 5797, 5805, 5873, 5880, 5963, 6092, 6138, 6164, 6171, 6193, 6289, 6296, 6298, 6344, 7258, 7469, 8083A, 8342, 9409T, 10100, 10264, 10504, 10514G, 11366, 11372, 11513, 11683, 11786, 11830, 11944, 12055, 12109, 12241, 12298, 12370, 12469, 12518, 12583, 12649, 12675, 12879, 12958, 12962, 13037, 13049, 13433A, 13581, 13601, 13605, 13838, 13997, 14044T, 14046, 14099, 14125A, 14209, 14213, 14313A, 14316, 14399, 14561C, 14639, 14738, 14812, 14839, 15064, 15196, 15713T, 16681, 16690+C | Cho,I.C., Han,S.H., Jeon,J.T., Ko,M.S., Fang,M. and Andersson,L. Unpublished |
| AY337045(Pig and wild boar) | Artificial recombination: missed mutations10517, 10680, 12649, 12675, 12958, 12962, 13049, 13433A, 13478, 13581, 13605, 13838, 13997, 14209, 14213, 14313A, 14399, 14561C, 14639 | E1a1, E1a1a1 | 109, 124T, 131, 136+C, 144, 152, 157, 180, 240, 293, 305, 322, 389, 391+G, 451, 500, 574, 691, 703, 705, 803-912d, 977d, 1168, 1175, 1225, 1304, 1315, 1412, 1639, 1959+C, 1990, 2064, 2065C, 2071, 2335, 2339, 2499d, 2591, 2613, 2846T, 3064, 3088, 3102, 3153, 3366, 3451, 3566T, 3594, 3640, 3873T, 3999, 4015, 4030, 4081, 4342, 4369, 4420, 4438, 4459, 4471, 4489A, 4711, 4737, 4754, 4846, 4939, 5168T, 5207, 5369, 5463A, 5474, 5552, 5628, 5636, 5672, 5678, 5708, 5753, 5797, 5873, 5880, 5948, 5963, 6092, 6138, 6164, 6171, 6296, 6298, 6344, 6508, 6925T, 6952, 6970, 7009, 7022, 7108, 7321, 7339, 7447, 7486, 7513, 7669, 7750, 7837, 8010, 8017, 8371, 8413, 8476C, 8498, 8581, 8605, 8713, 8743, 8761, 8773, 9058, 9156, 9157, 9225, 9234, 9435, 9553, 9605, 9752, 9789, 9973, 10070, 10100, 10484, 10529, 10753, 10816, 10944, | E1a1:137C, 142+XA, 10517, 10680, 12649, 12675, 12958, 12962, 13049, 13433A, 13478, 13581, 13605, 13838, 13997, 14209, 14213, 14313A, 14399, 14561C, 14639, 14738, 14812, 14839; E1a1a1:137C, 142+XA, 10680, 12649, 12675, | E1a1:136+C, 391+G, 691, 703, 705, 803-912d, 977d, 1959+C, 2499d, 2591, 2846T, 3153, 3566T, 5474, 6344, 8010, 8476C, 11973, 13197C, 13472C, 13745+TCA, 14135A, 14138G, 14163, 14272, 14316, 14336, 14607, 15179, 16681, 16690+C; E1a1a1:136+C, 180, 391+G, 691, 703, 705, 803-912d, 977d, 1959+C, 2499d, 2591, 2846T, 3153, 3566T, 5474, 6344, 8010, 8476C, 11973, 13197C, 13472C, 13745+TCA, 14135A, 14138G, 14163, 14272, 14316, 14336, 14607, 15179, 16681, 16690+C | Cho,I.C. Unpublished |
| | from haplogroup E1a; Phantom mutation 1304. | | 11018, 11071, 11162, 11184, 11189, 11259, 11289, 11366, 11372, 11683, 11786, 11830, 11944, 11973, 12109, 12241, 12298, 12370, 12469, 12518, 12583, 13197C, 13472C, 13745+TCA, 14135A, 14138G, 14163, 14272, 14316, 14336, 14607, 14710, 14947, 14998, 15025, 15109, 15127, 15151, 15179, 15277, 15362, 15625, 15661, 15685, 15721, 15838, 15961, 16111, 16258, 16292, 16294, 16297, 16358, 16456, 16492, 16552, 16564, 16608, 16681, 16690+C | 12958, 12962, 13049, 13433A, 13478, 13581, 13605, 13838, 13997, 14209, 14213, 14313A, 14399, 14561C, 14639, 14738, 14812, 14839 | | |
| AF486858(Pig and wild boar) | Artificial recombination: missing mutations 322, 1168, 1175, 1225, 1304, 1315, 1638T, 1640d of haplogroup E1a1b. | E1a_AF486858 | 109, 131, 136+C, 144, 152, 157, 180, 240, 293, 305, 389, 451, 500, 574, 608-1319d, 1412, 1638d, 1639, 1906, 1990, 2064, 2065C, 2335, 2339, 2613, 3064, 3088, 3102, 3366, 3451, 3594, 3640, 3873T, 3999, 4015, 4030, 4081, 4342, 4369, 4388T, 4420, 4438, 4459, 4471, 4489A, 4711, 4737, 4754, 4846, 4939, 5168T, 5207, 5369, 5463A, 5552, 5628, 5636, 5672, 5678, 5708, 5753, 5797, 5873, 5880, 5948, 5963, 6092, 6138, 6164, 6171, 6296, 6298, 6508, 6925T, 6952, 6970, 7009, 7022, 7108, 7321, 7339, 7447, 7486, 7513, 7669, 7750, 7837, 8017, 8371, 8413, 8498, 8581, 8605, 8713, 8743, 8761, 8773, 9058, 9156, 9157, 9225, 9234, 9435, 9553, 9605, 9752, 9789, 9973, 10070, 10100, 10484, 10529, 10680, 10753, 10816, 10944, 11018, 11071, 11162, 11184, 11189, 11259, 11289, 11366, 11372, 11683, 11786, 11830, 11944, 12061, 12109, 12241, 12298, 12370, 12469, 12518, 12583, 12649, 12675, 12958, 12962, 13049, 13433A, 13478, 13581, 13605, 13838, 13997, 14209, 14213, 14313A, 14399, 14561C, 14639, 14710, 14738, 14812, 14839, 14947, | E1a_AF486858:137C, 142+XA, 322, 1168, 1175, 1225, 1304, 1315, 1638T, 1640d | E1a_AF486858:136+C, 608-1319d, 1638d, 1639, 7447 | Yang,J., Wang,J., Kijas,J., Liu,B., Han,H., Yu,M., Yang,H., Zhao,S.and Li,K. J. Hered. 94 (5), 381-385 (2003) |
| | | | 14998, 15025, 15109, 15127, 15151, 15277, 15362, 15625, 15661, 15685, 15721, 15838, 15961, 16111, 16258, 16292, 16294, 16297, 16358, 16456, 16492, 16552, 16564, 16608 | | | |
| AF486866(Pig and wild boar) | Artificial recombination: missing mutations 1168, 1175, 1225, 1304, 1315, 10517 of haplogroup E1a1b. | E1a1b | 109, 124T, 131, 136+C, 144, 152, 157, 180, 240, 293, 305, 322, 389, 404, 451, 500, 574, 608-1319d, 1412, 1638d, 1639, 1990, 2064, 2065C, 2071, 2335, 2339, 2613, 3064, 3088, 3102, 3366, 3451, 3571, 3594, 3640, 3873T, 3999, 4015, 4030, 4081, 4342, 4369, 4420, 4438, 4459, 4489A, 4711, 4737, 4754, 4846, 4939, 5168T, 5207, 5369, 5463A, 5552, 5628, 5636, 5672, 5678, 5708, 5753, 5797, 5873, 5880, 5948, 5963, 6092, 6138, 6164, 6171, 6296, 6298, 6508, 6817, 6925T, 6952, 6970, 7009, 7022, 7108, 7321, 7339, 7447, 7486, 7513, 7669, 7750, 7837, 8017, 8371, 8413, 8498, 8581, 8605, 8713, 8743, 8761, 8773, 9058, 9156, 9157, 9225, 9234, 9435, 9553, 9605, 9752, 9789, 9973, 10070, 10100, 10484, 10529, 10680, 10753, 10816, 10944, 11018, 11071, 11162, 11184, 11189, 11259, 11289, 11366, 11372, 11683, 11786, 11830, 11944, 12109, 12241, 12298, 12370, 12469, 12518, 12583, 12649, 12675, 12958, 12962, 13049, 13433A, 13478, 13581, 13605, 13838, 13997, 14209, 14213, 14313A, 14399, 14561C, 14639, 14710, 14738, 14812, 14839, 14947, | E1a1b:137C, 142+XA, 1168, 1175, 1225, 1304, 1315, 10517 | E1a1b:136+C, 404, 608-1319d, 1638d, 3571, 6817, 15604 | Yang,J., Wang,J., Kijas,J., Liu,B., Han,H., Yu,M., Yang,H., Zhao,S.and Li,K. J. Hered. 94 (5), 381-385 (2003) |
| | | | 14998, 15025, 15109, 15127, 15151, 15277, 15362, 15604, 15625, 15661, 15685, 15721, 15838, 15961, 16111, 16258, 16292, 16294, 16297, 16358, 16456, 16492, 16552, 16564, 16608 | | | |
| AY574045(Pig and wild boar) | Surplus of mutations; Phantom mutation 1304. | D1a1 | 278, 451, 473, 500, 556d, 783-912d, 962d, 1304, 1315, 1638d, 2065C, 2374, 3163G, 3176G, 3179d, 3192d, 3594, 4342, 4369, 4399, 4420, 4438, 4459, 4489A, 4676C, 4800, 5483, 5702, 6011, 6210, 6344, 7022, 7233, 7252A, 7258, 7260T, 7263T, 7268C, 7341, 7611A, 7618A, 8125A, 8228d, 8234A, 8375A, 8377, 8378, 8380T, 8468C, 8581, 8898, 9122, 9143, 9423T, 10068, 10075, 10100, 10688, 10764T, 10771C, 11366, 11372, 11432, 11999C, 12010C, 12064, 12088C, 12199C, 12273C, 12337C, 12348C, 12428C, 12532C, 12583, 12942d, 12989d, 13840A, 14198, 14710, 14773, 15220, 15337, 15383, 15477, 15595T, 15838, 15907T, 16273, 16301, 16342, 16681 | D1a1:4471 | D1a1:473, 556d, 783-912d, 962d, 1304, 1315, 1638d, 3163G, 3176G, 3179d, 3192d, 4342, 4369, 4399, 4420, 4438, 4459, 4489A, 4676C, 4800, 5483, 5702, 6011, 6210, 6344, 7022, 7233, 7252A, 7258, 7260T, 7263T, 7268C, 7341, 7611A, 7618A, 8125A, 8228d, 8234A, 8375A, 8377, 8378, 8380T, 8468C, 8898, 9122, 9143, 9423T, 10068, 10075, 10100, 10688, 10764T, 10771C, 11366, 11372, 11999C, 12010C, 12088C, 12199C, 12273C, 12337C, 12348C, 12428C, 12532C, 12583, 12942d, 12989d, 13840A, 14773, 15383, 15477, 15595T, 15907T, 16273, 16342, 16681 | Cho,I.C., Park,J.J. and Jeon,J.T. Unpublished |
| AY574046(Pig and wild boar) | Surplus of mutations; Phantom mutation 1304. | E1a1b, E1a1b_JN601066_JN601067 | 28, 43, 109, 124T, 131, 136+C, 144, 152, 157, 180, 240, 293, 305, 322, 389, 451, 500, 574, 701, 703, 705, 772-921d, 949, 1168, 1175, 1225, 1304, 1315, 1412, 1638d, 1639, 1869+C, 1990, 2064, 2065C, 2071, 2335, 2339, 2341C, 2410, 2613, 3064, 3088, 3102, 3120, 3244, 3366, 3451, 3594, 3873T, 3999, 4015, 4030, 4081, 4342, 4369, 4399, 4420, 4438, 4459, 4489A, 4711, 4737, 4754, 4846, 4939, 5168T, 5207, 5369, 5377A, 5389, 5463A, 5470A, 5552, 5628, 5636, 5672, 5678, 5708, 5722T, 5753, 5797, 5873, 5880, 5948, 5963, 6092, 6138, 6164, 6171, 6296, 6298, 6344, 6508, 6814, 6925T, 6952, 6970, 7009, 7022, 7108, 7321, 7339, 7447, 7486, 7497, 7513, 7669, 7750, 7837, 8017, 8371, 8413, 8498, 8581, 8605, 8713, 8743, 8761, 8773, 9058, 9156, 9157, 9225, 9234, 9435, 9496C, 9521, 9524T, 9553, 9605, 9752, 9789, 9973, 10070, 10100, 10281A, 10421, 10434, 10484, 10529, 10646, 10680, 10753, 10816, 10944, 10972, 11018, 11071, 11162, 11184, 11189, 11259, 11289, 11366, 11372, 11683, 11705, 11786, 11823, 11830, 11880, 11944, 12109, 12241, 12298, 12370, 12469, 12518, 12583, 12603, 12649, 12675, 12850, 12958, 12962, 12997, 13049, 13187G, 13433A, 13478, 13581, 13605, 13722, 13838, 13931, 13997, 14149G, 14163G, 14209, 14213, 14259, 14313A, 14399, 14561C, 14639, 14710, 14738, 14739, 14812, 14839, 14947, 14998, 15025, 15109, 15127, 15151, 15277, 15362, 15625, 15661, 15685, 15721, 15838, 15961, 16111, 16217A, 16258, 16292, 16294, 16297, 16358, 16456, 16492, 16552, 16564, 16608, 16681 | E1a1b:137C, 142+XA, 3640, 10517; E1a1b_JN601066_JN601067:137C, 142+XA, 3640, 14477T, 15333 | E1a1b:28, 43, 136+C, 701, 703, 705, 772-921d, 949, 1638d, 1869+C, 2341C, 2410, 3120, 3244, 4399, 5377A, 5389, 5470A, 5722T, 6344, 6814, 7497, 9496C, 9521, 9524T, 10281A, 10421, 10434, 10646, 10972, 11705, 11823, 11880, 12603, 12850, 12997, 13187G, 13722, 13931, 14149G, 14163G, 14259, 14739, 16217A, 16681; E1a1b_JN601066_JN601067:28, 43, 136+C, 701, 703, 705, 772-921d, 1638d, 1869+C, 2341C, 2410, 3120, 3244, 4399, 5377A, 5389, 5470A, 5722T, 6344, 6814, 7497, 9496C, 9521, 9524T, 10281A, 10421, 10434, 10646, 10972, 11705, 11823, 11880, 12603, 12850, 12997, 13187G, 13722, 13931, 14149G, 14163G, 14259, 14739, 16217A, 16681 | Cho,I.C., Park,J.J. and Jeon,J.T. Unpublished |
| AY574047(Pig and wild boar) | Surplus of mutations; Phantom mutation 1304. | D, D3 | 180, 213, 240, 281, 305, 559, 577, 873-912d, 1304, 1315, 1943A, 2058+A, 2065C, 3140d, 3224, 3441+A, 3594, 4275, 4471, 4493, 4568, 4812, 5092A, 5118, 5170, 5369, 6053, 6344, 8088, 8098A, 8139T, 8148, 8152T, 8153, 8154G, 8155, 8157C, 8184d, 8209+T, 8521, 8581, 9352A, 9504, 9644A, 9648A, 9664, 9669G, 9987, 10100, 11366, 11372, 11729, 11848, 12530, 12583, 13047G, 13088G, 13126G, 13131, 13925A, 14552, 14613A, 14710, 15337, 15650, 15838, 16325G, 16681, 16686 | D3:451 | D:180, 213, 240, 281, 305, 559, 577, 873-912d, 1304, 1315, 1943A, 2058+A, 3140d, 3224, 3441+A, 4275, 4493, 4568, 4812, 5092A, 5118, 5170, 5369, 6053, 6344, 8088, 8098A, 8139T, 8148, 8152T, 8153, 8154G, 8155, 8157C, 8184d, 8209+T, 8521, 9352A, 9504, 9644A, 9648A, 9664, 9669G, 9987, 10100, 11366, 11372, 11729, 11848, 12530, 12583, 13047G, 13088G, 13126G, 13131, 13925A, 14552, 14613A, 15650, 16325G, 16681, 16686; D3:180, 213, 281, 305, 559, 577, 873-912d, 1304, 1315, 1943A, 2058+A, 3140d, 3224, 3441+A, 4275, 4493, 4568, 4812, 5092A, 5118, 5170, 5369, 6053, 6344, 8088, 8098A, 8139T, 8148, 8152T, 8153, 8154G, 8155, 8157C, 8184d, 8209+T, 8521, 9352A, 9504, 9644A, 9648A, 9664, 9669G, 9987, 10100, 11366, 11372, 11729, 11848, 12530, 12583, 13047G, 13088G, 13126G, 13131, 13925A, 14552, 14613A, 15650, 16325G, 16681, 16686 | Cho,I.C., Park,J.J. and Jeon,J.T. Unpublished |
| AY574048(Pig and wild boar) | Surplus of mutations; Phantom mutation 1304. | D1b1_JN601075 | 404, 480+T, 500, 664+A, 691, 772-921d, 945+A, 955-956d, 962, 964A, 1107+TAAAACACTTA, 1304, 1315, 2000d, 2065C, 2374, 2755, 3082, 3085T, 3144d, 3153d, 3434, 3590C, 3594, 3660C, 3822A, 3823C, 4291A, 4307, 4342, 4369, 4471, 5982, 6344, 6349-6360d, 6988, 7212, 8182, 8210T, 8267, 8545, 8581, 8875, 9372, 9553, 9605, 9633T, 10100, 10816, 10944, 11366, 11375T, 11432, 11918G, 12064, 12088C, 12229, 12297G, 12556, 13472C, 14361A, 14375, 14381, 14561C, 14710, 14747, 15185, 15337, 15838, 16301, 16552, 16681 | D1b1_JN601075: 1313 | D1b1_JN601075:480+T, 664+A, 691, 772-921d, 945+A, 955-956d, 962, 964A, 1107+TAAAACACTTA, 1304, 1315, 2000d, 2755, 3082, 3085T, 3144d, 3153d, 3590C, 3660C, 3822A, 3823C, 4291A, 4307, 4342, 4369, 5982, 6344, 6349-6360d, 7212, 8182, 8210T, 8875, 9553, 9605, 9633T, 10100, 10816, 10944, 11366, 11375T, 11918G, 12088C, 12229, 12297G, 12556, 14361A, 14375, 14381, 14561C, 14747, 15185, 16681 | Cho,I.C., Park,J.J. and Jeon,J.T. Unpublished |
| DQ207753 (Pig and wild boar) | Surplus of mutations; Phantom mutation | D, D3 | 180, 240, 266, 281, 322, 442, 458, 461, 500, 559, 691, 742-921d, 1304, 1315, 2058+A, 2065C, 2249T, 2864, 3166C, 3594, 3708, 4393, 4471, 4480, 4482, 4483, 4484, 4561, 4742, 4922, 5022, 5092, 5118, 5393, | D3:451 | D:180, 240, 266, 281, 322, 442, 458, 461, 500, 559, 691, 742-921d, 1304, 1315, 2058+A, 2249T, 2864, 3166C, 3708, 4393, 4480, 4482, 4483, 4484, 4561, 4742, 4922, 5022, 5092, 5118, 5393, 5688A, 5845, | Cho,I.C., Han,S.H., Ko,M.S. and Andersson,L. |
| | 1304. | | 5688A, 5845, 6344, 7122, 7171, 8177, 8242G, 8581, 9482A, 9485A, 9490, 10100, 10494, 10954, 11102C, 11104C, 11366, 11372, 11876, 12354A, 12507, 12583, 12957, 14455A, 14482G, 14492, 14536G, 14710, 14983, 15154, 15337, 15526, 15838, 15909, 15998T, 16166G, 16264, 16342 | | 6344, 7122, 7171, 8177, 8242G, 9482A, 9485A, 9490, 10100, 10494, 10954, 11102C, 11104C, 11366, 11372, 11876, 12354A, 12507, 12583, 12957, 14455A, 14482G, 14492, 14536G, 14983, 15154, 15526, 15909, 15998T, 16166G, 16264, 16342; D3:180, 266, 281, 322, 442, 458, 461, 500, 559, 691, 742-921d, 1304, 1315, 2058+A, 2249T, 2864, 3166C, 3708, 4393, 4480, 4482, 4483, 4484, 4561, 4742, 4922, 5022, 5092, 5118, 5393, 5688A, 5845, 6344, 7122, 7171, 8177, 8242G, 9482A, 9485A, 9490, 10100, 10494, 10954, 11102C, 11104C, 11366, 11372, 11876, 12354A, 12507, 12583, 12957, 14455A, 14482G, 14492, 14536G, 14983, 15154, 15526, 15909, 15998T, 16166G, 16264, 16342 | Unpublished |
| DQ207754 (Pig and wild boar) | Surplus of mutations; Phantom mutation 1304. | D, D3 | 180, 240, 266, 281, 322, 442, 458, 461, 500, 559, 691, 705-855d, 1304, 1315, 1348, 2065C, 2864, 2945C, 3594, 4266, 4393, 4471, 4480, 4575T, 4612N, 4646C, 4742, 5092, 5118, 5143G, 5154, 5393, 5797A, 5828A, 6344, 7171, 8261T, 8581, 9715, 9889C, 10100, 10494, 10736C, 10809T, 10826, 11366, 11372, 11876, 11944, 12105G, 12145T, 12507, 12572, 12583, 12957, 13496T, 13659A, 13781T, 14566T, 14710, 15064, 15154, 15321C, 15337, 15625, 15838, 16264, 16342 | D3:451 | D:180, 240, 266, 281, 322, 442, 458, 461, 500, 559, 691, 705-855d, 1304, 1315, 1348, 2864, 2945C, 4266, 4393, 4480, 4575T, 4612N, 4646C, 4742, 5092, 5118, 5143G, 5154, 5393, 5797A, 5828A, 6344, 7171, 8261T, 9715, 9889C, 10100, 10494, 10736C, 10809T, 10826, 11366, 11372, 11876, 11944, 12105G, 12145T, 12507, 12572, 12583, 12957, 13496T, 13659A, 13781T, 14566T, 15064, 15154, 15321C, 15625, 16264, 16342; D3:180, 266, 281, 322, 442, 458, 461, 500, 559, 691, 705-855d, 1304, 1315, 1348, 2864, 2945C, 4266, 4393, 4480, 4575T, 4612N, 4646C, 4742, 5092, 5118, 5143G, 5154, 5393, 5797A, 5828A, 6344, 7171, 8261T, 9715, 9889C, 10100, 10494, 10736C, 10809T, 10826, 11366, 11372, 11876, 11944, 12105G, 12145T, 12507, 12572, 12583, 12957, 13496T, 13659A, 13781T, 14566T, 15064, 15154, 15321C, 15625, 16264, 16342 | Cho,I.C., Han,S.H., Ko,M.S. and Andersson,L. Unpublished |
| DQ207755 (Pig and wild boar) | Surplus of mutations; Phantom mutation 1304. | D, D3 | 123d, 171, 180, 213, 240, 281, 305, 559, 663C, 691, 853-912d, 977d, 1024, 1304, 1315, 1390, 2045A, 2065C, 2089+A, 2704T, 3171C, 3594, 4471, 4561, 5092, 5118, 5154, 5369, 5688A, 6344, 8581, 9352A, 9433, 9485A, 9490, 9504, 10100, 11147, 11366, 11372, 11408, 11417, 11848, 12134T, 12298A, 12530, 12583, 14536G, 14552, 14553C, 14710, 15337, 15650 | D:15838; D3:451, 15838 | D:123d, 171, 180, 213, 240, 281, 305, 559, 663C, 691, 853-912d, 977d, 1024, 1304, 1315, 1390, 2045A, 2089+A, 2704T, 3171C, 4561, 5092, 5118, 5154, 5369, 5688A, 6344, 9352A, 9433, 9485A, 9490, 9504, 10100, 11147, 11366, 11372, 11408, 11417, 11848, 12134T, 12298A, 12530, 12583, 14536G, 14552, 14553C, 15650; D3:123d, 171, 180, 213, 281, 305, 559, 663C, 691, 853-912d, 977d, 1024, 1304, 1315, 1390, 2045A, 2089+A, 2704T, 3171C, 4561, 5092, 5118, 5154, 5369, 5688A, 6344, 9352A, 9433, 9485A, 9490, 9504, 10100, 11147, 11366, 11372, 11408, 11417, 11848, 12134T, 12298A, 12530, 12583, 14536G, 14552, 14553C, 15650 | Cho,I.C., Han,S.H., Ko,M.S. and Andersson,L. Unpublished |
| DQ268530 (Pig and wild boar) | Surplus of mutations; Phantom mutation 1304. | D, D3 | 180, 213, 240, 281, 305, 559, 752-921d, 1131, 1304, 1315, 1390+C, 1394T, 1395, 1399A, 1400+AAATGGCC, 1409+A, 1873+A, 1939C, 1964-1966d, 1968, 1972d, 2000d, 2033, 2034, 2037T, 2038G, 2042G, 2065C, 2197d, 2383A, 2384T, 2385A, 2386C, 3082d, 3156, 3171C, 3176+C, 3441+A, 3594, 3834d, 3943, 4471, 4532, 4561, 4721, 5022, 5092, 5118, 5369, 5688A, 5733C, 5774T, 5780T, 5787A, 5788C, 5789A, 5796T, 5798T, 6344, 8248, 8305G, 8377, 8581, 8965, 9352A, 9406, 9504, 10100, 10801, 10813T, 11001, 11366, 11372, 11681A, 11682, 11685C, 11686A, 11702C, 11703A, 11704C, 11706A, 11772G, 11848, 12134T, 12298A, 12583, 12797T, 14552, 14710, 15337, 15650, 15959, 16370, 16427 | D:15838; D3:451, 15838 | D:180, 213, 240, 281, 305, 559, 752-921d, 1131, 1304, 1315, 1390+C, 1394T, 1395, 1399A, 1400+AAATGGCC, 1409+A, 1873+A, 1939C, 1964-1966d, 1968, 1972d, 2000d, 2033, 2034, 2037T, 2038G, 2042G, 2197d, 2383A, 2384T, 2385A, 2386C, 3082d, 3156, 3171C, 3176+C, 3441+A, 3834d, 3943, 4532, 4561, 4721, 5022, 5092, 5118, 5369, 5688A, 5733C, 5774T, 5780T, 5787A, 5788C, 5789A, 5796T, 5798T, 6344, 8248, 8305G, 8377, 8965, 9352A, 9406, 9504, 10100, 10801, 10813T, 11001, 11366, 11372, 11681A, 11682, 11685C, 11686A, 11702C, 11703A, 11704C, 11706A, 11772G, 11848, 12134T, 12298A, 12583, 12797T, 14552, 15650, 15959, 16370, 16427; D3:180, 213, 281, 305, 559, 752-921d, 1131, 1304, 1315, 1390+C, 1394T, 1395, 1399A, 1400+AAATGGCC, 1409+A, 1873+A, 1939C, 1964-1966d, 1968, 1972d, 2000d, 2033, 2034, 2037T, 2038G, 2042G, 2197d, 2383A, 2384T, 2385A, 2386C, 3082d, 3156, 3171C, 3176+C, | Cho,I.C., Han,S.H., Ko,M.S. and Andersson,L. Unpublished |
| | | | | | 3441+A, 3834d, 3943, 4532, 4561, 4721, 5022, 5092, 5118, 5369, 5688A, 5733C, 5774T, 5780T, 5787A, 5788C, 5789A, 5796T, 5798T, 6344, 8248, 8305G, 8377, 8965, 9352A, 9406, 9504, 10100, 10801, 10813T, 11001, 11366, 11372, 11681A, 11682, 11685C, 11686A, 11702C, 11703A, 11704C, 11706A, 11772G, 11848, 12134T, 12298A, 12583, 12797T, 14552, 15650, 15959, 16370, 16427 | |
| DQ274110 (Pig and wild boar) | Surplus of mutations; Phantom mutation 1304. | D1a1 | 278, 302A, 451, 500, 546A, 559, 684A, 691C, 803-912d, 1304, 1307d, 1315, 2065C, 2374, 2382+A, 2390+A, 2952C, 2953A, 3140d, 3194T, 3195, 3224, 3241C, 3441+A, 3594, 4471, 4478N, 4484, 4511, 4561, 4737, 5022, 5089+CA, 5118, 5430T, 5745, 6275+T, 6344, 6823G, 6972A, 6982, 7003A, 7824, 8148d, 8160+C, 8209+T, 8581, 9316, 9485, 9556T, 9557T, 9560A, 9562C, 9727C, 9729A, 10100, 10504, 10857G, 10887, 11001, 11035A, 11281G, 11285C, 11294C, 11366, 11372, 11432, 11681A, 11682, 12064, 12190G, 12296C, 12298, 12583, 13003C, 13339C, 13414G, 13781T, 14198, 14209, 14710, 15220, 15337, 15838, 16301 | | D1a1:302A, 546A, 559, 684A, 691C, 803-912d, 1304, 1307d, 1315, 2382+A, 2390+A, 2952C, 2953A, 3140d, 3194T, 3195, 3224, 3241C, 3441+A, 4478N, 4484, 4511, 4561, 4737, 5022, 5089+CA, 5118, 5430T, 5745, 6275+T, 6344, 6823G, 6972A, 6982, 7003A, 7824, 8148d, 8160+C, 8209+T, 9316, 9485, 9556T, 9557T, 9560A, 9562C, 9727C, 9729A, 10100, 10504, 10857G, 10887, 11001, 11035A, 11281G, 11285C, 11294C, 11366, 11372, 11681A, 11682, 12190G, 12296C, 12298, 12583, 13003C, 13339C, 13414G, 13781T, 14209 | Cho,I.C., Han,S.H., Ko,M.S. and Andersson,L. Unpublished |
| DQ334860 (Pig and wild boar) | Surplus of mutations; Phantom mutation 1304. | D1a1 | 278, 451, 500, 557T, 1304, 1315, 2015+G, 2051, 2065C, 2249T, 2372G, 2374, 3084, 3085T, 3594, 3882T, 4471, 5092, 5118, 5133G, 5135C, 5146A, 5430T, 5688A, 5733C, 5774T, 5780T, 5787A, 5788C, 5789A, 5796T, 5798T, 6251T, 6252G, 6255T, 6256G, 6262T, 6263G, 6268, 6344, 8581, 9451, 9541, 10100, 10393, 10504, 10884, 11366, 11372, 11432, 12064, 12733, 12739, | | D1a1:557T, 1304, 1315, 2015+G, 2051, 2249T, 2372G, 3084, 3085T, 3882T, 5092, 5118, 5133G, 5135C, 5146A, 5430T, 5688A, 5733C, 5774T, 5780T, 5787A, 5788C, 5789A, 5796T, 5798T, 6251T, 6252G, 6255T, 6256G, 6262T, 6263G, 6268, 6344, 9451, 9541, 10100, 10393, 10504, 10884, 11366, 11372, 12733, 12739, 12742A, 13193, 13718T, 13769, | Han,S.-H., Ko,M.-S. and Cho,I.-C. Unpublished |
| | | | 12742A, 13193, 13718T, 13769, 13770, 13781T, 14198, 14710, 15220, 15337, 15838, 16301, 16448, 16681 | | 13770, 13781T, 16448, 16681 | |
| DQ334861 (Pig and wild boar) | Surplus of mutations; Phantom mutation 1304. | D1a1 | 268, 278, 323d, 451, 456, 500, 614+T, 631+T, 956, 1068A, 1069C, 1070A, 1073T, 1218, 1304, 1315, 1458A, 2015+G, 2051, 2065C, 2096d, 2270d, 2374, 2529, 2531, 2533C, 2535, 2539G, 2540C, 2541, 2553d, 2629+C, 2692+C, 3144d, 3594, 4132G, 4187A, 4471, 4484, 5092, 5118, 5430T, 5745, 5831, 5961T, 5964T, 6344, 6655, 6968, 8148, 8157C, 8581, 9322, 9397C, 9457, 10100, 10504, 10918, 11130, 11366, 11372, 11432, 11612, 11729, 12064, 12090C, 12583, 12733, 12739, 12742A, 13718T, 13779C, 14198, 14404G, 14536G, 14622G, 14710, 15220, 15337, 15838, 16301, 16681 | | D1a1:268, 323d, 456, 614+T, 631+T, 956, 1068A, 1069C, 1070A, 1073T, 1218, 1304, 1315, 1458A, 2015+G, 2051, 2096d, 2270d, 2529, 2531, 2533C, 2535, 2539G, 2540C, 2541, 2553d, 2629+C, 2692+C, 3144d, 4132G, 4187A, 4484, 5092, 5118, 5430T, 5745, 5831, 5961T, 5964T, 6344, 6655, 6968, 8148, 8157C, 9322, 9397C, 9457, 10100, 10504, 10918, 11130, 11366, 11372, 11612, 11729, 12090C, 12583, 12733, 12739, 12742A, 13718T, 13779C, 14404G, 14536G, 14622G, 16681 | Han,S.-H., Ko,M.-S. and Cho,I.-C. Unpublished |
| DQ466081 (Pig and wild boar) | Surplus of mutations. | D1a, D1c | 240, 442, 500, 691, 703, 713, 753, 903-912d, 1175, 1638d, 1670C, 1692T, 1693A, 1764, 1783d, 1897, 2065C, 2161, 2374, 3594, 3810T, 4471, 5105, 5599, 5765T, 7185, 8581, 9131, 10995, 11043, 11171, 11249, 11267T, 11271, 11432, 12061, 12064, 14198, 14290, 14618, 14710, 15337, 15601, 15838, 16301 | | D1a:240, 442, 691, 703, 713, 753, 903-912d, 1175, 1638d, 1670C, 1692T, 1693A, 1764, 1783d, 1897, 2161, 3810T, 5105, 5599, 5765T, 7185, 9131, 10995, 11043, 11171, 11249, 11267T, 11271, 12061, 14290, 14618, 15601; D1c:240, 691, 703, 713, 753, 903-912d, 1175, 1638d, 1670C, 1692T, 1693A, 1764, 1783d, 1897, 2161, 3810T, 5105, 5599, 5765T, 7185, 9131, 10995, 11043, 11171, 11249, 11267T, 11271, 12061, 14198, 14290, 14618, 15601 | Wang,J.F., Li,S. and Ran,X.Q. Direct Submission |
| DQ518915 (Pig and wild boar) | Surplus of mutations. The strand issue of *ND6* gene has been corrected. | A1c | 90, 180, 240, 278, 300, 301, 305, 322, 390, 451, 500, 534, 541, 574, 656, 947G, 949, 950T, 965T, 978A, 986, 992C, 1006A, 1032G, 1059C, 1081+CAAACCACACAAACCACACA, 1175, 1315, 1412, 1753, 1754, 1755-1756d, 1757, 1758C, 1764A, 1790, 1791d, 1797d, 1805d, 1817C, 1834, 1906, 2065C, 2339, 2514+A, 2726, 3013d, 3102, 3594, 3917,3920d, 4015, 4459, 4471, 4676, 4676+TGC, 4939, 5021d, 5405, 5463A, 5628, 5672, 5708, 5753, 5797, 6296, 6582, 6789, 6925, 6952, 6970, 7339, 7465, 7513, 7540, 7669, 7837, 8017, 8369C, 8398, 8581, 8668, 9052, 9156, 9309, 9414, 9435, 9480, 9699, 9727, 9822, 10070, 10100, 10205, 10457, 10582A, 10586, 10680, 11174, 11184, 11259, 11279, 11341, 11411, 11735, 11864A, 12064, 12649, 12787, 12909, 13049, 13159, 13175, 13227A, 13460, 13781, 14213, 14369, 14603, 14676, 14710, 14818A, 14819A, 14819+T, 15220,15337, 15499, 15502T, 15685, 15700, 15721, 15838, 16036, 16111, 16258, 16292, 16400 | A1c:1087+2ACAAACCAC, 11944, 14540, 14541, 14639, 15025 | A1c:947G, 949, 965T, 978A, 986, 992C, 1006A, 1032G, 1059C, 1081+CAAACCACACAAACCACACA, 1753, 1754, 1755-1756d, 1757, 1758C, 1764A, 1790, 1791d, 1797d, 1805d, 1817C, 3013d, 3917,3920d, 4676, 4676+TGC, 5021d, 6789, 8369C, 10582A, 10586, 11864A, 12064, 14603, 14676, 14818A, 14819+T,15220,15499,15502T | Chen,C.H., Huang,H.-L., Yang,H.-Y., Lai,S.-H., Yen,N.-T., Wu,M.-C. and Huang,M.C. Afr. J. Biotechnol. 10 (13), 2556-2561 (2011) |
| DQ534707 (Pig and wild boar) | Surplus of mutations. | D1a1a | 29A, 30, 32, 172G, 240, 278, 451, 500, 691, 921+TGCGTACACGTGCGTACACGTGCGTACACGTGCGTACACG, 925, 930C, 933, 934C, 944A, 945, 946T, 954T, 1022d, 1174, 1196C, 1318T, 1326C, 1486, 1609, 1641C, 1642+C, 1654, 1664, 1666G, 1681T, 1682G, 1688, 1691, 1692T, 1693A, 1695T, 1696+T, 1714T, 1715-1716d, 1728G, 1742d, 2065C, 2374, 2445, 2446, 2447A, 2448A, 2451G, 2591, 3488, 3594, 4471, 5847, 6097A, 6799, 6998C, 8291A, 8292, 8317, 8319, 8581, 8767C, 9014, 9388C, 9941A, 11366T, 11432, 11856, 11944, 14008T, 14057A, 14107A, 14198, 14476, 14710, 15025, 15337, 15838, 15961G, 16301 | D1a1a:12064, 15220 | D1a1a:29A, 30, 32, 172G, 691, 921+TGCGTACACGTGCGTACACGTGCGTACACGTGCGTACACG, 925, 930C, 933, 934C, 944A, 945, 946T, 954T, 1022d, 1174, 1196C, 1318T, 1326C, 1486, 1609, 1641C, 1642+C, 1654, 1664, 1666G, 1681T, 1682G, 1688, 1691, 1692T, 1693A, 1695T, 1696+T, 1714T, 1715-1716d, 1728G, 1742d, 2445, 2446, 2447A, 2448A, 2451G, 2591, 3488, 5847, 6097A, 6799, 6998C, 8291A, 8292, 8317, 8319, 8767C, 9014, 9388C, 9941A, 11366T, 11856, 11944, 14008T, 14057A, 14107A, 14476, 15025, 15961G | Chen,C.H., Yang,H.Y., Yen,N.T. and Huang,M.C. Direct Submission |
| DQ972936 (Pig and | Surplus of mutations. | D1b1 | 240, 404, 500, 691, 712-921d, 1107+TAAAACACTTA, 2065C, 2374, 3434, 3594, 4471, 6604G, 6660G, 8267, | | D1b1:240, 691, 712-921d, 1107+TAAAACACTTA, 6604G, 6660G, 8596, 9785, 11073C, 11506G, | Wu,C.Y., Jiang,Y.N., |
| wild boar) | | | 8581, 8596, 9372, 9785, 11073C, 11432, 11506G, 11575G, 11613, 11977, 12064, 12850, 13472C, 14105+A, 14124+T, 14129+G, 14710, 15337, 15838, 16301 | | 11575G, 11613, 11977, 12850, 14105+A, 14124+T, 14129+G | Chu,H.P., Li,S.H., Wang,Y., Li,Y.H., Chang,Y. and Ju,Y.T. Anim. Genet. 38 (5), 499-505 (2007) |
| EU090702(Pig and wild boar) | Surplus of mutations; Phantom mutation 1304. | D, D3 | 180, 240, 281, 305, 322, 442, 461, 500, 559, 691, 921+TGCGTACACGTGCGTACACGTGCGTACACG, 1187, 1304, 1315, 1577G, 1841C, 1861C, 1906, 1912C, 1914A, 1915C, 1920, 2030T, 2042, 2051, 2065C, 2089+A, 2362G, 2383G, 2864, 3113A, 3170C, 3173A, 3594, 3792G, 4331, 4375C, 4393, 4471, 4480, 4502, 4584, 4742, 5022, 5092, 5118, 5393, 6344, 6411, 7002, 7053, 7171, 7954C, 8080T, 8264, 8581, 8639, 8812, 9457, 9502, 9565, 9614G, 9727, 9890C, 10100, 10494, 10740A, 11366, 11372, 12035T, 12134T, 12185G, 12325G, 12331G, 12332C, 12507, 12532, 12572, 12583, 12632C, 12957, 13159, 13241, 13245G, 14275G, 14334T, 14386G, 14388T, 14409G, 14417G, 14589, 14710, 15154, 15337, 15838, 16055, 16264, 16342, 16681, 16686, 16690+CAGCACCCAAAGCTGAAATTCTAACTAAACTATTCCCTG | D3:451 | D:180, 240, 281, 305, 322, 442, 461, 500, 559, 691, 921+TGCGTACACGTGCGTACACGTGCGTACACG, 1187, 1304, 1315, 1577G, 1841C, 1861C, 1906, 1912C, 1914A, 1915C, 1920, 2030T, 2042, 2051, 2089+A, 2362G, 2383G, 2864, 3113A, 3170C, 3173A, 3792G, 4331, 4375C, 4393, 4480, 4502, 4584, 4742, 5022, 5092, 5118, 5393, 6344, 6411, 7002, 7053, 7171, 7954C, 8080T, 8264, 8639, 8812, 9457, 9502, 9565, 9614G, 9727, 9890C, 10100, 10494, 10740A, 11366, 11372, 12035T, 12134T, 12185G, 12325G, 12331G, 12332C, 12507, 12532, 12572, 12583, 12632C, 12957, 13159, 13241, 13245G, 14275G, 14334T, 14386G, 14388T, 14409G, 14417G, 14589, 15154, 16055, 16264, 16342, 16681, 16686, 16690+CAGCACCCAAAGCTGAAATTCTAACTAAACTATTCCCTG; D3:180, 281, 305, 322, 442, 461, 500, 559, 691, 921+TGCGTACACGTGCGTACACGTGCGTACACG, 1187, 1304, 1315, 1577G, 1841C, 1861C, 1906, 1912C, 1914A, 1915C, 1920, 2030T, 2042, 2051, 2089+A, 2362G, 2383G, 2864, 3113A, 3170C, 3173A, 3792G, 4331, 4375C, 4393, 4480, 4502, 4584, 4742, 5022, 5092, 5118, 5393, 6344, 6411, 7002, 7053, 7171, 7954C, 8080T, 8264, 8639, 8812, | Cho,I.C., Han,S.H., Lee,S.S., Ko,M.S., Lee,J.G. and Jeon,J.T. Unpublished |
| | | | | | 9457, 9502, 9565, 9614G, 9727, 9890C, 10100, 10494, 10740A, 11366, 11372, 12035T, 12134T, 12185G, 12325G, 12331G, 12332C, 12507, 12532, 12572, 12583, 12632C, 12957, 13159, 13241, 13245G, 14275G, 14334T, 14386G, 14388T, 14409G, 14417G, 14589, 15154, 16055, 16264, 16342, 16681, 16686, 16690+CAGCACCCAAAGCTGAAATTCTAACTAAACTATTCCCTG | |
| EU090703(Pig and wild boar) | Surplus of mutations; Phantom mutation 1304. | D1e'g'h | 67, 240, 266, 347A, 451, 500, 559, 873-912d, 1304, 1315, 1463C, 1576C, 1600C, 1868A, 1888T, 1891+ATC, 2065C, 2089+A, 2374, 2554, 2961C, 2992, 2999N, 3006C, 3075C, 3077C, 3162C, 3292, 3594, 4152, 4278, 4442T, 4471, 5022, 5092, 5118, 6077, 6344, 6829A, 6833G, 6842, 6850G, 6855, 6868, 6899G, 7013G, 7024, 7050T, 7051G, 7531, 8270, 8462T, 8581, 8853, 9193, 9559C, 9715, 9719G, 10100, 10583, 11366, 11372, 11432, 12064, 13202, 13742C, 14116, 14272, 14300G, 14428A, 14710, 14883, 15337, 15838, 16025, 16301, 16681, 16690+CAGCACCCAAAGCTGAAATTCTAACTAAACTATTCCCTG | | D1e'g'h:67, 266, 347A, 451, 873-912d, 1304, 1315, 1463C, 1576C, 1600C, 1868A, 1888T, 1891+ATC, 2089+A, 2554, 2961C, 2992, 2999N, 3006C, 3075C, 3077C, 3162C, 3292, 4152, 4278, 4442T, 5022, 5092, 5118, 6077, 6344, 6829A, 6833G, 6842, 6850G, 6855, 6868, 6899G, 7013G, 7024, 7050T, 7051G, 7531, 8270, 8462T, 8853, 9193, 9559C, 9715, 9719G, 10100, 10583, 11366, 11372, 13202, 13742C, 14116, 14272, 14300G, 14428A, 14883, 16025, 16681, 16690+CAGCACCCAAAGCTGAAATTCTAACTAAACTATTCCCTG | Cho,I.C., Han,S.H., Lee,S.S., Ko,M.S., Lee,J.G. and Jeon,J.T. Unpublished |
| EU333163 (Pig and wild boar) | Artificial recombination: missing mutations 2374, 11432 of haplogroup D1; mis-added mutations 109, 124T, 131, 137C, 142+A, 144, 152, 157, 1168, 1412, 2064, 2071, 2339, 2613, 4420, 4438, 4459, 4489A, 4737, 4754, 4939, 5168T, 5207, 5463A, 5552, 5628, 5636, 5672, 5678, 5708, 5753, 5880, 5948, 5963, 6138, 6171, 6296, 6298, 6508, 7486, 7750, 7837, 8017, 8371, 8413, 8605, 8743, 8761, 9058, 9156, 9157, 9225, | D1f | 109, 124T, 131, 136+C, 144, 152, 157, 240, 281, 299, 500, 803-912d, 977d, 1168, 1175, 1225, 1304, 1315, 1412, 1639, 1959+C, 1990, 2064, 2065C, 2071, 2335, 2339, 2499d, 2591, 2613, 3287, 3437+C, 3594, 4342, 4369, 4420, 4438, 4459, 4471, 4489A, 4711, 4737, 4754, 4846, 4939, 5168T, 5207, 5369, 5463A, 5474, 5552, 5628, 5636, 5672, 5678, 5708, 5753, 5797, 5873, 5880, 5948, 5963, 6092, 6138, 6164, 6171, 6296, 6298, 6344, 6508, 6658G, 6944, 7313, 7393G, 7447, 7486, 7613, 7624, 7627, 7750, 7837, 8010, 8017, 8371, 8413, 8476C, 8498, 8581, 8605, 8713, 8743, 8761, 8773, 9058, 9156, 9157, 9225, 9234, 10100, 10354, 10753, 10816, 10944, 11018, 11071, 11162, 11184, 11189, 11259, 11289, 11366, 11372, 11683, 11786, 11830, 12064, 12620G, 12713C, 13202, 14710, 15337, 15838, 16301 | D1f:2374, 11432 | D1f:109, 124T, 131, 136+C, 144, 152, 157, 281, 803-912d, 977d, 1168, 1175, 1225, 1304, 1315, 1412, 1639, 1959+C, 1990, 2064, 2071, 2335, 2339, 2499d, 2591, 2613, 3287, 3437+C, 4342, 4369, 4420, 4438, 4459, 4489A, 4711, 4737, 4754, 4846, 4939, 5168T, 5207, 5369, 5463A, 5474, 5552, 5628, 5636, 5672, 5678, 5708, 5753, 5797, 5873, 5880, 5948, 5963, 6092, 6138, 6164, 6171, 6296, 6298, 6344, 6508, 6658G, 6944, 7313, 7393G, 7447, 7486, 7613, 7624, 7627, 7750, 7837, 8010, 8017, 8371, 8413, 8476C, 8498, 8605, 8713, 8743, 8761, 8773, 9058, 9156, 9157, 9225, 9234, 10100, 10753, 10816, 10944, 11018, 11071, 11162, 11184, 11189, 11259, 11289, 11366, 11372, 11683, 11786, 11830, 12620G, 12713C | Yu,H., Li,L. and Liu,D. Unpublished |
| | 9234, 10100, 10753, 11081, 11162, 11189, 11289, 11372, 11683, 11786, 11830 of haplogroup E1a. | | | | | |
| GQ351599 (Pig and wild boar) | Artificial recombination: mis-added mutations 109, 124T, 131, 137C, 142+A, 144, 152, 157, 180, 293, 305, 322, 389, 500 of haplogroup E1a. | D3_EF545572 | 109, 124T, 131, 136+C, 144, 152, 157, 180, 240, 293, 305, 322, 389, 451, 500, 559, 691, 911-978d, 2065C, 3064Y, 3594, 4471, 4939Y, 5435, 5552R, 6841, 7309, 7540, 8581, 8686, 9752R, 9789Y, 10030Y, 11018Y, 11071R, 11259Y, 11432, 11539, 11552, 11830, 11944, 12109, 12649, 12675, 12719S, 12958, 12962, 13377S, 13433A, 13478, 14209, 14213, 14313A, 14399, 14561C, 14639R, 14710, 15109, 15127, 15151, 15277, 15362, 15420+AT, 15634, 15838, 15850, 16301, 16468, 16564, 16608, 16681 | D3_EF545572:278, 1311, 5393, 7468, 7615, 8533, 15154, 15337 | D3_EF545572:109, 124T, 131, 136+C, 144, 152, 157, 180, 293, 305, 322, 389, 500, 559, 691, 911-978d, 3064, 4939, 5552, 6841, 9752, 9789, 11018, 11071, 11259, 11432, 11830, 11944, 12109, 12649, 12675, 12719G, 12958, 12962, 13377G, 13433A, 13478, 14209, 14213, 14313A, 14399, 14561C, 14639, 15109, 15127, 15151, 15277, 15362, 15420+AT, 16301, 16564, 16608, 16681 | Chung,H.Y. Unpublished |
| KC250273(Pig and wild boar) | Artificial recombination: mis-added mutations 9553, 9605 from haplogroup E1a. | D1e | 240, 300, 500, 559, 691, 921+TGCGTACACGTGCGTACACG, 1638d, 2065C, 2374, 3594, 4471, 4842, 8135, 8581, 9553, 9605, 9917T, 11014A, 11432, 12064, 14710, 15337, 15838, 16301 | | D1e:691, 921+TGCGTACACGTGCGTACACG, 1638d, 4842, 9553, 9605, 9917T | Yu,G., Xiang,H., Wang,J. and Zhao,X. J Anim Sci Biotechnol 4 (1), 9 (2013) |
| KC469586(Pig and wild boar) | Artificial recombination: missing mutations 2374, 15220 of haplogroup D1a1; mis-added mutations 2613, 5672, 5678, 5708, 5753, 8017, 9156, 9157, 9225, 9234, 16111 from haplogroup E. | D1a1a | 240, 278, 451, 500, 833-912d, 1254+A, 1638d, 2065C, 2369d, 2613, 3594, 4471, 5672, 5678, 5708, 5753, 8017, 8581, 9156, 9157, 9225, 9234, 10616, 11432, 12064, 14198, 14710, 14812, 14839, 15337, 15838, 16111, 16301, 16690+G | D1a1a:2374, 15220 | D1a1a:833-912d, 1254+A, 1638d, 2369d, 2613, 5672, 5678, 5708, 5753, 8017, 9156, 9157, 9225, 9234, 10616, 14812, 14839, 16111, 16690+G | Yu,G., Xiang,H., Wang,J. and Zhao,X. J Anim Sci Biotechnol 4 (1), 9 (2013) |
| AY858379(Sheep) | Surplus of mutations | A2,A2_HM236175 | 281Y, 291Y, 566+G, 1099W, 1111M, 1261M, 1264S, 1277M, 1285S, 1288R, 1289R, 1290K, 1729+C, 2774Y, 2966Y, 3218R, 3431R, 3543W, 3662R, 6726R, 7216Y, 7435Y, 7500M, 7522R, 7983Y, 8039R, 8121Y, 8264C, 8376, 8651Y, 9284R, 9375R, 9756Y, 9996Y, 10549R, 10714R, 10852R, 11317R, 11668R, 11846Y, 12287Y, 12539S, 12571S, 13172Y, 13199R, 13576Y, 13813S, 14055Y, 14467Y, 14653R, 15459Y, 15484R, 15547R, 15583Y, 15594R, 15597Y, 15635R, 15638Y, 15639R, 15642+T, 15645R, 15656R, 15659K, 15708Y, 15710W, 15714R, 15716M, 15721Y, 15722M, 15731R, 15734K, 15789R, 15797M, 15800Y, 15806R, 15809K, 15818W, 15820Y, 15833W, 15858Y, 15872M, 15881R, 15882W, 15905W, 15920K, 15928Y, 15932S, 15944R, 15956Y, 15957R, 15958Y, 15978W, 15982R, 15986K, 15996W, 16003K, 16011R, 16038R, 16128Y, 16209Y, 16342+Y, 16343 | A2:538, 1112, 2443, 4182, 4215, 4443, 4839, 4915, 4935, 5565, 5784, 6267, 6510, 6555, 6615, 6628, 7141, 7719G, 8148, 8256, 9128, 9188, 10118, 10783, 11023, 11482, 11491, 11606, 11710d, 11783, 11834, 12023, 13097, 13436, 13837, | A2:281Y, 291Y, 1099W, 1111M, 1261M, 1264S, 1277M, 1285S, 1288R, 1289R, 1290K, 2774Y, 2966Y, 3218R, 3431R, 3543W, 3662R, 6726R, 7216Y, 7435Y, 7500M, 7522R, 7983Y, 8039R, 8121Y, 8651Y, 9284R, 9375R, 9756Y, 9996Y, 10549R, 10714R, 10852R, 11317R, 11668R, 11846Y, 12287Y, 12539S, 12571S, 13172Y, 13199R, 13576Y, 13813S, 14055Y, 14467Y, 14653R, 15459Y, 15484R, 15547R, 15583Y, 15594R, 15597Y, 15635R, 15638Y, 15639R, 15645R, 15656R, 15659K, 15708Y, 15710W, 15714R, 15716M, 15721Y, 15722M, 15731R, 15734K, 15789R, 15797M, 15800Y, 15806R, 15809K, 15818W, 15820Y, 15833W, 15858Y, 15872M, 15881R, 15882W, 15905W, 15920K, 15928Y, 15932S, 15944R, 15956Y, 15957R, 15958Y, 15978W, 15982R, 15986K, 15996W, 16003K, 16011R, 16038R, 16128Y, 16209Y, 16342+Y; A2_HM236175:281Y, 291Y, 1099W, 1111M, 1261M, | Ha,J.M. and Chung,H.Y. Unpublished. |

| | | | | | | | |
|---|---|---|---|---|---|---|---|
| | | | | 13855, 15864, 15939, 15972, 15978, 16020, 16022, 16036, 16042, 16048, 16096, 16097, 16217, 16342+C, 16440, 16453, 16472d, 16602; A2_HM236175:538, 1112, 2443, 4182, 4215, 4443, 4839, 4915, 4935, 5565, 5784, 6267, 6510, 6555, 6615, 6628, 7141, 7719G, 8148, 8256, 9128, 9188, 10118, 10783, 11023, 11482, 11491, 11606, 11710d, 11783, 11834, 12023, 13097, 13436, 13837, 13855, 15864, 15939, 15972, 15978, 16020, 16022, 16036, 16042, 16048, 16096, 16097, 16217, | 1264S, 1277M, 1285S, 1288R, 1289R, 1290K, 2774Y, 2966Y, 3218R, 3431R, 3543W, 3662R, 6726R, 7216Y, 7435Y, 7500M, 7522R, 7983Y, 8039R, 8121Y, 8651Y, 9284R, 9375R, 9756Y, 9996Y, 10549R, 10714R, 10852R, 11317R, 11668R, 11846Y, 12287Y, 12539S, 12571S, 13172Y, 13199R, 13576Y, 13813S, 14055Y, 14467Y, 14653R, 15459Y, 15484R, 15547R, 15583Y, 15594R, 15597Y, 15635R, 15638Y, 15639R, 15645R, 15656R, 15659K, 15708Y, 15710W, 15714R, 15716M, 15721Y, 15722M, 15731R, 15734K, 15789R, 15797M, 15800Y, 15806R, 15809K, 15818W, 15820Y, 15833W, 15858Y, 15872M, 15881R, 15882W, 15905W, 15920K, 15928Y, 15932S, 15944R, 15956Y, 15957R, 15958Y, 15978W, 15982R, 15986K, 15996W, 16003K, 16011R, 16038R, 16128Y, 16209Y, 16342+Y | |
| | | | | 16342+C, 16440, 16453, 16472d, 16602 | | |
| KF312238(Sheep) | Surplus of mutations | CE | 538, 566+G, 708-709d, 712+G, 713, 729C, 737, 744, 770, 787T, 788G, 1112, 1247, 1563, 1729+C, 1788, 2199, 2621, 2774, 2867, 3023, 3041G, 3218T, 3338, 3543A, 3662, 3930, 4008, 4182, 4208, 4428, 4443, 4524, 4839, 4915, 4935, 5137, 5147, 5763, 5784, 5850A, 5901, 5922, 5979, 6267, 6423, 6555, 6615, 6628, 6726, 6738, 6834, 7000, 7096, 7174, 7225, 7435, 7466, 7500A, 7639, 7645, 7755, 7773, 7918, 7978, 7980A, 7983, 7986, 7992, 8013, 8020, 8023, 8028, 8031, 8047, 8052A, 8082, 8089, 8091T, 8109, 8112, 8118, 8121, 8133, 8140, 8148, 8149, 8154, 8163, 8164, 8175, 8184, 8185, 8188, 8196A, 8220, 8227, 8238, 8239, 8250, 8256, 8264C, 8271, 8286, 8287, 8304A, 8310, 8346, 8352, 8355, 8382C, 8385, 8388, 8415, 8420, 8421, 8424, 8442T, 8443, 8452, 8457, 8476, 8480, 8487, 8489, 8505, 8508, 8510, 8514, 8520, 8524, 8526, 8527, 8535, 8536, 8571, 8577, 8583, 8607, 8633, 8663, 8666, 8693, 8705, 8708, 8738, 8744, 8747, 8750, 8783, 8789, 8798, 8799T, 8810, 8816, 8834, 8835, 8846, 8849, 8854, 8855, 8870, 8873, 8876, 8883, 8888, 8889, 9146, 9282, 9284, 9374, 9417, 9555, 9579, 9582, 9636, 9756, 9789, 10002, 10094, 10118, 10249, 10367, 10549, 10783, 10852, 10924, 10960, 11152, 11194T, 11203, 11204, 11212, 11215, 11317, 11425, 11454, 11482, 11491, 11668, 11710d, 11834, 11915, 12086, 12188, 12287, 12326, 12438, 12521, 12539C, 12571C, | CE:281, 711, 1413, 1469, 1683, 2903, 3218, 3431, 4224, 8376, 8413, 8651, 9056, 9375, 11338, 14551, 15597, 15607, 15956, 15963, 15965, 15967, 16020, 16133, 16156, 16472d | CE:708-709d, 712+G, 713, 729C, 737, 744, 770, 787T, 788G, 2199, 3041G, 3218T, 4443, 5147, 5922, 7225, 7978, 7980A, 7986, 7992, 8013, 8020, 8023, 8028, 8031, 8047, 8052A, 8082, 8089, 8091T, 8109, 8112, 8118, 8133, 8140, 8148, 8149, 8154, 8163, 8164, 8175, 8184, 8185, 8188, 8196A, 8220, 8227, 8238, 8239, 8250, 8256, 8271, 8286, 8287, 8304A, 8310, 8346, 8352, 8355, 8382C, 8385, 8388, 8415, 8420, 8421, 8424, 8442T, 8443, 8452, 8457, 8476, 8480, 8487, 8489, 8505, 8508, 8510, 8514, 8520, 8524, 8526, 8527, 8535, 8536, 8571, 8577, 8583, 8607, 8633, 8663, 8666, 8693, 8705, 8708, 8738, 8744, 8747, 8750, 8783, 8789, 8798, 8799T, 8810, 8834, 8835, 8846, 8849, 8854, 8855, 8870, 8873, 8876, 8883, 8888, 8889, 9146, 9417, 9555, 10249, 10960, 11194T, 11203, 11204, 11212, 11215, 11425, 12188, 12438, 13091, 13373, 13605, 14308, 14656, 15169, 15551, 15556, 15658, 15676, 15858, 15879+T, 15903, 15939, 15959, 15961, 15990, 16003, 16044, 16101, 16356, 16441 | Mereu,P. Unpublished |
| | | | 12887, 13091, 13151, 13154, 13172, 13199, 13361, 13373, 13436, 13552, 13576, 13605, 13675, 13806, 13810, 13813C, 13855, 14055, 14308, 14365, 14467, 14634, 14653, 14656, 14854, 14971, 15097, 15169, 15509, 15551, 15556, 15583, 15629, 15658, 15676, 15708, 15721, 15727, 15729+T, 15800, 15802, 15804+T, 15820, 15858, 15877, 15879+T, 15903, 15939, 15959, 15961, 15971, 15972, 15977, 15978, 15990, 16003, 16008, 16036, 16042, 16044, 16048, 16096, 16097, 16101, 16128, 16209, 16217, 16342+C, 16343, 16356, 16429, 16440, 16441, 16453, 16546, 16602 | | | |
| AY684273( Yak and wild yak) | Surplus of mutations. | B1b1 | 327, 328, 366G, 392, 406, 407+GC, 408, 409, 881G, 979G, 980G, 1016+T, 1617, 2073T, 2741T, 2866, 3681G, 3802, 4849A, 4852A, 4895, 5015, 5373, 5374G, 5375T, 5378d, 5386d, 5397d, 6136, 7537C, 7538, 7539, 7646G, 7808, 8366A, 8556, 8599, 8864, 8924, 9140, 9481, 9484, 9487, 10361, 11375, 11920G, 12267C, 13113, 13326, 13354, 13356G, 13363, 13370, 13476, 13477, 14037, 14684, 14685G, 14717T, 14718, 14719, 14942, 15074, 15093, 15397, 15719 | B1b1:5423 | B1b1:366G, 392, 406, 407+GC, 408, 409, 881G, 979G, 980G, 1016+T, 1617, 2073T, 2741T, 3681G, 3802, 4849A, 4852A, 4895, 5373, 5374G, 5375T, 5378d, 5386d, 5397d, 6136, 7537C, 7538, 7539, 7646G, 7808, 8366A, 8599, 8864, 8924, 9140, 9481, 9484, 9487, 10361, 11375, 11920G, 12267C, 13326, 13354, 13356G, 13363, 13370, 13476, 13477, 14037, 14684, 14685G, 14717T, 14718, 14719, 14942 | Gu,Z., Zhao,X., Li,N. and Wu,C. Mol. Phylogenet. Evol. 42 (1), 248-255 (2007) |
| GQ464251 (Yak and wild yak) | Artificial recombinatio: missing mutations 110T, 117, 135, 142, 187, 193, 218T, 220, 248, 249, 255, 275, 292, 308, 309, 316, 325, 333, 355, 390, 391, 500, 538, 693, 696, 719, 730, 810 of haplogroup C1b. | CD | 327, 328, 366G, 392, 406G, 881G, 987, 1082, 1612, 1656, 2007A, 2398, 2629, 2647, 2694A, 2852, 3498, 3502, 3676, 3760, 4421, 4881, 5015, 5390, 5693, 5748, 6491, 6500, 6761, 6764, 6866, 7271, 7409, 7941, 7950, 8400, 8415, 8481, 8484, 8523, 8556, 8577, 8848, 8875, 8938, 9756, 10083, 10338, 10477, 10498, 10534, 10645, 10824, 11315, 11711, 12131, 12252, 12314, 12320, 12323, 12383, 12675, 12804, 12823, 13086, 13113, 13530, 13569, 13786, 14636, 14645, 14675, 15093, 15287, 15332, 15398, 15419, 15503, 15719, 15833, 15950 | CD:135, 142, 187, 193, 248, 255, 275, 292, 308, 309, 316, 325, 355, 390, 391, 500, 538, 693, 696, 719, 730, 810 | CD:328, 366G, 392, 406G, 881G, 2694A, 3760, 4421, 7271, 13569, 14636 | Wang,Z., Shen,X., Liu,B., Su,J., Yonezawa,T., Yu,Y., Guo,S.,Ho,S.Y.W., Vila,C., Hasegawa,M. and Liu,J. J. Biogeogr. 37 (12), 2332-2344 (2010) |
| GQ464267 (Yak and wild yak) | Artificial recombination: missing mutations 6491, 6500, 6761, 6764, 6866, 7409 of haplogroup D1. | D1 | 135, 142, 187, 193, 248, 255, 275, 292, 308, 309, 316, 325, 327, 333, 355, 390, 500, 538, 693, 696, 719, 730, 810, 987, 1082, 1612, 1656, 2007A, 2398, 2629, 2647, 2850, 2852, 3498, 3502, 3676, 4881, 5015, 5390, 5693, 5748, 7941, 7950, 8400, 8415, 8481, 8484, 8523, 8556, 8577, 8655, 8848, 8875, 8938, 9756, 10050, 10083, 10338, 10477, 10498, 10534, 10645, 10824, 11315, 11554, 11711, 12131, 12252, 12314, 12320, 12323, 12383, 12490, 12675, 12804, 12823, 13086, 13113, 13256G, 13530, 13786, 14645, 14675, 15093, 15287, 15332, 15398, 15419, 15503, 15719, 15833, 15950 | D1:310, 391, 6491, 6500, 6761, 6764, 6866 | D1:333, 2850, 10050, 12490, 13256G | Wang,Z., Shen,X., Liu,B., Su,J., Yonezawa,T., Yu,Y., Guo,S.,Ho,S.Y.W., Vila,C., Hasegawa,M. and Liu,J. J. Biogeogr. 37 (12), 2332-2344 (2010) |
| GQ464276 (Yak and wild yak) | Artificial recombination: mis-added mutations333, | A1a3, A1a_GQ464290 | 160+A, 256, 295, 325, 327, 333, 355, 390, 391, 500, 538, 693, 696, 4895, 5015, 7808, 8382, 8556, 13113, 15089, 15093, 15719, 15833 | | A1a3:325, 355, 390, 391, 500, 538, 693, 696; A1a_GQ464290:325, 333, 355, 390, 391, 500, 693, 696 | Wang,Z., Shen,X., Liu,B., Su,J., Yonezawa,T., |
| | 355, 390, 391, 500, 538, 693, 696 from haplogroup CD. | | | | | Yu,Y., Guo,S.,Ho,S.Y.W., Vila,C., Hasegawa,M. and Liu,J. J. Biogeogr. 37 (12), 2332-2344 (2010) |
| GQ464288 (Yak and wild yak) | Artificial recombination: missing mutations 5693, 5748 of haplogroup CD. | C1a_GQ464287 | 110T, 117, 135, 142, 187, 193, 220, 248, 249, 255, 275, 292, 308, 309, 316, 325, 327, 333, 355, 390, 391, 500, 538, 693, 696, 719, 730, 810, 987, 1082, 1612, 1656, 2007A, 2398, 2629, 2647, 2852, 3498, 3502, 3676, 3760, 4493, 4881, 5015, 5390, 6491, 6500, 6761, 6764, 6866, 7271, 7409, 7941, 7950, 8400, 8415, 8481, 8484, 8523, 8556, 8577, 8848, 8875, 8938, 9756, 10083, 10338, 10477, 10498, 10534, 10645, 10824, 11084, 11315, 11635, 11711, 12131, 12252, 12314, 12320, 12323, 12383, 12675, 12804, 12823, 13086, 13113, 13530, 13786, 14645, 14675, 15093, 15287, 15332, 15398, 15419, 15503, 15719, 15833, 15950 | C1a_GQ464287:5693, 5748 | | Wang,Z., Shen,X., Liu,B., Su,J., Yonezawa,T., Yu,Y., Guo,S.,Ho,S.Y.W., Vila,C., Hasegawa,M. and Liu,J. J. Biogeogr. 37 (12), 2332-2344 (2010) |
| GQ464310 (Yak and wild yak) | Artificial recombination: missing mutations 160+A, 256, 295 of haplogroup A1a. | A1a1, A1a4, A1a_JQ846020 | 247, 248, 266, 327, 881G, 4895, 5015, 7808, 8382, 8556, 13113, 15089, 15093, 15719 | A1a1:160+A, 256, 295, 15833; A1a4:256, 295, 15833; A1a_JQ846020:160+A, 256, 295, 15833 | A1a1:247, 266, 881G; A1a4:247, 248, 266, 881G; A1a_JQ846020:247, 248, 881G | Wang,Z., Shen,X., Liu,B., Su,J., Yonezawa,T., Yu,Y., Guo,S.,Ho,S.Y.W., Vila,C., Hasegawa,M. and Liu,J. J. Biogeogr. 37 (12), |
| | | | | | | 2332-2344 (2010) |
| GQ464311 (Yak and wild yak) | Artificial recombination: missing mutations 255, 275, 292, 308, 309, 316, 325, 333, 355, 390, 391, 500, 538, 693, 696,7 19, 730 of haplogroup C; mis-added mutations 256, 295 from haplogroup A1. | C1a1_JQ846021 | 78, 100, 110T, 117, 135, 142, 187, 193, 220, 248, 249, 256, 295, 327, 810, 881G, 1082, 1612, 1656, 2007A, 2398, 2629, 2647, 2852, 3498, 3502, 3676, 3760, 4493, 4881, 5015, 5390, 5484, 5693, 5748, 5801, 6491, 6500, 6761, 6764, 6866, 7271, 7409, 7941, 7950, 8400, 8415, 8481, 8484, 8523, 8556, 8577, 8848, 8875, 8938, 9756, 10083, 10338, 10477, 10498, 10534, 10645, 10824, 11315, 11635, 11711, 12131, 12252, 12314, 12320, 12323, 12383, 12675, 12714, 12804, 12823, 13086, 13113, 13530, 13786, 14645, 14675, 14921, 15034, 15093, 15287, 15332, 15398, 15419, 15503, 15719, 15833, 15950 | C1a1_JQ846021: 255, 275, 292, 308, 309, 316, 325, 333, 355, 390, 391, 500, 538, 693, 696, 719, 730, 987 | C1a1_JQ846021:78, 100, 256, 295, 881G | Wang,Z., Shen,X., Liu,B., Su,J., Yonezawa,T., Yu,Y., Guo,S.,Ho,S.Y.W., Vila,C., Hasegawa,M. and Liu,J. J. Biogeogr. 37 (12), 2332-2344 (2010) |
| GQ464313 (Yak and wild yak) | Artificial recombination: missing mutations 333, 355, 390, 391, 500, 538, 693, 696, 719, 730 of haplogroup C. | C1a1 | 110T, 117, 135, 142, 187, 193, 220, 248, 249, 255, 275, 292, 308, 309, 316, 325, 327, 810, 987, 1082, 1612, 1656, 2007A, 2398, 2629, 2647, 2852, 3498, 3502, 3676, 3760, 4493, 4881, 5015, 5390, 5693, 5748, 6491, 6500, 6761, 6764, 6866, 7271, 7409, 7941, 7950, 8400, 8415, 8481, 8484, 8523, 8556, 8577, 8848, 8875, 8938, 9331, 9756, 10083, 10338, 10477, 10498, 10534, 10645, 10824, 11315, 11635, 11711, 12131, 12252, 12314, 12320, 12323, 12383, 12675, 12714, 12804, 12823, 13086, 13113, 13530, 13786, 14645, 14675, 15034, 15093, 15287, 15332, 15398, 15419, 15503, 15719, 15833 | C1a1:333, 355, 390, 391, 500, 538, 693, 696, 719, 730, 15950 | C1a1:9331 | Wang,Z., Shen,X., Liu,B., Su,J., Yonezawa,T., Yu,Y., Guo,S.,Ho,S.Y.W., Vila,C., Hasegawa,M. and Liu,J. J. Biogeogr. 37 (12), 2332-2344 (2010) |

Table S2. mtDNA haplogroup tree of dog and gray wolf.

Nucleotide position numbers are relative to the EU789787. Mutations are transitions unless an exact base change is specified.

Coding region mutations (np 1-15460) are shown in black; control region mutations (np 15461-16196) in blue.

Mutations toward a base identical-by-state to the EU789787 are indicated with the prefix @.

"W", "S", "M", "K", "R" and "Y" specify the heteroplasmic status of A/T, C/G, A/C, G/T, A/G and C/T at a certain site, respectively.

Undetermined mutations are noted by "N".

Deletions are indicated by a "d" after the deleted nucleotide position.

Insertions are indicated by a "+" followed by the position number and type of inserted nucleotide(s).

Repetitive and difficult-to-align regions np 15512-15513 and np 16019-16120 are not considered.

The length variations in the C tract scored at np 1691 (1691+C), the G tract scored at np 15918 (15918d) and the A tract scored at np 1493 (1493+A) were not considered for phylogenetic reconstruction and were therefore excluded from the tree.

EU442884, KF661078, KF661079, KF661080, KF661081, KF661088, KF661089 serve as an outgroups and are not considered in the tree.

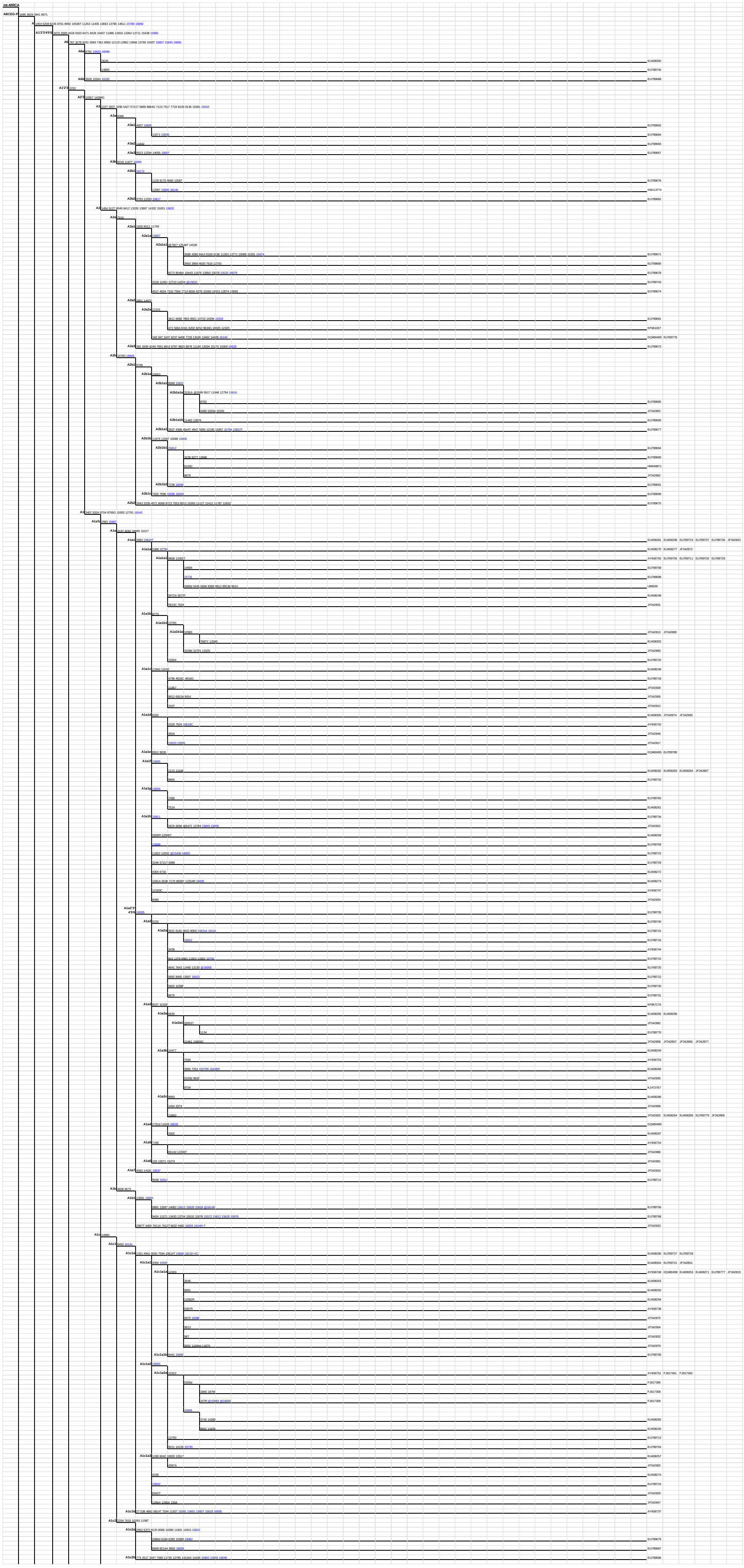

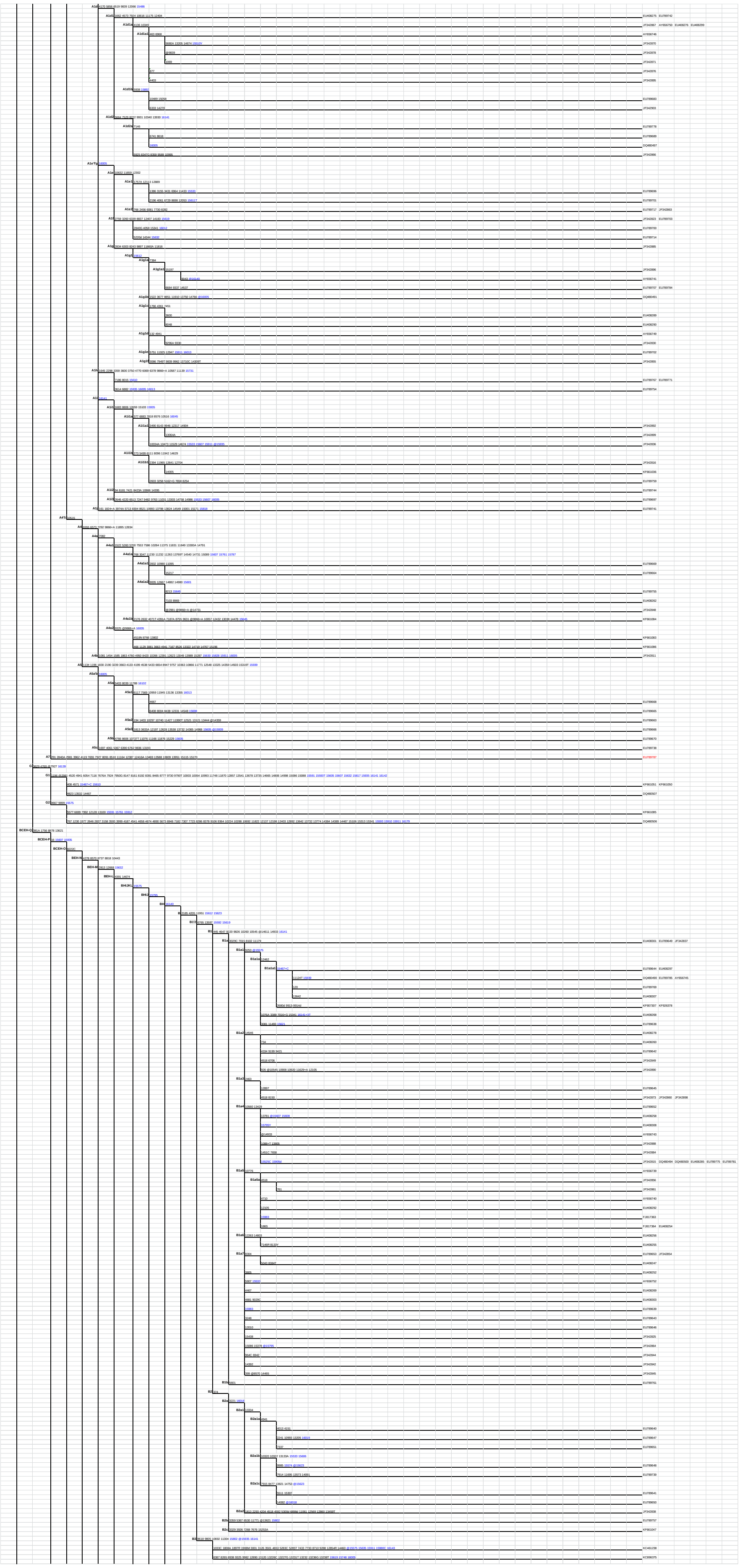

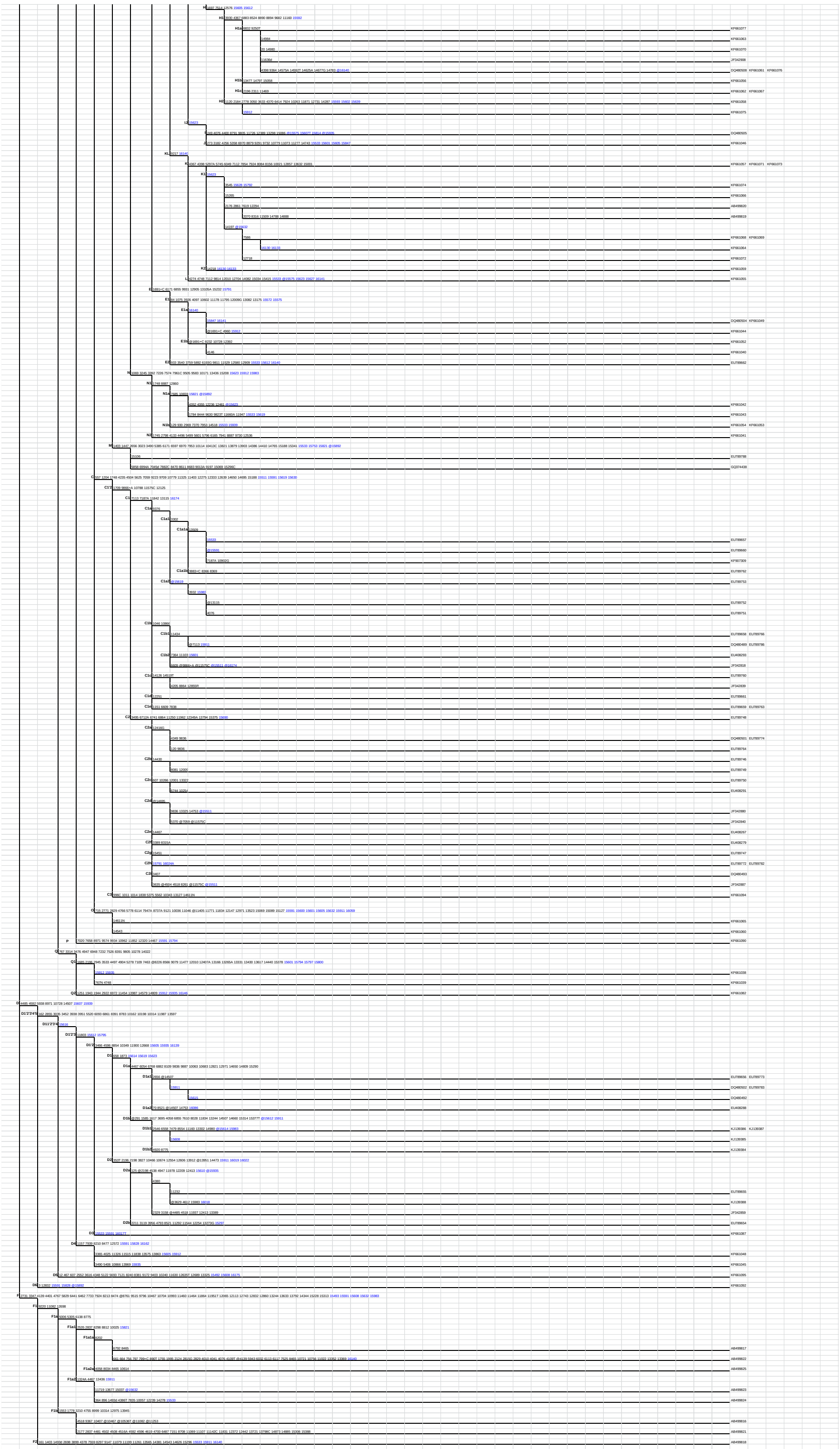

Table S3. mtDNA haplogroup tree of goat.

Nucleotide position numbers are relative to the GU068049. Mutations are transitions unless an exact base change is specified.

Coding region mutations (np 1-15430) are shown in black; control region mutations (np 15431-16642) in blue.

Mutations toward a base identical-by-state to the GU068049 are indicated with the prefix @.

"R" and "Y" specify the heteroplasmic status of A/G and C/T at a certain site, respectively.

Deletions are indicated by a "d" after the deleted nucleotide position.

Insertions are indicated by a "+" followed by the position number and type of inserted nucleotide(s).

Undetermined mutatiions are noted by "N".

The length variations scored at np 177 (177-178d),180 (180T) and 181 (181A) were not considered for phylogenetic reconstruction and were therefore excluded from the tree.

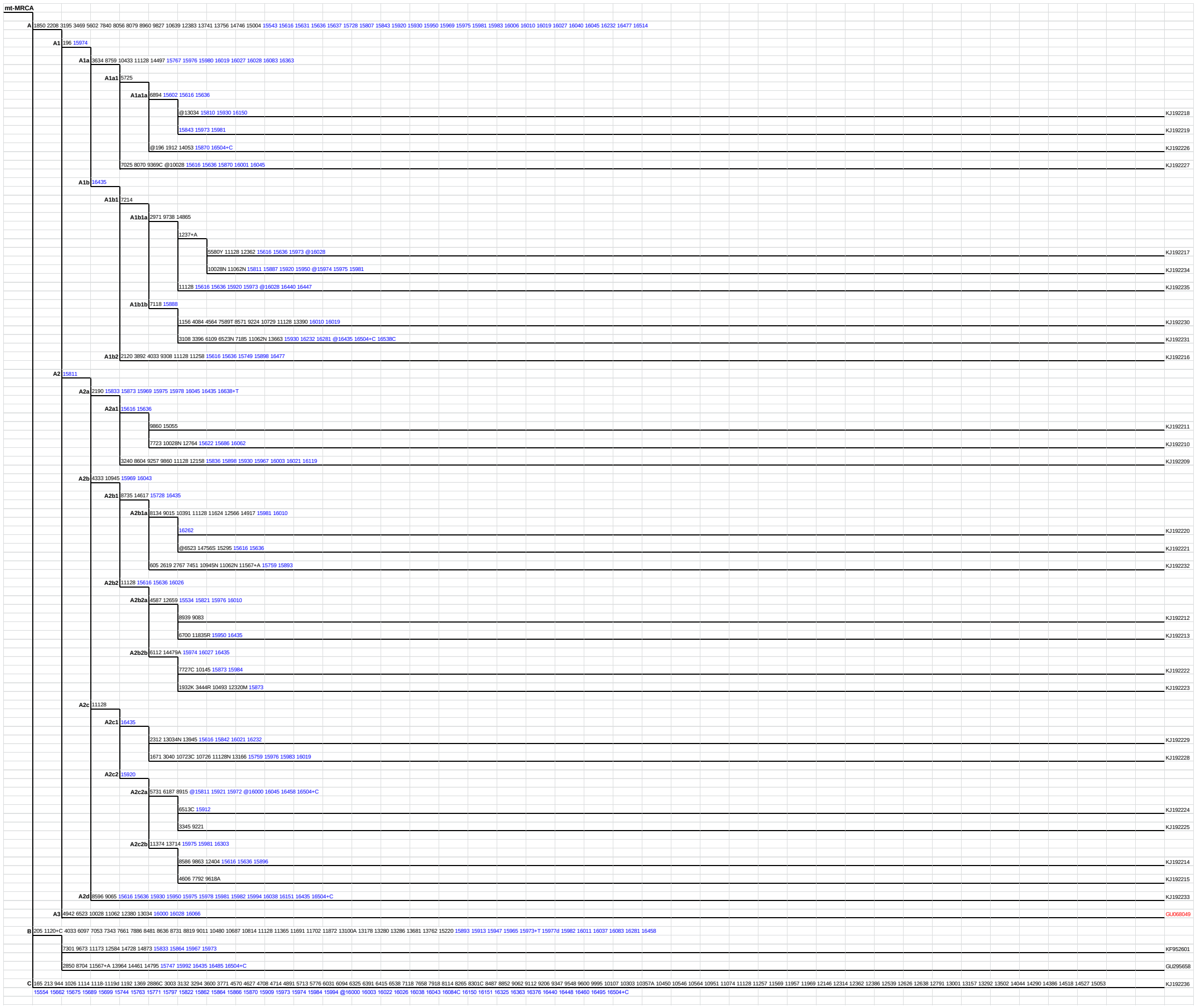

Table S4. mtDNA haplogroup tree of yak.

Nucleotide position numbers are relative to the GQ464259. Mutations are transitions unless an exact base change is specified.
Coding region mutations (np 893-16322) are shown in black; control region mutations (np 1-892) in blue.
Mutations toward a base identical-by-state to the GQ464259 are indicated with the prefix @.
"R" and "Y" specify the heteroplasmic status of A/G and C/T at a certain site, respectively.
Deletions are indicated by a "d" after the deleted nucleotide position.
Insertions are indicated by a "+" followed by the position number and type of inserted nucleotide(s).
The length variations in the C tract scored at np 751 (751+C, 751+2C) and 892 (881G, 892+C, 892+2C, 892d, 892+GGTGGG) were not considered for phylogenetic reconstruction and were therefore excluded from the tree.
GQ464260 appears remote to ABCD and serves as an outgroup sequence which is excluded from the tree.

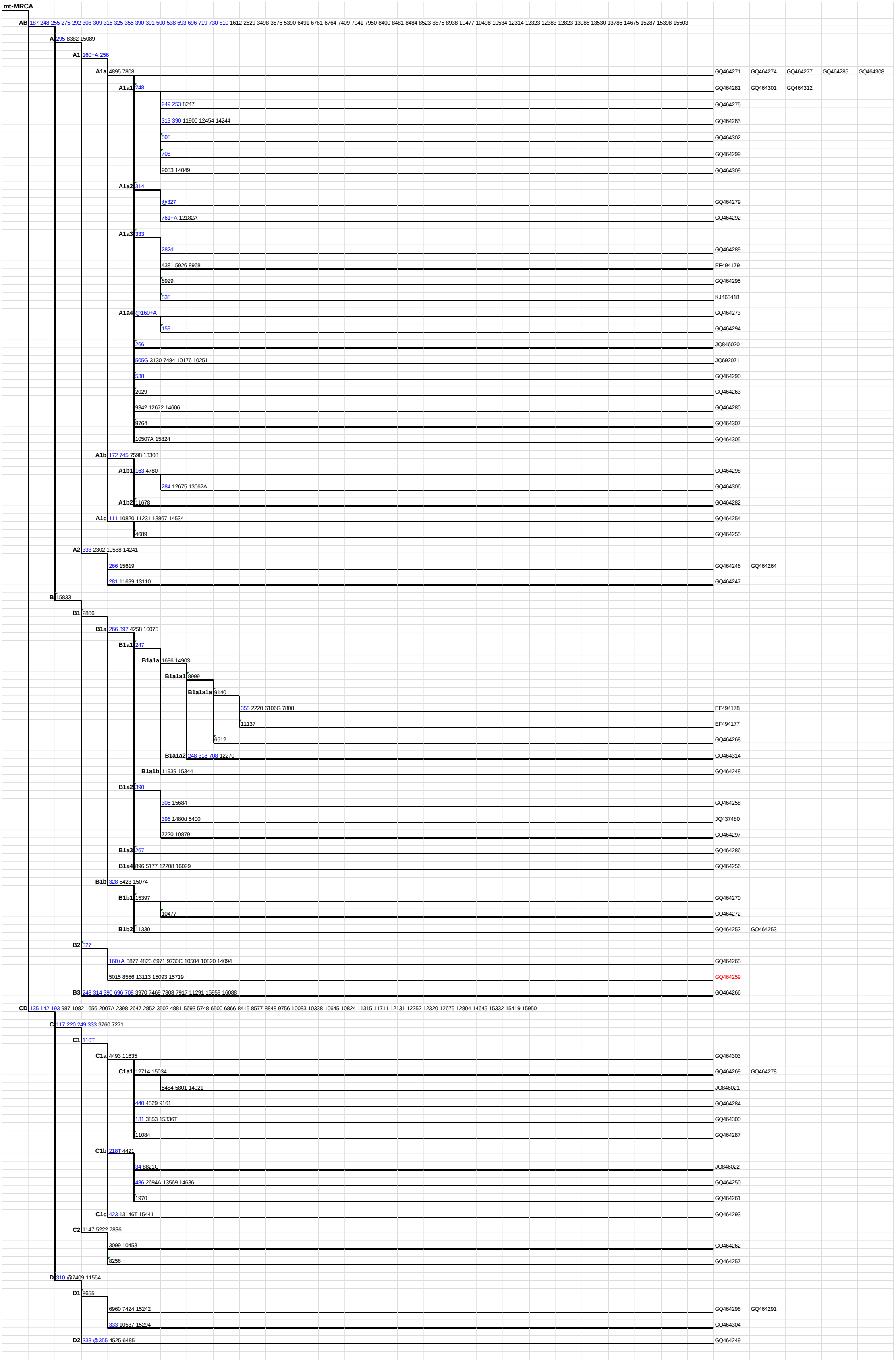
mt-MRCA
AB 187 248 255 275 292 308 309 316 325 355 390 391 500 538 693 696 719 730 810 1612 2629 3498 3676 5390 6491 6761 6764 7409 7941 7950 8400 8481 8484 8523 8875 8938 10477 10498 10534 12314 12323 12383 12823 13086 13530 13786 14675 15287 15398 15503
A 295 8382 15089
A1 160+A 256
A1a 4895 7808
GQ464271 GQ464274 GQ464277 GQ464285 GQ464308
A1a1 248
GQ464281 GQ464301 GQ464312
249 253 8247
GQ464275
313 390 11900 12454 14244
GQ464283
508
GQ464302
708
GQ464299
9033 14049
GQ464309
A1a2 314
@327
GQ464279
761+A 12182A
GQ464292
A1a3 333
282d
GQ464289
4381 5926 8968
EF494179
6929
GQ464295
538
KJ463418
A1a4 @160+A
GQ464273
159
GQ464294
266
JQ846020
505G 3130 7484 10176 10251
JQ692071
538
GQ464290
2029
GQ464263
9342 12672 14606
GQ464280
9764
GQ464307
10507A 15824
GQ464305
A1b 172 745 7598 13308
A1b1 163 4780
GQ464298
284 12675 13062A
GQ464306
A1b2 11678
GQ464282
A1c 111 10820 11231 13867 14534
GQ464254
4689
GQ464255
A2 333 2302 10588 14241
266 15619
GQ464246 GQ464264
281 11699 13110
GQ464247
B 15833
B1 2866
B1a 266 397 4258 10075
B1a1 247
B1a1a 1696 14903
B1a1a1 8999
B1a1a1a 9140
355 2220 6106G 7808
EF494178
11137
EF494177
6512
GQ464268
B1a1a2 248 318 708 12270
GQ464314
B1a1b 11939 15344
GQ464248
B1a2 390
305 15684
GQ464258
396 1480d 5400
JQ437480
7220 10879
GQ464297
B1a3 267
GQ464286
B1a4 896 5177 12208 16029
GQ464256
B1b 328 5423 15074
B1b1 15397
GQ464270
10477
GQ464272
B1b2 11330
GQ464252 GQ464253
B2 327
160+A 3877 4823 6971 9730C 10504 10820 14094
GQ464265
5015 8556 13113 15093 15719
GQ464259
B3 248 314 390 696 708 3970 7469 7808 7917 11291 15959 16088
GQ464266
CD 135 142 193 987 1082 1656 2007A 2398 2647 2852 3502 4881 5693 5748 6500 6866 8415 8577 8848 9756 10083 10338 10645 10824 11315 11711 12131 12252 12320 12675 12804 14645 15332 15419 15950
C 117 220 249 333 3760 7271
C1 110T
C1a 4493 11635
GQ464303
C1a1 12714 15034
GQ464269 GQ464278
5484 5801 14921
JQ846021
440 4529 9161
GQ464284
131 3853 15336T
GQ464300
11084
GQ464287
C1b 218T 4421
34 8821C
JQ846022
486 2694A 13569 14636
GQ464250
1970
GQ464261
C1c 423 13146T 15441
GQ464293
C2 1147 5222 7836
3099 10453
GQ464262
8256
GQ464257
D 310 @7409 11554
D1 8655
6960 7424 15242
GQ464296 GQ464291
333 10537 15294
GQ464304
D2 333 @355 4525 6485
GQ464249

Table S5. Potential errors detected in next-generation sequencing data.

| Sequence | Platform | Potential Error | Haplogroup Most Matched Sequence | Variants | Missing Diagnostic Mutations | Private | Sources |
|---|---|---|---|---|---|---|---|
| KF163061 (Cattle and Aurochs) | Roche/454 Life Sciences GS Junior instrument | Phantom mutations: 215+TCC, 231+C, 16200+A. | T1b1_KF163061_KF163076 | 169, 215+TCC, 231+C, 587+C, 1599-1600d, 2536A, 6050, 7542, 9682C, 12468, 12469, 13310C, 14348, 16022, 16050, 16084, 16085, 16113, 16200+A, 16255 | | T1b1_KF163061_KF163076:215+TCC, 234+T, 1599-1600d, 16084, 16085, 16200+A | Horsburgh,K.A., Prost,S., Gosling,A., Stanton,J.A., Rand,C. and Matisoo-Smith,E.A. PLoS ONE 8 (8), E71956 (2013) |
| KF163062 (Cattle and Aurochs) | Roche/454 Life Sciences GS Junior instrument | Phantom mutations: 215+TCC, 234+T, 16200+A. | T1b1b1b_KF163062_KF163086 | 169, 190, 215+TCC, 234+T, 497, 587+C, 1599-1600d, 2536A, 7542, 8710, 9682C, 12468, 12469, 12525, 13201, 13300, 13310C, 14523T, 16022, 16050, 16056, 16084, 16113, 16122, 16200+A, 16255, 16316 | | T1b1b1b_KF163062_KF163086:215+TCC, 234+T, 1599-1600d, 16084, 16200+A | Horsburgh,K.A., Prost,S., Gosling,A., Stanton,J.A., Rand,C. and Matisoo-Smith,E.A. PLoS ONE 8 (8), E71956 (2013) |
| KF163063 (Cattle and Aurochs) | Roche/454 Life Sciences GS Junior instrument | Phantom mutations: 215+TCC, 234+T, 16200+A. | T1b1, T1b1_KF163063 | 169, 215+TCC, 234+T, 587+C, 1599-1600d, 2536A, 7542, 9682C, 13310C, 16022, 16050, 16113, 16200+A, 16255, 16301 | | T1b1:215+TCC, 234+T, 1599-1600d, 16200+A, 16301; T1b1_KF163063:215+TCC, 234+T, 1599-1600d, 16200+A | Horsburgh,K.A., Prost,S., Gosling,A., Stanton,J.A., Rand,C. and Matisoo-Smith,E.A. PLoS ONE 8 (8), E71956 (2013) |
| KF163064 (Cattle and Aurochs) | Roche/454 Life Sciences GS Junior instrument | Phantom mutations: 215+TCC, 234+T, 16202+G. | T1b1b1b_KF163064 | 169, 190, 215+TCC, 234+T, 497, 587+C, 1599-1600d, 2536A, 3069, 7542, 8710, 9682C, 12525, 13201, 13300, 13310C, 14523T, 16022, 16050, 16056, 16057, 16084, 16113, 16202+G, 16316 | | T1b1b1b_KF163064:215+TCC, 234+T, 1599-1600d, 16084, 16202+G | Horsburgh,K.A., Prost,S., Gosling,A., Stanton,J.A., Rand,C. and Matisoo-Smith,E.A. PLoS ONE 8 (8), E71956 (2013) |
| KF163065 (Cattle and Aurochs) | Roche/454 Life Sciences GS Junior instrument | Phantom mutations: 215+TCC, 234+T, 16200+A. | T1b1b1b_KF163065 | 169, 190, 215+TCC, 234+T, 497, 587+C, 1599-1600d, 2536A, 3069, 4361, 7542, 8710, 9682C, 12468, 12469, 12525, 13201, 13300, 13310C, 14051, 14523T, 16022, 16050, 16056, 16084, 16085, 16113, 16200+A, 16316 | | T1b1b1b_KF163065:215+TCC, 234+T, 1599-1600d, 16084, 16085, 16200+A | Horsburgh,K.A., Prost,S., Gosling,A., Stanton,J.A., Rand,C. and Matisoo-Smith,E.A. PLoS ONE 8 (8), E71956 (2013) |
| KF163066 (Cattle and Aurochs) | Roche/454 Life Sciences GS Junior instrument | Phantom mutations: 215+TCC, 234+T, 16200+A. | T1b1b1b_KF163066 | 169, 190, 215+TCC, 234+T, 497, 587+C, 1599-1600d, 2536A, 3069, 7542, 8188, 8710, 9682C, 12468, 12469, 12525, 13201, 13300, 13310C, 14523T, 16022, 16050, 16056, 16085, 16113, 16200+A, 16316 | | T1b1b1b_KF163066:215+TCC, 234+T, 1599-1600d, 16085, 16200+A | Horsburgh,K.A., Prost,S., Gosling,A., Stanton,J.A., Rand,C. and Matisoo-Smith,E.A. PLoS ONE 8 (8), E71956 (2013) |
| KF163067 (Cattle and Aurochs) | Roche/454 Life Sciences GS Junior instrument | Phantom mutations: 215+TCC, 234+T, 16200+A. | T1b1b1b_KF163067 | 169, 173, 190, 215+TCC, 234+T, 497, 587+C, 1599-1600d, 2536A, 7542, 8710, 9682C, 12468, 12469, 12525, 13201, 13300, 13310C, 14523T, 16022, 16050, 16056, 16084, 16085, 16113, 16200+A, 16255, 16316 | | T1b1b1b_KF163067:215+TCC, 234+T, 1599-1600d, 16084, 16085, 16200+A | Horsburgh,K.A., Prost,S., Gosling,A., Stanton,J.A., Rand,C. and Matisoo-Smith,E.A. PLoS ONE 8 (8), E71956 (2013) |
| KF163068 | Roche/454 Life | Phantom | T1b1b1b_KF163069 | 169, 190, 215+TCC, 234+T, 497, 587+C, 1599-1600d, 2536A, 3270, 3413N, 3414N, 3415N, 3416N, 3417N, 3418N, 3419N, 3420N, | | T1b1b1b_KF163069:215+TCC, 234+T, 1599-1600d, 3413N, 3414N, 3415N, 3416N, 3417N, 3418N, | Horsburgh,K.A., Prost,S., |
| (Cattle and Aurochs) | Sciences GS Junior instrument | mutations: 215+TCC, 234+T, 16200+A. | | 3421N, 3422N, 3423N, 3424N, 3425N, 3426N, 3427N, 3428N, 3429N, 3430N, 3431N, 3432N, 3433N, 3434N, 3435N, 3436N, 3437N, 3438N, 3439N, 3440N, 3441N, 3442N, 3443N, 3444N, 3445N, 3446N, 3447N, 3448N, 3449N, 3450N, 3451N, 3452N, 4840Y, 7542, 8710, 9682C, 12361, 12468, 12469, 12525, 13201, 13300, 13310C, 14523T, 16022, 16050, 16056, 16084, 16085, 16113, 16200+A, 16255, 16316 | | 3419N, 3420N, 3421N, 3422N, 3423N, 3424N, 3425N, 3426N, 3427N, 3428N, 3429N, 3430N, 3431N, 3432N, 3433N, 3434N, 3435N, 3436N, 3437N, 3438N, 3439N, 3440N, 3441N, 3442N, 3443N, 3444N, 3445N, 3446N, 3447N, 3448N, 3449N, 3450N, 3451N, 3452N, 4840, 16084, 16085, 16200+A | Gosling,A., Stanton,J.A., Rand,C. and Matisoo-Smith,E.A. PLoS ONE 8 (8), E71956 (2013) |
| KF163069 (Cattle and Aurochs) | Roche/454 Life Sciences GS Junior instrument | Phantom mutations: 215+TCC, 234+T, 16200+A. | T1b1b1b_KF163069 | 169, 190, 215+TCC, 234+T, 497, 587+C, 1599-1600d, 2536A, 3270, 7542, 8710, 9682C, 12361, 12468, 12469, 12525, 13201, 13300, 13310C, 14523T, 16022, 16050, 16056, 16113, 16200+A, 16255, 16316 | | T1b1b1b_KF163069:215+TCC, 234+T, 1599-1600d, 16200+A | Horsburgh,K.A., Prost,S., Gosling,A., Stanton,J.A., Rand,C. and Matisoo-Smith,E.A. PLoS ONE 8 (8), E71956 (2013) |
| KF163070 (Cattle and Aurochs) | Roche/454 Life Sciences GS Junior instrument | Phantom mutations: 215+TCC, 234+T, 16200+A. | T1b1b1b_KF163070_KF163091 | 169, 190, 215+TCC, 234+T, 497, 587+C, 1599-1600d, 2536A, 7542, 7622, 8710, 9682C, 12468, 12469, 12525, 13201, 13260, 13300, 13310C, 14523T, 16022, 16050, 16056, 16084, 16085, 16113, 16200+A, 16255, 16316 | | T1b1b1b_KF163070_KF163091:215+TCC, 234+T, 1599-1600d, 16084, 16085, 16200+A | Horsburgh,K.A., Prost,S., Gosling,A., Stanton,J.A., Rand,C. and Matisoo-Smith,E.A. PLoS ONE 8 (8), E71956 (2013) |
| KF163071 (Cattle and Aurochs) | Roche/454 Life Sciences GS Junior instrument | Phantom mutations: 215+TCC, 234+T, 16202+G. | T1b1b1b_KF163071 | 169, 190, 216Y, 221+CCC, 234+T, 497, 587+C, 1599-1600d, 2536A, 3270, 7542, 8710, 9682C, 12361, 12468, 12525, 13201, 13300, 13310C, 14523T, 16022, 16050, 16056, 16113, 16202+G, 16255, 16316 | | T1b1b1b_KF163071:221+CCC, 234+T, 1599-1600d, 16202+G | Horsburgh,K.A., Prost,S., Gosling,A., Stanton,J.A., Rand,C. and Matisoo-Smith,E.A. PLoS ONE 8 (8), E71956 (2013) |
| KF163072 (Cattle and Aurochs) | Roche/454 Life Sciences GS Junior instrument | Phantom mutations: 215+TCC, 234+T, 16200+A. | T1c_KF163072 | 169, 215+TCC, 234+T, 587+C, 1599-1600d, 2536A, 4586, 6025, 9682C, 9906, 10907, 13310C, 15460, 16050, 16084, 16085, 16113, 16121, 16122, 16200+A, 16255 | | T1c_KF163072:215+TCC, 234+T, 1599-1600d, 16084, 16085, 16200+A | Horsburgh,K.A., Prost,S., Gosling,A., Stanton,J.A., Rand,C. and Matisoo-Smith,E.A. PLoS ONE 8 (8), E71956 (2013) |
| KF163073 (Cattle and Aurochs) | Roche/454 Life Sciences GS Junior instrument | Phantom mutations: 215+TCC, 234+T, 16200+A. | T1b1b1c | 169, 215+TCC, 234+T, 587+C, 1599-1600d, 2536A, 7542, 9682C, 12468, 12469, 13201, 13310C, 14523T, 15156, 16022, 16050, 16113, 16200+A, 16255 | T1b1b1c:16148 | T1b1b1c:215+TCC, 234+T, 1599-1600d, 15156, 16200+A | Horsburgh,K.A., Prost,S., Gosling,A., Stanton,J.A., Rand,C. and Matisoo-Smith,E.A. PLoS ONE 8 (8), E71956 (2013) |
| KF163074 (Cattle and Aurochs) | Roche/454 Life Sciences GS Junior instrument | Phantom mutations: 215+TCC, 234+T, 16200+A. | T1b1b1c | 169, 215+TCC, 234+T, 587+C, 1599-1600d, 2536A, 7542, 9682C, 13201, 13310C, 14523T, 16022, 16050, 16113, 16148, 16200+A, 16255 | | T1b1b1c:215+TCC, 234+T, 1599-1600d, 16200+A | Horsburgh,K.A., Prost,S., Gosling,A., Stanton,J.A., Rand,C. and Matisoo-Smith,E.A. PLoS ONE 8 (8), E71956 (2013) |
| KF163075 (Cattle and Aurochs) | Roche/454 Life Sciences GS Junior instrument | Phantom mutations: 215+TCC, 234+T, 16200+A. | T1b1b1a2_KF163075 | 169, 215+TCC, 234+T, 587+C, 1599-1600d, 2536A, 5225, 7542, 9682C, 12402, 12469, 13201, 13310C, 14523T, 14771, 16022, 16050, 16056, 16057, 16113, 16200+A, 16255 | | T1b1b1a2_KF163075:215+TCC, 234+T, 1599-1600d, 16200+A | Horsburgh,K.A., Prost,S., Gosling,A., Stanton,J.A., Rand,C. and Matisoo-Smith,E.A. PLoS ONE 8 (8), E71956 (2013) |
| KF163076 | Roche/454 Life | Phantom | T1b1_KF163061_KF16 | 169, 215+TCC, 234+T, 587+C, 1599-1600d, 2536A, 6050, 7542, 9682C, 12468, 12469, 13310C, 14348, 16022, 16050, 16084, 16085, | | T1b1_KF163061_KF163076:215+TCC, 234+T, 1599-1600d, 16084, 16085, 16200+A | Horsburgh,K.A., Prost,S., |
| (Cattle and Aurochs) | Sciences GS Junior instrument | mutations: 215+TCC, 234+T, 16200+A. | 3076 | 16113, 16200+A, 16255 | | | Gosling,A., Stanton,J.A., Rand,C. and Matisoo-Smith,E.A. PLoS ONE 8 (8), E71956 (2013) |
| KF163077 (Cattle and Aurochs) | Roche/454 Life Sciences GS Junior instrument | Phantom mutations: 215+TCC, 234+T, 16200+A. | T1b1b1b_KF163077 | 169, 190, 215+TCC, 234+T, 497, 587+C, 1075C, 1599-1600d, 2536A, 6391, 7542, 8710, 9095, 9682C, 12525, 13201, 13300, 13310C, 14523T, 16022, 16050, 16056, 16084, 16085, 16113, 16121, 16200+A, 16255, 16316 | | T1b1b1b_KF163077:215+TCC, 234+T, 1599-1600d, 16084, 16085, 16200+A | Horsburgh,K.A., Prost,S., Gosling,A., Stanton,J.A., Rand,C. and Matisoo-Smith,E.A. PLoS ONE 8 (8), E71956 (2013) |
| KF163078 (Cattle and Aurochs) | Roche/454 Life Sciences GS Junior instrument | Phantom mutations: 215+TCC, 234+T, 16200+A. | T1b1b1b_KF163078 | 169, 190, 215+TCC, 234+T, 497, 587+C, 1599-1600d, 2536A, 3270, 7542, 8710, 9682C, 12361, 12468, 12469, 12525, 13201, 13300, 13310C, 14523T, 16022, 16050, 16056, 16085, 16113, 16121, 16122, 16200+A, 16255, 16316 | | T1b1b1b_KF163078:215+TCC, 234+T, 1599-1600d, 16085, 16200+A | Horsburgh,K.A., Prost,S., Gosling,A., Stanton,J.A., Rand,C. and Matisoo-Smith,E.A. PLoS ONE 8 (8), E71956 (2013) |
| KF163079 (Cattle and Aurochs) | Roche/454 Life Sciences GS Junior instrument | Phantom mutations: 215+TCC, 234+T, 16200+A. | T1b1b1b_KF163079 | 169, 173, 190, 215+TCC, 234+T, 497, 587+C, 1599-1600d, 2536A, 7542, 8710, 9682C, 12525, 13201, 13300, 13310C, 14523T, 16022, 16050, 16056, 16113, 16200+A, 16255, 16316 | | T1b1b1b_KF163079:215+TCC, 234+T, 1599-1600d, 16200+A | Horsburgh,K.A., Prost,S., Gosling,A., Stanton,J.A., Rand,C. and Matisoo-Smith,E.A. PLoS ONE 8 (8), E71956 (2013) |
| KF163080 (Cattle and Aurochs) | Roche/454 Life Sciences GS Junior instrument | Phantom mutations: 215+TCC, 234+T, 16200+A. | T1b1b1bc_KF163080 | 169, 190, 215+TCC, 234+T, 497, 587+C, 1599-1600d, 2536A, 6475, 7542, 8710, 9682C, 12468, 12469, 12525, 13201, 13300, 13310C, 14523T, 16022, 16050, 16056, 16084, 16085, 16113, 16122, 16200+A, 16255, 16316 | | T1b1b1b_KF163080:215+TCC, 234+T, 1599-1600d, 16084, 16085, 16200+A | Horsburgh,K.A., Prost,S., Gosling,A., Stanton,J.A., Rand,C. and Matisoo-Smith,E.A. PLoS ONE 8 (8), E71956 (2013) |
| KF163081 (Cattle and Aurochs) | Roche/454 Life Sciences GS Junior instrument | Phantom mutations: 215+TCC, 234+T, 16200+A. | T1d_KF163081 | 169, 215+TCC, 234+T, 587+C, 1599-1600d, 2536A, 6235, 9682C, 10347, 10657, 12468, 13310C, 16050, 16084, 16085, 16113, 16121, 16200+A, 16255 | | T1d_KF163081:215+TCC, 234+T, 1599-1600d, 16084, 16085, 16200+A | Horsburgh,K.A., Prost,S., Gosling,A., Stanton,J.A., Rand,C. and Matisoo-Smith,E.A. PLoS ONE 8 (8), E71956 (2013) |
| KF163082 (Cattle and Aurochs) | Roche/454 Life Sciences GS Junior instrument | Phantom mutations: 215+TCC, 234+T, 16200+A. | T1_KF163082 | 169, 215+TCC, 234+T, 299, 587+C, 1599-1600d, 2536A, 3682, 9201, 9682C, 12468, 12469, 13310C, 16050, 16113, 16200+A, 16248, 16255, 16260 | | T1_KF163082:215+TCC, 234+T, 1599-1600d, 16200+A | Horsburgh,K.A., Prost,S., Gosling,A., Stanton,J.A., Rand,C. and Matisoo-Smith,E.A. PLoS ONE 8 (8), E71956 (2013) |
| KF163083 (Cattle and Aurochs) | Roche/454 Life Sciences GS Junior instrument | Phantom mutations: 215+TCC, 234+T, 16200+A. | T1b1b1b_KF163083 | 169, 190, 215+TCC, 234+T, 497, 587+C, 1599-1600d, 2536A, 3270, 7514, 8710, 9682C, 12361, 12525, 13201, 13300, 13310C, 14523T, 16022, 16050, 16056, 16084, 16085, 16113, 16200+A, 16255, 16316 | | T1b1b1b_KF163083:215+TCC, 234+T, 1599-1600d, 16084, 16085, 16200+A | Horsburgh,K.A., Prost,S., Gosling,A., Stanton,J.A., Rand,C. and Matisoo-Smith,E.A. PLoS ONE 8 (8), E71956 (2013) |
| KF163084 (Cattle and | Roche/454 Life Sciences GS | Phantom mutations: | T1b1_KF163084 | 169, 215+TCC, 234+T, 587+C, 1599-1600d, 2536A, 7542, 9682C, 13310C, 13395, 16022, 16050, 16057, 16068, 16084, 16085, 16113, 16200+A, 16255 | | T1b1_KF163084:215+TCC, 234+T, 1599-1600d, 16084, 16085, 16200+A | Horsburgh,K.A., Prost,S., Gosling,A., Stanton,J.A., Rand,C. |
| Aurochs) | Junior instrument | 215+TCC, 234+T, 16200+A. | | | | | and Matisoo-Smith,E.A. PLoS ONE 8 (8), E71956 (2013) |
| KF163085 (Cattle and Aurochs) | Roche/454 Life Sciences GS Junior instrument | Phantom mutations: 215+TCC, 234+T, 16200+A. | T1b1b1b_KF163085 | 169, 190, 215+TCC, 234+T, 249, 296N, 297N, 497, 587+C, 1599-1600d, 2536A, 7542, 8710, 9682C, 12468, 12469, 12525, 13005, 13201, 13300, 13310C, 14523T, 15951, 15953G, 16022, 16050, 16056, 16085, 16113, 16200+A, 16255, 16316 | | T1b1b1b_KF163085:215+TCC, 234+T, 296N, 297N, 1599-1600d, 16085, 16200+A | Horsburgh,K.A., Prost,S., Gosling,A., Stanton,J.A., Rand,C. and Matisoo-Smith,E.A. PLoS ONE 8 (8), E71956 (2013) |
| KF163086 (Cattle and Aurochs) | Roche/454 Life Sciences GS Junior instrument | Phantom mutations: 215+TCC, 234+T, 16200+A. | T1b1b1c | 169, 190, 215+TCC, 234+T, 497, 587+C, 1599-1600d, 2536A, 7542, 8710, 9682C, 12525, 13201, 13300, 13310C, 14523T, 16022, 16050, 16056, 16113, 16122, 16200+A, 16255, 16316 | | T1b1b1b:215+TCC, 234+T, 1599-1600d, 16122, 16200+A | Horsburgh,K.A., Prost,S., Gosling,A., Stanton,J.A., Rand,C. and Matisoo-Smith,E.A. PLoS ONE 8 (8), E71956 (2013) |
| KF163087 (Cattle and Aurochs) | Roche/454 Life Sciences GS Junior instrument | Phantom mutations: 215+TCC, 234+T, 16200+A. | T1b1_KF163087 | 169, 215+TCC, 234+T, 587+C, 1599-1600d, 2536A, 6050, 7542, 9682C, 13310C, 14348, 16022, 16050, 16084, 16085, 16113, 16200+A, 16255 | | T1b1_KF163087:215+TCC, 234+T, 1599-1600d, 16084, 16085, 16200+A | Horsburgh,K.A., Prost,S., Gosling,A., Stanton,J.A., Rand,C. and Matisoo-Smith,E.A. PLoS ONE 8 (8), E71956 (2013) |
| KF163088 (Cattle and Aurochs) | Roche/454 Life Sciences GS Junior instrument | Phantom mutations: 215+TCC, 234+T, 16200+A. | T1b1b1c_KF163088 | 169, 215+TCC, 234+T, 587+C, 1599-1600d, 2536A, 7542, 9682C, 12468, 12469, 13201, 13310C, 14523T, 16022, 16050, 16084, 16085, 16113, 16148, 16200+A, 16255 | | T1b1b1c_KF163088:215+TCC, 234+T, 1599-1600d, 16084, 16085, 16200+A | Horsburgh,K.A., Prost,S., Gosling,A., Stanton,J.A., Rand,C. and Matisoo-Smith,E.A. PLoS ONE 8 (8), E71956 (2013) |
| KF163089 (Cattle and Aurochs) | Roche/454 Life Sciences GS Junior instrument | Phantom mutations: 215+TCC, 234+T, 16200+A. | T1b1b1a2_KF163089 | 169, 179, 215+TCC, 234+T, 587+C, 1481, 1599-1600d, 2536A, 7542, 9682C, 10460, 12468, 12469, 13201, 13310C, 14523T, 16022, 16050, 16056, 16084, 16085, 16113, 16200+A, 16255 | | T1b1b1a2_KF163089:215+TCC, 234+T, 1599-1600d, 16084, 16085, 16200+A | Horsburgh,K.A., Prost,S., Gosling,A., Stanton,J.A., Rand,C. and Matisoo-Smith,E.A. PLoS ONE 8 (8), E71956 (2013) |
| KF163090 (Cattle and Aurochs) | Roche/454 Life Sciences GS Junior instrument | Phantom mutations: 215+TCC, 234+T, 16200+A. | T1b1b1b_KF163090 | 169, 190, 215+TCC, 234+T, 497, 587+C, 1599-1600d, 2536A, 3270, 3481, 7542, 8710, 9682C, 12468, 12469, 12525, 13201, 13300, 13310C, 14523T, 16022, 16050, 16056, 16084, 16085, 16113, 16200+A, 16255, 16316 | | T1b1b1b_KF163090:215+TCC, 234+T, 1599-1600d, 16084, 16085, 16200+A | Horsburgh,K.A., Prost,S., Gosling,A., Stanton,J.A., Rand,C. and Matisoo-Smith,E.A. PLoS ONE 8 (8), E71956 (2013) |
| KF163091 (Cattle and Aurochs) | Roche/454 Life Sciences GS Junior instrument | Phantom mutations: 215+TCC, 234+T, 16200+A. | T1b1b1b_KF163070_KF163091 | 169, 190, 221+CCC, 234+T, 497, 587+C, 1599-1600d, 2536A, 7542, 7622, 8710, 9682C, 12468, 12469, 12525, 13201, 13260, 13300, 13310C, 14523T, 16022, 16050, 16056, 16113, 16200+A, 16255, 16316 | | T1b1b1b_KF163070_KF163091:221+CCC, 234+T, 1599-1600d, 16200+A | Horsburgh,K.A., Prost,S., Gosling,A., Stanton,J.A., Rand,C. and Matisoo-Smith,E.A. PLoS ONE 8 (8), E71956 (2013) |
| KF163092 (Cattle and Aurochs) | Roche/454 Life Sciences GS Junior | Phantom mutations: 215+TCC, 234+T, | T1b1_KF163092 | 169, 215+TCC, 234+T, 587+C, 1599-1600d, 2536A, 7542, 9501, 9682C, 13260, 13310C, 16022, 16050, 16084, 16085, 16113, 16200+A, 16255 | | T1b1_KF163092:215+TCC, 234+T, 1599-1600d, 16084, 16085, 16200+A | Horsburgh,K.A., Prost,S., Gosling,A., Stanton,J.A., Rand,C. and Matisoo-Smith,E.A. PLoS |
| | instrument | 16200+A. | | | | | ONE 8 (8), E71956 (2013) |
| KF163093 (Cattle and Aurochs) | Roche/454 Life Sciences GS Junior instrument | Phantom mutations: 215+TCC, 234+T, 16200+A. | T1_KF163093 | 169, 215+TCC, 234+T, 299, 587+C, 1599-1600d, 2536A, 3682, 9201, 9682C, 12468, 12469, 13310C, 16050, 16113, 16200+A, 16247, 16255, 16260 | | T1_KF163093:215+TCC, 234+T, 1599-1600d, 16200+A | Horsburgh,K.A., Prost,S., Gosling,A., Stanton,J.A., Rand,C. and Matisoo-Smith,E.A. PLoS ONE 8 (8), E71956 (2013) |
| KF163094 (Cattle and Aurochs) | Roche/454 Life Sciences GS Junior instrument | Phantom mutations: 215+TCC, 234+T, 16200+A. | T1b1b1a2_KF163089 | 169, 179, 215+TCC, 234+T, 587+C, 1481, 1599-1600d, 2536A, 3153Y, 7542, 9682C, 10460, 12468, 12469, 13201, 13310C, 14523T, 16022, 16050, 16056, 16113, 16200+A, 16255 | | T1b1b1a2_KF163089:215+TCC, 234+T, 1599-1600d, 3153, 16200+A | Horsburgh,K.A., Prost,S., Gosling,A., Stanton,J.A., Rand,C. and Matisoo-Smith,E.A. PLoS ONE 8 (8), E71956 (2013) |
| **Sequence** | **Platform** | **Potential Error** | **Haplogroup_Most Matched Sequence** | **Variants** | **Missing Diagnostic Mutations** | **Private** | **Sources** |
| HQ439441 (Horse and Przewalski's horse) | Illumina/Solexa Genome Analyzer II | Phantom mutaions: 358d, 2226+C, 5239+A, 5274+G, 15385+T. | A1_HQ439441 | 358d, 2226+C, 5239+A, 5274+G, 6784, 8807, 15385+T, 15492, 15823, 16126N, 16127N, 16128N, 16129N, 16130N, 16131N, 16132N, 16133N, 16134N, 16135N, 16136N, 16137N, 16138N, 16139N, 16140N, 16141N, 16142N, 16143N, 16144N, 16145N, 16146N, 16147N, 16148N, 16149N, 16150N, 16151N, 16152N, 16153N, 16154N, 16155N, 16156N, 16157N, 16158N, 16159N, 16160N, 16161N, 16162N, 16163N, 16164N, 16165N, 16166N, 16167N, 16168N, 16169N, 16170N, 16171N, 16172N, 16173N, 16174N, 16175N, 16176N, 16177N, 16178N, 16179N, 16180N, 16181N, 16182N, 16183N, 16184N, 16185N, 16186N, 16187N, 16188N, 16189N, 16190N, 16191N, 16192N, 16193N, 16194N, 16195N, 16196N, 16197N, 16198N, 16199N, 16200N, 16201N, 16202N, 16203N, 16204N, 16205N, 16206N, 16207N, 16208N, 16209N, 16210N, 16211N, 16212N, 16213N, 16214N, 16215N, 16216N, 16217N, 16218N, 16219N, 16220N, 16221N, 16222N, 16223N, 16224N, 16225N, 16226N, 16227N, 16228N, 16229N, 16230N, 16231N, 16232N, 16233N, 16234N, 16235N, 16236N, 16237N, 16238N, 16239N, 16240N, 16241N, 16242N, 16243N, 16244N, 16245N, 16246N, 16247N, 16248N, 16249N, 16250N, 16251N, 16252N, 16253N, 16254N, 16255N, 16256N, 16257N, 16258N, 16259N, 16260N, 16261N, 16262N, 16263N, 16264N, 16265N, 16266N, 16267N, 16268N, 16269N, 16270N, 16271N, 16272N, 16273N, 16274N, 16275N, 16276N, 16277N, 16278N, 16279N, 16280N, 16281N, 16282N, 16283N, 16284N, 16285N, 16286N, 16287N, 16288N, 16289N, 16290N, 16291N, 16292N, 16293N, 16294N, 16295N, 16296N, 16297N, 16298N, 16299N, 16300N, 16301N, 16302N, 16303N, 16304N, 16305N, 16306N, 16307N, 16308N, 16309N, 16310N, 16311N, 16312N, 16313N, 16314N, 16315N, 16316N, 16317N, 16318N, 16319N, 16320N, 16321N, 16322N, 16323N, 16324N, 16325N, 16326N, 16327N, 16328N, 16329N, 16330N, 16331N, 16332N, 16333N, 16334N, 16335N, 16336N, 16337N, 16338N, 16339N, 16340N, 16341N, 16342N, 16343N, 16344N, 16345N, 16346N, 16347N, 16348N, 16349N, 16350N, 16351N, 16352N, 16353N, 16354N, 16355N, 16356N, 16357N | | A1_HQ439441:358d, 2226+C, 5239+A, 5274+G, 15385+T, 16126N, 16127N, 16128N, 16129N, 16130N, 16131N, 16132N, 16133N, 16134N, 16135N, 16136N, 16137N, 16138N, 16139N, 16140N, 16141N, 16142N, 16143N, 16144N, 16145N, 16146N, 16147N, 16148N, 16149N, 16150N, 16151N, 16152N, 16153N, 16154N, 16155N, 16156N, 16157N, 16158N, 16159N, 16160N, 16161N, 16162N, 16163N, 16164N, 16165N, 16166N, 16167N, 16168N, 16169N, 16170N, 16171N, 16172N, 16173N, 16174N, 16175N, 16176N, 16177N, 16178N, 16179N, 16180N, 16181N, 16182N, 16183N, 16184N, 16185N, 16186N, 16187N, 16188N, 16189N, 16190N, 16191N, 16192N, 16193N, 16194N, 16195N, 16196N, 16197N, 16198N, 16199N, 16200N, 16201N, 16202N, 16203N, 16204N, 16205N, 16206N, 16207N, 16208N, 16209N, 16210N, 16211N, 16212N, 16213N, 16214N, 16215N, 16216N, 16217N, 16218N, 16219N, 16220N, 16221N, 16222N, 16223N, 16224N, 16225N, 16226N, 16227N, 16228N, 16229N, 16230N, 16231N, 16232N, 16233N, 16234N, 16235N, 16236N, 16237N, 16238N, 16239N, 16240N, 16241N, 16242N, 16243N, 16244N, 16245N, 16246N, 16247N, 16248N, 16249N, 16250N, 16251N, 16252N, 16253N, 16254N, 16255N, 16256N, 16257N, 16258N, 16259N, 16260N, 16261N, 16262N, 16263N, 16264N, 16265N, 16266N, 16267N, 16268N, 16269N, 16270N, 16271N, 16272N, 16273N, 16274N, 16275N, 16276N, 16277N, 16278N, 16279N, 16280N, 16281N, 16282N, 16283N, 16284N, 16285N, 16286N, 16287N, 16288N, 16289N, 16290N, 16291N, 16292N, 16293N, 16294N, 16295N, 16296N, 16297N, 16298N, 16299N, 16300N, 16301N, 16302N, 16303N, 16304N, 16305N, 16306N, 16307N, 16308N, 16309N, 16310N, 16311N, 16312N, 16313N, 16314N, 16315N, 16316N, 16317N, 16318N, 16319N, 16320N, 16321N, 16322N, 16323N, 16324N, 16325N, 16326N, 16327N, 16328N, 16329N, 16330N, 16331N, 16332N, 16333N, 16334N, 16335N, 16336N, 16337N, 16338N, 16339N, 16340N, 16341N, 16342N, 16343N, 16344N, 16345N, 16346N, 16347N, 16348N, 16349N, 16350N, 16351N, 16352N, 16353N, 16354N, 16355N, 16356N, 16357N | Lippold,S., Matzke,N., Reissman,M., Burbano,H. and Hofreiter,M. BMC Evolutionary Biology. 328 (11), 1471-2148 (2011) |
| HQ439442 (Horse and Przewalski's horse) | Illumina/Solexa Genome Analyzer II | Phantom mutaions: 358d, 1387A, 2227+T, 5239+A, 5277+A, 15385+T. | M1 | 158, 356d, 427, 961, 1387A, 1609T, 2227+T, 2339A, 2788, 3070, 3100, 3475, 3800, 4062, 4526, 4536, 4599G, 4605, 4646, 4669, 4898, 5103, 5239+A, 5277+A, 5527, 5827, 5884, 6004, 6076, 6307, 6712, 6784, 7001, 7432, 7666, 7900, 8005, 8043, 8076, 8150, 8175, 8238, 8358, 8556T, 8565, 8798, 9086, 9332, 9540, 10110, 10173, 10214, 10292, 10376, 10448, 10460, 10646, 10859, 11240, 11394, 11492, 11543, 11842, 11879, 11966, 12029, 12095, 12332, 12767, 13049, 13100, 13333, 13356, 13502, 13615, 13629, 13720, 13920, 13933, 14422, 14626, 14671, 14803, 14815, 15052A, 15133, 15342, 15385+T, 15492, 15599, 15614, 15656, 15717, 15768, 15803, 15824, 15866, 15953, 16065, 16077, 16118, 16126N, 16127N, 16128N, 16129N, 16130N, 16131N, 16132N, 16133N, 16134N, 16135N, 16136N, 16137N, 16138N, 16139N, 16140N, 16141N, 16142N, 16143N, 16144N, 16145N, 16146N, 16147N, 16148N, 16149N, 16150N, 16151N, 16152N, 16153N, 16154N, 16155N, 16156N, 16157N, 16158N, 16159N, 16160N, 16161N, 16162N, 16163N, 16164N, 16165N, 16166N, 16167N, 16168N, 16169N, 16170N, 16171N, 16172N, 16173N, 16174N, 16175N, 16176N, 16177N, 16178N, 16179N, 16180N, 16181N, 16182N, 16183N, 16184N, 16185N, 16186N, 16187N, 16188N, 16189N, 16190N, 16191N, 16192N, 16193N, 16194N, 16195N, 16196N, 16197N, 16198N, 16199N, 16200N, 16201N, 16202N, 16203N, 16204N, 16205N, 16206N, 16207N, 16208N, 16209N, 16210N, 16211N, 16212N, 16213N, 16214N, 16215N, 16216N, 16217N, 16218N, 16219N, 16220N, 16221N, 16222N, 16223N, 16224N, 16225N, 16226N, 16227N, 16228N, 16229N, 16230N, 16231N, 16232N, 16233N, 16234N, 16235N, 16236N, 16237N, 16238N, 16239N, 16240N, 16241N, 16242N, 16243N, 16244N, 16245N, 16246N, 16247N, 16248N, 16249N, 16250N, 16251N, 16252N, 16253N, 16254N, 16255N, 16256N, 16257N, 16258N, 16259N, 16260N, 16261N, 16262N, 16263N, 16264N, 16265N, 16266N, 16267N, 16268N, 16269N, 16270N, 16271N, 16272N, 16273N, 16274N, 16275N, 16276N, 16277N, 16278N, 16279N, 16280N, 16281N, 16282N, 16283N, 16284N, 16285N, 16286N, 16287N, 16288N, 16289N, 16290N, 16291N, 16292N, 16293N, 16294N, 16295N, 16296N, 16297N, 16298N, 16299N, 16300N, 16301N, 16302N, 16303N, 16304N, 16305N, 16306N, 16307N, 16308N, 16309N, 16310N, 16311N, 16312N, 16313N, 16314N, 16315N, 16316N, 16317N, 16318N, 16319N, 16320N, 16321N, 16322N, 16323N, 16324N, 16325N, 16326N, 16327N, 16328N, 16329N, 16330N, 16331N, 16332N, 16333N, 16334N, 16335N, 16336N, 16337N, 16338N, 16339N, 16340N, 16341N, 16342N, 16343N, 16344N, 16345N, 16346N, 16347N, 16348N, 16349N, 16350N, 16351N, 16352N, 16353N, 16354N, 16355N, 16356N, 16357N, 16368, 16540A, 16543, 16556, 16626 | M1:356 | M1:356d, 2227+T, 5239+A, 5277+A, 15385+T, 16126N, 16127N, 16128N, 16129N, 16130N, 16131N, 16132N, 16133N, 16134N, 16135N, 16136N, 16137N, 16138N, 16139N, 16140N, 16141N, 16142N, 16143N, 16144N, 16145N, 16146N, 16147N, 16148N, 16149N, 16150N, 16151N, 16152N, 16153N, 16154N, 16155N, 16156N, 16157N, 16158N, 16159N, 16160N, 16161N, 16162N, 16163N, 16164N, 16165N, 16166N, 16167N, 16168N, 16169N, 16170N, 16171N, 16172N, 16173N, 16174N, 16175N, 16176N, 16177N, 16178N, 16179N, 16180N, 16181N, 16182N, 16183N, 16184N, 16185N, 16186N, 16187N, 16188N, 16189N, 16190N, 16191N, 16192N, 16193N, 16194N, 16195N, 16196N, 16197N, 16198N, 16199N, 16200N, 16201N, 16202N, 16203N, 16204N, 16205N, 16206N, 16207N, 16208N, 16209N, 16210N, 16211N, 16212N, 16213N, 16214N, 16215N, 16216N, 16217N, 16218N, 16219N, 16220N, 16221N, 16222N, 16223N, 16224N, 16225N, 16226N, 16227N, 16228N, 16229N, 16230N, 16231N, 16232N, 16233N, 16234N, 16235N, 16236N, 16237N, 16238N, 16239N, 16240N, 16241N, 16242N, 16243N, 16244N, 16245N, 16246N, 16247N, 16248N, 16249N, 16250N, 16251N, 16252N, 16253N, 16254N, 16255N, 16256N, 16257N, 16258N, 16259N, 16260N, 16261N, 16262N, 16263N, 16264N, 16265N, 16266N, 16267N, 16268N, 16269N, 16270N, 16271N, 16272N, 16273N, 16274N, 16275N, 16276N, 16277N, 16278N, 16279N, 16280N, 16281N, 16282N, 16283N, 16284N, 16285N, 16286N, 16287N, 16288N, 16289N, 16290N, 16291N, 16292N, 16293N, 16294N, 16295N, 16296N, 16297N, 16298N, 16299N, 16300N, 16301N, 16302N, 16303N, 16304N, 16305N, 16306N, 16307N, 16308N, 16309N, 16310N, 16311N, 16312N, 16313N, 16314N, 16315N, 16316N, 16317N, 16318N, 16319N, 16320N, 16321N, 16322N, 16323N, 16324N, 16325N, 16326N, 16327N, 16328N, 16329N, 16330N, 16331N, 16332N, 16333N, 16334N, 16335N, 16336N, 16337N, 16338N, 16339N, 16340N, 16341N, 16342N, 16343N, 16344N, 16345N, 16346N, 16347N, 16348N, 16349N, 16350N, 16351N, 16352N, 16353N, 16354N, 16355N, 16356N, 16357N | Lippold,S., Matzke,N., Reissman,M., Burbano,H. and Hofreiter,M. BMC Evolutionary Biology. 328 (11), 1471-2148 (2011) |
| HQ439443 (Horse and Przewalski's horse) | Illumina/Solexa Genome Analyzer II | Phantom mutaions: 358d, 1387A, 2227+T, 5239+A, 5277+A, 15385+T. | G1'2'3_HQ439443 | 158, 222, 356d, 382, 387, 416, 1387A, 2227+T, 2788, 2940, 3053, 3576, 4062, 4646, 4669, 4830, 5239+A, 5277+A, 5498, 5830, 5881, 5884, 6004, 6307, 6688, 6784, 7001, 8005, 8037, 9239, 9402, 9669, 9741A, 10214, 10376, 10471, 11165, 11210, 11240, 11543, 11552, 11842, 12767, 12860, 13049, 13175, 13223, 13333, 13502, 14158, 14626, 14734, 15385+T, 15492, 15539, 15594, 15599, 15632, 15647, 15663, 15700, 15717, 15867, 15868, 16100, 16110, 16126N, 16127N, 16128N, 16129N, 16130N, 16131N, 16132N, 16133N, 16134N, 16135N, 16136N, 16137N, 16138N, 16139N, 16140N, 16141N, 16142N, 16143N, 16144N, 16145N, 16146N, 16147N, 16148N, 16149N, 16150N, 16151N, 16152N, 16153N, 16154N, 16155N, 16156N, 16157N, 16158N, 16159N, 16160N, 16161N, 16162N, 16163N, 16164N, 16165N, 16166N, 16167N, 16168N, 16169N, 16170N, 16171N, 16172N, 16173N, 16174N, 16175N, 16176N, 16177N, 16178N, 16179N, 16180N, 16181N, 16182N, 16183N, 16184N, 16185N, 16186N, 16187N, 16188N, 16189N, 16190N, 16191N, 16192N, 16193N, 16194N, 16195N, 16196N, 16197N, 16198N, 16199N, 16200N, 16201N, 16202N, 16203N, 16204N, 16205N, 16206N, 16207N, 16208N, 16209N, 16210N, 16211N, 16212N, 16213N, 16214N, 16215N, 16216N, 16217N, 16218N, 16219N, 16220N, 16221N, 16222N, 16223N, 16224N, 16225N, 16226N, 16227N, 16228N, 16229N, 16230N, 16231N, 16232N, 16233N, 16234N, 16235N, 16236N, 16237N, 16238N, 16239N, | G1'2'3_HQ439443:356 | G1'2'3_HQ439443:356d, 2227+T, 5239+A, 5277+A, 15385+T, 15868, 16100, 16126N, 16127N, 16128N, 16129N, 16130N, 16131N, 16132N, 16133N, 16134N, 16135N, 16136N, 16137N, 16138N, 16139N, 16140N, 16141N, 16142N, 16143N, 16144N, 16145N, 16146N, 16147N, 16148N, 16149N, 16150N, 16151N, 16152N, 16153N, 16154N, 16155N, 16156N, 16157N, 16158N, 16159N, 16160N, 16161N, 16162N, 16163N, 16164N, 16165N, 16166N, 16167N, 16168N, 16169N, 16170N, 16171N, 16172N, 16173N, 16174N, 16175N, 16176N, 16177N, 16178N, 16179N, 16180N, 16181N, 16182N, 16183N, 16184N, 16185N, 16186N, 16187N, 16188N, 16189N, 16190N, 16191N, 16192N, 16193N, 16194N, 16195N, 16196N, 16197N, 16198N, 16199N, 16200N, 16201N, 16202N, 16203N, 16204N, 16205N, 16206N, 16207N, 16208N, 16209N, 16210N, 16211N, 16212N, 16213N, 16214N, 16215N, 16216N, 16217N, 16218N, 16219N, 16220N, 16221N, 16222N, 16223N, 16224N, 16225N, 16226N, 16227N, 16228N, 16229N, 16230N, 16231N, 16232N, 16233N, 16234N, 16235N, 16236N, 16237N, 16238N, | Lippold,S., Matzke,N., Reissman,M., Burbano,H. and Hofreiter,M. BMC Evolutionary Biology. 328 (11), 1471-2148 (2011) |
| | | | | 16240N, 16241N, 16242N, 16243N, 16244N, 16245N, 16246N, 16247N, 16248N, 16249N, 16250N, 16251N, 16252N, 16253N, 16254N, 16255N, 16256N, 16257N, 16258N, 16259N, 16260N, 16261N, 16262N, 16263N, 16264N, 16265N, 16266N, 16267N, 16268N, 16269N, 16270N, 16271N, 16272N, 16273N, 16274N, 16275N, 16276N, 16277N, 16278N, 16279N, 16280N, 16281N, 16282N, 16283N, 16284N, 16285N, 16286N, 16287N, 16288N, 16289N, 16290N, 16291N, 16292N, 16293N, 16294N, 16295N, 16296N, 16297N, 16298N, 16299N, 16300N, 16301N, 16302N, 16303N, 16304N, 16305N, 16306N, 16307N, 16308N, 16309N, 16310N, 16311N, 16312N, 16313N, 16314N, 16315N, 16316N, 16317N, 16318N, 16319N, 16320N, 16321N, 16322N, 16323N, 16324N, 16325N, 16326N, 16327N, 16328N, 16329N, 16330N, 16331N, 16332N, 16333N, 16334N, 16335N, 16336N, 16337N, 16338N, 16339N, 16340N, 16341N, 16342N, 16343N, 16344N, 16345N, 16346N, 16347N, 16348N, 16349N, 16350N, 16351N, 16352N, 16353N, 16354N, 16355N, 16356N, 16357N, 16368 | | 16239N, 16240N, 16241N, 16242N, 16243N, 16244N, 16245N, 16246N, 16247N, 16248N, 16249N, 16250N, 16251N, 16252N, 16253N, 16254N, 16255N, 16256N, 16257N, 16258N, 16259N, 16260N, 16261N, 16262N, 16263N, 16264N, 16265N, 16266N, 16267N, 16268N, 16269N, 16270N, 16271N, 16272N, 16273N, 16274N, 16275N, 16276N, 16277N, 16278N, 16279N, 16280N, 16281N, 16282N, 16283N, 16284N, 16285N, 16286N, 16287N, 16288N, 16289N, 16290N, 16291N, 16292N, 16293N, 16294N, 16295N, 16296N, 16297N, 16298N, 16299N, 16300N, 16301N, 16302N, 16303N, 16304N, 16305N, 16306N, 16307N, 16308N, 16309N, 16310N, 16311N, 16312N, 16313N, 16314N, 16315N, 16316N, 16317N, 16318N, 16319N, 16320N, 16321N, 16322N, 16323N, 16324N, 16325N, 16326N, 16327N, 16328N, 16329N, 16330N, 16331N, 16332N, 16333N, 16334N, 16335N, 16336N, 16337N, 16338N, 16339N, 16340N, 16341N, 16342N, 16343N, 16344N, 16345N, 16346N, 16347N, 16348N, 16349N, 16350N, 16351N, 16352N, 16353N, 16354N, 16355N, 16356N, 16357N | |
| HQ439444 (Horse and Przewalski's horse) | Illumina/Solexa Genome Analyzer II | Phantom mutaions: 358d, 1387A, 2227+T, 5239+A, 5277+A, 15385+T. | P4a_HQ439444 | 158, 356d, 739, 860, 961, 1383, 1387A, 1684C, 2227+T, 2788, 3070, 3259T, 3271, 3557, 3616, 3800, 3942, 4062, 4536, 4605, 4646, 4669, 5239+A, 5277+A, 5286, 5527, 5827, 5884, 6004, 6307, 6505, 6529, 6712, 6784, 7001, 7243, 7427, 7612T, 7666, 7898, 7900, 8005, 8076, 8150, 8238, 8358, 8361, 8556T, 8565, 8855, 9053, 9086, 9239, 9775, 10110, 10173, 10214, 10292, 10376, 10448, 10859, 11240, 11378, 11394, 11424, 11492, 11543, 11842, 11879, 11966, 12167, 12230, 12287, 12299, 12332, 12404, 12767, 13049, 13333, 13370, 13463, 13466, 13629, 13920, 13933, 14626, 14803, 15202, 15342, 15385+T, 15492, 15594, 15599, 15601, 15664, 15700, 15717, 15768, 15774, 15806, 15953, 16035, 16118, 16126N, 16127N, 16128N, 16129N, 16130N, 16131N, 16132N, 16133N, 16134N, 16135N, 16136N, 16137N, 16138N, 16139N, 16140N, 16141N, 16142N, 16143N, 16144N, 16145N, 16146N, 16147N, 16148N, 16149N, 16150N, 16151N, 16152N, 16153N, 16154N, 16155N, 16156N, 16157N, 16158N, 16159N, 16160N, 16161N, 16162N, 16163N, 16164N, 16165N, 16166N, 16167N, 16168N, 16169N, 16170N, 16171N, 16172N, 16173N, 16174N, 16175N, 16176N, 16177N, 16178N, 16179N, 16180N, 16181N, 16182N, 16183N, 16184N, 16185N, 16186N, 16187N, 16188N, 16189N, 16190N, 16191N, 16192N, 16193N, 16194N, 16195N, 16196N, 16197N, 16198N, 16199N, 16200N, 16201N, 16202N, 16203N, 16204N, 16205N, 16206N, 16207N, 16208N, 16209N, 16210N, 16211N, 16212N, 16213N, 16214N, 16215N, 16216N, 16217N, 16218N, 16219N, 16220N, 16221N, 16222N, 16223N, 16224N, 16225N, 16226N, 16227N, 16228N, 16229N, 16230N, 16231N, 16232N, 16233N, 16234N, 16235N, 16236N, 16237N, 16238N, 16239N, 16240N, 16241N, 16242N, 16243N, 16244N, 16245N, 16246N, 16247N, 16248N, 16249N, 16250N, 16251N, 16252N, 16253N, 16254N, 16255N, 16256N, 16257N, 16258N, 16259N, 16260N, 16261N, 16262N, 16263N, 16264N, 16265N, 16266N, 16267N, 16268N, 16269N, 16270N, 16271N, 16272N, 16273N, 16274N, 16275N, 16276N, 16277N, 16278N, 16279N, 16280N, 16281N, 16282N, 16283N, 16284N, 16285N, 16286N, 16287N, 16288N, 16289N, 16290N, 16291N, 16292N, 16293N, 16294N, 16295N, 16296N, 16297N, 16298N, 16299N, 16300N, 16301N, 16302N, 16303N, 16304N, 16305N, 16306N, 16307N, 16308N, 16309N, 16310N, 16311N, 16312N, 16313N, 16314N, 16315N, 16316N, 16317N, 16318N, 16319N, 16320N, 16321N, 16322N, 16323N, 16324N, 16325N, 16326N, 16327N, 16328N, 16329N, 16330N, 16331N, 16332N, 16333N, 16334N, 16335N, 16336N, 16337N, 16338N, 16339N, 16340N, 16341N, 16342N, 16343N, 16344N, 16345N, 16346N, 16347N, 16348N, 16349N, 16350N, 16351N, 16352N, 16353N, 16354N, 16355N, 16356N, 16357N, 16368, 16540A, 16626 | P4a_HQ439444:356 | P4a_HQ439444:356d, 2227+T, 5239+A, 5277+A, 15385+T, 16126N, 16127N, 16128N, 16129N, 16130N, 16131N, 16132N, 16133N, 16134N, 16135N, 16136N, 16137N, 16138N, 16139N, 16140N, 16141N, 16142N, 16143N, 16144N, 16145N, 16146N, 16147N, 16148N, 16149N, 16150N, 16151N, 16152N, 16153N, 16154N, 16155N, 16156N, 16157N, 16158N, 16159N, 16160N, 16161N, 16162N, 16163N, 16164N, 16165N, 16166N, 16167N, 16168N, 16169N, 16170N, 16171N, 16172N, 16173N, 16174N, 16175N, 16176N, 16177N, 16178N, 16179N, 16180N, 16181N, 16182N, 16183N, 16184N, 16185N, 16186N, 16187N, 16188N, 16189N, 16190N, 16191N, 16192N, 16193N, 16194N, 16195N, 16196N, 16197N, 16198N, 16199N, 16200N, 16201N, 16202N, 16203N, 16204N, 16205N, 16206N, 16207N, 16208N, 16209N, 16210N, 16211N, 16212N, 16213N, 16214N, 16215N, 16216N, 16217N, 16218N, 16219N, 16220N, 16221N, 16222N, 16223N, 16224N, 16225N, 16226N, 16227N, 16228N, 16229N, 16230N, 16231N, 16232N, 16233N, 16234N, 16235N, 16236N, 16237N, 16238N, 16239N, 16240N, 16241N, 16242N, 16243N, 16244N, 16245N, 16246N, 16247N, 16248N, 16249N, 16250N, 16251N, 16252N, 16253N, 16254N, 16255N, 16256N, 16257N, 16258N, 16259N, 16260N, 16261N, 16262N, 16263N, 16264N, 16265N, 16266N, 16267N, 16268N, 16269N, 16270N, 16271N, 16272N, 16273N, 16274N, 16275N, 16276N, 16277N, 16278N, 16279N, 16280N, 16281N, 16282N, 16283N, 16284N, 16285N, 16286N, 16287N, 16288N, 16289N, 16290N, 16291N, 16292N, 16293N, 16294N, 16295N, 16296N, 16297N, 16298N, 16299N, 16300N, 16301N, 16302N, 16303N, 16304N, 16305N, 16306N, 16307N, 16308N, 16309N, 16310N, 16311N, 16312N, 16313N, 16314N, 16315N, 16316N, 16317N, 16318N, 16319N, 16320N, 16321N, 16322N, 16323N, 16324N, 16325N, 16326N, 16327N, 16328N, 16329N, 16330N, 16331N, 16332N, 16333N, 16334N, 16335N, 16336N, 16337N, 16338N, 16339N, 16340N, 16341N, 16342N, 16343N, 16344N, 16345N, 16346N, 16347N, 16348N, 16349N, 16350N, 16351N, 16352N, 16353N, 16354N, 16355N, 16356N, 16357N | Lippold,S., Matzke,N., Reissman,M., Burbano,H. and Hofreiter,M. BMC Evolutionary Biology. 328 (11), 1471-2148 (2011) |
| HQ439445 (Horse and Przewalski's horse) | Illumina/Solexa Genome Analyzer II | Phantom mutaions: 358d, 1387G, 2226+C, 5239+A, 5277+A, 15382+C. | P4a_HQ439445 | 158, 172T, 356d, 739, 860, 961, 1383, 1387G, 1684C, 2226+C, 2788, 3070, 3259T, 3271, 3557, 3616, 3800, 3942, 4062, 4536, 4605, 4646, 4669, 5239+A, 5277+A, 5527, 5827, 5884, 6004, 6307, 6505, 6529, 6712, 6784, 7001, 7243, 7427, 7612T, 7666, 7898, 7900, 8005, 8076, 8150, 8238, 8358, 8361, 8556T, 8565, 8855, 9053, 9086, 9239, 9696, 9775, 10110, 10173, 10214, 10292, 10376, 10448, 10859, 11240, 11378, 11394, 11424, 11492, 11543, 11842, 11879, 11966, 12167, 12230, 12287, 12332, 12404, 12767, 13049, 13333, 13370, 13463, 13466, 13629, 13920, 13933, 14193T, 14626, 14803, 14825, 15202, 15342, 15382+C, 15492, 15594, 15599, 15601, 15664, 15700, 15717, 15768, 15774, 15806, 15953, 16035, 16054, 16118, 16126N, 16127N, 16128N, 16129N, 16130N, 16131N, 16132N, 16133N, 16134N, 16135N, 16136N, 16137N, 16138N, 16139N, 16140N, 16141N, 16142N, 16143N, 16144N, 16145N, 16146N, 16147N, 16148N, 16149N, 16150N, 16151N, 16152N, 16153N, 16154N, 16155N, 16156N, 16157N, 16158N, 16159N, 16160N, 16161N, 16162N, 16163N, 16164N, 16165N, 16166N, 16167N, 16168N, 16169N, 16170N, 16171N, 16172N, 16173N, 16174N, 16175N, 16176N, 16177N, 16178N, 16179N, 16180N, 16181N, 16182N, 16183N, 16184N, 16185N, 16186N, 16187N, 16188N, 16189N, 16190N, 16191N, 16192N, 16193N, 16194N, 16195N, 16196N, 16197N, 16198N, 16199N, 16200N, 16201N, 16202N, 16203N, 16204N, 16205N, 16206N, 16207N, 16208N, 16209N, 16210N, 16211N, 16212N, 16213N, 16214N, 16215N, 16216N, 16217N, 16218N, 16219N, 16220N, 16221N, 16222N, 16223N, 16224N, 16225N, 16226N, 16227N, 16228N, 16229N, 16230N, 16231N, 16232N, 16233N, 16234N, 16235N, 16236N, 16237N, 16238N, 16239N, 16240N, 16241N, 16242N, 16243N, 16244N, 16245N, 16246N, 16247N, 16248N, 16249N, 16250N, 16251N, 16252N, 16253N, 16254N, 16255N, 16256N, 16257N, 16258N, 16259N, 16260N, 16261N, 16262N, 16263N, 16264N, 16265N, 16266N, 16267N, 16268N, 16269N, 16270N, 16271N, 16272N, 16273N, 16274N, 16275N, 16276N, 16277N, 16278N, 16279N, 16280N, 16281N, 16282N, 16283N, 16284N, 16285N, 16286N, 16287N, 16288N, 16289N, 16290N, 16291N, 16292N, 16293N, 16294N, 16295N, 16296N, 16297N, 16298N, 16299N, 16300N, 16301N, 16302N, 16303N, 16304N, 16305N, 16306N, 16307N, 16308N, 16309N, 16310N, 16311N, 16312N, 16313N, 16314N, 16315N, 16316N, 16317N, 16318N, 16319N, 16320N, 16321N, 16322N, 16323N, 16324N, 16325N, 16326N, 16327N, 16328N, 16329N, 16330N, 16331N, 16332N, 16333N, 16334N, 16335N, 16336N, 16337N, 16338N, 16339N, 16340N, 16341N, 16342N, 16343N, 16344N, 16345N, 16346N, 16347N, 16348N, 16349N, 16350N, 16351N, 16352N, 16353N, 16354N, 16355N, 16356N, 16357N, 16368, 16398A, 16540A, 16626 | P4a_HQ439445:356 | P4a_HQ439445:356d, 2226+C, 5239+A, 5277+A, 15382+C, 16126N, 16127N, 16128N, 16129N, 16130N, 16131N, 16132N, 16133N, 16134N, 16135N, 16136N, 16137N, 16138N, 16139N, 16140N, 16141N, 16142N, 16143N, 16144N, 16145N, 16146N, 16147N, 16148N, 16149N, 16150N, 16151N, 16152N, 16153N, 16154N, 16155N, 16156N, 16157N, 16158N, 16159N, 16160N, 16161N, 16162N, 16163N, 16164N, 16165N, 16166N, 16167N, 16168N, 16169N, 16170N, 16171N, 16172N, 16173N, 16174N, 16175N, 16176N, 16177N, 16178N, 16179N, 16180N, 16181N, 16182N, 16183N, 16184N, 16185N, 16186N, 16187N, 16188N, 16189N, 16190N, 16191N, 16192N, 16193N, 16194N, 16195N, 16196N, 16197N, 16198N, 16199N, 16200N, 16201N, 16202N, 16203N, 16204N, 16205N, 16206N, 16207N, 16208N, 16209N, 16210N, 16211N, 16212N, 16213N, 16214N, 16215N, 16216N, 16217N, 16218N, 16219N, 16220N, 16221N, 16222N, 16223N, 16224N, 16225N, 16226N, 16227N, 16228N, 16229N, 16230N, 16231N, 16232N, 16233N, 16234N, 16235N, 16236N, 16237N, 16238N, 16239N, 16240N, 16241N, 16242N, 16243N, 16244N, 16245N, 16246N, 16247N, 16248N, 16249N, 16250N, 16251N, 16252N, 16253N, 16254N, 16255N, 16256N, 16257N, 16258N, 16259N, 16260N, 16261N, 16262N, 16263N, 16264N, 16265N, 16266N, 16267N, 16268N, 16269N, 16270N, 16271N, 16272N, 16273N, 16274N, 16275N, 16276N, 16277N, 16278N, 16279N, 16280N, 16281N, 16282N, 16283N, 16284N, 16285N, 16286N, 16287N, 16288N, 16289N, 16290N, 16291N, 16292N, 16293N, 16294N, 16295N, 16296N, 16297N, 16298N, 16299N, 16300N, 16301N, 16302N, 16303N, 16304N, 16305N, 16306N, 16307N, 16308N, 16309N, 16310N, 16311N, 16312N, 16313N, 16314N, 16315N, 16316N, 16317N, 16318N, 16319N, 16320N, 16321N, 16322N, 16323N, 16324N, 16325N, 16326N, 16327N, 16328N, 16329N, 16330N, 16331N, 16332N, 16333N, 16334N, 16335N, 16336N, 16337N, 16338N, 16339N, 16340N, 16341N, 16342N, 16343N, 16344N, 16345N, 16346N, 16347N, 16348N, 16349N, 16350N, 16351N, 16352N, 16353N, 16354N, 16355N, 16356N, 16357N | Lippold,S., Matzke,N., Reissman,M., Burbano,H. and Hofreiter,M. BMC Evolutionary Biology. 328 (11), 1471-2148 (2011) |
| HQ439446 (Horse and Przewalski's horse) | Illumina/Solexa Genome Analyzer II | Phantom mutaions: 358d, 1387A, 2227+T, 5239+A, 5277+A, 15385+T. | L1a1a_HQ439446 | 158, 356d, 961, 1375, 1387A, 2227+T, 2788, 2899, 3517, 3942, 4062, 4536, 4646, 4669, 5239+A, 5277+A, 5527, 5815, 5884, 6004, 6307, 6784, 7001, 7516, 7666, 7900, 8005, 8058, 8142A, 8301, 8319, 8358, 8565, 9239, 9684, 9951, 10110, 10214, 10292, 10376, 10421, 10613, 11240, 11543, 11682, 11693, 11842, 11879, 12119, 12200, 12767, 12896, 12950, 13040, 13049, 13333, 13520, 14803, 14995, 15313, 15385+T, 15491, 15492, 15493, 15531, 15599, 15600, 15646, 15717, 15768, 15867, 15868, 15953, 15971, 16065, 16100, 16126N, 16127N, 16128N, 16129N, 16130N, 16131N, 16132N, 16133N, 16134N, 16135N, 16136N, 16137N, 16138N, 16139N, 16140N, 16141N, 16142N, 16143N, 16144N, 16145N, 16146N, 16147N, 16148N, 16149N, 16150N, 16151N, 16152N, 16153N, 16154N, 16155N, 16156N, 16157N, 16158N, 16159N, 16160N, 16161N, 16162N, 16163N, 16164N, 16165N, 16166N, 16167N, 16168N, 16169N, 16170N, 16171N, 16172N, 16173N, 16174N, 16175N, 16176N, 16177N, 16178N, 16179N, 16180N, 16181N, 16182N, 16183N, 16184N, 16185N, 16186N, 16187N, 16188N, 16189N, 16190N, 16191N, 16192N, 16193N, 16194N, 16195N, 16196N, 16197N, 16198N, 16199N, 16200N, 16201N, 16202N, 16203N, 16204N, 16205N, 16206N, 16207N, 16208N, 16209N, 16210N, 16211N, 16212N, 16213N, 16214N, 16215N, 16216N, 16217N, 16218N, 16219N, 16220N, 16221N, 16222N, 16223N, 16224N, 16225N, 16226N, 16227N, 16228N, 16229N, 16230N, | L1a1a_HQ439446:356 | L1a1a_HQ439446:356d, 2227+T, 5239+A, 5277+A, 15385+T, 16126N, 16127N, 16128N, 16129N, 16130N, 16131N, 16132N, 16133N, 16134N, 16135N, 16136N, 16137N, 16138N, 16139N, 16140N, 16141N, 16142N, 16143N, 16144N, 16145N, 16146N, 16147N, 16148N, 16149N, 16150N, 16151N, 16152N, 16153N, 16154N, 16155N, 16156N, 16157N, 16158N, 16159N, 16160N, 16161N, 16162N, 16163N, 16164N, 16165N, 16166N, 16167N, 16168N, 16169N, 16170N, 16171N, 16172N, 16173N, 16174N, 16175N, 16176N, 16177N, 16178N, 16179N, 16180N, 16181N, 16182N, 16183N, 16184N, 16185N, 16186N, 16187N, 16188N, 16189N, 16190N, 16191N, 16192N, 16193N, 16194N, 16195N, 16196N, 16197N, 16198N, 16199N, 16200N, 16201N, 16202N, 16203N, 16204N, 16205N, 16206N, 16207N, 16208N, 16209N, 16210N, 16211N, 16212N, 16213N, 16214N, 16215N, 16216N, 16217N, 16218N, 16219N, 16220N, 16221N, 16222N, 16223N, 16224N, 16225N, 16226N, 16227N, 16228N, 16229N, 16230N, 16231N, 16232N, 16233N, 16234N, 16235N, 16236N, 16237N, 16238N, 16239N, | Lippold,S., Matzke,N., Reissman,M., Burbano,H. and Hofreiter,M. BMC Evolutionary Biology. 328 (11), 1471-2148 (2011) |
| | | | | 16231N, 16232N, 16233N, 16234N, 16235N, 16236N, 16237N, 16238N, 16239N, 16240N, 16241N, 16242N, 16243N, 16244N, 16245N, 16246N, 16247N, 16248N, 16249N, 16250N, 16251N, 16252N, 16253N, 16254N, 16255N, 16256N, 16257N, 16258N, 16259N, 16260N, 16261N, 16262N, 16263N, 16264N, 16265N, 16266N, 16267N, 16268N, 16269N, 16270N, 16271N, 16272N, 16273N, 16274N, 16275N, 16276N, 16277N, 16278N, 16279N, 16280N, 16281N, 16282N, 16283N, 16284N, 16285N, 16286N, 16287N, 16288N, 16289N, 16290N, 16291N, 16292N, 16293N, 16294N, 16295N, 16296N, 16297N, 16298N, 16299N, 16300N, 16301N, 16302N, 16303N, 16304N, 16305N, 16306N, 16307N, 16308N, 16309N, 16310N, 16311N, 16312N, 16313N, 16314N, 16315N, 16316N, 16317N, 16318N, 16319N, 16320N, 16321N, 16322N, 16323N, 16324N, 16325N, 16326N, 16327N, 16328N, 16329N, 16330N, 16331N, 16332N, 16333N, 16334N, 16335N, 16336N, 16337N, 16338N, 16339N, 16340N, 16341N, 16342N, 16343N, 16344N, 16345N, 16346N, 16347N, 16348N, 16349N, 16350N, 16351N, 16352N, 16353N, 16354N, 16355N, 16356N, 16357N, 16368, 16398A, 16626 | | 16240N, 16241N, 16242N, 16243N, 16244N, 16245N, 16246N, 16247N, 16248N, 16249N, 16250N, 16251N, 16252N, 16253N, 16254N, 16255N, 16256N, 16257N, 16258N, 16259N, 16260N, 16261N, 16262N, 16263N, 16264N, 16265N, 16266N, 16267N, 16268N, 16269N, 16270N, 16271N, 16272N, 16273N, 16274N, 16275N, 16276N, 16277N, 16278N, 16279N, 16280N, 16281N, 16282N, 16283N, 16284N, 16285N, 16286N, 16287N, 16288N, 16289N, 16290N, 16291N, 16292N, 16293N, 16294N, 16295N, 16296N, 16297N, 16298N, 16299N, 16300N, 16301N, 16302N, 16303N, 16304N, 16305N, 16306N, 16307N, 16308N, 16309N, 16310N, 16311N, 16312N, 16313N, 16314N, 16315N, 16316N, 16317N, 16318N, 16319N, 16320N, 16321N, 16322N, 16323N, 16324N, 16325N, 16326N, 16327N, 16328N, 16329N, 16330N, 16331N, 16332N, 16333N, 16334N, 16335N, 16336N, 16337N, 16338N, 16339N, 16340N, 16341N, 16342N, 16343N, 16344N, 16345N, 16346N, 16347N, 16348N, 16349N, 16350N, 16351N, 16352N, 16353N, 16354N, 16355N, 16356N, 16357N | |
| HQ439447 (Horse and Przewalski's horse) | Illumina/Solexa Genome Analyzer II | Phantom mutaions: 358d, 1387G, 2227+T, 5237+G, 5277+A, 15385+T. | P4a_HQ439447 | 158, 356d, 739, 860, 961, 1383, 1387G, 1684C, 2227+T, 2788, 3070, 3259T, 3271, 3557, 3616, 3800, 3942, 4062, 4536, 4605, 4646, 4669, 5237+G, 5277+A, 5527, 5827, 5884, 6004, 6307, 6505, 6529, 6712, 6784, 7001, 7243, 7427, 7612T, 7666, 7898, 7900, 8005, 8076, 8150, 8238, 8358, 8361, 8556T, 8565, 8855, 9053, 9086, 9239, 9703, 9775, 10110, 10173, 10214, 10292, 10376, 10448, 10859, 11240, 11378, 11394, 11424, 11492, 11543, 11842, 11879, 11966, 12167, 12230, 12287, 12332, 12404, 12767, 13049, 13333, 13370, 13463, 13466, 13629, 13920, 13933, 14626, 14803, 15202, 15342, 15385+T, 15492, 15594, 15599, 15601, 15664, 15700, 15717, 15768, 15774, 15806, 15953, 16035, 16118, 16126N, 16127N, 16128N, 16129N, 16130N, 16131N, 16132N, 16133N, 16134N, 16135N, 16136N, 16137N, 16138N, 16139N, 16140N, 16141N, 16142N, 16143N, 16144N, 16145N, 16146N, 16147N, 16148N, 16149N, 16150N, 16151N, 16152N, 16153N, 16154N, 16155N, 16156N, 16157N, 16158N, 16159N, 16160N, 16161N, 16162N, 16163N, 16164N, 16165N, 16166N, 16167N, 16168N, 16169N, 16170N, 16171N, 16172N, 16173N, 16174N, 16175N, 16176N, 16177N, 16178N, 16179N, 16180N, 16181N, 16182N, 16183N, 16184N, 16185N, 16186N, 16187N, 16188N, 16189N, 16190N, 16191N, 16192N, 16193N, 16194N, 16195N, 16196N, 16197N, 16198N, 16199N, 16200N, 16201N, 16202N, 16203N, 16204N, 16205N, 16206N, 16207N, 16208N, 16209N, 16210N, 16211N, 16212N, 16213N, 16214N, 16215N, 16216N, 16217N, 16218N, 16219N, 16220N, 16221N, 16222N, 16223N, 16224N, 16225N, 16226N, 16227N, 16228N, 16229N, 16230N, 16231N, 16232N, 16233N, 16234N, 16235N, 16236N, 16237N, 16238N, 16239N, 16240N, 16241N, 16242N, 16243N, 16244N, 16245N, 16246N, 16247N, 16248N, 16249N, 16250N, 16251N, 16252N, 16253N, 16254N, 16255N, 16256N, 16257N, 16258N, 16259N, 16260N, 16261N, 16262N, 16263N, 16264N, 16265N, 16266N, 16267N, 16268N, 16269N, 16270N, 16271N, 16272N, 16273N, 16274N, 16275N, 16276N, 16277N, 16278N, 16279N, 16280N, 16281N, 16282N, 16283N, 16284N, 16285N, 16286N, 16287N, 16288N, 16289N, 16290N, 16291N, 16292N, 16293N, 16294N, 16295N, 16296N, 16297N, 16298N, 16299N, 16300N, 16301N, 16302N, 16303N, 16304N, 16305N, 16306N, 16307N, 16308N, 16309N, 16310N, 16311N, 16312N, 16313N, 16314N, 16315N, 16316N, 16317N, 16318N, 16319N, 16320N, 16321N, 16322N, 16323N, 16324N, 16325N, 16326N, 16327N, 16328N, 16329N, 16330N, 16331N, 16332N, 16333N, 16334N, 16335N, 16336N, 16337N, 16338N, 16339N, 16340N, 16341N, 16342N, 16343N, 16344N, 16345N, 16346N, 16347N, 16348N, 16349N, 16350N, 16351N, 16352N, 16353N, 16354N, 16355N, 16356N, 16357N, 16368, 16540A, 16626 | P4a_HQ439447:356 | P4a_HQ439447:356d, 2227+T, 5237+G, 5277+A, 15385+T, 16126N, 16127N, 16128N, 16129N, 16130N, 16131N, 16132N, 16133N, 16134N, 16135N, 16136N, 16137N, 16138N, 16139N, 16140N, 16141N, 16142N, 16143N, 16144N, 16145N, 16146N, 16147N, 16148N, 16149N, 16150N, 16151N, 16152N, 16153N, 16154N, 16155N, 16156N, 16157N, 16158N, 16159N, 16160N, 16161N, 16162N, 16163N, 16164N, 16165N, 16166N, 16167N, 16168N, 16169N, 16170N, 16171N, 16172N, 16173N, 16174N, 16175N, 16176N, 16177N, 16178N, 16179N, 16180N, 16181N, 16182N, 16183N, 16184N, 16185N, 16186N, 16187N, 16188N, 16189N, 16190N, 16191N, 16192N, 16193N, 16194N, 16195N, 16196N, 16197N, 16198N, 16199N, 16200N, 16201N, 16202N, 16203N, 16204N, 16205N, 16206N, 16207N, 16208N, 16209N, 16210N, 16211N, 16212N, 16213N, 16214N, 16215N, 16216N, 16217N, 16218N, 16219N, 16220N, 16221N, 16222N, 16223N, 16224N, 16225N, 16226N, 16227N, 16228N, 16229N, 16230N, 16231N, 16232N, 16233N, 16234N, 16235N, 16236N, 16237N, 16238N, 16239N, 16240N, 16241N, 16242N, 16243N, 16244N, 16245N, 16246N, 16247N, 16248N, 16249N, 16250N, 16251N, 16252N, 16253N, 16254N, 16255N, 16256N, 16257N, 16258N, 16259N, 16260N, 16261N, 16262N, 16263N, 16264N, 16265N, 16266N, 16267N, 16268N, 16269N, 16270N, 16271N, 16272N, 16273N, 16274N, 16275N, 16276N, 16277N, 16278N, 16279N, 16280N, 16281N, 16282N, 16283N, 16284N, 16285N, 16286N, 16287N, 16288N, 16289N, 16290N, 16291N, 16292N, 16293N, 16294N, 16295N, 16296N, 16297N, 16298N, 16299N, 16300N, 16301N, 16302N, 16303N, 16304N, 16305N, 16306N, 16307N, 16308N, 16309N, 16310N, 16311N, 16312N, 16313N, 16314N, 16315N, 16316N, 16317N, 16318N, 16319N, 16320N, 16321N, 16322N, 16323N, 16324N, 16325N, 16326N, 16327N, 16328N, 16329N, 16330N, 16331N, 16332N, 16333N, 16334N, 16335N, 16336N, 16337N, 16338N, 16339N, 16340N, 16341N, 16342N, 16343N, 16344N, 16345N, 16346N, 16347N, 16348N, 16349N, 16350N, 16351N, 16352N, 16353N, 16354N, 16355N, 16356N, 16357N | Lippold,S., Matzke,N., Reissman,M., Burbano,H. and Hofreiter,M. BMC Evolutionary Biology. 328 (11), 1471-2148 (2011) |
| HQ439448 (Horse and Przewalski's horse) | Illumina/Solexa Genome Analyzer II | Phantom mutaions: 358d, 1387G, 2226+C, 5239+A, 5274+G, 15383+G. | Q1b1_HQ439448 | 158, 302, 341, 356d, 739, 860, 961, 1387G, 2226+C, 2788, 3070, 3259T, 3271, 3616, 3800, 3942, 4062, 4201, 4536, 4605, 4646, 4669, 5239+A, 5274+G, 5527, 5827, 5884, 6004, 6307, 6529, 6688, 6712, 6784, 7001, 7243, 7294, 7576, 7612T, 7666, 7898, 7900, 8005, 8076, 8150, 8238, 8358, 8361, 8556T, 8565, 8855, 9071, 9086, 9203, 9239, 9775, 10110, 10173, 10214, 10292, 10376, 10448, 10859, 11240, 11378, 11394, 11424, 11492, 11543, 11842, 11879, 11966, 12167, 12230, 12332, 12404, 12767, 13049, 13333, 13463, 13629, 13920, 13933, 14626, 14803, 15202, 15342, 15383+G, 15492, 15599, 15700, 15717, 15723, 15737, 15768, 15774, 15808, 15953, 16034, 16035, 16054, 16060C, 16118N, 16126N, 16127N, 16128N, 16129N, 16130N, 16131N, 16132N, 16133N, 16134N, 16135N, 16136N, 16137N, 16138N, 16139N, 16140N, 16141N, 16142N, 16143N, 16144N, 16145N, 16146N, 16147N, 16148N, 16149N, 16150N, 16151N, 16152N, 16153N, 16154N, 16155N, 16156N, 16157N, 16158N, 16159N, 16160N, 16161N, 16162N, 16163N, 16164N, 16165N, 16166N, 16167N, 16168N, 16169N, 16170N, 16171N, 16172N, 16173N, 16174N, 16175N, 16176N, 16177N, 16178N, 16179N, 16180N, 16181N, 16182N, 16183N, 16184N, 16185N, 16186N, 16187N, 16188N, 16189N, 16190N, 16191N, 16192N, 16193N, 16194N, 16195N, 16196N, 16197N, 16198N, 16199N, 16200N, 16201N, 16202N, 16203N, 16204N, 16205N, 16206N, 16207N, 16208N, 16209N, 16210N, 16211N, 16212N, 16213N, 16214N, 16215N, 16216N, 16217N, 16218N, 16219N, 16220N, 16221N, 16222N, 16223N, 16224N, 16225N, 16226N, 16227N, 16228N, 16229N, 16230N, 16231N, 16232N, 16233N, 16234N, 16235N, 16236N, 16237N, 16238N, 16239N, 16240N, 16241N, 16242N, 16243N, 16244N, 16245N, 16246N, 16247N, 16248N, 16249N, 16250N, 16251N, 16252N, 16253N, 16254N, 16255N, 16256N, 16257N, 16258N, 16259N, 16260N, 16261N, 16262N, 16263N, 16264N, 16265N, 16266N, 16267N, 16268N, 16269N, 16270N, 16271N, 16272N, 16273N, 16274N, 16275N, 16276N, 16277N, 16278N, 16279N, 16280N, 16281N, 16282N, 16283N, 16284N, 16285N, 16286N, 16287N, 16288N, 16289N, 16290N, 16291N, 16292N, 16293N, 16294N, 16295N, 16296N, 16297N, 16298N, 16299N, 16300N, 16301N, 16302N, 16303N, 16304N, 16305N, 16306N, 16307N, 16308N, 16309N, 16310N, 16311N, 16312N, 16313N, 16314N, 16315N, 16316N, 16317N, 16318N, 16319N, 16320N, 16321N, 16322N, 16323N, 16324N, 16325N, 16326N, 16327N, 16328N, 16329N, 16330N, 16331N, 16332N, 16333N, 16334N, 16335N, 16336N, 16337N, 16338N, 16339N, 16340N, 16341N, 16342N, 16343N, 16344N, 16345N, 16346N, 16347N, 16348N, 16349N, 16350N, 16351N, 16352N, 16353N, 16354N, 16355N, 16356N, 16357N, 16368, 16540A, 16626 | Q1b1_HQ439448:356, 1387d | Q1b1_HQ439448:356d, 1387G, 2226+C, 5239+A, 5274+G, 15383+G, 16118N, 16126N, 16127N, 16128N, 16129N, 16130N, 16131N, 16132N, 16133N, 16134N, 16135N, 16136N, 16137N, 16138N, 16139N, 16140N, 16141N, 16142N, 16143N, 16144N, 16145N, 16146N, 16147N, 16148N, 16149N, 16150N, 16151N, 16152N, 16153N, 16154N, 16155N, 16156N, 16157N, 16158N, 16159N, 16160N, 16161N, 16162N, 16163N, 16164N, 16165N, 16166N, 16167N, 16168N, 16169N, 16170N, 16171N, 16172N, 16173N, 16174N, 16175N, 16176N, 16177N, 16178N, 16179N, 16180N, 16181N, 16182N, 16183N, 16184N, 16185N, 16186N, 16187N, 16188N, 16189N, 16190N, 16191N, 16192N, 16193N, 16194N, 16195N, 16196N, 16197N, 16198N, 16199N, 16200N, 16201N, 16202N, 16203N, 16204N, 16205N, 16206N, 16207N, 16208N, 16209N, 16210N, 16211N, 16212N, 16213N, 16214N, 16215N, 16216N, 16217N, 16218N, 16219N, 16220N, 16221N, 16222N, 16223N, 16224N, 16225N, 16226N, 16227N, 16228N, 16229N, 16230N, 16231N, 16232N, 16233N, 16234N, 16235N, 16236N, 16237N, 16238N, 16239N, 16240N, 16241N, 16242N, 16243N, 16244N, 16245N, 16246N, 16247N, 16248N, 16249N, 16250N, 16251N, 16252N, 16253N, 16254N, 16255N, 16256N, 16257N, 16258N, 16259N, 16260N, 16261N, 16262N, 16263N, 16264N, 16265N, 16266N, 16267N, 16268N, 16269N, 16270N, 16271N, 16272N, 16273N, 16274N, 16275N, 16276N, 16277N, 16278N, 16279N, 16280N, 16281N, 16282N, 16283N, 16284N, 16285N, 16286N, 16287N, 16288N, 16289N, 16290N, 16291N, 16292N, 16293N, 16294N, 16295N, 16296N, 16297N, 16298N, 16299N, 16300N, 16301N, 16302N, 16303N, 16304N, 16305N, 16306N, 16307N, 16308N, 16309N, 16310N, 16311N, 16312N, 16313N, 16314N, 16315N, 16316N, 16317N, 16318N, 16319N, 16320N, 16321N, 16322N, 16323N, 16324N, 16325N, 16326N, 16327N, 16328N, 16329N, 16330N, 16331N, 16332N, 16333N, 16334N, 16335N, 16336N, 16337N, 16338N, 16339N, 16340N, 16341N, 16342N, 16343N, 16344N, 16345N, 16346N, 16347N, 16348N, 16349N, 16350N, 16351N, 16352N, 16353N, 16354N, 16355N, 16356N, 16357N | Lippold,S., Matzke,N., Reissman,M., Burbano,H. and Hofreiter,M. BMC Evolutionary Biology. 328 (11), 1471-2148 (2011) |
| HQ439449 (Horse and Przewalski's horse) | Illumina/Solexa Genome Analyzer II | Phantom mutaions: 58d, 1387G, 2226+C, 5239+A, 5277+A, 15383+G. | C1a_HQ439449 | 158, 356d, 957, 1387G, 2226+C, 2238, 2788, 3800, 4062, 4599, 4884, 4993, 5239+A, 5277+A, 5500, 6076, 6784, 7001, 7942, 8723, 9239, 9278, 9664, 10214, 10217, 11046, 11129, 11240, 11543, 12352, 13079, 15383+G, 15492, 15599, 15647, 15717, 15823, 15867, 15953, 15971, 16110, 16126N, 16127N, 16128N, 16129N, 16130N, 16131N, 16132N, 16133N, 16134N, 16135N, 16136N, 16137N, 16138N, 16139N, 16140N, 16141N, 16142N, 16143N, 16144N, 16145N, 16146N, 16147N, 16148N, 16149N, 16150N, 16151N, 16152N, 16153N, 16154N, 16155N, 16156N, 16157N, 16158N, 16159N, 16160N, 16161N, 16162N, 16163N, 16164N, 16165N, 16166N, 16167N, 16168N, 16169N, 16170N, 16171N, 16172N, 16173N, 16174N, 16175N, 16176N, 16177N, 16178N, 16179N, 16180N, 16181N, 16182N, 16183N, 16184N, 16185N, 16186N, 16187N, 16188N, 16189N, 16190N, 16191N, 16192N, 16193N, 16194N, 16195N, 16196N, 16197N, 16198N, 16199N, 16200N, 16201N, 16202N, 16203N, 16204N, 16205N, 16206N, 16207N, 16208N, 16209N, 16210N, 16211N, 16212N, 16213N, 16214N, 16215N, 16216N, 16217N, 16218N, 16219N, 16220N, 16221N, 16222N, 16223N, 16224N, 16225N, 16226N, 16227N, 16228N, 16229N, 16230N, 16231N, 16232N, 16233N, 16234N, 16235N, 16236N, 16237N, 16238N, 16239N, 16240N, 16241N, 16242N, 16243N, 16244N, 16245N, 16246N, 16247N, 16248N, 16249N, 16250N, 16251N, 16252N, 16253N, 16254N, 16255N, 16256N, 16257N, | C1a_HQ439449:356 | C1a_HQ439449:356d, 2226+C, 5239+A, 5277+A, 15383+G, 16126N, 16127N, 16128N, 16129N, 16130N, 16131N, 16132N, 16133N, 16134N, 16135N, 16136N, 16137N, 16138N, 16139N, 16140N, 16141N, 16142N, 16143N, 16144N, 16145N, 16146N, 16147N, 16148N, 16149N, 16150N, 16151N, 16152N, 16153N, 16154N, 16155N, 16156N, 16157N, 16158N, 16159N, 16160N, 16161N, 16162N, 16163N, 16164N, 16165N, 16166N, 16167N, 16168N, 16169N, 16170N, 16171N, 16172N, 16173N, 16174N, 16175N, 16176N, 16177N, 16178N, 16179N, 16180N, 16181N, 16182N, 16183N, 16184N, 16185N, 16186N, 16187N, 16188N, 16189N, 16190N, 16191N, 16192N, 16193N, 16194N, 16195N, 16196N, 16197N, 16198N, 16199N, 16200N, 16201N, 16202N, 16203N, 16204N, 16205N, 16206N, 16207N, 16208N, 16209N, 16210N, 16211N, 16212N, 16213N, 16214N, 16215N, 16216N, 16217N, 16218N, 16219N, 16220N, 16221N, 16222N, 16223N, 16224N, 16225N, 16226N, 16227N, 16228N, 16229N, 16230N, 16231N, 16232N, 16233N, 16234N, 16235N, 16236N, 16237N, 16238N, 16239N, 16240N, | Lippold,S., Matzke,N., Reissman,M., Burbano,H. and Hofreiter,M. BMC Evolutionary Biology. 328 (11), 1471-2148 (2011) |
| | | | | 16258N, 16259N, 16260N, 16261N, 16262N, 16263N, 16264N, 16265N, 16266N, 16267N, 16268N, 16269N, 16270N, 16271N, 16272N, 16273N, 16274N, 16275N, 16276N, 16277N, 16278N, 16279N, 16280N, 16281N, 16282N, 16283N, 16284N, 16285N, 16286N, 16287N, 16288N, 16289N, 16290N, 16291N, 16292N, 16293N, 16294N, 16295N, 16296N, 16297N, 16298N, 16299N, 16300N, 16301N, 16302N, 16303N, 16304N, 16305N, 16306N, 16307N, 16308N, 16309N, 16310N, 16311N, 16312N, 16313N, 16314N, 16315N, 16316N, 16317N, 16318N, 16319N, 16320N, 16321N, 16322N, 16323N, 16324N, 16325N, 16326N, 16327N, 16328N, 16329N, 16330N, 16331N, 16332N, 16333N, 16334N, 16335N, 16336N, 16337N, 16338N, 16339N, 16340N, 16341N, 16342N, 16343N, 16344N, 16345N, 16346N, 16347N, 16348N, 16349N, 16350N, 16351N, 16352N, 16353N, 16354N, 16355N, 16356N, 16357N, 16368 | | 16241N, 16242N, 16243N, 16244N, 16245N, 16246N, 16247N, 16248N, 16249N, 16250N, 16251N, 16252N, 16253N, 16254N, 16255N, 16256N, 16257N, 16258N, 16259N, 16260N, 16261N, 16262N, 16263N, 16264N, 16265N, 16266N, 16267N, 16268N, 16269N, 16270N, 16271N, 16272N, 16273N, 16274N, 16275N, 16276N, 16277N, 16278N, 16279N, 16280N, 16281N, 16282N, 16283N, 16284N, 16285N, 16286N, 16287N, 16288N, 16289N, 16290N, 16291N, 16292N, 16293N, 16294N, 16295N, 16296N, 16297N, 16298N, 16299N, 16300N, 16301N, 16302N, 16303N, 16304N, 16305N, 16306N, 16307N, 16308N, 16309N, 16310N, 16311N, 16312N, 16313N, 16314N, 16315N, 16316N, 16317N, 16318N, 16319N, 16320N, 16321N, 16322N, 16323N, 16324N, 16325N, 16326N, 16327N, 16328N, 16329N, 16330N, 16331N, 16332N, 16333N, 16334N, 16335N, 16336N, 16337N, 16338N, 16339N, 16340N, 16341N, 16342N, 16343N, 16344N, 16345N, 16346N, 16347N, 16348N, 16349N, 16350N, 16351N, 16352N, 16353N, 16354N, 16355N, 16356N, 16357N | |
| HQ439450 (Horse and Przewalski's horse) | Illumina/Solexa Genome Analyzer II | Phantom mutaions: 358d, 1387A, 2227+T, 5237+G, 5274+G, 15385+T. | Q2a_HQ439450 | 158, 302, 341, 356d, 739, 854, 860, 961, 1387A, 2227+T, 2788, 3070, 3259T, 3271, 3616, 3800, 3942, 4062, 4201, 4536, 4605, 4646, 4669, 5237+G, 5274+G, 5527, 5827, 5875, 5884, 6004, 6307, 6529, 6712, 6784, 7001, 7243, 7294, 7612T, 7666, 7898, 7900, 8005, 8076, 8150, 8238, 8358, 8361, 8556T, 8565, 8855, 9086, 9203, 9239, 9775, 10110, 10173, 10214, 10292, 10376, 10448, 10859, 11240, 11378, 11394, 11424, 11492, 11543, 11842, 11879, 11966, 12167, 12230, 12332, 12404, 12767, 13049, 13333, 13463, 13629, 13920, 13933, 14626, 14803, 15202, 15342, 15385+T, 15492, 15599, 15601, 15647, 15700, 15717, 15737, 15768, 15774, 15808, 15953, 15992, 16034, 16035, 16110, 16118, 16124N, 16126N, 16127N, 16128N, 16129N, 16130N, 16131N, 16132N, 16133N, 16134N, 16135N, 16136N, 16137N, 16138N, 16139N, 16140N, 16141N, 16142N, 16143N, 16144N, 16145N, 16146N, 16147N, 16148N, 16149N, 16150N, 16151N, 16152N, 16153N, 16154N, 16155N, 16156N, 16157N, 16158N, 16159N, 16160N, 16161N, 16162N, 16163N, 16164N, 16165N, 16166N, 16167N, 16168N, 16169N, 16170N, 16171N, 16172N, 16173N, 16174N, 16175N, 16176N, 16177N, 16178N, 16179N, 16180N, 16181N, 16182N, 16183N, 16184N, 16185N, 16186N, 16187N, 16188N, 16189N, 16190N, 16191N, 16192N, 16193N, 16194N, 16195N, 16196N, 16197N, 16198N, 16199N, 16200N, 16201N, 16202N, 16203N, 16204N, 16205N, 16206N, 16207N, 16208N, 16209N, 16210N, 16211N, 16212N, 16213N, 16214N, 16215N, 16216N, 16217N, 16218N, 16219N, 16220N, 16221N, 16222N, 16223N, 16224N, 16225N, 16226N, 16227N, 16228N, 16229N, 16230N, 16231N, 16232N, 16233N, 16234N, 16235N, 16236N, 16237N, 16238N, 16239N, 16240N, 16241N, 16242N, 16243N, 16244N, 16245N, 16246N, 16247N, 16248N, 16249N, 16250N, 16251N, 16252N, 16253N, 16254N, 16255N, 16256N, 16257N, 16258N, 16259N, 16260N, 16261N, 16262N, 16263N, 16264N, 16265N, 16266N, 16267N, 16268N, 16269N, 16270N, 16271N, 16272N, 16273N, 16274N, 16275N, 16276N, 16277N, 16278N, 16279N, 16280N, 16281N, 16282N, 16283N, 16284N, 16285N, 16286N, 16287N, 16288N, 16289N, 16290N, 16291N, 16292N, 16293N, 16294N, 16295N, 16296N, 16297N, 16298N, 16299N, 16300N, 16301N, 16302N, 16303N, 16304N, 16305N, 16306N, 16307N, 16308N, 16309N, 16310N, 16311N, 16312N, 16313N, 16314N, 16315N, 16316N, 16317N, 16318N, 16319N, 16320N, 16321N, 16322N, 16323N, 16324N, 16325N, 16326N, 16327N, 16328N, 16329N, 16330N, 16331N, 16332N, 16333N, 16334N, 16335N, 16336N, 16337N, 16338N, 16339N, 16340N, 16341N, 16342N, 16343N, 16344N, 16345N, 16346N, 16347N, 16348N, 16349N, 16350N, 16351N, 16352N, 16353N, 16354N, 16355N, 16356N, 16357N, 16368, 16540A, 16626 | Q2a_HQ439450:356 | Q2a_HQ439450:356d, 2227+T, 5237+G, 5274+G, 15385+T, 16124N, 16126N, 16127N, 16128N, 16129N, 16130N, 16131N, 16132N, 16133N, 16134N, 16135N, 16136N, 16137N, 16138N, 16139N, 16140N, 16141N, 16142N, 16143N, 16144N, 16145N, 16146N, 16147N, 16148N, 16149N, 16150N, 16151N, 16152N, 16153N, 16154N, 16155N, 16156N, 16157N, 16158N, 16159N, 16160N, 16161N, 16162N, 16163N, 16164N, 16165N, 16166N, 16167N, 16168N, 16169N, 16170N, 16171N, 16172N, 16173N, 16174N, 16175N, 16176N, 16177N, 16178N, 16179N, 16180N, 16181N, 16182N, 16183N, 16184N, 16185N, 16186N, 16187N, 16188N, 16189N, 16190N, 16191N, 16192N, 16193N, 16194N, 16195N, 16196N, 16197N, 16198N, 16199N, 16200N, 16201N, 16202N, 16203N, 16204N, 16205N, 16206N, 16207N, 16208N, 16209N, 16210N, 16211N, 16212N, 16213N, 16214N, 16215N, 16216N, 16217N, 16218N, 16219N, 16220N, 16221N, 16222N, 16223N, 16224N, 16225N, 16226N, 16227N, 16228N, 16229N, 16230N, 16231N, 16232N, 16233N, 16234N, 16235N, 16236N, 16237N, 16238N, 16239N, 16240N, 16241N, 16242N, 16243N, 16244N, 16245N, 16246N, 16247N, 16248N, 16249N, 16250N, 16251N, 16252N, 16253N, 16254N, 16255N, 16256N, 16257N, 16258N, 16259N, 16260N, 16261N, 16262N, 16263N, 16264N, 16265N, 16266N, 16267N, 16268N, 16269N, 16270N, 16271N, 16272N, 16273N, 16274N, 16275N, 16276N, 16277N, 16278N, 16279N, 16280N, 16281N, 16282N, 16283N, 16284N, 16285N, 16286N, 16287N, 16288N, 16289N, 16290N, 16291N, 16292N, 16293N, 16294N, 16295N, 16296N, 16297N, 16298N, 16299N, 16300N, 16301N, 16302N, 16303N, 16304N, 16305N, 16306N, 16307N, 16308N, 16309N, 16310N, 16311N, 16312N, 16313N, 16314N, 16315N, 16316N, 16317N, 16318N, 16319N, 16320N, 16321N, 16322N, 16323N, 16324N, 16325N, 16326N, 16327N, 16328N, 16329N, 16330N, 16331N, 16332N, 16333N, 16334N, 16335N, 16336N, 16337N, 16338N, 16339N, 16340N, 16341N, 16342N, 16343N, 16344N, 16345N, 16346N, 16347N, 16348N, 16349N, 16350N, 16351N, 16352N, 16353N, 16354N, 16355N, 16356N, 16357N | Lippold,S., Matzke,N., Reissman,M., Burbano,H. and Hofreiter,M. BMC Evolutionary Biology. 328 (11), 1471-2148 (2011) |
| HQ439451 (Horse and Przewalski's horse) | Illumina/Solexa Genome Analyzer II | Phantom mutaions: 358d, 1387A, 2227+T, 5239+A, 5274+G, 15385+T. | L1a1a_HQ439451 | 158, 356d, 961, 1375, 1387A, 2227+T, 2788, 2899, 3517, 3942, 4062, 4536, 4646, 4669, 5239+A, 5274+G, 5527, 5815, 5884, 6004, 6307, 6784, 7001, 7516, 7666, 7900, 8005, 8058, 8301, 8319, 8358, 8565, 9239, 9684, 9951, 10110, 10214, 10292, 10376, 10421, 10613, 11240, 11543, 11682, 11693, 11842, 11879, 12119, 12200, 12767, 12896, 12950, 13049, 13333, 13520, 14252, 14803, 14995, 15313, 15385+T, 15491, 15492, 15493, 15531, 15599, 15600, 15646, 15717, 15768, 15867, 15868, 15953, 15971, 16065, 16100, 16126N, 16127N, 16128N, 16129N, 16130N, 16131N, 16132N, 16133N, 16134N, 16135N, 16136N, 16137N, 16138N, 16139N, 16140N, 16141N, 16142N, 16143N, 16144N, 16145N, 16146N, 16147N, 16148N, 16149N, 16150N, 16151N, 16152N, 16153N, 16154N, 16155N, 16156N, 16157N, 16158N, 16159N, 16160N, 16161N, 16162N, 16163N, 16164N, 16165N, 16166N, 16167N, 16168N, 16169N, 16170N, 16171N, 16172N, 16173N, 16174N, 16175N, 16176N, 16177N, 16178N, 16179N, 16180N, 16181N, 16182N, 16183N, 16184N, 16185N, 16186N, 16187N, 16188N, 16189N, 16190N, 16191N, 16192N, 16193N, 16194N, 16195N, 16196N, 16197N, 16198N, 16199N, 16200N, 16201N, 16202N, 16203N, 16204N, 16205N, 16206N, 16207N, 16208N, 16209N, 16210N, 16211N, 16212N, 16213N, 16214N, 16215N, 16216N, 16217N, 16218N, 16219N, 16220N, 16221N, 16222N, 16223N, 16224N, 16225N, 16226N, 16227N, 16228N, 16229N, 16230N, 16231N, 16232N, 16233N, 16234N, 16235N, 16236N, 16237N, 16238N, 16239N, 16240N, 16241N, 16242N, 16243N, 16244N, 16245N, 16246N, 16247N, 16248N, 16249N, 16250N, 16251N, 16252N, 16253N, 16254N, 16255N, 16256N, 16257N, 16258N, 16259N, 16260N, 16261N, 16262N, 16263N, 16264N, 16265N, 16266N, 16267N, 16268N, 16269N, 16270N, 16271N, 16272N, 16273N, 16274N, 16275N, 16276N, 16277N, 16278N, 16279N, 16280N, 16281N, 16282N, 16283N, 16284N, 16285N, 16286N, 16287N, 16288N, 16289N, 16290N, 16291N, 16292N, 16293N, 16294N, 16295N, 16296N, 16297N, 16298N, 16299N, 16300N, 16301N, 16302N, 16303N, 16304N, 16305N, 16306N, 16307N, 16308N, 16309N, 16310N, 16311N, 16312N, 16313N, 16314N, 16315N, 16316N, 16317N, 16318N, 16319N, 16320N, 16321N, 16322N, 16323N, 16324N, 16325N, 16326N, 16327N, 16328N, 16329N, 16330N, 16331N, 16332N, 16333N, 16334N, 16335N, 16336N, 16337N, 16338N, 16339N, 16340N, 16341N, 16342N, 16343N, 16344N, 16345N, 16346N, 16347N, 16348N, 16349N, 16350N, 16351N, 16352N, 16353N, 16354N, 16355N, 16356N, 16357N, 16368, 16398A, 16626 | L1a1a_HQ439451:356 | L1a1a_HQ439451:356d, 2227+T, 5239+A, 5274+G, 15385+T, 16126N, 16127N, 16128N, 16129N, 16130N, 16131N, 16132N, 16133N, 16134N, 16135N, 16136N, 16137N, 16138N, 16139N, 16140N, 16141N, 16142N, 16143N, 16144N, 16145N, 16146N, 16147N, 16148N, 16149N, 16150N, 16151N, 16152N, 16153N, 16154N, 16155N, 16156N, 16157N, 16158N, 16159N, 16160N, 16161N, 16162N, 16163N, 16164N, 16165N, 16166N, 16167N, 16168N, 16169N, 16170N, 16171N, 16172N, 16173N, 16174N, 16175N, 16176N, 16177N, 16178N, 16179N, 16180N, 16181N, 16182N, 16183N, 16184N, 16185N, 16186N, 16187N, 16188N, 16189N, 16190N, 16191N, 16192N, 16193N, 16194N, 16195N, 16196N, 16197N, 16198N, 16199N, 16200N, 16201N, 16202N, 16203N, 16204N, 16205N, 16206N, 16207N, 16208N, 16209N, 16210N, 16211N, 16212N, 16213N, 16214N, 16215N, 16216N, 16217N, 16218N, 16219N, 16220N, 16221N, 16222N, 16223N, 16224N, 16225N, 16226N, 16227N, 16228N, 16229N, 16230N, 16231N, 16232N, 16233N, 16234N, 16235N, 16236N, 16237N, 16238N, 16239N, 16240N, 16241N, 16242N, 16243N, 16244N, 16245N, 16246N, 16247N, 16248N, 16249N, 16250N, 16251N, 16252N, 16253N, 16254N, 16255N, 16256N, 16257N, 16258N, 16259N, 16260N, 16261N, 16262N, 16263N, 16264N, 16265N, 16266N, 16267N, 16268N, 16269N, 16270N, 16271N, 16272N, 16273N, 16274N, 16275N, 16276N, 16277N, 16278N, 16279N, 16280N, 16281N, 16282N, 16283N, 16284N, 16285N, 16286N, 16287N, 16288N, 16289N, 16290N, 16291N, 16292N, 16293N, 16294N, 16295N, 16296N, 16297N, 16298N, 16299N, 16300N, 16301N, 16302N, 16303N, 16304N, 16305N, 16306N, 16307N, 16308N, 16309N, 16310N, 16311N, 16312N, 16313N, 16314N, 16315N, 16316N, 16317N, 16318N, 16319N, 16320N, 16321N, 16322N, 16323N, 16324N, 16325N, 16326N, 16327N, 16328N, 16329N, 16330N, 16331N, 16332N, 16333N, 16334N, 16335N, 16336N, 16337N, 16338N, 16339N, 16340N, 16341N, 16342N, 16343N, 16344N, 16345N, 16346N, 16347N, 16348N, 16349N, 16350N, 16351N, 16352N, 16353N, 16354N, 16355N, 16356N, 16357N | Lippold,S., Matzke,N., Reissman,M., Burbano,H. and Hofreiter,M. BMC Evolutionary Biology. 328 (11), 1471-2148 (2011) |
| HQ439452 (Horse and Przewalski's horse) | Illumina/Solexa Genome Analyzer II | Phantom mutaions: 358d, 1387A, 2227+T, 5237+G, 5277+A, 15385+T. | L2a2a_HQ439452 | 158, 356d, 961, 1375, 1387A, 1837, 2227+T, 2607, 2788, 2899, 3517, 3942, 4062, 4536, 4646, 4669, 5237+G, 5277+A, 5527, 5815, 5884, 6004, 6307, 6784, 7001, 7516, 7666, 7900, 8005, 8058, 8199, 8301, 8319, 8358, 8403, 8565, 9239, 9951, 10110, 10214, 10292, 10376, 10421, 10613, 11240, 11543, 11693, 11842, 11879, 12119, 12200, 12767, 12896, 12950, 13049, 13333, 13520, 14803, 14825, 14995, 15313, 15385+T, 15491, 15492, 15493, 15531, 15582, 15600, 15646, 15717, 15768, 15867, 15868, 15953, 15971, 16065, 16100, 16126N, 16127N, 16128N, 16129N, 16130N, 16131N, 16132N, 16133N, 16134N, 16135N, 16136N, 16137N, 16138N, 16139N, 16140N, 16141N, 16142N, 16143N, 16144N, 16145N, 16146N, 16147N, 16148N, 16149N, 16150N, 16151N, 16152N, 16153N, 16154N, 16155N, 16156N, 16157N, 16158N, 16159N, 16160N, 16161N, 16162N, 16163N, 16164N, 16165N, 16166N, 16167N, 16168N, 16169N, 16170N, 16171N, 16172N, 16173N, 16174N, 16175N, 16176N, 16177N, 16178N, 16179N, 16180N, 16181N, 16182N, 16183N, 16184N, 16185N, 16186N, 16187N, 16188N, 16189N, 16190N, 16191N, 16192N, 16193N, 16194N, 16195N, 16196N, 16197N, 16198N, 16199N, 16200N, 16201N, 16202N, 16203N, 16204N, 16205N, 16206N, 16207N, 16208N, 16209N, 16210N, 16211N, 16212N, 16213N, 16214N, 16215N, 16216N, 16217N, 16218N, 16219N, 16220N, 16221N, 16222N, 16223N, 16224N, 16225N, 16226N, 16227N, 16228N, 16229N, 16230N, | L2a2a_HQ439452:356 | L2a2a_HQ439452:356d, 2227+T, 5237+G, 5277+A, 15385+T, 16126N, 16127N, 16128N, 16129N, 16130N, 16131N, 16132N, 16133N, 16134N, 16135N, 16136N, 16137N, 16138N, 16139N, 16140N, 16141N, 16142N, 16143N, 16144N, 16145N, 16146N, 16147N, 16148N, 16149N, 16150N, 16151N, 16152N, 16153N, 16154N, 16155N, 16156N, 16157N, 16158N, 16159N, 16160N, 16161N, 16162N, 16163N, 16164N, 16165N, 16166N, 16167N, 16168N, 16169N, 16170N, 16171N, 16172N, 16173N, 16174N, 16175N, 16176N, 16177N, 16178N, 16179N, 16180N, 16181N, 16182N, 16183N, 16184N, 16185N, 16186N, 16187N, 16188N, 16189N, 16190N, 16191N, 16192N, 16193N, 16194N, 16195N, 16196N, 16197N, 16198N, 16199N, 16200N, 16201N, 16202N, 16203N, 16204N, 16205N, 16206N, 16207N, 16208N, 16209N, 16210N, 16211N, 16212N, 16213N, 16214N, 16215N, 16216N, 16217N, 16218N, 16219N, 16220N, 16221N, 16222N, 16223N, 16224N, 16225N, 16226N, 16227N, 16228N, 16229N, 16230N, 16231N, 16232N, 16233N, 16234N, 16235N, 16236N, 16237N, 16238N, 16239N, | Lippold,S., Matzke,N., Reissman,M., Burbano,H. and Hofreiter,M. BMC Evolutionary Biology. 328 (11), 1471-2148 (2011) |
| | | | | 16231N, 16232N, 16233N, 16234N, 16235N, 16236N, 16237N, 16238N, 16239N, 16240N, 16241N, 16242N, 16243N, 16244N, 16245N, 16246N, 16247N, 16248N, 16249N, 16250N, 16251N, 16252N, 16253N, 16254N, 16255N, 16256N, 16257N, 16258N, 16259N, 16260N, 16261N, 16262N, 16263N, 16264N, 16265N, 16266N, 16267N, 16268N, 16269N, 16270N, 16271N, 16272N, 16273N, 16274N, 16275N, 16276N, 16277N, 16278N, 16279N, 16280N, 16281N, 16282N, 16283N, 16284N, 16285N, 16286N, 16287N, 16288N, 16289N, 16290N, 16291N, 16292N, 16293N, 16294N, 16295N, 16296N, 16297N, 16298N, 16299N, 16300N, 16301N, 16302N, 16303N, 16304N, 16305N, 16306N, 16307N, 16308N, 16309N, 16310N, 16311N, 16312N, 16313N, 16314N, 16315N, 16316N, 16317N, 16318N, 16319N, 16320N, 16321N, 16322N, 16323N, 16324N, 16325N, 16326N, 16327N, 16328N, 16329N, 16330N, 16331N, 16332N, 16333N, 16334N, 16335N, 16336N, 16337N, 16338N, 16339N, 16340N, 16341N, 16342N, 16343N, 16344N, 16345N, 16346N, 16347N, 16348N, 16349N, 16350N, 16351N, 16352N, 16353N, 16354N, 16355N, 16356N, 16357N, 16368, 16398A, 16626 | | 16240N, 16241N, 16242N, 16243N, 16244N, 16245N, 16246N, 16247N, 16248N, 16249N, 16250N, 16251N, 16252N, 16253N, 16254N, 16255N, 16256N, 16257N, 16258N, 16259N, 16260N, 16261N, 16262N, 16263N, 16264N, 16265N, 16266N, 16267N, 16268N, 16269N, 16270N, 16271N, 16272N, 16273N, 16274N, 16275N, 16276N, 16277N, 16278N, 16279N, 16280N, 16281N, 16282N, 16283N, 16284N, 16285N, 16286N, 16287N, 16288N, 16289N, 16290N, 16291N, 16292N, 16293N, 16294N, 16295N, 16296N, 16297N, 16298N, 16299N, 16300N, 16301N, 16302N, 16303N, 16304N, 16305N, 16306N, 16307N, 16308N, 16309N, 16310N, 16311N, 16312N, 16313N, 16314N, 16315N, 16316N, 16317N, 16318N, 16319N, 16320N, 16321N, 16322N, 16323N, 16324N, 16325N, 16326N, 16327N, 16328N, 16329N, 16330N, 16331N, 16332N, 16333N, 16334N, 16335N, 16336N, 16337N, 16338N, 16339N, 16340N, 16341N, 16342N, 16343N, 16344N, 16345N, 16346N, 16347N, 16348N, 16349N, 16350N, 16351N, 16352N, 16353N, 16354N, 16355N, 16356N, 16357N | |
| HQ439453 (Horse and Przewalski's horse) | Illumina/Solexa Genome Analyzer II | Phantom mutaions: 358d, 1387A, 2227+T, 5239+A, 5277+A, 15383+A. | L3a2_HQ439453 | 77, 158, 356d, 961, 1375, 1387A, 2227+T, 2788, 2899, 3517, 3942, 4062, 4536, 4646, 4669, 5239+A, 5277+A, 5527, 5815, 5884, 6004, 6307, 6784, 7001, 7516, 7666, 7840, 7900, 8005, 8058, 8301, 8319, 8358, 8565, 9239, 9398, 9951, 10110, 10214, 10292, 10376, 10421, 10613, 11240, 11543, 11693, 11842, 11879, 12119, 12200, 12767, 12896, 12950, 13049, 13333, 13520, 14803, 14825, 14995, 15313, 15383+A, 15491, 15492, 15493, 15531, 15599, 15600, 15601, 15646, 15717, 15768, 15867, 15868, 15953, 15971, 16065, 16100, 16126N, 16127N, 16128N, 16129N, 16130N, 16131N, 16132N, 16133N, 16134N, 16135N, 16136N, 16137N, 16138N, 16139N, 16140N, 16141N, 16142N, 16143N, 16144N, 16145N, 16146N, 16147N, 16148N, 16149N, 16150N, 16151N, 16152N, 16153N, 16154N, 16155N, 16156N, 16157N, 16158N, 16159N, 16160N, 16161N, 16162N, 16163N, 16164N, 16165N, 16166N, 16167N, 16168N, 16169N, 16170N, 16171N, 16172N, 16173N, 16174N, 16175N, 16176N, 16177N, 16178N, 16179N, 16180N, 16181N, 16182N, 16183N, 16184N, 16185N, 16186N, 16187N, 16188N, 16189N, 16190N, 16191N, 16192N, 16193N, 16194N, 16195N, 16196N, 16197N, 16198N, 16199N, 16200N, 16201N, 16202N, 16203N, 16204N, 16205N, 16206N, 16207N, 16208N, 16209N, 16210N, 16211N, 16212N, 16213N, 16214N, 16215N, 16216N, 16217N, 16218N, 16219N, 16220N, 16221N, 16222N, 16223N, 16224N, 16225N, 16226N, 16227N, 16228N, 16229N, 16230N, 16231N, 16232N, 16233N, 16234N, 16235N, 16236N, 16237N, 16238N, 16239N, 16240N, 16241N, 16242N, 16243N, 16244N, 16245N, 16246N, 16247N, 16248N, 16249N, 16250N, 16251N, 16252N, 16253N, 16254N, 16255N, 16256N, 16257N, 16258N, 16259N, 16260N, 16261N, 16262N, 16263N, 16264N, 16265N, 16266N, 16267N, 16268N, 16269N, 16270N, 16271N, 16272N, 16273N, 16274N, 16275N, 16276N, 16277N, 16278N, 16279N, 16280N, 16281N, 16282N, 16283N, 16284N, 16285N, 16286N, 16287N, 16288N, 16289N, 16290N, 16291N, 16292N, 16293N, 16294N, 16295N, 16296N, 16297N, 16298N, 16299N, 16300N, 16301N, 16302N, 16303N, 16304N, 16305N, 16306N, 16307N, 16308N, 16309N, 16310N, 16311N, 16312N, 16313N, 16314N, 16315N, 16316N, 16317N, 16318N, 16319N, 16320N, 16321N, 16322N, 16323N, 16324N, 16325N, 16326N, 16327N, 16328N, 16329N, 16330N, 16331N, 16332N, 16333N, 16334N, 16335N, 16336N, 16337N, 16338N, 16339N, 16340N, 16341N, 16342N, 16343N, 16344N, 16345N, 16346N, 16347N, 16348N, 16349N, 16350N, 16351N, 16352N, 16353N, 16354N, 16355N, 16356N, 16357N, 16368, 16553, 16554, 16626, 16654-16657d | L3a2_HQ439453:356 | L3a2_HQ439453:356d, 2227+T, 5239+A, 5277+A, 15383+A, 16126N, 16127N, 16128N, 16129N, 16130N, 16131N, 16132N, 16133N, 16134N, 16135N, 16136N, 16137N, 16138N, 16139N, 16140N, 16141N, 16142N, 16143N, 16144N, 16145N, 16146N, 16147N, 16148N, 16149N, 16150N, 16151N, 16152N, 16153N, 16154N, 16155N, 16156N, 16157N, 16158N, 16159N, 16160N, 16161N, 16162N, 16163N, 16164N, 16165N, 16166N, 16167N, 16168N, 16169N, 16170N, 16171N, 16172N, 16173N, 16174N, 16175N, 16176N, 16177N, 16178N, 16179N, 16180N, 16181N, 16182N, 16183N, 16184N, 16185N, 16186N, 16187N, 16188N, 16189N, 16190N, 16191N, 16192N, 16193N, 16194N, 16195N, 16196N, 16197N, 16198N, 16199N, 16200N, 16201N, 16202N, 16203N, 16204N, 16205N, 16206N, 16207N, 16208N, 16209N, 16210N, 16211N, 16212N, 16213N, 16214N, 16215N, 16216N, 16217N, 16218N, 16219N, 16220N, 16221N, 16222N, 16223N, 16224N, 16225N, 16226N, 16227N, 16228N, 16229N, 16230N, 16231N, 16232N, 16233N, 16234N, 16235N, 16236N, 16237N, 16238N, 16239N, 16240N, 16241N, 16242N, 16243N, 16244N, 16245N, 16246N, 16247N, 16248N, 16249N, 16250N, 16251N, 16252N, 16253N, 16254N, 16255N, 16256N, 16257N, 16258N, 16259N, 16260N, 16261N, 16262N, 16263N, 16264N, 16265N, 16266N, 16267N, 16268N, 16269N, 16270N, 16271N, 16272N, 16273N, 16274N, 16275N, 16276N, 16277N, 16278N, 16279N, 16280N, 16281N, 16282N, 16283N, 16284N, 16285N, 16286N, 16287N, 16288N, 16289N, 16290N, 16291N, 16292N, 16293N, 16294N, 16295N, 16296N, 16297N, 16298N, 16299N, 16300N, 16301N, 16302N, 16303N, 16304N, 16305N, 16306N, 16307N, 16308N, 16309N, 16310N, 16311N, 16312N, 16313N, 16314N, 16315N, 16316N, 16317N, 16318N, 16319N, 16320N, 16321N, 16322N, 16323N, 16324N, 16325N, 16326N, 16327N, 16328N, 16329N, 16330N, 16331N, 16332N, 16333N, 16334N, 16335N, 16336N, 16337N, 16338N, 16339N, 16340N, 16341N, 16342N, 16343N, 16344N, 16345N, 16346N, 16347N, 16348N, 16349N, 16350N, 16351N, 16352N, 16353N, 16354N, 16355N, 16356N, 16357N | Lippold,S., Matzke,N., Reissman,M., Burbano,H. and Hofreiter,M. BMC Evolutionary Biology. 328 (11), 1471-2148 (2011) |
| HQ439454 (Horse and Przewalski's horse) | Illumina/Solexa Genome Analyzer II | Phantom mutaions: 358d, 1387A, 2226+C, 5237+G, 5277+A, 15385+T. | G1b_HQ439454 | 158, 222, 356d, 382, 387, 416, 1387A, 2226+C, 2788, 2940, 3053, 3576, 4062, 4646, 4669, 4830, 5237+G, 5277+A, 5498, 5669, 5830, 5881, 5884, 6004, 6307, 6688, 6784, 7001, 7831, 8005, 8037, 9239, 9402, 9669, 9741A, 10214, 10376, 10471, 11165, 11240, 11453, 11543, 11552, 11842, 12767, 12860, 13049, 13223, 13333, 13502, 14350, 14476, 14626, 14734, 14825, 15385+T, 15492, 15539, 15582, 15594, 15599, 15632, 15647, 15663, 15700, 15717, 15867, 16028, 16110, 16126N, 16127N, 16128N, 16129N, 16130N, 16131N, 16132N, 16133N, 16134N, 16135N, 16136N, 16137N, 16138N, 16139N, 16140N, 16141N, 16142N, 16143N, 16144N, 16145N, 16146N, 16147N, 16148N, 16149N, 16150N, 16151N, 16152N, 16153N, 16154N, 16155N, 16156N, 16157N, 16158N, 16159N, 16160N, 16161N, 16162N, 16163N, 16164N, 16165N, 16166N, 16167N, 16168N, 16169N, 16170N, 16171N, 16172N, 16173N, 16174N, 16175N, 16176N, 16177N, 16178N, 16179N, 16180N, 16181N, 16182N, 16183N, 16184N, 16185N, 16186N, 16187N, 16188N, 16189N, 16190N, 16191N, 16192N, 16193N, 16194N, 16195N, 16196N, 16197N, 16198N, 16199N, 16200N, 16201N, 16202N, 16203N, 16204N, 16205N, 16206N, 16207N, 16208N, 16209N, 16210N, 16211N, 16212N, 16213N, 16214N, 16215N, 16216N, 16217N, 16218N, 16219N, 16220N, 16221N, 16222N, 16223N, 16224N, 16225N, 16226N, 16227N, 16228N, 16229N, 16230N, 16231N, 16232N, 16233N, 16234N, 16235N, 16236N, 16237N, 16238N, 16239N, 16240N, 16241N, 16242N, 16243N, 16244N, 16245N, 16246N, 16247N, 16248N, 16249N, 16250N, 16251N, 16252N, 16253N, 16254N, 16255N, 16256N, 16257N, 16258N, 16259N, 16260N, 16261N, 16262N, 16263N, 16264N, 16265N, 16266N, 16267N, 16268N, 16269N, 16270N, 16271N, 16272N, 16273N, 16274N, 16275N, 16276N, 16277N, 16278N, 16279N, 16280N, 16281N, 16282N, 16283N, 16284N, 16285N, 16286N, 16287N, 16288N, 16289N, 16290N, 16291N, 16292N, 16293N, 16294N, 16295N, 16296N, 16297N, 16298N, 16299N, 16300N, 16301N, 16302N, 16303N, 16304N, 16305N, 16306N, 16307N, 16308N, 16309N, 16310N, 16311N, 16312N, 16313N, 16314N, 16315N, 16316N, 16317N, 16318N, 16319N, 16320N, 16321N, 16322N, 16323N, 16324N, 16325N, 16326N, 16327N, 16328N, 16329N, 16330N, 16331N, 16332N, 16333N, 16334N, 16335N, 16336N, 16337N, 16338N, 16339N, 16340N, 16341N, 16342N, 16343N, 16344N, 16345N, 16346N, 16347N, 16348N, 16349N, 16350N, 16351N, 16352N, 16353N, 16354N, 16355N, 16356N, 16357N, 16368 | G1b_HQ439454:356 | G1b_HQ439454:356d, 2226+C, 5237+G, 5277+A, 15385+T, 16126N, 16127N, 16128N, 16129N, 16130N, 16131N, 16132N, 16133N, 16134N, 16135N, 16136N, 16137N, 16138N, 16139N, 16140N, 16141N, 16142N, 16143N, 16144N, 16145N, 16146N, 16147N, 16148N, 16149N, 16150N, 16151N, 16152N, 16153N, 16154N, 16155N, 16156N, 16157N, 16158N, 16159N, 16160N, 16161N, 16162N, 16163N, 16164N, 16165N, 16166N, 16167N, 16168N, 16169N, 16170N, 16171N, 16172N, 16173N, 16174N, 16175N, 16176N, 16177N, 16178N, 16179N, 16180N, 16181N, 16182N, 16183N, 16184N, 16185N, 16186N, 16187N, 16188N, 16189N, 16190N, 16191N, 16192N, 16193N, 16194N, 16195N, 16196N, 16197N, 16198N, 16199N, 16200N, 16201N, 16202N, 16203N, 16204N, 16205N, 16206N, 16207N, 16208N, 16209N, 16210N, 16211N, 16212N, 16213N, 16214N, 16215N, 16216N, 16217N, 16218N, 16219N, 16220N, 16221N, 16222N, 16223N, 16224N, 16225N, 16226N, 16227N, 16228N, 16229N, 16230N, 16231N, 16232N, 16233N, 16234N, 16235N, 16236N, 16237N, 16238N, 16239N, 16240N, 16241N, 16242N, 16243N, 16244N, 16245N, 16246N, 16247N, 16248N, 16249N, 16250N, 16251N, 16252N, 16253N, 16254N, 16255N, 16256N, 16257N, 16258N, 16259N, 16260N, 16261N, 16262N, 16263N, 16264N, 16265N, 16266N, 16267N, 16268N, 16269N, 16270N, 16271N, 16272N, 16273N, 16274N, 16275N, 16276N, 16277N, 16278N, 16279N, 16280N, 16281N, 16282N, 16283N, 16284N, 16285N, 16286N, 16287N, 16288N, 16289N, 16290N, 16291N, 16292N, 16293N, 16294N, 16295N, 16296N, 16297N, 16298N, 16299N, 16300N, 16301N, 16302N, 16303N, 16304N, 16305N, 16306N, 16307N, 16308N, 16309N, 16310N, 16311N, 16312N, 16313N, 16314N, 16315N, 16316N, 16317N, 16318N, 16319N, 16320N, 16321N, 16322N, 16323N, 16324N, 16325N, 16326N, 16327N, 16328N, 16329N, 16330N, 16331N, 16332N, 16333N, 16334N, 16335N, 16336N, 16337N, 16338N, 16339N, 16340N, 16341N, 16342N, 16343N, 16344N, 16345N, 16346N, 16347N, 16348N, 16349N, 16350N, 16351N, 16352N, 16353N, 16354N, 16355N, 16356N, 16357N | Lippold,S., Matzke,N., Reissman,M., Burbano,H. and Hofreiter,M. BMC Evolutionary Biology. 328 (11), 1471-2148 (2011) |
| HQ439455 (Horse and Przewalski's horse) | Illumina/Solexa Genome Analyzer II | Phantom mutaions: 358d, 1387A, 2227+T, 5237+G, 5274+G, 15385+T. | M1a, M1a1 | 158, 356d, 427, 874, 961, 1387A, 1609T, 2227+T, 2339A, 2788, 3070, 3100, 3475, 3800, 4062, 4526, 4536, 4599G, 4605, 4646, 4669, 4898, 5103, 5237+G, 5274+G, 5527, 5827, 5884, 6004, 6076, 6307, 6712, 6784, 7001, 7432, 7666, 7900, 8005, 8043, 8076, 8150, 8175, 8238, 8358, 8556T, 8565, 8798, 9086, 9332, 9540, 10110, 10173, 10214, 10292, 10376, 10448, 10460, 10646, 10859, 11240, 11394, 11492, 11543, 11842, 11879, 11966, 12029, 12095, 12332, 12767, 13049, 13100, 13333, 13356, 13502, 13615, 13629, 13720, 13920, 13933, 14422, 14626, 14671, 14803, 14815, 15052A, 15133, 15342, 15385+T, 15492, 15599, 15614, 15656, 15717, 15768, 15803, 15824, 15866, 15953, 16065, 16077, 16110, 16118, 16126N, 16127N, 16128N, 16129N, 16130N, 16131N, 16132N, 16133N, 16134N, 16135N, 16136N, 16137N, 16138N, 16139N, 16140N, 16141N, 16142N, 16143N, 16144N, 16145N, 16146N, 16147N, 16148N, 16149N, 16150N, 16151N, 16152N, 16153N, 16154N, 16155N, 16156N, 16157N, 16158N, 16159N, 16160N, 16161N, 16162N, 16163N, 16164N, 16165N, 16166N, 16167N, 16168N, 16169N, 16170N, 16171N, 16172N, 16173N, 16174N, 16175N, 16176N, 16177N, 16178N, 16179N, 16180N, 16181N, 16182N, 16183N, 16184N, 16185N, 16186N, 16187N, 16188N, 16189N, 16190N, 16191N, 16192N, 16193N, 16194N, 16195N, 16196N, 16197N, 16198N, 16199N, 16200N, 16201N, 16202N, 16203N, 16204N, 16205N, 16206N, 16207N, 16208N, 16209N, | M1a:356; M1a1:356 | M1a:356d, 2227+T, 5237+G, 5274+G, 15385+T, 16110, 16126N, 16127N, 16128N, 16129N, 16130N, 16131N, 16132N, 16133N, 16134N, 16135N, 16136N, 16137N, 16138N, 16139N, 16140N, 16141N, 16142N, 16143N, 16144N, 16145N, 16146N, 16147N, 16148N, 16149N, 16150N, 16151N, 16152N, 16153N, 16154N, 16155N, 16156N, 16157N, 16158N, 16159N, 16160N, 16161N, 16162N, 16163N, 16164N, 16165N, 16166N, 16167N, 16168N, 16169N, 16170N, 16171N, 16172N, 16173N, 16174N, 16175N, 16176N, 16177N, 16178N, 16179N, 16180N, 16181N, 16182N, 16183N, 16184N, 16185N, 16186N, 16187N, 16188N, 16189N, 16190N, 16191N, 16192N, 16193N, 16194N, 16195N, 16196N, 16197N, 16198N, 16199N, 16200N, 16201N, 16202N, 16203N, 16204N, 16205N, 16206N, 16207N, 16208N, 16209N, 16210N, 16211N, 16212N, 16213N, 16214N, 16215N, 16216N, 16217N, 16218N, 16219N, 16220N, 16221N, 16222N, 16223N, 16224N, 16225N, 16226N, 16227N, 16228N, 16229N, 16230N, 16231N, 16232N, 16233N, 16234N, 16235N, 16236N, 16237N, 16238N, 16239N, 16240N, | Lippold,S., Matzke,N., Reissman,M., Burbano,H. and Hofreiter,M. BMC Evolutionary Biology. 328 (11), 1471-2148 (2011) |
| | | | | 16210N, 16211N, 16212N, 16213N, 16214N, 16215N, 16216N, 16217N, 16218N, 16219N, 16220N, 16221N, 16222N, 16223N, 16224N, 16225N, 16226N, 16227N, 16228N, 16229N, 16230N, 16231N, 16232N, 16233N, 16234N, 16235N, 16236N, 16237N, 16238N, 16239N, 16240N, 16241N, 16242N, 16243N, 16244N, 16245N, 16246N, 16247N, 16248N, 16249N, 16250N, 16251N, 16252N, 16253N, 16254N, 16255N, 16256N, 16257N, 16258N, 16259N, 16260N, 16261N, 16262N, 16263N, 16264N, 16265N, 16266N, 16267N, 16268N, 16269N, 16270N, 16271N, 16272N, 16273N, 16274N, 16275N, 16276N, 16277N, 16278N, 16279N, 16280N, 16281N, 16282N, 16283N, 16284N, 16285N, 16286N, 16287N, 16288N, 16289N, 16290N, 16291N, 16292N, 16293N, 16294N, 16295N, 16296N, 16297N, 16298N, 16299N, 16300N, 16301N, 16302N, 16303N, 16304N, 16305N, 16306N, 16307N, 16308N, 16309N, 16310N, 16311N, 16312N, 16313N, 16314N, 16315N, 16316N, 16317N, 16318N, 16319N, 16320N, 16321N, 16322N, 16323N, 16324N, 16325N, 16326N, 16327N, 16328N, 16329N, 16330N, 16331N, 16332N, 16333N, 16334N, 16335N, 16336N, 16337N, 16338N, 16339N, 16340N, 16341N, 16342N, 16343N, 16344N, 16345N, 16346N, 16347N, 16348N, 16349N, 16350N, 16351N, 16352N, 16353N, 16354N, 16355N, 16356N, 16357N, 16368, 16540A, 16543, 16556, 16626 | | 16241N, 16242N, 16243N, 16244N, 16245N, 16246N, 16247N, 16248N, 16249N, 16250N, 16251N, 16252N, 16253N, 16254N, 16255N, 16256N, 16257N, 16258N, 16259N, 16260N, 16261N, 16262N, 16263N, 16264N, 16265N, 16266N, 16267N, 16268N, 16269N, 16270N, 16271N, 16272N, 16273N, 16274N, 16275N, 16276N, 16277N, 16278N, 16279N, 16280N, 16281N, 16282N, 16283N, 16284N, 16285N, 16286N, 16287N, 16288N, 16289N, 16290N, 16291N, 16292N, 16293N, 16294N, 16295N, 16296N, 16297N, 16298N, 16299N, 16300N, 16301N, 16302N, 16303N, 16304N, 16305N, 16306N, 16307N, 16308N, 16309N, 16310N, 16311N, 16312N, 16313N, 16314N, 16315N, 16316N, 16317N, 16318N, 16319N, 16320N, 16321N, 16322N, 16323N, 16324N, 16325N, 16326N, 16327N, 16328N, 16329N, 16330N, 16331N, 16332N, 16333N, 16334N, 16335N, 16336N, 16337N, 16338N, 16339N, 16340N, 16341N, 16342N, 16343N, 16344N, 16345N, 16346N, 16347N, 16348N, 16349N, 16350N, 16351N, 16352N, 16353N, 16354N, 16355N, 16356N, 16357N; M1a1:356d, 1387A, 2227+T, 5237+G, 5274+G, 15385+T, 16126N, 16127N, 16128N, 16129N, 16130N, 16131N, 16132N, 16133N, 16134N, 16135N, 16136N, 16137N, 16138N, 16139N, 16140N, 16141N, 16142N, 16143N, 16144N, 16145N, 16146N, 16147N, 16148N, 16149N, 16150N, 16151N, 16152N, 16153N, 16154N, 16155N, 16156N, 16157N, 16158N, 16159N, 16160N, 16161N, 16162N, 16163N, 16164N, 16165N, 16166N, 16167N, 16168N, 16169N, 16170N, 16171N, 16172N, 16173N, 16174N, 16175N, 16176N, 16177N, 16178N, 16179N, 16180N, 16181N, 16182N, 16183N, 16184N, 16185N, 16186N, 16187N, 16188N, 16189N, 16190N, 16191N, 16192N, 16193N, 16194N, 16195N, 16196N, 16197N, 16198N, 16199N, 16200N, 16201N, 16202N, 16203N, 16204N, 16205N, 16206N, 16207N, 16208N, 16209N, 16210N, 16211N, 16212N, 16213N, 16214N, 16215N, 16216N, 16217N, 16218N, 16219N, 16220N, 16221N, 16222N, 16223N, 16224N, 16225N, 16226N, 16227N, 16228N, 16229N, 16230N, 16231N, 16232N, 16233N, 16234N, 16235N, 16236N, 16237N, 16238N, 16239N, 16240N, 16241N, 16242N, 16243N, 16244N, 16245N, 16246N, 16247N, 16248N, 16249N, 16250N, 16251N, 16252N, 16253N, 16254N, 16255N, 16256N, 16257N, 16258N, 16259N, 16260N, 16261N, 16262N, 16263N, 16264N, 16265N, 16266N, 16267N, 16268N, 16269N, 16270N, 16271N, 16272N, 16273N, 16274N, 16275N, 16276N, 16277N, 16278N, 16279N, 16280N, 16281N, 16282N, 16283N, 16284N, 16285N, 16286N, 16287N, 16288N, 16289N, 16290N, 16291N, 16292N, 16293N, 16294N, 16295N, 16296N, 16297N, 16298N, 16299N, 16300N, 16301N, 16302N, 16303N, 16304N, 16305N, 16306N, 16307N, 16308N, 16309N, 16310N, 16311N, 16312N, 16313N, 16314N, 16315N, 16316N, 16317N, 16318N, 16319N, 16320N, 16321N, 16322N, 16323N, 16324N, 16325N, 16326N, 16327N, 16328N, 16329N, 16330N, 16331N, 16332N, 16333N, 16334N, 16335N, 16336N, 16337N, 16338N, 16339N, 16340N, 16341N, 16342N, 16343N, 16344N, 16345N, 16346N, 16347N, 16348N, 16349N, 16350N, 16351N, 16352N, 16353N, 16354N, 16355N, 16356N, 16357N | |
| HQ439456 (Horse and Przewalski's horse) | Illumina/Solexa Genome Analyzer II | Phantom mutaions: 358d, 1387G, 2226+C, 5241+T, 5277+A, 15385+T. | L3a1_HQ439456 | 1d, 77, 158, 356d, 961, 1375, 1387G, 1891N, 1892N, 1893N, 1894N, 1895N, 1896N, 1897N, 1898N, 1899N, 1900N, 1901N, 1902N, 2226+C, 2788, 2899, 3517, 3557, 3942, 4062, 4536, 4646, 4669, 5241+T, 5277+A, 5527, 5815, 5884, 6004, 6307, 6784, 7001, 7516, 7666, 7900, 8005, 8058, 8301, 8319, 8358, 8565, 9239, 9951, 10110, 10214, 10292, 10376, 10421, 10613, 11084, 11240, 11543, 11693, 11842, 11879, 12119, 12200, 12398, 12767, 12896, 12950, 13049, 13333, 13520, 13996N, 14803, 14995, 15313, 15385+T, 15491, 15492, 15493, 15531, 15582, 15599, 15600, 15601, 15646, 15717, 15768, 15867, 15868, 15953, 15971, 16065, 16100, 16108, 16117N, 16118N, 16119N, 16120N, 16121N, 16122N, 16123N, 16124N, 16125N, 16126N, 16127N, 16128N, 16129N, 16130N, 16131N, 16132N, 16133N, 16134N, 16135N, 16136N, 16137N, 16138N, 16139N, 16140N, 16141N, 16142N, 16143N, 16144N, 16145N, 16146N, 16147N, 16148N, 16149N, 16150N, 16151N, 16152N, 16153N, 16154N, 16155N, 16156N, 16157N, 16158N, 16159N, 16160N, 16161N, 16162N, 16163N, 16164N, 16165N, 16166N, 16167N, 16168N, 16169N, 16170N, 16171N, 16172N, 16173N, 16174N, 16175N, 16176N, 16177N, 16178N, 16179N, 16180N, 16181N, 16182N, 16183N, 16184N, 16185N, 16186N, 16187N, 16188N, 16189N, 16190N, 16191N, 16192N, 16193N, 16194N, 16195N, 16196N, 16197N, 16198N, 16199N, 16200N, 16201N, 16202N, 16203N, 16204N, 16205N, 16206N, 16207N, 16208N, 16209N, 16210N, 16211N, 16212N, 16213N, 16214N, 16215N, 16216N, 16217N, 16218N, 16219N, 16220N, 16221N, 16222N, 16223N, 16224N, 16225N, 16226N, 16227N, 16228N, 16229N, 16230N, 16231N, 16232N, 16233N, 16234N, 16235N, 16236N, 16237N, 16238N, 16239N, 16240N, 16241N, 16242N, 16243N, 16244N, 16245N, 16246N, 16247N, 16248N, 16249N, 16250N, 16251N, 16252N, 16253N, 16254N, 16255N, 16256N, 16257N, 16258N, 16259N, 16260N, 16261N, 16262N, 16263N, 16264N, 16265N, 16266N, 16267N, 16268N, 16269N, 16270N, 16271N, 16272N, 16273N, 16274N, 16275N, 16276N, 16277N, 16278N, 16279N, 16280N, 16281N, 16282N, 16283N, 16284N, 16285N, 16286N, 16287N, 16288N, 16289N, 16290N, 16291N, 16292N, 16293N, 16294N, 16295N, 16296N, 16297N, 16298N, 16299N, 16300N, 16301N, 16302N, 16303N, 16304N, 16305N, 16306N, 16307N, 16308N, 16309N, 16310N, 16311N, 16312N, 16313N, 16314N, 16315N, 16316N, 16317N, 16318N, 16319N, 16320N, 16321N, 16322N, 16323N, 16324N, 16325N, 16326N, 16327N, 16328N, 16329N, 16330N, 16331N, 16332N, 16333N, 16334N, 16335N, 16336N, 16337N, 16338N, 16339N, 16340N, 16341N, 16342N, 16343N, 16344N, 16345N, 16346N, 16347N, 16348N, 16349N, 16350N, 16351N, 16352N, 16353N, 16354N, 16355N, 16356N, 16357N, 16368, 16554, 16626 | L3a1_HQ439456:356 | L3a1_HQ439456:1d, 356d, 1891N, 1892N, 1893N, 1894N, 1895N, 1896N, 1897N, 1898N, 1899N, 1900N, 1901N, 1902N, 2226+C, 3557, 5277+A, 13996N, 15385+T, 16117N, 16118N, 16119N, 16120N, 16121N, 16122N, 16123N, 16124N, 16125N, 16126N, 16127N, 16128N, 16129N, 16130N, 16131N, 16132N, 16133N, 16134N, 16135N, 16136N, 16137N, 16138N, 16139N, 16140N, 16141N, 16142N, 16143N, 16144N, 16145N, 16146N, 16147N, 16148N, 16149N, 16150N, 16151N, 16152N, 16153N, 16154N, 16155N, 16156N, 16157N, 16158N, 16159N, 16160N, 16161N, 16162N, 16163N, 16164N, 16165N, 16166N, 16167N, 16168N, 16169N, 16170N, 16171N, 16172N, 16173N, 16174N, 16175N, 16176N, 16177N, 16178N, 16179N, 16180N, 16181N, 16182N, 16183N, 16184N, 16185N, 16186N, 16187N, 16188N, 16189N, 16190N, 16191N, 16192N, 16193N, 16194N, 16195N, 16196N, 16197N, 16198N, 16199N, 16200N, 16201N, 16202N, 16203N, 16204N, 16205N, 16206N, 16207N, 16208N, 16209N, 16210N, 16211N, 16212N, 16213N, 16214N, 16215N, 16216N, 16217N, 16218N, 16219N, 16220N, 16221N, 16222N, 16223N, 16224N, 16225N, 16226N, 16227N, 16228N, 16229N, 16230N, 16231N, 16232N, 16233N, 16234N, 16235N, 16236N, 16237N, 16238N, 16239N, 16240N, 16241N, 16242N, 16243N, 16244N, 16245N, 16246N, 16247N, 16248N, 16249N, 16250N, 16251N, 16252N, 16253N, 16254N, 16255N, 16256N, 16257N, 16258N, 16259N, 16260N, 16261N, 16262N, 16263N, 16264N, 16265N, 16266N, 16267N, 16268N, 16269N, 16270N, 16271N, 16272N, 16273N, 16274N, 16275N, 16276N, 16277N, 16278N, 16279N, 16280N, 16281N, 16282N, 16283N, 16284N, 16285N, 16286N, 16287N, 16288N, 16289N, 16290N, 16291N, 16292N, 16293N, 16294N, 16295N, 16296N, 16297N, 16298N, 16299N, 16300N, 16301N, 16302N, 16303N, 16304N, 16305N, 16306N, 16307N, 16308N, 16309N, 16310N, 16311N, 16312N, 16313N, 16314N, 16315N, 16316N, 16317N, 16318N, 16319N, 16320N, 16321N, 16322N, 16323N, 16324N, 16325N, 16326N, 16327N, 16328N, 16329N, 16330N, 16331N, 16332N, 16333N, 16334N, 16335N, 16336N, 16337N, 16338N, 16339N, 16340N, 16341N, 16342N, 16343N, 16344N, 16345N, 16346N, 16347N, 16348N, 16349N, 16350N, 16351N, 16352N, 16353N, 16354N, 16355N, 16356N, 16357N | Lippold,S., Matzke,N., Reissman,M., Burbano,H. and Hofreiter,M. BMC Evolutionary Biology. 328 (11), 1471-2148 (2011) |
| HQ439457 (Horse and Przewalski's horse) | Illumina/Solexa Genome Analyzer II | Phantom mutaions: 358d, 1387A, 2227+T, 5239+A, 5277+A, 15385+T. | P_HQ439457 | 158, 356d, 739, 860, 961, 1383, 1387A, 1581N, 1582N, 1583N, 1584N, 1585N, 1586N, 1587N, 1588N, 1589N, 1590N, 1591N, 1592N, 1593N, 1594N, 1595N, 1596N, 1597N, 1598N, 1599N, 1684C, 1876N, 1877N, 1878N, 1879N, 1880N, 1881N, 1882N, 1883N, 1884N, 1885N, 1886N, 1887N, 1888N, 1889N, 1890N, 1891N, 1892N, 1893N, 1894N, 1895N, 1896N, 1897N, 2227+T, 2788, 2945N, 2946N, 2947N, 2948N, 2949N, 2950N, 2951N, 2952N, 2953N, 2954N, 2955N, 2956N, 2957N, 2958N, 2959N, 2960N, 2961N, 2962N, 3070, 3259T, 3271, 3557, 3616, 3800N, 3815N, 3816N, 3817N, 3818N, 3819N, 3820N, 3821N, 3822N, 3823N, 3942, 4062, 4536, 4605, 4646N, 4669N, 5094N, 5098N, 5239+A, 5277+A, 5527, 5827, 5884, 6004, 6307, 6505, 6529, 6712, 6784, 7001, 7243, 7290N, 7291N, 7292N, 7293N, 7294N, 7295N, 7296N, 7427, 7612T, 7666, 7898, 7900, 8005, 8076, 8150, 8238, 8358, 8361, 8556T, 8565, 8855, 9053, 9086, 9239, 9753, 9775, 10110, 10173, 10214, 10292, 10376, 10448N, 10826N, 10827N, 10859, 11045, 11240, 11378, 11394, 11424, 11492, 11543, 11842, 11879, 11966, 12167, 12230, 12287, 12332, 12404, 12767, 13049, 13333, 13463, 13466, 13629, 13685N, 13747, | P_HQ439457:356, 3800, 4646, 4669, 10448, 16118, 16626 | P_HQ439457:356d, 1581N, 1582N, 1583N, 1584N, 1585N, 1586N, 1587N, 1588N, 1589N, 1590N, 1591N, 1592N, 1593N, 1594N, 1595N, 1596N, 1597N, 1598N, 1599N, 1876N, 1877N, 1878N, 1879N, 1880N, 1881N, 1882N, 1883N, 1884N, 1885N, 1886N, 1887N, 1888N, 1889N, 1890N, 1891N, 1892N, 1893N, 1894N, 1895N, 1896N, 1897N, 2227+T, 2945N, 2946N, 2947N, 2948N, 2949N, 2950N, 2951N, 2952N, 2953N, 2954N, 2955N, 2956N, 2957N, 2958N, 2959N, 2960N, 2961N, 2962N, 3800N, 3815N, 3816N, 3817N, 3818N, 3819N, 3820N, 3821N, 3822N, 3823N, 4646N, 4669N, 5094N, 5098N, 5239+A, 5277+A, 7290N, 7291N, 7292N, 7293N, 7294N, 7295N, 7296N, 10448N, 10826N, 10827N, 13685N, 15385+T, 16118N, 16126N, 16127N, 16128N, 16129N, 16130N, 16131N, 16132N, 16133N, 16134N, 16135N, 16136N, 16137N, 16138N, 16139N, 16140N, 16141N, 16142N, 16143N, 16144N, 16145N, | Lippold,S., Matzke,N., Reissman,M., Burbano,H. and Hofreiter,M. BMC Evolutionary Biology. 328 (11), 1471-2148 (2011) |
| | | | | 13761, 13920, 13933, 14626, 14803, 15202, 15342, 15385+T, 15492, 15594, 15599, 15664, 15700, 15717, 15768, 15774, 15806, 15953, 16035, 16118N, 16126N, 16127N, 16128N, 16129N, 16130N, 16131N, 16132N, 16133N, 16134N, 16135N, 16136N, 16137N, 16138N, 16139N, 16140N, 16141N, 16142N, 16143N, 16144N, 16145N, 16146N, 16147N, 16148N, 16149N, 16150N, 16151N, 16152N, 16153N, 16154N, 16155N, 16156N, 16157N, 16158N, 16159N, 16160N, 16161N, 16162N, 16163N, 16164N, 16165N, 16166N, 16167N, 16168N, 16169N, 16170N, 16171N, 16172N, 16173N, 16174N, 16175N, 16176N, 16177N, 16178N, 16179N, 16180N, 16181N, 16182N, 16183N, 16184N, 16185N, 16186N, 16187N, 16188N, 16189N, 16190N, 16191N, 16192N, 16193N, 16194N, 16195N, 16196N, 16197N, 16198N, 16199N, 16200N, 16201N, 16202N, 16203N, 16204N, 16205N, 16206N, 16207N, 16208N, 16209N, 16210N, 16211N, 16212N, 16213N, 16214N, 16215N, 16216N, 16217N, 16218N, 16219N, 16220N, 16221N, 16222N, 16223N, 16224N, 16225N, 16226N, 16227N, 16228N, 16229N, 16230N, 16231N, 16232N, 16233N, 16234N, 16235N, 16236N, 16237N, 16238N, 16239N, 16240N, 16241N, 16242N, 16243N, 16244N, 16245N, 16246N, 16247N, 16248N, 16249N, 16250N, 16251N, 16252N, 16253N, 16254N, 16255N, 16256N, 16257N, 16258N, 16259N, 16260N, 16261N, 16262N, 16263N, 16264N, 16265N, 16266N, 16267N, 16268N, 16269N, 16270N, 16271N, 16272N, 16273N, 16274N, 16275N, 16276N, 16277N, 16278N, 16279N, 16280N, 16281N, 16282N, 16283N, 16284N, 16285N, 16286N, 16287N, 16288N, 16289N, 16290N, 16291N, 16292N, 16293N, 16294N, 16295N, 16296N, 16297N, 16298N, 16299N, 16300N, 16301N, 16302N, 16303N, 16304N, 16305N, 16306N, 16307N, 16308N, 16309N, 16310N, 16311N, 16312N, 16313N, 16314N, 16315N, 16316N, 16317N, 16318N, 16319N, 16320N, 16321N, 16322N, 16323N, 16324N, 16325N, 16326N, 16327N, 16328N, 16329N, 16330N, 16331N, 16332N, 16333N, 16334N, 16335N, 16336N, 16337N, 16338N, 16339N, 16340N, 16341N, 16342N, 16343N, 16344N, 16345N, 16346N, 16347N, 16348N, 16349N, 16350N, 16351N, 16352N, 16353N, 16354N, 16355N, 16356N, 16357N, 16368, 16540A, 16626N, 16651-16657d | | 16146N, 16147N, 16148N, 16149N, 16150N, 16151N, 16152N, 16153N, 16154N, 16155N, 16156N, 16157N, 16158N, 16159N, 16160N, 16161N, 16162N, 16163N, 16164N, 16165N, 16166N, 16167N, 16168N, 16169N, 16170N, 16171N, 16172N, 16173N, 16174N, 16175N, 16176N, 16177N, 16178N, 16179N, 16180N, 16181N, 16182N, 16183N, 16184N, 16185N, 16186N, 16187N, 16188N, 16189N, 16190N, 16191N, 16192N, 16193N, 16194N, 16195N, 16196N, 16197N, 16198N, 16199N, 16200N, 16201N, 16202N, 16203N, 16204N, 16205N, 16206N, 16207N, 16208N, 16209N, 16210N, 16211N, 16212N, 16213N, 16214N, 16215N, 16216N, 16217N, 16218N, 16219N, 16220N, 16221N, 16222N, 16223N, 16224N, 16225N, 16226N, 16227N, 16228N, 16229N, 16230N, 16231N, 16232N, 16233N, 16234N, 16235N, 16236N, 16237N, 16238N, 16239N, 16240N, 16241N, 16242N, 16243N, 16244N, 16245N, 16246N, 16247N, 16248N, 16249N, 16250N, 16251N, 16252N, 16253N, 16254N, 16255N, 16256N, 16257N, 16258N, 16259N, 16260N, 16261N, 16262N, 16263N, 16264N, 16265N, 16266N, 16267N, 16268N, 16269N, 16270N, 16271N, 16272N, 16273N, 16274N, 16275N, 16276N, 16277N, 16278N, 16279N, 16280N, 16281N, 16282N, 16283N, 16284N, 16285N, 16286N, 16287N, 16288N, 16289N, 16290N, 16291N, 16292N, 16293N, 16294N, 16295N, 16296N, 16297N, 16298N, 16299N, 16300N, 16301N, 16302N, 16303N, 16304N, 16305N, 16306N, 16307N, 16308N, 16309N, 16310N, 16311N, 16312N, 16313N, 16314N, 16315N, 16316N, 16317N, 16318N, 16319N, 16320N, 16321N, 16322N, 16323N, 16324N, 16325N, 16326N, 16327N, 16328N, 16329N, 16330N, 16331N, 16332N, 16333N, 16334N, 16335N, 16336N, 16337N, 16338N, 16339N, 16340N, 16341N, 16342N, 16343N, 16344N, 16345N, 16346N, 16347N, 16348N, 16349N, 16350N, 16351N, 16352N, 16353N, 16354N, 16355N, 16356N, 16357N, 16626N, 16651-16657d | |
| HQ439458 (Horse and Przewalski's horse) | Illumina/Solexa Genome Analyzer II | Phantom mutaions: 358d, 2227+T, 5239+A, 5277+A, 15383+A. | B1a1_HQ439458 | 1-9d, 158, 356d, 859, 2227+T, 4062, 4440N, 5239+A, 5277+A, 5566N, 5567N, 6784, 7204, 7627, 9570, 9961, 10764, 11240, 11896, 11942, 11951, 13502, 13816, 13884N, 15383+A, 15492, 15582, 15647, 15663, 15717, 15807, 15823, 16052, 16110, 16126N, 16127N, 16128N, 16129N, 16130N, 16131N, 16132N, 16133N, 16134N, 16135N, 16136N, 16137N, 16138N, 16139N, 16140N, 16141N, 16142N, 16143N, 16144N, 16145N, 16146N, 16147N, 16148N, 16149N, 16150N, 16151N, 16152N, 16153N, 16154N, 16155N, 16156N, 16157N, 16158N, 16159N, 16160N, 16161N, 16162N, 16163N, 16164N, 16165N, 16166N, 16167N, 16168N, 16169N, 16170N, 16171N, 16172N, 16173N, 16174N, 16175N, 16176N, 16177N, 16178N, 16179N, 16180N, 16181N, 16182N, 16183N, 16184N, 16185N, 16186N, 16187N, 16188N, 16189N, 16190N, 16191N, 16192N, 16193N, 16194N, 16195N, 16196N, 16197N, 16198N, 16199N, 16200N, 16201N, 16202N, 16203N, 16204N, 16205N, 16206N, 16207N, 16208N, 16209N, 16210N, 16211N, 16212N, 16213N, 16214N, 16215N, 16216N, 16217N, 16218N, 16219N, 16220N, 16221N, 16222N, 16223N, 16224N, 16225N, 16226N, 16227N, 16228N, 16229N, 16230N, 16231N, 16232N, 16233N, 16234N, 16235N, 16236N, 16237N, 16238N, 16239N, 16240N, 16241N, 16242N, 16243N, 16244N, 16245N, 16246N, 16247N, 16248N, 16249N, 16250N, 16251N, 16252N, 16253N, 16254N, 16255N, 16256N, 16257N, 16258N, 16259N, 16260N, 16261N, 16262N, 16263N, 16264N, 16265N, 16266N, 16267N, 16268N, 16269N, 16270N, 16271N, 16272N, 16273N, 16274N, 16275N, 16276N, 16277N, 16278N, 16279N, 16280N, 16281N, 16282N, 16283N, 16284N, 16285N, 16286N, 16287N, 16288N, 16289N, 16290N, 16291N, 16292N, | B1a1_HQ439458:356 | B1a1_HQ439458:1-9d, 356d, 2227+T, 4440N, 5239+A, 5277+A, 5566N, 5567N, 13884N, 15383+A, 16126N, 16127N, 16128N, 16129N, 16130N, 16131N, 16132N, 16133N, 16134N, 16135N, 16136N, 16137N, 16138N, 16139N, 16140N, 16141N, 16142N, 16143N, 16144N, 16145N, 16146N, 16147N, 16148N, 16149N, 16150N, 16151N, 16152N, 16153N, 16154N, 16155N, 16156N, 16157N, 16158N, 16159N, 16160N, 16161N, 16162N, 16163N, 16164N, 16165N, 16166N, 16167N, 16168N, 16169N, 16170N, 16171N, 16172N, 16173N, 16174N, 16175N, 16176N, 16177N, 16178N, 16179N, 16180N, 16181N, 16182N, 16183N, 16184N, 16185N, 16186N, 16187N, 16188N, 16189N, 16190N, 16191N, 16192N, 16193N, 16194N, 16195N, 16196N, 16197N, 16198N, 16199N, 16200N, 16201N, 16202N, 16203N, 16204N, 16205N, 16206N, 16207N, 16208N, 16209N, 16210N, 16211N, 16212N, 16213N, 16214N, 16215N, 16216N, 16217N, 16218N, 16219N, 16220N, 16221N, 16222N, 16223N, 16224N, 16225N, 16226N, 16227N, 16228N, 16229N, 16230N, 16231N, 16232N, 16233N, 16234N, 16235N, 16236N, 16237N, 16238N, 16239N, 16240N, 16241N, 16242N, 16243N, 16244N, 16245N, 16246N, 16247N, 16248N, 16249N, 16250N, 16251N, 16252N, 16253N, 16254N, 16255N, 16256N, 16257N, | Lippold,S., Matzke,N., Reissman,M., Burbano,H. and Hofreiter,M. BMC Evolutionary Biology. 328 (11), 1471-2148 (2011) |
| | | | | 16293N, 16294N, 16295N, 16296N, 16297N, 16298N, 16299N, 16300N, 16301N, 16302N, 16303N, 16304N, 16305N, 16306N, 16307N, 16308N, 16309N, 16310N, 16311N, 16312N, 16313N, 16314N, 16315N, 16316N, 16317N, 16318N, 16319N, 16320N, 16321N, 16322N, 16323N, 16324N, 16325N, 16326N, 16327N, 16328N, 16329N, 16330N, 16331N, 16332N, 16333N, 16334N, 16335N, 16336N, 16337N, 16338N, 16339N, 16340N, 16341N, 16342N, 16343N, 16344N, 16345N, 16346N, 16347N, 16348N, 16349N, 16350N, 16351N, 16352N, 16353N, 16354N, 16355N, 16356N, 16357N, 16368, 16653-16657d | | 16258N, 16259N, 16260N, 16261N, 16262N, 16263N, 16264N, 16265N, 16266N, 16267N, 16268N, 16269N, 16270N, 16271N, 16272N, 16273N, 16274N, 16275N, 16276N, 16277N, 16278N, 16279N, 16280N, 16281N, 16282N, 16283N, 16284N, 16285N, 16286N, 16287N, 16288N, 16289N, 16290N, 16291N, 16292N, 16293N, 16294N, 16295N, 16296N, 16297N, 16298N, 16299N, 16300N, 16301N, 16302N, 16303N, 16304N, 16305N, 16306N, 16307N, 16308N, 16309N, 16310N, 16311N, 16312N, 16313N, 16314N, 16315N, 16316N, 16317N, 16318N, 16319N, 16320N, 16321N, 16322N, 16323N, 16324N, 16325N, 16326N, 16327N, 16328N, 16329N, 16330N, 16331N, 16332N, 16333N, 16334N, 16335N, 16336N, 16337N, 16338N, 16339N, 16340N, 16341N, 16342N, 16343N, 16344N, 16345N, 16346N, 16347N, 16348N, 16349N, 16350N, 16351N, 16352N, 16353N, 16354N, 16355N, 16356N, 16357N | |
| HQ439459 (Horse and Przewalski's horse) | Illumina/Solexa Genome Analyzer II | Phantom mutaions: 358d, 1387A, 2226+C, 5239+A, 5277+A, 15385+T. | L2a2_HQ439459 | 158, 356d, 961, 1375, 1387A, 1521, 2226+C, 2607, 2788, 2899, 3517, 3942, 4062N, 4536N, 4548N, 4549N, 4550N, 4551N, 4552N, 4553N, 4554N, 4555N, 4556N, 4557N, 4558N, 4559N, 4560N, 4561N, 4562N, 4563N, 4564N, 4565N, 4566N, 4567N, 4646, 4669, 5239+A, 5275N, 5277+A, 5527N, 5815, 5884, 5998, 6004, 6307, 6784, 7001, 7516, 7666, 7900, 8005, 8058, 8199, 8301, 8319, 8358, 8403, 8565, 9239, 9716, 9951, 10110, 10214, 10292, 10376, 10421, 10613, 11240, 11543, 11693, 11842, 11879, 12119, 12200, 12767, 12896, 12950, 13049, 13333, 13520, 14803, 14995, 15313, 15385+T, 15491, 15492, 15493, 15531, 15582, 15600, 15646, 15717, 15768, 15867, 15868, 15953, 15971, 16065, 16100, 16126N, 16127N, 16128N, 16129N, 16130N, 16131N, 16132N, 16133N, 16134N, 16135N, 16136N, 16137N, 16138N, 16139N, 16140N, 16141N, 16142N, 16143N, 16144N, 16145N, 16146N, 16147N, 16148N, 16149N, 16150N, 16151N, 16152N, 16153N, 16154N, 16155N, 16156N, 16157N, 16158N, 16159N, 16160N, 16161N, 16162N, 16163N, 16164N, 16165N, 16166N, 16167N, 16168N, 16169N, 16170N, 16171N, 16172N, 16173N, 16174N, 16175N, 16176N, 16177N, 16178N, 16179N, 16180N, 16181N, 16182N, 16183N, 16184N, 16185N, 16186N, 16187N, 16188N, 16189N, 16190N, 16191N, 16192N, 16193N, 16194N, 16195N, 16196N, 16197N, 16198N, 16199N, 16200N, 16201N, 16202N, 16203N, 16204N, 16205N, 16206N, 16207N, 16208N, 16209N, 16210N, 16211N, 16212N, 16213N, 16214N, 16215N, 16216N, 16217N, 16218N, 16219N, 16220N, 16221N, 16222N, 16223N, 16224N, 16225N, 16226N, 16227N, 16228N, 16229N, 16230N, 16231N, 16232N, 16233N, 16234N, 16235N, 16236N, 16237N, 16238N, 16239N, 16240N, 16241N, 16242N, 16243N, 16244N, 16245N, 16246N, 16247N, 16248N, 16249N, 16250N, 16251N, 16252N, 16253N, 16254N, 16255N, 16256N, 16257N, 16258N, 16259N, 16260N, 16261N, 16262N, 16263N, 16264N, 16265N, 16266N, 16267N, 16268N, 16269N, 16270N, 16271N, 16272N, 16273N, 16274N, 16275N, 16276N, 16277N, 16278N, 16279N, 16280N, 16281N, 16282N, 16283N, 16284N, 16285N, 16286N, 16287N, 16288N, 16289N, 16290N, 16291N, 16292N, 16293N, 16294N, 16295N, 16296N, 16297N, 16298N, 16299N, 16300N, 16301N, 16302N, 16303N, 16304N, 16305N, 16306N, 16307N, 16308N, 16309N, 16310N, 16311N, 16312N, 16313N, 16314N, 16315N, 16316N, 16317N, 16318N, 16319N, 16320N, 16321N, 16322N, 16323N, 16324N, 16325N, 16326N, 16327N, 16328N, 16329N, 16330N, 16331N, 16332N, 16333N, 16334N, 16335N, 16336N, 16337N, 16338N, 16339N, 16340N, 16341N, 16342N, 16343N, 16344N, 16345N, 16346N, 16347N, 16348N, 16349N, 16350N, 16351N, 16352N, 16353N, 16354N, 16355N, 16356N, 16357N, 16358N, 16359N, 16360N, 16361N, 16362N, 16363N, 16364N, 16365N, 16366N, 16367N, 16368N, 16369N, 16370N, 16371N, 16372N, 16373N, 16626 | L2a2_HQ439459:356, 4062, 4536, 5527, 16368 | L2a2_HQ439459:356d, 2226+C, 4062N, 4536N, 4548N, 4549N, 4550N, 4551N, 4552N, 4553N, 4554N, 4555N, 4556N, 4557N, 4558N, 4559N, 4560N, 4561N, 4562N, 4563N, 4564N, 4565N, 4566N, 4567N, 5239+A, 5275N, 5277+A, 5527N, 5998, 8565, 15385+T, 16126N, 16127N, 16128N, 16129N, 16130N, 16131N, 16132N, 16133N, 16134N, 16135N, 16136N, 16137N, 16138N, 16139N, 16140N, 16141N, 16142N, 16143N, 16144N, 16145N, 16146N, 16147N, 16148N, 16149N, 16150N, 16151N, 16152N, 16153N, 16154N, 16155N, 16156N, 16157N, 16158N, 16159N, 16160N, 16161N, 16162N, 16163N, 16164N, 16165N, 16166N, 16167N, 16168N, 16169N, 16170N, 16171N, 16172N, 16173N, 16174N, 16175N, 16176N, 16177N, 16178N, 16179N, 16180N, 16181N, 16182N, 16183N, 16184N, 16185N, 16186N, 16187N, 16188N, 16189N, 16190N, 16191N, 16192N, 16193N, 16194N, 16195N, 16196N, 16197N, 16198N, 16199N, 16200N, 16201N, 16202N, 16203N, 16204N, 16205N, 16206N, 16207N, 16208N, 16209N, 16210N, 16211N, 16212N, 16213N, 16214N, 16215N, 16216N, 16217N, 16218N, 16219N, 16220N, 16221N, 16222N, 16223N, 16224N, 16225N, 16226N, 16227N, 16228N, 16229N, 16230N, 16231N, 16232N, 16233N, 16234N, 16235N, 16236N, 16237N, 16238N, 16239N, 16240N, 16241N, 16242N, 16243N, 16244N, 16245N, 16246N, 16247N, 16248N, 16249N, 16250N, 16251N, 16252N, 16253N, 16254N, 16255N, 16256N, 16257N, 16258N, 16259N, 16260N, 16261N, 16262N, 16263N, 16264N, 16265N, 16266N, 16267N, 16268N, 16269N, 16270N, 16271N, 16272N, 16273N, 16274N, 16275N, 16276N, 16277N, 16278N, 16279N, 16280N, 16281N, 16282N, 16283N, 16284N, 16285N, 16286N, 16287N, 16288N, 16289N, 16290N, 16291N, 16292N, 16293N, 16294N, 16295N, 16296N, 16297N, 16298N, 16299N, 16300N, 16301N, 16302N, 16303N, 16304N, 16305N, 16306N, 16307N, 16308N, 16309N, 16310N, 16311N, 16312N, 16313N, 16314N, 16315N, 16316N, 16317N, 16318N, 16319N, 16320N, 16321N, 16322N, 16323N, 16324N, 16325N, 16326N, 16327N, 16328N, 16329N, 16330N, 16331N, 16332N, 16333N, 16334N, 16335N, 16336N, 16337N, 16338N, 16339N, 16340N, 16341N, 16342N, 16343N, 16344N, 16345N, 16346N, 16347N, 16348N, 16349N, 16350N, | Lippold,S., Matzke,N., Reissman,M., Burbano,H. and Hofreiter,M. BMC Evolutionary Biology. 328 (11), 1471-2148 (2011) |
| | | | | | | 16351N, 16352N, 16353N, 16354N, 16355N, 16356N, 16357N, 16358N, 16359N, 16360N, 16361N, 16362N, 16363N, 16364N, 16365N, 16366N, 16367N, 16368N, 16369N, 16370N, 16371N, 16372N, 16373N | |
| HQ439460 (Horse and Przewalski's horse) | Illumina/Solexa Genome Analyzer II | Phantom mutaions: 358d, 2226+C, 5239+A, 5277+A, 15383+G. | B1c1_HQ439460 | 19N, 21N, 32N, 33N, 34N, 35N, 36N, 37N, 158, 356d, 669, 859, 1563N, 1564N, 1565N, 1566N, 1567N, 1568N, 1569N, 1580N, 2226+C, 4062N, 4743, 5239+A, 5277+A, 5929, 6784, 7176N, 7177N, 7178N, 7179N, 7180N, 7627, 9961, 10212N, 10213N, 10214N, 10215N, 10216N, 10217N, 10218N, 10219N, 10220N, 10221N, 10222N, 10223N, 10224N, 10225N, 10226N, 10227N, 10228N, 10229N, 10230N, 10231N, 10232N, 10233N, 10234N, 10235N, 10236N, 10237N, 10238N, 10239N, 10240N, 10241N, 10242N, 10243N, 10244N, 10245N, 10246N, 10247N, 10248N, 10249N, 10250N, 10251N, 10252N, 10764, 11240, 12314, 12332, 14689, 15383+G, 15492, 15647, 15663, 15717, 15823, 16052, 16108, 16110, 16124N, 16125N, 16126N, 16127N, 16128N, 16129N, 16130N, 16131N, 16132N, 16133N, 16134N, 16135N, 16136N, 16137N, 16138N, 16139N, 16140N, 16141N, 16142N, 16143N, 16144N, 16145N, 16146N, 16147N, 16148N, 16149N, 16150N, 16151N, 16152N, 16153N, 16154N, 16155N, 16156N, 16157N, 16158N, 16159N, 16160N, 16161N, 16162N, 16163N, 16164N, 16165N, 16166N, 16167N, 16168N, 16169N, 16170N, 16171N, 16172N, 16173N, 16174N, 16175N, 16176N, 16177N, 16178N, 16179N, 16180N, 16181N, 16182N, 16183N, 16184N, 16185N, 16186N, 16187N, 16188N, 16189N, 16190N, 16191N, 16192N, 16193N, 16194N, 16195N, 16196N, 16197N, 16198N, 16199N, 16200N, 16201N, 16202N, 16203N, 16204N, 16205N, 16206N, 16207N, 16208N, 16209N, 16210N, 16211N, 16212N, 16213N, 16214N, 16215N, 16216N, 16217N, 16218N, 16219N, 16220N, 16221N, 16222N, 16223N, 16224N, 16225N, 16226N, 16227N, 16228N, 16229N, 16230N, 16231N, 16232N, 16233N, 16234N, 16235N, 16236N, 16237N, 16238N, 16239N, 16240N, 16241N, 16242N, 16243N, 16244N, 16245N, 16246N, 16247N, 16248N, 16249N, 16250N, 16251N, 16252N, 16253N, 16254N, 16255N, 16256N, 16257N, 16258N, 16259N, 16260N, 16261N, 16262N, 16263N, 16264N, 16265N, 16266N, 16267N, 16268N, 16269N, 16270N, 16271N, 16272N, 16273N, 16274N, 16275N, 16276N, 16277N, 16278N, 16279N, 16280N, 16281N, 16282N, 16283N, 16284N, 16285N, 16286N, 16287N, 16288N, 16289N, 16290N, 16291N, 16292N, 16293N, 16294N, 16295N, 16296N, 16297N, 16298N, 16299N, 16300N, 16301N, 16302N, 16303N, 16304N, 16305N, 16306N, 16307N, 16308N, 16309N, 16310N, 16311N, 16312N, 16313N, 16314N, 16315N, 16316N, 16317N, 16318N, 16319N, 16320N, 16321N, 16322N, 16323N, 16324N, 16325N, 16326N, 16327N, 16328N, 16329N, 16330N, 16331N, 16332N, 16333N, 16334N, 16335N, 16336N, 16337N, 16338N, 16339N, 16340N, 16341N, 16342N, 16343N, 16344N, 16345N, 16346N, 16347N, 16348N, 16349N, 16350N, 16351N, 16352N, 16353N, 16354N, 16355N, 16356N, 16357N, 16358N, 16359N, 16360N, 16361N, 16362N, 16363N, 16364N, 16365N, 16366N, 16367N, 16368N, 16369N, 16370N, 16371N, 16372N, 16373N, 16374N, 16375N, 16376N, 16377N, 16378N, 16379N, 16380N, 16381N, 16382N, 16383N, 16384N, 16385N, 16386N | B1c1_HQ439460:356, 4062, 16368 | B1c1_HQ439460:19N, 21N, 32N, 33N, 34N, 35N, 36N, 37N, 356d, 1563N, 1564N, 1565N, 1566N, 1567N, 1568N, 1569N, 1580N, 2226+C, 4062N, 5239+A, 5277+A, 7176N, 7177N, 7178N, 7179N, 7180N, 10212N, 10213N, 10214N, 10215N, 10216N, 10217N, 10218N, 10219N, 10220N, 10221N, 10222N, 10223N, 10224N, 10225N, 10226N, 10227N, 10228N, 10229N, 10230N, 10231N, 10232N, 10233N, 10234N, 10235N, 10236N, 10237N, 10238N, 10239N, 10240N, 10241N, 10242N, 10243N, 10244N, 10245N, 10246N, 10247N, 10248N, 10249N, 10250N, 10251N, 10252N, 15383+G, 16124N, 16125N, 16126N, 16127N, 16128N, 16129N, 16130N, 16131N, 16132N, 16133N, 16134N, 16135N, 16136N, 16137N, 16138N, 16139N, 16140N, 16141N, 16142N, 16143N, 16144N, 16145N, 16146N, 16147N, 16148N, 16149N, 16150N, 16151N, 16152N, 16153N, 16154N, 16155N, 16156N, 16157N, 16158N, 16159N, 16160N, 16161N, 16162N, 16163N, 16164N, 16165N, 16166N, 16167N, 16168N, 16169N, 16170N, 16171N, 16172N, 16173N, 16174N, 16175N, 16176N, 16177N, 16178N, 16179N, 16180N, 16181N, 16182N, 16183N, 16184N, 16185N, 16186N, 16187N, 16188N, 16189N, 16190N, 16191N, 16192N, 16193N, 16194N, 16195N, 16196N, 16197N, 16198N, 16199N, 16200N, 16201N, 16202N, 16203N, 16204N, 16205N, 16206N, 16207N, 16208N, 16209N, 16210N, 16211N, 16212N, 16213N, 16214N, 16215N, 16216N, 16217N, 16218N, 16219N, 16220N, 16221N, 16222N, 16223N, 16224N, 16225N, 16226N, 16227N, 16228N, 16229N, 16230N, 16231N, 16232N, 16233N, 16234N, 16235N, 16236N, 16237N, 16238N, 16239N, 16240N, 16241N, 16242N, 16243N, 16244N, 16245N, 16246N, 16247N, 16248N, 16249N, 16250N, 16251N, 16252N, 16253N, 16254N, 16255N, 16256N, 16257N, 16258N, 16259N, 16260N, 16261N, 16262N, 16263N, 16264N, 16265N, 16266N, 16267N, 16268N, 16269N, 16270N, 16271N, 16272N, 16273N, 16274N, 16275N, 16276N, 16277N, 16278N, 16279N, 16280N, 16281N, 16282N, 16283N, 16284N, 16285N, 16286N, 16287N, 16288N, 16289N, 16290N, 16291N, 16292N, 16293N, 16294N, 16295N, 16296N, 16297N, 16298N, 16299N, 16300N, 16301N, 16302N, 16303N, 16304N, 16305N, 16306N, 16307N, 16308N, 16309N, 16310N, 16311N, 16312N, 16313N, 16314N, 16315N, 16316N, 16317N, 16318N, 16319N, 16320N, 16321N, 16322N, 16323N, 16324N, 16325N, 16326N, 16327N, 16328N, 16329N, 16330N, 16331N, 16332N, 16333N, 16334N, 16335N, 16336N, 16337N, 16338N, 16339N, 16340N, 16341N, 16342N, 16343N, 16344N, 16345N, 16346N, 16347N, 16348N, 16349N, 16350N, 16351N, 16352N, 16353N, 16354N, 16355N, 16356N, 16357N, 16358N, 16359N, 16360N, 16361N, 16362N, 16363N, 16364N, 16365N, 16366N, 16367N, 16368N, 16369N, 16370N, 16371N, 16372N, 16373N, 16374N, 16375N, 16376N, 16377N, 16378N, 16379N, 16380N, 16381N, 16382N, 16383N, 16384N, 16385N, 16386N | Lippold,S., Matzke,N., Reissman,M., Burbano,H. and Hofreiter,M. BMC Evolutionary Biology. 328 (11), 1471-2148 (2011) |
| HQ439461 (Horse and Przewalski's horse) | Illumina/Solexa Genome Analyzer II | Phantom mutaions: 358d, 1387A, 2227+T, 5239+A, 5274+G, 15385+T. | L2b2 | 158, 356d, 961, 1375, 1387A, 2227+T, 2607, 2788, 2899, 3517, 3727, 3942, 4062, 4536, 4646, 4669, 5239+A, 5274+G, 5527, 5815, 5884, 6004, 6307, 6784, 7001, 7516, 7666, 7900, 8005, 8058, 8301, 8319, 8358, 8565, 9239, 9951, 10110, 10214, 10292, 10376, 10421, 10613, 11240, 11543, 11693, 11842, 11879, 12119, 12200, 12494, 12767, 12896, 12950, 13049, 13333, 13520, 14803, 14995, 15313, 15385+T, 15491, 15492, 15493, 15531, 15582, 15600, 15646, 15717, 15768, 15867, 15868, 15953, 15971, 16065, 16100, 16108N, 16118N, 16126N, 16127N, 16128N, 16129N, 16130N, 16131N, 16132N, 16133N, 16134N, 16135N, 16136N, 16137N, 16138N, 16139N, 16140N, 16141N, 16142N, 16143N, 16144N, 16145N, 16146N, 16147N, 16148N, 16149N, 16150N, 16151N, 16152N, 16153N, 16154N, 16155N, 16156N, 16157N, 16158N, 16159N, 16160N, 16161N, 16162N, 16163N, 16164N, 16165N, 16166N, 16167N, 16168N, 16169N, 16170N, 16171N, 16172N, 16173N, 16174N, 16175N, 16176N, 16177N, 16178N, 16179N, 16180N, 16181N, 16182N, 16183N, 16184N, 16185N, 16186N, 16187N, 16188N, 16189N, 16190N, 16191N, 16192N, 16193N, 16194N, 16195N, 16196N, 16197N, 16198N, 16199N, 16200N, 16201N, 16202N, 16203N, 16204N, 16205N, 16206N, 16207N, 16208N, 16209N, 16210N, 16211N, 16212N, 16213N, 16214N, 16215N, 16216N, 16217N, 16218N, 16219N, 16220N, 16221N, 16222N, 16223N, 16224N, 16225N, 16226N, 16227N, 16228N, 16229N, 16230N, 16231N, 16232N, 16233N, 16234N, 16235N, 16236N, 16237N, 16238N, 16239N, 16240N, 16241N, 16242N, 16243N, 16244N, 16245N, 16246N, 16247N, 16248N, 16249N, 16250N, 16251N, 16252N, 16253N, 16254N, 16255N, 16256N, 16257N, 16258N, 16259N, 16260N, 16261N, 16262N, 16263N, 16264N, 16265N, 16266N, 16267N, 16268N, 16269N, 16270N, 16271N, 16272N, 16273N, 16274N, 16275N, 16276N, 16277N, 16278N, 16279N, 16280N, 16281N, 16282N, 16283N, 16284N, 16285N, 16286N, 16287N, 16288N, 16289N, 16290N, 16291N, 16292N, 16293N, 16294N, 16295N, 16296N, 16297N, 16298N, 16299N, 16300N, 16301N, 16302N, 16303N, 16304N, 16305N, 16306N, 16307N, 16308N, 16309N, 16310N, 16311N, 16312N, 16313N, 16314N, 16315N, 16316N, 16317N, 16318N, 16319N, 16320N, 16321N, 16322N, 16323N, 16324N, 16325N, 16326N, 16327N, 16328N, 16329N, 16330N, 16331N, 16332N, 16333N, 16334N, 16335N, 16336N, 16337N, 16338N, 16339N, 16340N, 16341N, 16342N, 16343N, 16344N, 16345N, 16346N, 16347N, 16348N, 16349N, 16350N, 16351N, 16352N, 16353N, 16354N, 16355N, 16356N, 16357N, 16368, 16398A, 16626 | L2b2:356 | L2b2:356d, 2227+T, 5239+A, 5274+G, 15385+T, 16108N, 16118N, 16126N, 16127N, 16128N, 16129N, 16130N, 16131N, 16132N, 16133N, 16134N, 16135N, 16136N, 16137N, 16138N, 16139N, 16140N, 16141N, 16142N, 16143N, 16144N, 16145N, 16146N, 16147N, 16148N, 16149N, 16150N, 16151N, 16152N, 16153N, 16154N, 16155N, 16156N, 16157N, 16158N, 16159N, 16160N, 16161N, 16162N, 16163N, 16164N, 16165N, 16166N, 16167N, 16168N, 16169N, 16170N, 16171N, 16172N, 16173N, 16174N, 16175N, 16176N, 16177N, 16178N, 16179N, 16180N, 16181N, 16182N, 16183N, 16184N, 16185N, 16186N, 16187N, 16188N, 16189N, 16190N, 16191N, 16192N, 16193N, 16194N, 16195N, 16196N, 16197N, 16198N, 16199N, 16200N, 16201N, 16202N, 16203N, 16204N, 16205N, 16206N, 16207N, 16208N, 16209N, 16210N, 16211N, 16212N, 16213N, 16214N, 16215N, 16216N, 16217N, 16218N, 16219N, 16220N, 16221N, 16222N, 16223N, 16224N, 16225N, 16226N, 16227N, 16228N, 16229N, 16230N, 16231N, 16232N, 16233N, 16234N, 16235N, 16236N, 16237N, 16238N, 16239N, 16240N, 16241N, 16242N, 16243N, 16244N, 16245N, 16246N, 16247N, 16248N, 16249N, 16250N, 16251N, 16252N, 16253N, 16254N, 16255N, 16256N, 16257N, 16258N, 16259N, 16260N, 16261N, 16262N, 16263N, 16264N, 16265N, 16266N, 16267N, 16268N, 16269N, 16270N, 16271N, 16272N, 16273N, 16274N, 16275N, 16276N, 16277N, 16278N, 16279N, 16280N, 16281N, 16282N, 16283N, 16284N, 16285N, 16286N, 16287N, 16288N, 16289N, 16290N, 16291N, 16292N, 16293N, 16294N, 16295N, 16296N, 16297N, 16298N, 16299N, 16300N, 16301N, 16302N, 16303N, 16304N, 16305N, 16306N, 16307N, 16308N, 16309N, 16310N, 16311N, 16312N, 16313N, 16314N, 16315N, 16316N, 16317N, 16318N, 16319N, 16320N, 16321N, 16322N, 16323N, 16324N, 16325N, 16326N, 16327N, 16328N, 16329N, 16330N, 16331N, 16332N, 16333N, 16334N, 16335N, 16336N, 16337N, 16338N, 16339N, 16340N, 16341N, 16342N, 16343N, 16344N, 16345N, 16346N, 16347N, 16348N, 16349N, 16350N, 16351N, 16352N, 16353N, 16354N, 16355N, 16356N, 16357N | Lippold,S., Matzke,N., Reissman,M., Burbano,H. and Hofreiter,M. BMC Evolutionary Biology. 328 (11), 1471-2148 (2011) |
| HQ439462 (Horse and Przewalski's horse) | Illumina/Solexa Genome Analyzer II | Phantom mutaions: 358d, 1387A, 2227+T, 5239+A, 5277+A, 15383+A. | D, D1_JN398400, D2 | 1-28d, 158, 356d, 1387A, 1479N, 1694, 2217, 2227+T, 2788, 4062, 4609N, 4610N, 4611N, 4612N, 4613N, 4614N, 4615N, 4616N, 4617N, 4618N, 4619N, 4646N, 5239+A, 5277+A, 5884, 6004, 6532, 6784, 6970N, 7001N, 7143N, 7144N, 7145N, 7146N, 7147N, 7148N, 7149N, 7150N, 7151N, 7152N, 7153N, 7154N, 7155N, 7156N, 7157N, 7158N, 7159N, 7160N, 7161N, 7162N, 7163N, 7164N, 7165N, 7166N, 7167N, 7168N, 7169N, 7170N, 7171N, 7172N, 7173N, 7174N, 7175N, 7176N, 7177N, 7178N, 7179N, 7180N, 7181N, 7666, 7896, 8110, 8777N, 8867, 9239, 9384, 9402, 10214, 10376, 10398N, 10412N, 10413N, 10414N, 10415N, 10416N, 10429N, 10826N, 10827N, 11240, 11417, 11543, 12139N, 12140N, 12141N, 12142N, 12143N, 12144N, 12145N, 12146N, 12147N, 12191N, 12218, 12791, 13049, 13333, 13920N, 13948N, 14200, 14731, 15202, 15383+A, 15492, 15582, 15593, 15599, 15717, 15734, 15767, 15824, 15867, 16004, 16108N, 16110N, 16115N, 16116N, 16117N, 16118N, 16119N, 16120N, 16121N, 16122N, 16123N, 16124N, 16125N, 16126N, 16127N, 16128N, 16129N, 16130N, 16131N, 16132N, 16133N, 16134N, 16135N, 16136N, 16137N, 16138N, 16139N, 16140N, 16141N, 16142N, 16143N, 16144N, 16145N, 16146N, 16147N, 16148N, 16149N, 16150N, 16151N, 16152N, 16153N, 16154N, 16155N, 16156N, 16157N, 16158N, 16159N, 16160N, 16161N, 16162N, 16163N, 16164N, 16165N, 16166N, 16167N, 16168N, 16169N, 16170N, | D:356, 1387d, 4646, 5497, 7001, 11457, 12191, 13948, 14825, 15518, 15807, 16084, 16108, 16110, 16368; D1_JN398400:356, 1387d, 3557, 4646, 5497, 7001, 11457, 12191, 13948, 15518, 15807, 16084, 16108, 16110, 16368; D2:356, 1387d, 4646, 5497, 7001, 11457, 12191, 13948, 14825, 15518, 15807, | D:1-28d, 356d, 1387A, 1479N, 2227+T, 4609N, 4610N, 4611N, 4612N, 4613N, 4614N, 4615N, 4616N, 4617N, 4618N, 4619N, 4646N, 5239+A, 5277+A, 6970N, 7001N, 7143N, 7144N, 7145N, 7146N, 7147N, 7148N, 7149N, 7150N, 7151N, 7152N, 7153N, 7154N, 7155N, 7156N, 7157N, 7158N, 7159N, 7160N, 7161N, 7162N, 7163N, 7164N, 7165N, 7166N, 7167N, 7168N, 7169N, 7170N, 7171N, 7172N, 7173N, 7174N, 7175N, 7176N, 7177N, 7178N, 7179N, 7180N, 7181N, 7666, 7896, 8110, 8777N, 9384, 10398N, 10412N, 10413N, 10414N, 10415N, 10416N, 10429N, 10826N, 10827N, 12139N, 12140N, 12141N, 12142N, 12143N, 12144N, 12145N, 12146N, 12147N, 12191N, 13920N, 13948N, 14731, 15202, 15383+A, 15582, 15593, 15824, 16004, 16108N, 16110N, 16115N, 16116N, 16117N, 16118N, 16119N, 16120N, 16121N, 16122N, 16123N, 16124N, 16125N, 16126N, 16127N, 16128N, 16129N, 16130N, 16131N, 16132N, 16133N, 16134N, 16135N, 16136N, 16137N, 16138N, 16139N, 16140N, 16141N, 16142N, 16143N, 16144N, 16145N, 16146N, 16147N, 16148N, 16149N, 16150N, 16151N, 16152N, | Lippold,S., Matzke,N., Reissman,M., Burbano,H. and Hofreiter,M. BMC Evolutionary Biology. 328 (11), 1471-2148 (2011) |
| | | | | 16171N, 16172N, 16173N, 16174N, 16175N, 16176N, 16177N, 16178N, 16179N, 16180N, 16181N, 16182N, 16183N, 16184N, 16185N, 16186N, 16187N, 16188N, 16189N, 16190N, 16191N, 16192N, 16193N, 16194N, 16195N, 16196N, 16197N, 16198N, 16199N, 16200N, 16201N, 16202N, 16203N, 16204N, 16205N, 16206N, 16207N, 16208N, 16209N, 16210N, 16211N, 16212N, 16213N, 16214N, 16215N, 16216N, 16217N, 16218N, 16219N, 16220N, 16221N, 16222N, 16223N, 16224N, 16225N, 16226N, 16227N, 16228N, 16229N, 16230N, 16231N, 16232N, 16233N, 16234N, 16235N, 16236N, 16237N, 16238N, 16239N, 16240N, 16241N, 16242N, 16243N, 16244N, 16245N, 16246N, 16247N, 16248N, 16249N, 16250N, 16251N, 16252N, 16253N, 16254N, 16255N, 16256N, 16257N, 16258N, 16259N, 16260N, 16261N, 16262N, 16263N, 16264N, 16265N, 16266N, 16267N, 16268N, 16269N, 16270N, 16271N, 16272N, 16273N, 16274N, 16275N, 16276N, 16277N, 16278N, 16279N, 16280N, 16281N, 16282N, 16283N, 16284N, 16285N, 16286N, 16287N, 16288N, 16289N, 16290N, 16291N, 16292N, 16293N, 16294N, 16295N, 16296N, 16297N, 16298N, 16299N, 16300N, 16301N, 16302N, 16303N, 16304N, 16305N, 16306N, 16307N, 16308N, 16309N, 16310N, 16311N, 16312N, 16313N, 16314N, 16315N, 16316N, 16317N, 16318N, 16319N, 16320N, 16321N, 16322N, 16323N, 16324N, 16325N, 16326N, 16327N, 16328N, 16329N, 16330N, 16331N, 16332N, 16333N, 16334N, 16335N, 16336N, 16337N, 16338N, 16339N, 16340N, 16341N, 16342N, 16343N, 16344N, 16345N, 16346N, 16347N, 16348N, 16349N, 16350N, 16351N, 16352N, 16353N, 16354N, 16355N, 16356N, 16357N, 16368N, 16391, 16652-16657d | 15865, 16084, 16108, 16110, 16368 | 16153N, 16154N, 16155N, 16156N, 16157N, 16158N, 16159N, 16160N, 16161N, 16162N, 16163N, 16164N, 16165N, 16166N, 16167N, 16168N, 16169N, 16170N, 16171N, 16172N, 16173N, 16174N, 16175N, 16176N, 16177N, 16178N, 16179N, 16180N, 16181N, 16182N, 16183N, 16184N, 16185N, 16186N, 16187N, 16188N, 16189N, 16190N, 16191N, 16192N, 16193N, 16194N, 16195N, 16196N, 16197N, 16198N, 16199N, 16200N, 16201N, 16202N, 16203N, 16204N, 16205N, 16206N, 16207N, 16208N, 16209N, 16210N, 16211N, 16212N, 16213N, 16214N, 16215N, 16216N, 16217N, 16218N, 16219N, 16220N, 16221N, 16222N, 16223N, 16224N, 16225N, 16226N, 16227N, 16228N, 16229N, 16230N, 16231N, 16232N, 16233N, 16234N, 16235N, 16236N, 16237N, 16238N, 16239N, 16240N, 16241N, 16242N, 16243N, 16244N, 16245N, 16246N, 16247N, 16248N, 16249N, 16250N, 16251N, 16252N, 16253N, 16254N, 16255N, 16256N, 16257N, 16258N, 16259N, 16260N, 16261N, 16262N, 16263N, 16264N, 16265N, 16266N, 16267N, 16268N, 16269N, 16270N, 16271N, 16272N, 16273N, 16274N, 16275N, 16276N, 16277N, 16278N, 16279N, 16280N, 16281N, 16282N, 16283N, 16284N, 16285N, 16286N, 16287N, 16288N, 16289N, 16290N, 16291N, 16292N, 16293N, 16294N, 16295N, 16296N, 16297N, 16298N, 16299N, 16300N, 16301N, 16302N, 16303N, 16304N, 16305N, 16306N, 16307N, 16308N, 16309N, 16310N, 16311N, 16312N, 16313N, 16314N, 16315N, 16316N, 16317N, 16318N, 16319N, 16320N, 16321N, 16322N, 16323N, 16324N, 16325N, 16326N, 16327N, 16328N, 16329N, 16330N, 16331N, 16332N, 16333N, 16334N, 16335N, 16336N, 16337N, 16338N, 16339N, 16340N, 16341N, 16342N, 16343N, 16344N, 16345N, 16346N, 16347N, 16348N, 16349N, 16350N, 16351N, 16352N, 16353N, 16354N, 16355N, 16356N, 16357N, 16368N, 16652-16657d; D1_JN398400:1-28d, 356d, 1387A, 1479N, 2227+T, 4609N, 4610N, 4611N, 4612N, 4613N, 4614N, 4615N, 4616N, 4617N, 4618N, 4619N, 4646N, 5239+A, 5277+A, 6970N, 7001N, 7143N, 7144N, 7145N, 7146N, 7147N, 7148N, 7149N, 7150N, 7151N, 7152N, 7153N, 7154N, 7155N, 7156N, 7157N, 7158N, 7159N, 7160N, 7161N, 7162N, 7163N, 7164N, 7165N, 7166N, 7167N, 7168N, 7169N, 7170N, 7171N, 7172N, 7173N, 7174N, 7175N, 7176N, 7177N, 7178N, 7179N, 7180N, 7181N, 7666, 7896, 8110, 8777N, 9384, 10398N, 10412N, 10413N, 10414N, 10415N, 10416N, 10429N, 10826N, 10827N, 12139N, 12140N, 12141N, 12142N, 12143N, 12144N, 12145N, 12146N, 12147N, 12191N, 13920N, 13948N, 14731, 15202, 15383+A, 15582, 15593, 15824, 16004, 16108N, 16110N, 16115N, 16116N, 16117N, 16118N, 16119N, 16120N, 16121N, 16122N, 16123N, 16124N, 16125N, 16126N, 16127N, 16128N, 16129N, 16130N, 16131N, 16132N, 16133N, 16134N, 16135N, 16136N, 16137N, 16138N, 16139N, 16140N, 16141N, 16142N, 16143N, 16144N, 16145N, 16146N, 16147N, 16148N, 16149N, 16150N, 16151N, 16152N, 16153N, 16154N, 16155N, 16156N, 16157N, 16158N, 16159N, 16160N, 16161N, 16162N, 16163N, 16164N, 16165N, 16166N, 16167N, 16168N, 16169N, 16170N, 16171N, 16172N, 16173N, 16174N, 16175N, 16176N, 16177N, 16178N, 16179N, 16180N, 16181N, 16182N, 16183N, | |
| | | | | | | 16184N, 16185N, 16186N, 16187N, 16188N, 16189N, 16190N, 16191N, 16192N, 16193N, 16194N, 16195N, 16196N, 16197N, 16198N, 16199N, 16200N, 16201N, 16202N, 16203N, 16204N, 16205N, 16206N, 16207N, 16208N, 16209N, 16210N, 16211N, 16212N, 16213N, 16214N, 16215N, 16216N, 16217N, 16218N, 16219N, 16220N, 16221N, 16222N, 16223N, 16224N, 16225N, 16226N, 16227N, 16228N, 16229N, 16230N, 16231N, 16232N, 16233N, 16234N, 16235N, 16236N, 16237N, 16238N, 16239N, 16240N, 16241N, 16242N, 16243N, 16244N, 16245N, 16246N, 16247N, 16248N, 16249N, 16250N, 16251N, 16252N, 16253N, 16254N, 16255N, 16256N, 16257N, 16258N, 16259N, 16260N, 16261N, 16262N, 16263N, 16264N, 16265N, 16266N, 16267N, 16268N, 16269N, 16270N, 16271N, 16272N, 16273N, 16274N, 16275N, 16276N, 16277N, 16278N, 16279N, 16280N, 16281N, 16282N, 16283N, 16284N, 16285N, 16286N, 16287N, 16288N, 16289N, 16290N, 16291N, 16292N, 16293N, 16294N, 16295N, 16296N, 16297N, 16298N, 16299N, 16300N, 16301N, 16302N, 16303N, 16304N, 16305N, 16306N, 16307N, 16308N, 16309N, 16310N, 16311N, 16312N, 16313N, 16314N, 16315N, 16316N, 16317N, 16318N, 16319N, 16320N, 16321N, 16322N, 16323N, 16324N, 16325N, 16326N, 16327N, 16328N, 16329N, 16330N, 16331N, 16332N, 16333N, 16334N, 16335N, 16336N, 16337N, 16338N, 16339N, 16340N, 16341N, 16342N, 16343N, 16344N, 16345N, 16346N, 16347N, 16348N, 16349N, 16350N, 16351N, 16352N, 16353N, 16354N, 16355N, 16356N, 16357N, 16368N, 16652-16657d; D2:1-28d, 356d, 1387A, 1479N, 2227+T, 4609N, 4610N, 4611N, 4612N, 4613N, 4614N, 4615N, 4616N, 4617N, 4618N, 4619N, 4646N, 5239+A, 5277+A, 6970N, 7001N, 7143N, 7144N, 7145N, 7146N, 7147N, 7148N, 7149N, 7150N, 7151N, 7152N, 7153N, 7154N, 7155N, 7156N, 7157N, 7158N, 7159N, 7160N, 7161N, 7162N, 7163N, 7164N, 7165N, 7166N, 7167N, 7168N, 7169N, 7170N, 7171N, 7172N, 7173N, 7174N, 7175N, 7176N, 7177N, 7178N, 7179N, 7180N, 7181N, 7666, 7896, 8110, 8777N, 9384, 10398N, 10412N, 10413N, 10414N, 10415N, 10416N, 10429N, 10826N, 10827N, 12139N, 12140N, 12141N, 12142N, 12143N, 12144N, 12145N, 12146N, 12147N, 12191N, 13920N, 13948N, 14731, 15202, 15383+A, 15582, 15824, 16004, 16108N, 16110N, 16115N, 16116N, 16117N, 16118N, 16119N, 16120N, 16121N, 16122N, 16123N, 16124N, 16125N, 16126N, 16127N, 16128N, 16129N, 16130N, 16131N, 16132N, 16133N, 16134N, 16135N, 16136N, 16137N, 16138N, 16139N, 16140N, 16141N, 16142N, 16143N, 16144N, 16145N, 16146N, 16147N, 16148N, 16149N, 16150N, 16151N, 16152N, 16153N, 16154N, 16155N, 16156N, 16157N, 16158N, 16159N, 16160N, 16161N, 16162N, 16163N, 16164N, 16165N, 16166N, 16167N, 16168N, 16169N, 16170N, 16171N, 16172N, 16173N, 16174N, 16175N, 16176N, 16177N, 16178N, 16179N, 16180N, 16181N, 16182N, 16183N, 16184N, 16185N, 16186N, 16187N, 16188N, 16189N, 16190N, 16191N, 16192N, 16193N, 16194N, 16195N, 16196N, 16197N, 16198N, 16199N, 16200N, 16201N, 16202N, 16203N, 16204N, 16205N, 16206N, 16207N, 16208N, 16209N, 16210N, 16211N, 16212N, 16213N, 16214N, 16215N, 16216N, 16217N, | |
| | | | | | | 16218N, 16219N, 16220N, 16221N, 16222N, 16223N, 16224N, 16225N, 16226N, 16227N, 16228N, 16229N, 16230N, 16231N, 16232N, 16233N, 16234N, 16235N, 16236N, 16237N, 16238N, 16239N, 16240N, 16241N, 16242N, 16243N, 16244N, 16245N, 16246N, 16247N, 16248N, 16249N, 16250N, 16251N, 16252N, 16253N, 16254N, 16255N, 16256N, 16257N, 16258N, 16259N, 16260N, 16261N, 16262N, 16263N, 16264N, 16265N, 16266N, 16267N, 16268N, 16269N, 16270N, 16271N, 16272N, 16273N, 16274N, 16275N, 16276N, 16277N, 16278N, 16279N, 16280N, 16281N, 16282N, 16283N, 16284N, 16285N, 16286N, 16287N, 16288N, 16289N, 16290N, 16291N, 16292N, 16293N, 16294N, 16295N, 16296N, 16297N, 16298N, 16299N, 16300N, 16301N, 16302N, 16303N, 16304N, 16305N, 16306N, 16307N, 16308N, 16309N, 16310N, 16311N, 16312N, 16313N, 16314N, 16315N, 16316N, 16317N, 16318N, 16319N, 16320N, 16321N, 16322N, 16323N, 16324N, 16325N, 16326N, 16327N, 16328N, 16329N, 16330N, 16331N, 16332N, 16333N, 16334N, 16335N, 16336N, 16337N, 16338N, 16339N, 16340N, 16341N, 16342N, 16343N, 16344N, 16345N, 16346N, 16347N, 16348N, 16349N, 16350N, 16351N, 16352N, 16353N, 16354N, 16355N, 16356N, 16357N, 16368N, 16652-16657d | |
| HQ439463 (Horse and Przewalski's horse) | Illumina/Solexa Genome Analyzer II | Phantom mutaions: 358d, 1387A, 2227+T, 5239+A, 5277+A, 15385+T. | I2a4_HQ439463 | 158, 356d, 1387A, 1587, 1791, 2227+T, 2614, 2770, 2788, 4062, 4063, 4392, 4646, 4669, 4830, 5061, 5210, 5239+A, 5277+A, 5884, 6004, 6175, 6247, 6307, 6784, 7001, 8005, 8187, 8358, 8379, 8792, 9239, 9251, 9694, 9948, 10083, 10214, 10238, 10376, 11240, 11543, 11827, 11842, 12404, 12443, 12683, 12767, 13049, 13333, 13502, 13761, 14554, 15385+T, 15492, 15535, 15593, 15599, 15647, 15706, 15717, 15768, 15823, 15867, 15971, 16110, 16119, 16126N, 16127N, 16128N, 16129N, 16130N, 16131N, 16132N, 16133N, 16134N, 16135N, 16136N, 16137N, 16138N, 16139N, 16140N, 16141N, 16142N, 16143N, 16144N, 16145N, 16146N, 16147N, 16148N, 16149N, 16150N, 16151N, 16152N, 16153N, 16154N, 16155N, 16156N, 16157N, 16158N, 16159N, 16160N, 16161N, 16162N, 16163N, 16164N, 16165N, 16166N, 16167N, 16168N, 16169N, 16170N, 16171N, 16172N, 16173N, 16174N, 16175N, 16176N, 16177N, 16178N, 16179N, 16180N, 16181N, 16182N, 16183N, 16184N, 16185N, 16186N, 16187N, 16188N, 16189N, 16190N, 16191N, 16192N, 16193N, 16194N, 16195N, 16196N, 16197N, 16198N, 16199N, 16200N, 16201N, 16202N, 16203N, 16204N, 16205N, 16206N, 16207N, 16208N, 16209N, 16210N, 16211N, 16212N, 16213N, 16214N, 16215N, 16216N, 16217N, 16218N, 16219N, 16220N, 16221N, 16222N, 16223N, 16224N, 16225N, 16226N, 16227N, 16228N, 16229N, 16230N, 16231N, 16232N, 16233N, 16234N, 16235N, 16236N, 16237N, 16238N, 16239N, 16240N, 16241N, 16242N, 16243N, 16244N, 16245N, 16246N, 16247N, 16248N, 16249N, 16250N, 16251N, 16252N, 16253N, 16254N, 16255N, 16256N, 16257N, 16258N, 16259N, 16260N, 16261N, 16262N, 16263N, 16264N, 16265N, 16266N, 16267N, 16268N, 16269N, 16270N, 16271N, 16272N, 16273N, 16274N, 16275N, 16276N, 16277N, 16278N, 16279N, 16280N, 16281N, 16282N, 16283N, 16284N, 16285N, 16286N, 16287N, 16288N, 16289N, 16290N, 16291N, 16292N, 16293N, 16294N, 16295N, 16296N, 16297N, 16298N, 16299N, 16300N, 16301N, 16302N, 16303N, 16304N, 16305N, 16306N, 16307N, 16308N, 16309N, 16310N, 16311N, 16312N, 16313N, 16314N, 16315N, 16316N, 16317N, 16318N, 16319N, 16320N, 16321N, 16322N, 16323N, 16324N, 16325N, 16326N, 16327N, 16328N, 16329N, 16330N, 16331N, 16332N, 16333N, 16334N, 16335N, 16336N, 16337N, 16338N, 16339N, 16340N, 16341N, 16342N, 16343N, 16344N, 16345N, 16346N, 16347N, 16348N, 16349N, 16350N, 16351N, 16352N, 16353N, 16354N, 16355N, 16356N, 16357N, 16368, 16436 | I2a4_HQ439463:356 | I2a4_HQ439463:356d, 2227+T, 5239+A, 5277+A, 15385+T, 15647, 16126N, 16127N, 16128N, 16129N, 16130N, 16131N, 16132N, 16133N, 16134N, 16135N, 16136N, 16137N, 16138N, 16139N, 16140N, 16141N, 16142N, 16143N, 16144N, 16145N, 16146N, 16147N, 16148N, 16149N, 16150N, 16151N, 16152N, 16153N, 16154N, 16155N, 16156N, 16157N, 16158N, 16159N, 16160N, 16161N, 16162N, 16163N, 16164N, 16165N, 16166N, 16167N, 16168N, 16169N, 16170N, 16171N, 16172N, 16173N, 16174N, 16175N, 16176N, 16177N, 16178N, 16179N, 16180N, 16181N, 16182N, 16183N, 16184N, 16185N, 16186N, 16187N, 16188N, 16189N, 16190N, 16191N, 16192N, 16193N, 16194N, 16195N, 16196N, 16197N, 16198N, 16199N, 16200N, 16201N, 16202N, 16203N, 16204N, 16205N, 16206N, 16207N, 16208N, 16209N, 16210N, 16211N, 16212N, 16213N, 16214N, 16215N, 16216N, 16217N, 16218N, 16219N, 16220N, 16221N, 16222N, 16223N, 16224N, 16225N, 16226N, 16227N, 16228N, 16229N, 16230N, 16231N, 16232N, 16233N, 16234N, 16235N, 16236N, 16237N, 16238N, 16239N, 16240N, 16241N, 16242N, 16243N, 16244N, 16245N, 16246N, 16247N, 16248N, 16249N, 16250N, 16251N, 16252N, 16253N, 16254N, 16255N, 16256N, 16257N, 16258N, 16259N, 16260N, 16261N, 16262N, 16263N, 16264N, 16265N, 16266N, 16267N, 16268N, 16269N, 16270N, 16271N, 16272N, 16273N, 16274N, 16275N, 16276N, 16277N, 16278N, 16279N, 16280N, 16281N, 16282N, 16283N, 16284N, 16285N, 16286N, 16287N, 16288N, 16289N, 16290N, 16291N, 16292N, 16293N, 16294N, 16295N, 16296N, 16297N, 16298N, 16299N, 16300N, 16301N, 16302N, 16303N, 16304N, 16305N, 16306N, 16307N, 16308N, 16309N, 16310N, 16311N, 16312N, 16313N, 16314N, 16315N, 16316N, 16317N, 16318N, 16319N, 16320N, 16321N, 16322N, 16323N, 16324N, 16325N, 16326N, 16327N, 16328N, 16329N, 16330N, 16331N, 16332N, 16333N, 16334N, 16335N, 16336N, 16337N, 16338N, | Lippold,S., Matzke,N., Reissman,M., Burbano,H. and Hofreiter,M. BMC Evolutionary Biology. 328 (11), 1471-2148 (2011) |
| | | | | | | 16339N, 16340N, 16341N, 16342N, 16343N, 16344N, 16345N, 16346N, 16347N, 16348N, 16349N, 16350N, 16351N, 16352N, 16353N, 16354N, 16355N, 16356N, 16357N | |
| HQ439464 (Horse and Przewalski's horse) | Illumina/Solexa Genome Analyzer II | Phantom mutaions: 358d, 1387G, 2226+C, 5239+A, 5277+A, 15383+G. | N1_HQ439464 | 158, 356d, 961, 1387G, 1609T, 2226+C, 2339A, 2788, 2802, 3070, 3100, 3475, 3557, 3800, 4062, 4526, 4536, 4605, 4646, 4669, 4898, 4917, 5239+A, 5277+A, 5527, 5827, 5884, 6004, 6076, 6307, 6712, 6784, 7001, 7432, 7666, 7900, 8005, 8043, 8076, 8150, 8175, 8238, 8280, 8358, 8556T, 8565, 8798, 9071, 9086, 9239, 9332, 9402, 9540, 10110, 10173, 10214, 10292, 10376, 10404, 10448, 10460, 10517, 10646, 10859, 11240, 11394, 11492, 11543, 11842, 11879, 11966, 12029, 12095, 12332, 12767, 13049, 13100, 13309, 13333, 13356, 13502, 13567, 13615, 13629, 13720, 13920, 13933, 14422, 14626, 14671, 14803, 14815, 15052A, 15133, 15342, 15383+G, 15492, 15541, 15598, 15599, 15717, 15768, 15803, 15824, 15835, 15866, 15953, 16004, 16065, 16110, 16118, 16126N, 16127N, 16128N, 16129N, 16130N, 16131N, 16132N, 16133N, 16134N, 16135N, 16136N, 16137N, 16138N, 16139N, 16140N, 16141N, 16142N, 16143N, 16144N, 16145N, 16146N, 16147N, 16148N, 16149N, 16150N, 16151N, 16152N, 16153N, 16154N, 16155N, 16156N, 16157N, 16158N, 16159N, 16160N, 16161N, 16162N, 16163N, 16164N, 16165N, 16166N, 16167N, 16168N, 16169N, 16170N, 16171N, 16172N, 16173N, 16174N, 16175N, 16176N, 16177N, 16178N, 16179N, 16180N, 16181N, 16182N, 16183N, 16184N, 16185N, 16186N, 16187N, 16188N, 16189N, 16190N, 16191N, 16192N, 16193N, 16194N, 16195N, 16196N, 16197N, 16198N, 16199N, 16200N, 16201N, 16202N, 16203N, 16204N, 16205N, 16206N, 16207N, 16208N, 16209N, 16210N, 16211N, 16212N, 16213N, 16214N, 16215N, 16216N, 16217N, 16218N, 16219N, 16220N, 16221N, 16222N, 16223N, 16224N, 16225N, 16226N, 16227N, 16228N, 16229N, 16230N, 16231N, 16232N, 16233N, 16234N, 16235N, 16236N, 16237N, 16238N, 16239N, 16240N, 16241N, 16242N, 16243N, 16244N, 16245N, 16246N, 16247N, 16248N, 16249N, 16250N, 16251N, 16252N, 16253N, 16254N, 16255N, 16256N, 16257N, 16258N, 16259N, 16260N, 16261N, 16262N, 16263N, 16264N, 16265N, 16266N, 16267N, 16268N, 16269N, 16270N, 16271N, 16272N, 16273N, 16274N, 16275N, 16276N, 16277N, 16278N, 16279N, 16280N, 16281N, 16282N, 16283N, 16284N, 16285N, 16286N, 16287N, 16288N, 16289N, 16290N, 16291N, 16292N, 16293N, 16294N, 16295N, 16296N, 16297N, 16298N, 16299N, 16300N, 16301N, 16302N, 16303N, 16304N, 16305N, 16306N, 16307N, 16308N, 16309N, 16310N, 16311N, 16312N, 16313N, 16314N, 16315N, 16316N, 16317N, 16318N, 16319N, 16320N, 16321N, 16322N, 16323N, 16324N, 16325N, 16326N, 16327N, 16328N, 16329N, 16330N, 16331N, 16332N, 16333N, 16334N, 16335N, 16336N, 16337N, 16338N, 16339N, 16340N, 16341N, 16342N, 16343N, 16344N, 16345N, 16346N, 16347N, 16348N, 16349N, 16350N, 16351N, 16352N, 16353N, 16354N, 16355N, 16356N, 16357N, 16368, 16398A, 16540A, 16543, 16556, 16626 | N1_HQ439464:356 | N1_HQ439464:356d, 2226+C, 5239+A, 5277+A, 15383+G, 16126N, 16127N, 16128N, 16129N, 16130N, 16131N, 16132N, 16133N, 16134N, 16135N, 16136N, 16137N, 16138N, 16139N, 16140N, 16141N, 16142N, 16143N, 16144N, 16145N, 16146N, 16147N, 16148N, 16149N, 16150N, 16151N, 16152N, 16153N, 16154N, 16155N, 16156N, 16157N, 16158N, 16159N, 16160N, 16161N, 16162N, 16163N, 16164N, 16165N, 16166N, 16167N, 16168N, 16169N, 16170N, 16171N, 16172N, 16173N, 16174N, 16175N, 16176N, 16177N, 16178N, 16179N, 16180N, 16181N, 16182N, 16183N, 16184N, 16185N, 16186N, 16187N, 16188N, 16189N, 16190N, 16191N, 16192N, 16193N, 16194N, 16195N, 16196N, 16197N, 16198N, 16199N, 16200N, 16201N, 16202N, 16203N, 16204N, 16205N, 16206N, 16207N, 16208N, 16209N, 16210N, 16211N, 16212N, 16213N, 16214N, 16215N, 16216N, 16217N, 16218N, 16219N, 16220N, 16221N, 16222N, 16223N, 16224N, 16225N, 16226N, 16227N, 16228N, 16229N, 16230N, 16231N, 16232N, 16233N, 16234N, 16235N, 16236N, 16237N, 16238N, 16239N, 16240N, 16241N, 16242N, 16243N, 16244N, 16245N, 16246N, 16247N, 16248N, 16249N, 16250N, 16251N, 16252N, 16253N, 16254N, 16255N, 16256N, 16257N, 16258N, 16259N, 16260N, 16261N, 16262N, 16263N, 16264N, 16265N, 16266N, 16267N, 16268N, 16269N, 16270N, 16271N, 16272N, 16273N, 16274N, 16275N, 16276N, 16277N, 16278N, 16279N, 16280N, 16281N, 16282N, 16283N, 16284N, 16285N, 16286N, 16287N, 16288N, 16289N, 16290N, 16291N, 16292N, 16293N, 16294N, 16295N, 16296N, 16297N, 16298N, 16299N, 16300N, 16301N, 16302N, 16303N, 16304N, 16305N, 16306N, 16307N, 16308N, 16309N, 16310N, 16311N, 16312N, 16313N, 16314N, 16315N, 16316N, 16317N, 16318N, 16319N, 16320N, 16321N, 16322N, 16323N, 16324N, 16325N, 16326N, 16327N, 16328N, 16329N, 16330N, 16331N, 16332N, 16333N, 16334N, 16335N, 16336N, 16337N, 16338N, 16339N, 16340N, 16341N, 16342N, 16343N, 16344N, 16345N, 16346N, 16347N, 16348N, 16349N, 16350N, 16351N, 16352N, 16353N, 16354N, 16355N, 16356N, 16357N | Lippold,S., Matzke,N., Reissman,M., Burbano,H. and Hofreiter,M. BMC Evolutionary Biology. 328 (11), 1471-2148 (2011) |
| HQ439465 (Horse and Przewalski's horse) | Illumina/Solexa Genome Analyzer II | Phantom mutaions: 358d, 1387G, 2227+T, 5239+A, 5277+A, 15385+T. | Q1b1a_HQ439465 | 158, 302, 341, 356d, 739, 860, 961, 1387G, 2227+T, 2788, 3070, 3259T, 3271, 3616, 3800, 3942, 4062, 4201, 4536, 4605, 4646, 4669, 5239+A, 5277+A, 5527, 5827, 5884, 6004, 6307, 6529, 6688, 6712, 6784, 7001, 7243, 7294, 7612T, 7666, 7898, 7900, 8005, 8076, 8150, 8238, 8358, 8361, 8556T, 8565, 8855, 9086, 9203, 9239, 9775, 10110, 10173, 10214, 10292, 10376, 10448, 10859, 11240, 11378, 11394, 11424, 11492, 11543, 11842, 11879, 11966, 12167, 12230, 12332, 12404, 12767, 13049, 13307, 13333, 13463, 13629, 13920, 13933, 14626, 14803, 15202, 15342, 15385+T, 15492, 15599, 15601, 15700, 15717, 15723, 15737, 15768, 15808, 15866, 15953, 16035, 16054, 16060C, 16110, 16118, 16126N, 16127N, 16128N, 16129N, 16130N, 16131N, 16132N, 16133N, 16134N, 16135N, 16136N, 16137N, 16138N, 16139N, 16140N, 16141N, 16142N, 16143N, 16144N, 16145N, 16146N, 16147N, 16148N, 16149N, 16150N, 16151N, 16152N, 16153N, 16154N, 16155N, 16156N, 16157N, 16158N, 16159N, 16160N, 16161N, 16162N, 16163N, 16164N, 16165N, 16166N, 16167N, 16168N, 16169N, 16170N, 16171N, 16172N, 16173N, 16174N, 16175N, 16176N, 16177N, 16178N, 16179N, 16180N, 16181N, | Q1b1a_HQ439465:356 | Q1b1a_HQ439465:356d, 2227+T, 5239+A, 5277+A, 15385+T, 16110, 16126N, 16127N, 16128N, 16129N, 16130N, 16131N, 16132N, 16133N, 16134N, 16135N, 16136N, 16137N, 16138N, 16139N, 16140N, 16141N, 16142N, 16143N, 16144N, 16145N, 16146N, 16147N, 16148N, 16149N, 16150N, 16151N, 16152N, 16153N, 16154N, 16155N, 16156N, 16157N, 16158N, 16159N, 16160N, 16161N, 16162N, 16163N, 16164N, 16165N, 16166N, 16167N, 16168N, 16169N, 16170N, 16171N, 16172N, 16173N, 16174N, 16175N, 16176N, 16177N, 16178N, 16179N, 16180N, 16181N, 16182N, 16183N, 16184N, 16185N, 16186N, 16187N, 16188N, 16189N, 16190N, 16191N, 16192N, 16193N, 16194N, 16195N, 16196N, 16197N, 16198N, 16199N, 16200N, 16201N, 16202N, 16203N, 16204N, 16205N, 16206N, 16207N, 16208N, 16209N, 16210N, 16211N, 16212N, 16213N, 16214N, 16215N, 16216N, | Lippold,S., Matzke,N., Reissman,M., Burbano,H. and Hofreiter,M. BMC Evolutionary Biology. 328 (11), 1471-2148 (2011) |
| | | | | 16182N, 16183N, 16184N, 16185N, 16186N, 16187N, 16188N, 16189N, 16190N, 16191N, 16192N, 16193N, 16194N, 16195N, 16196N, 16197N, 16198N, 16199N, 16200N, 16201N, 16202N, 16203N, 16204N, 16205N, 16206N, 16207N, 16208N, 16209N, 16210N, 16211N, 16212N, 16213N, 16214N, 16215N, 16216N, 16217N, 16218N, 16219N, 16220N, 16221N, 16222N, 16223N, 16224N, 16225N, 16226N, 16227N, 16228N, 16229N, 16230N, 16231N, 16232N, 16233N, 16234N, 16235N, 16236N, 16237N, 16238N, 16239N, 16240N, 16241N, 16242N, 16243N, 16244N, 16245N, 16246N, 16247N, 16248N, 16249N, 16250N, 16251N, 16252N, 16253N, 16254N, 16255N, 16256N, 16257N, 16258N, 16259N, 16260N, 16261N, 16262N, 16263N, 16264N, 16265N, 16266N, 16267N, 16268N, 16269N, 16270N, 16271N, 16272N, 16273N, 16274N, 16275N, 16276N, 16277N, 16278N, 16279N, 16280N, 16281N, 16282N, 16283N, 16284N, 16285N, 16286N, 16287N, 16288N, 16289N, 16290N, 16291N, 16292N, 16293N, 16294N, 16295N, 16296N, 16297N, 16298N, 16299N, 16300N, 16301N, 16302N, 16303N, 16304N, 16305N, 16306N, 16307N, 16308N, 16309N, 16310N, 16311N, 16312N, 16313N, 16314N, 16315N, 16316N, 16317N, 16318N, 16319N, 16320N, 16321N, 16322N, 16323N, 16324N, 16325N, 16326N, 16327N, 16328N, 16329N, 16330N, 16331N, 16332N, 16333N, 16334N, 16335N, 16336N, 16337N, 16338N, 16339N, 16340N, 16341N, 16342N, 16343N, 16344N, 16345N, 16346N, 16347N, 16348N, 16349N, 16350N, 16351N, 16352N, 16353N, 16354N, 16355N, 16356N, 16357N, 16368, 16540A, 16626 | | 16217N, 16218N, 16219N, 16220N, 16221N, 16222N, 16223N, 16224N, 16225N, 16226N, 16227N, 16228N, 16229N, 16230N, 16231N, 16232N, 16233N, 16234N, 16235N, 16236N, 16237N, 16238N, 16239N, 16240N, 16241N, 16242N, 16243N, 16244N, 16245N, 16246N, 16247N, 16248N, 16249N, 16250N, 16251N, 16252N, 16253N, 16254N, 16255N, 16256N, 16257N, 16258N, 16259N, 16260N, 16261N, 16262N, 16263N, 16264N, 16265N, 16266N, 16267N, 16268N, 16269N, 16270N, 16271N, 16272N, 16273N, 16274N, 16275N, 16276N, 16277N, 16278N, 16279N, 16280N, 16281N, 16282N, 16283N, 16284N, 16285N, 16286N, 16287N, 16288N, 16289N, 16290N, 16291N, 16292N, 16293N, 16294N, 16295N, 16296N, 16297N, 16298N, 16299N, 16300N, 16301N, 16302N, 16303N, 16304N, 16305N, 16306N, 16307N, 16308N, 16309N, 16310N, 16311N, 16312N, 16313N, 16314N, 16315N, 16316N, 16317N, 16318N, 16319N, 16320N, 16321N, 16322N, 16323N, 16324N, 16325N, 16326N, 16327N, 16328N, 16329N, 16330N, 16331N, 16332N, 16333N, 16334N, 16335N, 16336N, 16337N, 16338N, 16339N, 16340N, 16341N, 16342N, 16343N, 16344N, 16345N, 16346N, 16347N, 16348N, 16349N, 16350N, 16351N, 16352N, 16353N, 16354N, 16355N, 16356N, 16357N | |
| HQ439466 (Horse and Przewalski's horse) | Illumina/Solexa Genome Analyzer II | Phantom mutaions: 358d, 1387A, 2227+T, 5239+A, 5274+G, 15383+G. | M1b_HQ439466 | 158, 356d, 427, 961, 1387A, 1609T, 2227+T, 2339A, 2788, 3070, 3100, 3211T, 3475, 3800, 4062, 4526, 4536, 4599G, 4646, 4669, 4898, 5103, 5239+A, 5274+G, 5527, 5827, 5884, 6004, 6076, 6307, 6712, 6784, 7001, 7432, 7666, 7900, 8005, 8043, 8076, 8150, 8175, 8238, 8358, 8556T, 8565, 8798, 9086, 9332, 9540, 10110, 10173, 10214, 10292, 10376, 10448, 10460, 10515, 10646, 10859, 11240, 11394, 11492, 11513, 11543, 11789, 11842, 11879, 11966, 12029, 12095, 12332, 12767, 13049, 13100, 13333, 13356, 13502, 13615, 13629, 13720, 13920, 13933, 14422, 14626, 14671, 14803, 14815, 15052A, 15133, 15342, 15383+G, 15492, 15599, 15614, 15656, 15717, 15768, 15803, 15824, 15866, 15953, 16065, 16077, 16118, 16126N, 16127N, 16128N, 16129N, 16130N, 16131N, 16132N, 16133N, 16134N, 16135N, 16136N, 16137N, 16138N, 16139N, 16140N, 16141N, 16142N, 16143N, 16144N, 16145N, 16146N, 16147N, 16148N, 16149N, 16150N, 16151N, 16152N, 16153N, 16154N, 16155N, 16156N, 16157N, 16158N, 16159N, 16160N, 16161N, 16162N, 16163N, 16164N, 16165N, 16166N, 16167N, 16168N, 16169N, 16170N, 16171N, 16172N, 16173N, 16174N, 16175N, 16176N, 16177N, 16178N, 16179N, 16180N, 16181N, 16182N, 16183N, 16184N, 16185N, 16186N, 16187N, 16188N, 16189N, 16190N, 16191N, 16192N, 16193N, 16194N, 16195N, 16196N, 16197N, 16198N, 16199N, 16200N, 16201N, 16202N, 16203N, 16204N, 16205N, 16206N, 16207N, 16208N, 16209N, 16210N, 16211N, 16212N, 16213N, 16214N, 16215N, 16216N, 16217N, 16218N, 16219N, 16220N, 16221N, 16222N, 16223N, 16224N, 16225N, 16226N, 16227N, 16228N, 16229N, 16230N, 16231N, 16232N, 16233N, 16234N, 16235N, 16236N, 16237N, 16238N, 16239N, 16240N, 16241N, 16242N, 16243N, 16244N, 16245N, 16246N, 16247N, 16248N, 16249N, 16250N, 16251N, 16252N, 16253N, 16254N, 16255N, 16256N, 16257N, 16258N, 16259N, 16260N, 16261N, 16262N, 16263N, 16264N, 16265N, 16266N, 16267N, 16268N, 16269N, 16270N, 16271N, 16272N, 16273N, 16274N, 16275N, 16276N, 16277N, 16278N, 16279N, 16280N, 16281N, 16282N, 16283N, 16284N, 16285N, 16286N, 16287N, 16288N, 16289N, 16290N, 16291N, 16292N, 16293N, 16294N, 16295N, 16296N, 16297N, 16298N, 16299N, 16300N, 16301N, 16302N, 16303N, 16304N, 16305N, 16306N, 16307N, 16308N, 16309N, 16310N, 16311N, 16312N, 16313N, 16314N, 16315N, 16316N, 16317N, 16318N, 16319N, 16320N, 16321N, 16322N, 16323N, 16324N, 16325N, 16326N, 16327N, 16328N, 16329N, 16330N, 16331N, 16332N, 16333N, 16334N, 16335N, 16336N, 16337N, 16338N, 16339N, 16340N, 16341N, 16342N, 16343N, | M1b_HQ439466:356 | M1b_HQ439466:356d, 1387A, 2227+T, 5239+A, 5274+G, 15383+G, 16126N, 16127N, 16128N, 16129N, 16130N, 16131N, 16132N, 16133N, 16134N, 16135N, 16136N, 16137N, 16138N, 16139N, 16140N, 16141N, 16142N, 16143N, 16144N, 16145N, 16146N, 16147N, 16148N, 16149N, 16150N, 16151N, 16152N, 16153N, 16154N, 16155N, 16156N, 16157N, 16158N, 16159N, 16160N, 16161N, 16162N, 16163N, 16164N, 16165N, 16166N, 16167N, 16168N, 16169N, 16170N, 16171N, 16172N, 16173N, 16174N, 16175N, 16176N, 16177N, 16178N, 16179N, 16180N, 16181N, 16182N, 16183N, 16184N, 16185N, 16186N, 16187N, 16188N, 16189N, 16190N, 16191N, 16192N, 16193N, 16194N, 16195N, 16196N, 16197N, 16198N, 16199N, 16200N, 16201N, 16202N, 16203N, 16204N, 16205N, 16206N, 16207N, 16208N, 16209N, 16210N, 16211N, 16212N, 16213N, 16214N, 16215N, 16216N, 16217N, 16218N, 16219N, 16220N, 16221N, 16222N, 16223N, 16224N, 16225N, 16226N, 16227N, 16228N, 16229N, 16230N, 16231N, 16232N, 16233N, 16234N, 16235N, 16236N, 16237N, 16238N, 16239N, 16240N, 16241N, 16242N, 16243N, 16244N, 16245N, 16246N, 16247N, 16248N, 16249N, 16250N, 16251N, 16252N, 16253N, 16254N, 16255N, 16256N, 16257N, 16258N, 16259N, 16260N, 16261N, 16262N, 16263N, 16264N, 16265N, 16266N, 16267N, 16268N, 16269N, 16270N, 16271N, 16272N, 16273N, 16274N, 16275N, 16276N, 16277N, 16278N, 16279N, 16280N, 16281N, 16282N, 16283N, 16284N, 16285N, 16286N, 16287N, 16288N, 16289N, 16290N, 16291N, 16292N, 16293N, 16294N, 16295N, 16296N, 16297N, 16298N, 16299N, 16300N, 16301N, 16302N, 16303N, 16304N, 16305N, 16306N, 16307N, 16308N, 16309N, 16310N, 16311N, 16312N, 16313N, 16314N, 16315N, 16316N, 16317N, 16318N, 16319N, 16320N, 16321N, 16322N, 16323N, 16324N, 16325N, 16326N, 16327N, 16328N, 16329N, 16330N, 16331N, 16332N, 16333N, 16334N, 16335N, 16336N, 16337N, 16338N, | Lippold,S., Matzke,N., Reissman,M., Burbano,H. and Hofreiter,M. BMC Evolutionary Biology. 328 (11), 1471-2148 (2011) |
| | | | | 16344N, 16345N, 16346N, 16347N, 16348N, 16349N, 16350N, 16351N, 16352N, 16353N, 16354N, 16355N, 16356N, 16357N, 16368, 16540A, 16543, 16556, 16626 | | 16339N, 16340N, 16341N, 16342N, 16343N, 16344N, 16345N, 16346N, 16347N, 16348N, 16349N, 16350N, 16351N, 16352N, 16353N, 16354N, 16355N, 16356N, 16357N | |
| HQ439467 (Horse and Przewalski's horse) | Illumina/Solexa Genome Analyzer II | Phantom mutaions: 358d, 1387A, 2227+T, 5237+A, 5277+A, 15385+T. | O_HQ439467 | 158, 356d, 739, 860, 961, 965, 1387A, 1684C, 2227+T, 2788, 3070, 3259T, 3271, 3557, 3616, 3800, 3942, 4062, 4536, 4605, 4646, 4669, 5237+G, 5277+A, 5527, 5821, 5827, 5884, 6004, 6307, 6529, 6712, 6745, 6784, 7001, 7243, 7375, 7427, 7612T, 7666, 7898, 7900, 8005, 8040, 8076, 8150, 8238, 8358, 8361, 8556T, 8565, 8855, 9053, 9086, 9239, 9775, 10110, 10173, 10214, 10292, 10361T, 10376, 10501, 10859, 11240, 11378, 11394, 11424, 11492, 11543, 11842, 11879, 11966, 12167, 12230, 12287, 12305, 12332, 12404, 12767, 12927, 13049, 13333, 13463, 13466, 13629, 13920, 13933, 13951, 14626, 14650, 14803, 15202, 15342, 15385+T, 15492, 15594, 15599, 15632, 15664, 15700, 15717, 15768, 15774, 15806, 15953, 16035, 16110, 16118, 16126N, 16127N, 16128N, 16129N, 16130N, 16131N, 16132N, 16133N, 16134N, 16135N, 16136N, 16137N, 16138N, 16139N, 16140N, 16141N, 16142N, 16143N, 16144N, 16145N, 16146N, 16147N, 16148N, 16149N, 16150N, 16151N, 16152N, 16153N, 16154N, 16155N, 16156N, 16157N, 16158N, 16159N, 16160N, 16161N, 16162N, 16163N, 16164N, 16165N, 16166N, 16167N, 16168N, 16169N, 16170N, 16171N, 16172N, 16173N, 16174N, 16175N, 16176N, 16177N, 16178N, 16179N, 16180N, 16181N, 16182N, 16183N, 16184N, 16185N, 16186N, 16187N, 16188N, 16189N, 16190N, 16191N, 16192N, 16193N, 16194N, 16195N, 16196N, 16197N, 16198N, 16199N, 16200N, 16201N, 16202N, 16203N, 16204N, 16205N, 16206N, 16207N, 16208N, 16209N, 16210N, 16211N, 16212N, 16213N, 16214N, 16215N, 16216N, 16217N, 16218N, 16219N, 16220N, 16221N, 16222N, 16223N, 16224N, 16225N, 16226N, 16227N, 16228N, 16229N, 16230N, 16231N, 16232N, 16233N, 16234N, 16235N, 16236N, 16237N, 16238N, 16239N, 16240N, 16241N, 16242N, 16243N, 16244N, 16245N, 16246N, 16247N, 16248N, 16249N, 16250N, 16251N, 16252N, 16253N, 16254N, 16255N, 16256N, 16257N, 16258N, 16259N, 16260N, 16261N, 16262N, 16263N, 16264N, 16265N, 16266N, 16267N, 16268N, 16269N, 16270N, 16271N, 16272N, 16273N, 16274N, 16275N, 16276N, 16277N, 16278N, 16279N, 16280N, 16281N, 16282N, 16283N, 16284N, 16285N, 16286N, 16287N, 16288N, 16289N, 16290N, 16291N, 16292N, 16293N, 16294N, 16295N, 16296N, 16297N, 16298N, 16299N, 16300N, 16301N, 16302N, 16303N, 16304N, 16305N, 16306N, 16307N, 16308N, 16309N, 16310N, 16311N, 16312N, 16313N, 16314N, 16315N, 16316N, 16317N, 16318N, 16319N, 16320N, 16321N, 16322N, 16323N, 16324N, 16325N, 16326N, 16327N, 16328N, 16329N, 16330N, 16331N, 16332N, 16333N, 16334N, 16335N, 16336N, 16337N, 16338N, 16339N, 16340N, 16341N, 16342N, 16343N, 16344N, 16345N, 16346N, 16347N, 16348N, 16349N, 16350N, 16351N, 16352N, 16353N, 16354N, 16355N, 16356N, 16357N, 16368, 16398A, 16540A, 16626 | O_HQ439467:356 | O_HQ439467:356d, 2227+T, 5237+G, 5277+A, 15385+T, 16126N, 16127N, 16128N, 16129N, 16130N, 16131N, 16132N, 16133N, 16134N, 16135N, 16136N, 16137N, 16138N, 16139N, 16140N, 16141N, 16142N, 16143N, 16144N, 16145N, 16146N, 16147N, 16148N, 16149N, 16150N, 16151N, 16152N, 16153N, 16154N, 16155N, 16156N, 16157N, 16158N, 16159N, 16160N, 16161N, 16162N, 16163N, 16164N, 16165N, 16166N, 16167N, 16168N, 16169N, 16170N, 16171N, 16172N, 16173N, 16174N, 16175N, 16176N, 16177N, 16178N, 16179N, 16180N, 16181N, 16182N, 16183N, 16184N, 16185N, 16186N, 16187N, 16188N, 16189N, 16190N, 16191N, 16192N, 16193N, 16194N, 16195N, 16196N, 16197N, 16198N, 16199N, 16200N, 16201N, 16202N, 16203N, 16204N, 16205N, 16206N, 16207N, 16208N, 16209N, 16210N, 16211N, 16212N, 16213N, 16214N, 16215N, 16216N, 16217N, 16218N, 16219N, 16220N, 16221N, 16222N, 16223N, 16224N, 16225N, 16226N, 16227N, 16228N, 16229N, 16230N, 16231N, 16232N, 16233N, 16234N, 16235N, 16236N, 16237N, 16238N, 16239N, 16240N, 16241N, 16242N, 16243N, 16244N, 16245N, 16246N, 16247N, 16248N, 16249N, 16250N, 16251N, 16252N, 16253N, 16254N, 16255N, 16256N, 16257N, 16258N, 16259N, 16260N, 16261N, 16262N, 16263N, 16264N, 16265N, 16266N, 16267N, 16268N, 16269N, 16270N, 16271N, 16272N, 16273N, 16274N, 16275N, 16276N, 16277N, 16278N, 16279N, 16280N, 16281N, 16282N, 16283N, 16284N, 16285N, 16286N, 16287N, 16288N, 16289N, 16290N, 16291N, 16292N, 16293N, 16294N, 16295N, 16296N, 16297N, 16298N, 16299N, 16300N, 16301N, 16302N, 16303N, 16304N, 16305N, 16306N, 16307N, 16308N, 16309N, 16310N, 16311N, 16312N, 16313N, 16314N, 16315N, 16316N, 16317N, 16318N, 16319N, 16320N, 16321N, 16322N, 16323N, 16324N, 16325N, 16326N, 16327N, 16328N, 16329N, 16330N, 16331N, 16332N, 16333N, 16334N, 16335N, 16336N, 16337N, 16338N, 16339N, 16340N, 16341N, 16342N, 16343N, 16344N, 16345N, 16346N, 16347N, 16348N, 16349N, 16350N, 16351N, 16352N, 16353N, 16354N, 16355N, 16356N, 16357N | Lippold,S., Matzke,N., Reissman,M., Burbano,H. and Hofreiter,M. BMC Evolutionary Biology. 328 (11), 1471-2148 (2011) |
| HQ439468 (Horse and Przewalski's horse) | Illumina/Solexa Genome Analyzer II | Phantom mutaions: 358d, 1387G, 2226+C, 5237+G, 5277+A, 15385+T. | G1a_HQ439468 | 158, 222, 356d, 382, 387, 416, 1387G, 1603, 2226+C, 2788, 2940, 3053, 3576, 4062, 4646, 4669, 4830, 5237+G, 5277+A, 5498, 5669, 5830, 5881, 5884, 6004, 6307, 6688, 6784, 7001, 8005, 8037, 8230, 9239, 9402, 9669, 9741A, 10214, 10376, 10471, 11165, 11240, 11543, 11552, 11842, 12767, 12860, 13049, 13079, 13223, 13333, 13502, 14350, 14626, 14651, 14734, 15385+T, 15492, 15539, 15582, 15594, 15599, 15632, 15647, 15663, 15700, 15717, 15867, 16110, 16126N, 16127N, 16128N, 16129N, 16130N, 16131N, 16132N, 16133N, 16134N, 16135N, 16136N, 16137N, 16138N, 16139N, 16140N, 16141N, 16142N, 16143N, 16144N, 16145N, 16146N, 16147N, 16148N, 16149N, 16150N, 16151N, 16152N, 16153N, 16154N, 16155N, 16156N, 16157N, 16158N, 16159N, 16160N, 16161N, 16162N, 16163N, 16164N, 16165N, 16166N, 16167N, 16168N, 16169N, 16170N, 16171N, 16172N, 16173N, 16174N, 16175N, 16176N, 16177N, 16178N, 16179N, 16180N, 16181N, 16182N, 16183N, 16184N, 16185N, 16186N, 16187N, 16188N, 16189N, 16190N, 16191N, 16192N, 16193N, 16194N, 16195N, 16196N, 16197N, 16198N, 16199N, 16200N, 16201N, 16202N, 16203N, 16204N, 16205N, 16206N, 16207N, | G1a_HQ439468:356 | G1a_HQ439468:356d, 2226+C, 5237+G, 5277+A, 15385+T, 16126N, 16127N, 16128N, 16129N, 16130N, 16131N, 16132N, 16133N, 16134N, 16135N, 16136N, 16137N, 16138N, 16139N, 16140N, 16141N, 16142N, 16143N, 16144N, 16145N, 16146N, 16147N, 16148N, 16149N, 16150N, 16151N, 16152N, 16153N, 16154N, 16155N, 16156N, 16157N, 16158N, 16159N, 16160N, 16161N, 16162N, 16163N, 16164N, 16165N, 16166N, 16167N, 16168N, 16169N, 16170N, 16171N, 16172N, 16173N, 16174N, 16175N, 16176N, 16177N, 16178N, 16179N, 16180N, 16181N, 16182N, 16183N, 16184N, 16185N, 16186N, 16187N, 16188N, 16189N, 16190N, 16191N, 16192N, 16193N, 16194N, 16195N, 16196N, 16197N, 16198N, 16199N, 16200N, 16201N, 16202N, 16203N, 16204N, 16205N, 16206N, 16207N, 16208N, 16209N, 16210N, 16211N, 16212N, 16213N, 16214N, 16215N, 16216N, 16217N, 16218N, | Lippold,S., Matzke,N., Reissman,M., Burbano,H. and Hofreiter,M. BMC Evolutionary Biology. 328 (11), 1471-2148 (2011) |
| | | | | 16208N, 16209N, 16210N, 16211N, 16212N, 16213N, 16214N, 16215N, 16216N, 16217N, 16218N, 16219N, 16220N, 16221N, 16222N, 16223N, 16224N, 16225N, 16226N, 16227N, 16228N, 16229N, 16230N, 16231N, 16232N, 16233N, 16234N, 16235N, 16236N, 16237N, 16238N, 16239N, 16240N, 16241N, 16242N, 16243N, 16244N, 16245N, 16246N, 16247N, 16248N, 16249N, 16250N, 16251N, 16252N, 16253N, 16254N, 16255N, 16256N, 16257N, 16258N, 16259N, 16260N, 16261N, 16262N, 16263N, 16264N, 16265N, 16266N, 16267N, 16268N, 16269N, 16270N, 16271N, 16272N, 16273N, 16274N, 16275N, 16276N, 16277N, 16278N, 16279N, 16280N, 16281N, 16282N, 16283N, 16284N, 16285N, 16286N, 16287N, 16288N, 16289N, 16290N, 16291N, 16292N, 16293N, 16294N, 16295N, 16296N, 16297N, 16298N, 16299N, 16300N, 16301N, 16302N, 16303N, 16304N, 16305N, 16306N, 16307N, 16308N, 16309N, 16310N, 16311N, 16312N, 16313N, 16314N, 16315N, 16316N, 16317N, 16318N, 16319N, 16320N, 16321N, 16322N, 16323N, 16324N, 16325N, 16326N, 16327N, 16328N, 16329N, 16330N, 16331N, 16332N, 16333N, 16334N, 16335N, 16336N, 16337N, 16338N, 16339N, 16340N, 16341N, 16342N, 16343N, 16344N, 16345N, 16346N, 16347N, 16348N, 16349N, 16350N, 16351N, 16352N, 16353N, 16354N, 16355N, 16356N, 16357N, 16368 | | 16219N, 16220N, 16221N, 16222N, 16223N, 16224N, 16225N, 16226N, 16227N, 16228N, 16229N, 16230N, 16231N, 16232N, 16233N, 16234N, 16235N, 16236N, 16237N, 16238N, 16239N, 16240N, 16241N, 16242N, 16243N, 16244N, 16245N, 16246N, 16247N, 16248N, 16249N, 16250N, 16251N, 16252N, 16253N, 16254N, 16255N, 16256N, 16257N, 16258N, 16259N, 16260N, 16261N, 16262N, 16263N, 16264N, 16265N, 16266N, 16267N, 16268N, 16269N, 16270N, 16271N, 16272N, 16273N, 16274N, 16275N, 16276N, 16277N, 16278N, 16279N, 16280N, 16281N, 16282N, 16283N, 16284N, 16285N, 16286N, 16287N, 16288N, 16289N, 16290N, 16291N, 16292N, 16293N, 16294N, 16295N, 16296N, 16297N, 16298N, 16299N, 16300N, 16301N, 16302N, 16303N, 16304N, 16305N, 16306N, 16307N, 16308N, 16309N, 16310N, 16311N, 16312N, 16313N, 16314N, 16315N, 16316N, 16317N, 16318N, 16319N, 16320N, 16321N, 16322N, 16323N, 16324N, 16325N, 16326N, 16327N, 16328N, 16329N, 16330N, 16331N, 16332N, 16333N, 16334N, 16335N, 16336N, 16337N, 16338N, 16339N, 16340N, 16341N, 16342N, 16343N, 16344N, 16345N, 16346N, 16347N, 16348N, 16349N, 16350N, 16351N, 16352N, 16353N, 16354N, 16355N, 16356N, 16357N | |
| HQ439469 (Horse and Przewalski's horse) | Illumina/Solexa Genome Analyzer II | Phantom mutaions: 358d, 2227+T, 5237+G, 5277+A, 15385+T. | B1a1_HQ439469 | 158, 356d, 859, 2227+T, 4062, 4245, 5237+G, 5277+A, 6784, 7627, 9961, 10764, 11240, 11399, 13166, 13816, 15385+T, 15492, 15582, 15647, 15663, 15717, 15807, 15823, 16052, 16110, 16126N, 16127N, 16128N, 16129N, 16130N, 16131N, 16132N, 16133N, 16134N, 16135N, 16136N, 16137N, 16138N, 16139N, 16140N, 16141N, 16142N, 16143N, 16144N, 16145N, 16146N, 16147N, 16148N, 16149N, 16150N, 16151N, 16152N, 16153N, 16154N, 16155N, 16156N, 16157N, 16158N, 16159N, 16160N, 16161N, 16162N, 16163N, 16164N, 16165N, 16166N, 16167N, 16168N, 16169N, 16170N, 16171N, 16172N, 16173N, 16174N, 16175N, 16176N, 16177N, 16178N, 16179N, 16180N, 16181N, 16182N, 16183N, 16184N, 16185N, 16186N, 16187N, 16188N, 16189N, 16190N, 16191N, 16192N, 16193N, 16194N, 16195N, 16196N, 16197N, 16198N, 16199N, 16200N, 16201N, 16202N, 16203N, 16204N, 16205N, 16206N, 16207N, 16208N, 16209N, 16210N, 16211N, 16212N, 16213N, 16214N, 16215N, 16216N, 16217N, 16218N, 16219N, 16220N, 16221N, 16222N, 16223N, 16224N, 16225N, 16226N, 16227N, 16228N, 16229N, 16230N, 16231N, 16232N, 16233N, 16234N, 16235N, 16236N, 16237N, 16238N, 16239N, 16240N, 16241N, 16242N, 16243N, 16244N, 16245N, 16246N, 16247N, 16248N, 16249N, 16250N, 16251N, 16252N, 16253N, 16254N, 16255N, 16256N, 16257N, 16258N, 16259N, 16260N, 16261N, 16262N, 16263N, 16264N, 16265N, 16266N, 16267N, 16268N, 16269N, 16270N, 16271N, 16272N, 16273N, 16274N, 16275N, 16276N, 16277N, 16278N, 16279N, 16280N, 16281N, 16282N, 16283N, 16284N, 16285N, 16286N, 16287N, 16288N, 16289N, 16290N, 16291N, 16292N, 16293N, 16294N, 16295N, 16296N, 16297N, 16298N, 16299N, 16300N, 16301N, 16302N, 16303N, 16304N, 16305N, 16306N, 16307N, 16308N, 16309N, 16310N, 16311N, 16312N, 16313N, 16314N, 16315N, 16316N, 16317N, 16318N, 16319N, 16320N, 16321N, 16322N, 16323N, 16324N, 16325N, 16326N, 16327N, 16328N, 16329N, 16330N, 16331N, 16332N, 16333N, 16334N, 16335N, 16336N, 16337N, 16338N, 16339N, 16340N, 16341N, 16342N, 16343N, 16344N, 16345N, 16346N, 16347N, 16348N, 16349N, 16350N, 16351N, 16352N, 16353N, 16354N, 16355N, 16356N, 16357N, 16368 | B1a1_HQ439469:356 | B1a1_HQ439469:356d, 2227+T, 5237+G, 5277+A, 15385+T, 16126N, 16127N, 16128N, 16129N, 16130N, 16131N, 16132N, 16133N, 16134N, 16135N, 16136N, 16137N, 16138N, 16139N, 16140N, 16141N, 16142N, 16143N, 16144N, 16145N, 16146N, 16147N, 16148N, 16149N, 16150N, 16151N, 16152N, 16153N, 16154N, 16155N, 16156N, 16157N, 16158N, 16159N, 16160N, 16161N, 16162N, 16163N, 16164N, 16165N, 16166N, 16167N, 16168N, 16169N, 16170N, 16171N, 16172N, 16173N, 16174N, 16175N, 16176N, 16177N, 16178N, 16179N, 16180N, 16181N, 16182N, 16183N, 16184N, 16185N, 16186N, 16187N, 16188N, 16189N, 16190N, 16191N, 16192N, 16193N, 16194N, 16195N, 16196N, 16197N, 16198N, 16199N, 16200N, 16201N, 16202N, 16203N, 16204N, 16205N, 16206N, 16207N, 16208N, 16209N, 16210N, 16211N, 16212N, 16213N, 16214N, 16215N, 16216N, 16217N, 16218N, 16219N, 16220N, 16221N, 16222N, 16223N, 16224N, 16225N, 16226N, 16227N, 16228N, 16229N, 16230N, 16231N, 16232N, 16233N, 16234N, 16235N, 16236N, 16237N, 16238N, 16239N, 16240N, 16241N, 16242N, 16243N, 16244N, 16245N, 16246N, 16247N, 16248N, 16249N, 16250N, 16251N, 16252N, 16253N, 16254N, 16255N, 16256N, 16257N, 16258N, 16259N, 16260N, 16261N, 16262N, 16263N, 16264N, 16265N, 16266N, 16267N, 16268N, 16269N, 16270N, 16271N, 16272N, 16273N, 16274N, 16275N, 16276N, 16277N, 16278N, 16279N, 16280N, 16281N, 16282N, 16283N, 16284N, 16285N, 16286N, 16287N, 16288N, 16289N, 16290N, 16291N, 16292N, 16293N, 16294N, 16295N, 16296N, 16297N, 16298N, 16299N, 16300N, 16301N, 16302N, 16303N, 16304N, 16305N, 16306N, 16307N, 16308N, 16309N, 16310N, 16311N, 16312N, 16313N, 16314N, 16315N, 16316N, 16317N, 16318N, 16319N, 16320N, 16321N, 16322N, 16323N, 16324N, 16325N, 16326N, 16327N, 16328N, 16329N, 16330N, 16331N, 16332N, 16333N, 16334N, 16335N, 16336N, 16337N, 16338N, | Lippold,S., Matzke,N., Reissman,M., Burbano,H. and Hofreiter,M. BMC Evolutionary Biology. 328 (11), 1471-2148 (2011) |
| | | | | | | 16339N, 16340N, 16341N, 16342N, 16343N, 16344N, 16345N, 16346N, 16347N, 16348N, 16349N, 16350N, 16351N, 16352N, 16353N, 16354N, 16355N, 16356N, 16357N | |
| HQ439470 (Horse and Przewalski's horse) | Illumina/Solexa Genome Analyzer II | Phantom mutaions: 358d, 1387G, 2227+T, 5239+A, 5277+A, 15383+G. | I2_HQ439470 | 158, 356d, 961, 1375, 1387G, 2227+T, 2607, 2788, 2899, 3517, 3727, 3942, 4062, 4536, 4646, 4669, 4757, 5239+A, 5277+A, 5527, 5815, 5884, 6004, 6307, 6784, 7001, 7516, 7666, 7900, 8005, 8058, 8301, 8319, 8358, 8565, 9239, 9951, 10087, 10110, 10214, 10292, 10376, 10421, 10613, 11240, 11543, 11693, 11842, 11879, 12119, 12200, 12767, 12896, 12950, 13049, 13333, 13520, 14803, 14995, 15313, 15383+G, 15475, 15491, 15492, 15493, 15531, 15582, 15600, 15646, 15717, 15768, 15824, 15867, 15868, 15953, 15971, 16065, 16100, 16108, 16126N, 16127N, 16128N, 16129N, 16130N, 16131N, 16132N, 16133N, 16134N, 16135N, 16136N, 16137N, 16138N, 16139N, 16140N, 16141N, 16142N, 16143N, 16144N, 16145N, 16146N, 16147N, 16148N, 16149N, 16150N, 16151N, 16152N, 16153N, 16154N, 16155N, 16156N, 16157N, 16158N, 16159N, 16160N, 16161N, 16162N, 16163N, 16164N, 16165N, 16166N, 16167N, 16168N, 16169N, 16170N, 16171N, 16172N, 16173N, 16174N, 16175N, 16176N, 16177N, 16178N, 16179N, 16180N, 16181N, 16182N, 16183N, 16184N, 16185N, 16186N, 16187N, 16188N, 16189N, 16190N, 16191N, 16192N, 16193N, 16194N, 16195N, 16196N, 16197N, 16198N, 16199N, 16200N, 16201N, 16202N, 16203N, 16204N, 16205N, 16206N, 16207N, 16208N, 16209N, 16210N, 16211N, 16212N, 16213N, 16214N, 16215N, 16216N, 16217N, 16218N, 16219N, 16220N, 16221N, 16222N, 16223N, 16224N, 16225N, 16226N, 16227N, 16228N, 16229N, 16230N, 16231N, 16232N, 16233N, 16234N, 16235N, 16236N, 16237N, 16238N, 16239N, 16240N, 16241N, 16242N, 16243N, 16244N, 16245N, 16246N, 16247N, 16248N, 16249N, 16250N, 16251N, 16252N, 16253N, 16254N, 16255N, 16256N, 16257N, 16258N, 16259N, 16260N, 16261N, 16262N, 16263N, 16264N, 16265N, 16266N, 16267N, 16268N, 16269N, 16270N, 16271N, 16272N, 16273N, 16274N, 16275N, 16276N, 16277N, 16278N, 16279N, 16280N, 16281N, 16282N, 16283N, 16284N, 16285N, 16286N, 16287N, 16288N, 16289N, 16290N, 16291N, 16292N, 16293N, 16294N, 16295N, 16296N, 16297N, 16298N, 16299N, 16300N, 16301N, 16302N, 16303N, 16304N, 16305N, 16306N, 16307N, 16308N, 16309N, 16310N, 16311N, 16312N, 16313N, 16314N, 16315N, 16316N, 16317N, 16318N, 16319N, 16320N, 16321N, 16322N, 16323N, 16324N, 16325N, 16326N, 16327N, 16328N, 16329N, 16330N, 16331N, 16332N, 16333N, 16334N, 16335N, 16336N, 16337N, 16338N, 16339N, 16340N, 16341N, 16342N, 16343N, 16344N, 16345N, 16346N, 16347N, 16348N, 16349N, 16350N, 16351N, 16352N, 16353N, 16354N, 16355N, 16356N, 16357N, 16368, 16626 | I2_HQ439470:356 | I2_HQ439470:356d, 2227+T, 5239+A, 5277+A, 15383+G, 16126N, 16127N, 16128N, 16129N, 16130N, 16131N, 16132N, 16133N, 16134N, 16135N, 16136N, 16137N, 16138N, 16139N, 16140N, 16141N, 16142N, 16143N, 16144N, 16145N, 16146N, 16147N, 16148N, 16149N, 16150N, 16151N, 16152N, 16153N, 16154N, 16155N, 16156N, 16157N, 16158N, 16159N, 16160N, 16161N, 16162N, 16163N, 16164N, 16165N, 16166N, 16167N, 16168N, 16169N, 16170N, 16171N, 16172N, 16173N, 16174N, 16175N, 16176N, 16177N, 16178N, 16179N, 16180N, 16181N, 16182N, 16183N, 16184N, 16185N, 16186N, 16187N, 16188N, 16189N, 16190N, 16191N, 16192N, 16193N, 16194N, 16195N, 16196N, 16197N, 16198N, 16199N, 16200N, 16201N, 16202N, 16203N, 16204N, 16205N, 16206N, 16207N, 16208N, 16209N, 16210N, 16211N, 16212N, 16213N, 16214N, 16215N, 16216N, 16217N, 16218N, 16219N, 16220N, 16221N, 16222N, 16223N, 16224N, 16225N, 16226N, 16227N, 16228N, 16229N, 16230N, 16231N, 16232N, 16233N, 16234N, 16235N, 16236N, 16237N, 16238N, 16239N, 16240N, 16241N, 16242N, 16243N, 16244N, 16245N, 16246N, 16247N, 16248N, 16249N, 16250N, 16251N, 16252N, 16253N, 16254N, 16255N, 16256N, 16257N, 16258N, 16259N, 16260N, 16261N, 16262N, 16263N, 16264N, 16265N, 16266N, 16267N, 16268N, 16269N, 16270N, 16271N, 16272N, 16273N, 16274N, 16275N, 16276N, 16277N, 16278N, 16279N, 16280N, 16281N, 16282N, 16283N, 16284N, 16285N, 16286N, 16287N, 16288N, 16289N, 16290N, 16291N, 16292N, 16293N, 16294N, 16295N, 16296N, 16297N, 16298N, 16299N, 16300N, 16301N, 16302N, 16303N, 16304N, 16305N, 16306N, 16307N, 16308N, 16309N, 16310N, 16311N, 16312N, 16313N, 16314N, 16315N, 16316N, 16317N, 16318N, 16319N, 16320N, 16321N, 16322N, 16323N, 16324N, 16325N, 16326N, 16327N, 16328N, 16329N, 16330N, 16331N, 16332N, 16333N, 16334N, 16335N, 16336N, 16337N, 16338N, 16339N, 16340N, 16341N, 16342N, 16343N, 16344N, 16345N, 16346N, 16347N, 16348N, 16349N, 16350N, 16351N, 16352N, 16353N, 16354N, 16355N, 16356N, 16357N | Lippold,S., Matzke,N., Reissman,M., Burbano,H. and Hofreiter,M. BMC Evolutionary Biology. 328 (11), 1471-2148 (2011) |
| HQ439471 (Horse and Przewalski's horse) | Illumina/Solexa Genome Analyzer II | Phantom mutaions: 358d, 1387G, 2226+C, 5237+G, 5277+A, 15385+T. | C1_HQ439471 | 158, 356d, 957, 1387G, 2226+C, 2238, 2788, 3800, 4062, 4599, 4884, 4993, 5237+G, 5277+A, 5500, 5914, 6076, 6784, 7001, 7942, 9239, 9278, 9664, 10214, 10217, 11046, 11129, 11240, 11543, 12352, 13079, 15385+T, 15492, 15582, 15599, 15647, 15717, 15823, 15867, 15953, 15971, 16110, 16126N, 16127N, 16128N, 16129N, 16130N, 16131N, 16132N, 16133N, 16134N, 16135N, 16136N, 16137N, 16138N, 16139N, 16140N, 16141N, 16142N, 16143N, 16144N, 16145N, 16146N, 16147N, 16148N, 16149N, 16150N, 16151N, 16152N, 16153N, 16154N, 16155N, 16156N, 16157N, 16158N, 16159N, 16160N, 16161N, 16162N, 16163N, 16164N, 16165N, 16166N, 16167N, 16168N, 16169N, 16170N, 16171N, 16172N, 16173N, 16174N, 16175N, 16176N, 16177N, 16178N, 16179N, 16180N, 16181N, 16182N, 16183N, 16184N, 16185N, 16186N, 16187N, 16188N, 16189N, 16190N, 16191N, 16192N, 16193N, 16194N, 16195N, 16196N, 16197N, 16198N, 16199N, 16200N, 16201N, 16202N, 16203N, 16204N, 16205N, 16206N, 16207N, 16208N, 16209N, 16210N, 16211N, 16212N, 16213N, 16214N, 16215N, 16216N, 16217N, 16218N, 16219N, 16220N, 16221N, 16222N, 16223N, 16224N, 16225N, 16226N, | C1_HQ439471:356, 1387A | C1_HQ439471:356d, 1387G, 2226+C, 5237+G, 5277+A, 15385+T, 16126N, 16127N, 16128N, 16129N, 16130N, 16131N, 16132N, 16133N, 16134N, 16135N, 16136N, 16137N, 16138N, 16139N, 16140N, 16141N, 16142N, 16143N, 16144N, 16145N, 16146N, 16147N, 16148N, 16149N, 16150N, 16151N, 16152N, 16153N, 16154N, 16155N, 16156N, 16157N, 16158N, 16159N, 16160N, 16161N, 16162N, 16163N, 16164N, 16165N, 16166N, 16167N, 16168N, 16169N, 16170N, 16171N, 16172N, 16173N, 16174N, 16175N, 16176N, 16177N, 16178N, 16179N, 16180N, 16181N, 16182N, 16183N, 16184N, 16185N, 16186N, 16187N, 16188N, 16189N, 16190N, 16191N, 16192N, 16193N, 16194N, 16195N, 16196N, 16197N, 16198N, 16199N, 16200N, 16201N, 16202N, 16203N, 16204N, 16205N, 16206N, 16207N, 16208N, 16209N, 16210N, 16211N, 16212N, 16213N, 16214N, 16215N, 16216N, 16217N, | Lippold,S., Matzke,N., Reissman,M., Burbano,H. and Hofreiter,M. BMC Evolutionary Biology. 328 (11), 1471-2148 (2011) |
| | | | | 16227N, 16228N, 16229N, 16230N, 16231N, 16232N, 16233N, 16234N, 16235N, 16236N, 16237N, 16238N, 16239N, 16240N, 16241N, 16242N, 16243N, 16244N, 16245N, 16246N, 16247N, 16248N, 16249N, 16250N, 16251N, 16252N, 16253N, 16254N, 16255N, 16256N, 16257N, 16258N, 16259N, 16260N, 16261N, 16262N, 16263N, 16264N, 16265N, 16266N, 16267N, 16268N, 16269N, 16270N, 16271N, 16272N, 16273N, 16274N, 16275N, 16276N, 16277N, 16278N, 16279N, 16280N, 16281N, 16282N, 16283N, 16284N, 16285N, 16286N, 16287N, 16288N, 16289N, 16290N, 16291N, 16292N, 16293N, 16294N, 16295N, 16296N, 16297N, 16298N, 16299N, 16300N, 16301N, 16302N, 16303N, 16304N, 16305N, 16306N, 16307N, 16308N, 16309N, 16310N, 16311N, 16312N, 16313N, 16314N, 16315N, 16316N, 16317N, 16318N, 16319N, 16320N, 16321N, 16322N, 16323N, 16324N, 16325N, 16326N, 16327N, 16328N, 16329N, 16330N, 16331N, 16332N, 16333N, 16334N, 16335N, 16336N, 16337N, 16338N, 16339N, 16340N, 16341N, 16342N, 16343N, 16344N, 16345N, 16346N, 16347N, 16348N, 16349N, 16350N, 16351N, 16352N, 16353N, 16354N, 16355N, 16356N, 16357N, 16368 | | 16218N, 16219N, 16220N, 16221N, 16222N, 16223N, 16224N, 16225N, 16226N, 16227N, 16228N, 16229N, 16230N, 16231N, 16232N, 16233N, 16234N, 16235N, 16236N, 16237N, 16238N, 16239N, 16240N, 16241N, 16242N, 16243N, 16244N, 16245N, 16246N, 16247N, 16248N, 16249N, 16250N, 16251N, 16252N, 16253N, 16254N, 16255N, 16256N, 16257N, 16258N, 16259N, 16260N, 16261N, 16262N, 16263N, 16264N, 16265N, 16266N, 16267N, 16268N, 16269N, 16270N, 16271N, 16272N, 16273N, 16274N, 16275N, 16276N, 16277N, 16278N, 16279N, 16280N, 16281N, 16282N, 16283N, 16284N, 16285N, 16286N, 16287N, 16288N, 16289N, 16290N, 16291N, 16292N, 16293N, 16294N, 16295N, 16296N, 16297N, 16298N, 16299N, 16300N, 16301N, 16302N, 16303N, 16304N, 16305N, 16306N, 16307N, 16308N, 16309N, 16310N, 16311N, 16312N, 16313N, 16314N, 16315N, 16316N, 16317N, 16318N, 16319N, 16320N, 16321N, 16322N, 16323N, 16324N, 16325N, 16326N, 16327N, 16328N, 16329N, 16330N, 16331N, 16332N, 16333N, 16334N, 16335N, 16336N, 16337N, 16338N, 16339N, 16340N, 16341N, 16342N, 16343N, 16344N, 16345N, 16346N, 16347N, 16348N, 16349N, 16350N, 16351N, 16352N, 16353N, 16354N, 16355N, 16356N, 16357N | |
| HQ439472 (Horse and Przewalski's horse) | Illumina/Solexa Genome Analyzer II | Phantom mutaions: 358d, 1387A, 2227+T, 5237+G, 5274+G, 15385+T. | G1a_HQ439472 | 158, 222, 356d, 382, 387, 416, 1387A, 2227+T, 2788, 2940, 3053, 3576, 4062, 4646, 4669, 4830, 5237+G, 5274+G, 5498, 5669, 5830, 5881, 5884, 6004, 6307, 6688, 6784, 7001, 8005, 8037, 9239, 9402, 9669, 9741A, 10214, 10376, 10471, 11165, 11240, 11543, 11552, 11842, 12767, 12860, 13049, 13223, 13333, 13502, 13812, 14350, 14626, 14651, 14734, 15297, 15385+T, 15492, 15539, 15582, 15599, 15632, 15647, 15663, 15717, 15867, 15971, 16028, 16110, 16126N, 16127N, 16128N, 16129N, 16130N, 16131N, 16132N, 16133N, 16134N, 16135N, 16136N, 16137N, 16138N, 16139N, 16140N, 16141N, 16142N, 16143N, 16144N, 16145N, 16146N, 16147N, 16148N, 16149N, 16150N, 16151N, 16152N, 16153N, 16154N, 16155N, 16156N, 16157N, 16158N, 16159N, 16160N, 16161N, 16162N, 16163N, 16164N, 16165N, 16166N, 16167N, 16168N, 16169N, 16170N, 16171N, 16172N, 16173N, 16174N, 16175N, 16176N, 16177N, 16178N, 16179N, 16180N, 16181N, 16182N, 16183N, 16184N, 16185N, 16186N, 16187N, 16188N, 16189N, 16190N, 16191N, 16192N, 16193N, 16194N, 16195N, 16196N, 16197N, 16198N, 16199N, 16200N, 16201N, 16202N, 16203N, 16204N, 16205N, 16206N, 16207N, 16208N, 16209N, 16210N, 16211N, 16212N, 16213N, 16214N, 16215N, 16216N, 16217N, 16218N, 16219N, 16220N, 16221N, 16222N, 16223N, 16224N, 16225N, 16226N, 16227N, 16228N, 16229N, 16230N, 16231N, 16232N, 16233N, 16234N, 16235N, 16236N, 16237N, 16238N, 16239N, 16240N, 16241N, 16242N, 16243N, 16244N, 16245N, 16246N, 16247N, 16248N, 16249N, 16250N, 16251N, 16252N, 16253N, 16254N, 16255N, 16256N, 16257N, 16258N, 16259N, 16260N, 16261N, 16262N, 16263N, 16264N, 16265N, 16266N, 16267N, 16268N, 16269N, 16270N, 16271N, 16272N, 16273N, 16274N, 16275N, 16276N, 16277N, 16278N, 16279N, 16280N, 16281N, 16282N, 16283N, 16284N, 16285N, 16286N, 16287N, 16288N, 16289N, 16290N, 16291N, 16292N, 16293N, 16294N, 16295N, 16296N, 16297N, 16298N, 16299N, 16300N, 16301N, 16302N, 16303N, 16304N, 16305N, 16306N, 16307N, 16308N, 16309N, 16310N, 16311N, 16312N, 16313N, 16314N, 16315N, 16316N, 16317N, 16318N, 16319N, 16320N, 16321N, 16322N, 16323N, 16324N, 16325N, 16326N, 16327N, 16328N, 16329N, 16330N, 16331N, 16332N, 16333N, 16334N, 16335N, 16336N, 16337N, 16338N, 16339N, 16340N, 16341N, 16342N, 16343N, 16344N, 16345N, 16346N, 16347N, 16348N, 16349N, 16350N, 16351N, 16352N, 16353N, 16354N, 16355N, 16356N, 16357N, 16368 | G1a_HQ439472:356 | G1a_HQ439472:356d, 2227+T, 5237+G, 5274+G, 15385+T, 16126N, 16127N, 16128N, 16129N, 16130N, 16131N, 16132N, 16133N, 16134N, 16135N, 16136N, 16137N, 16138N, 16139N, 16140N, 16141N, 16142N, 16143N, 16144N, 16145N, 16146N, 16147N, 16148N, 16149N, 16150N, 16151N, 16152N, 16153N, 16154N, 16155N, 16156N, 16157N, 16158N, 16159N, 16160N, 16161N, 16162N, 16163N, 16164N, 16165N, 16166N, 16167N, 16168N, 16169N, 16170N, 16171N, 16172N, 16173N, 16174N, 16175N, 16176N, 16177N, 16178N, 16179N, 16180N, 16181N, 16182N, 16183N, 16184N, 16185N, 16186N, 16187N, 16188N, 16189N, 16190N, 16191N, 16192N, 16193N, 16194N, 16195N, 16196N, 16197N, 16198N, 16199N, 16200N, 16201N, 16202N, 16203N, 16204N, 16205N, 16206N, 16207N, 16208N, 16209N, 16210N, 16211N, 16212N, 16213N, 16214N, 16215N, 16216N, 16217N, 16218N, 16219N, 16220N, 16221N, 16222N, 16223N, 16224N, 16225N, 16226N, 16227N, 16228N, 16229N, 16230N, 16231N, 16232N, 16233N, 16234N, 16235N, 16236N, 16237N, 16238N, 16239N, 16240N, 16241N, 16242N, 16243N, 16244N, 16245N, 16246N, 16247N, 16248N, 16249N, 16250N, 16251N, 16252N, 16253N, 16254N, 16255N, 16256N, 16257N, 16258N, 16259N, 16260N, 16261N, 16262N, 16263N, 16264N, 16265N, 16266N, 16267N, 16268N, 16269N, 16270N, 16271N, 16272N, 16273N, 16274N, 16275N, 16276N, 16277N, 16278N, 16279N, 16280N, 16281N, 16282N, 16283N, 16284N, 16285N, 16286N, 16287N, 16288N, 16289N, 16290N, 16291N, 16292N, 16293N, 16294N, 16295N, 16296N, 16297N, 16298N, 16299N, 16300N, 16301N, 16302N, 16303N, 16304N, 16305N, 16306N, 16307N, 16308N, 16309N, 16310N, 16311N, 16312N, 16313N, 16314N, 16315N, 16316N, 16317N, 16318N, 16319N, 16320N, 16321N, 16322N, 16323N, 16324N, 16325N, 16326N, 16327N, 16328N, 16329N, 16330N, 16331N, 16332N, 16333N, 16334N, 16335N, 16336N, 16337N, 16338N, 16339N, | Lippold,S., Matzke,N., Reissman,M., Burbano,H. and Hofreiter,M. BMC Evolutionary Biology. 328 (11), 1471-2148 (2011) |
| | | | | | | 16340N, 16341N, 16342N, 16343N, 16344N, 16345N, 16346N, 16347N, 16348N, 16349N, 16350N, 16351N, 16352N, 16353N, 16354N, 16355N, 16356N, 16357N | |
| HQ439473 (Horse and Przewalski's horse) | Illumina/Solexa Genome Analyzer II | Phantom mutaions: 358d, 1387A, 2226+C, 5239+A, 5277+A, 15381+A. | L3a2_HQ439473 | 77, 158, 356d, 961, 1375, 1387A, 2226+C, 2788, 2899, 3517, 3942, 4062, 4536, 4646, 4669, 5239+A, 5277+A, 5527, 5815, 5884, 6004, 6307, 6784, 7001, 7202T, 7516, 7666, 7900, 8005, 8058, 8301, 8319, 8358, 8565, 9239, 9951, 10110, 10214, 10292, 10376, 10421, 10613, 11240, 11288, 11543, 11693, 11842, 11879, 12119, 12200, 12767, 12896, 12950, 13049, 13333, 13520, 14803, 14995, 15313, 15381+A, 15491, 15492, 15493, 15531, 15582, 15599, 15600, 15601, 15646, 15717, 15768, 15867, 15868, 15953, 15971, 16065, 16100, 16126N, 16127N, 16128N, 16129N, 16130N, 16131N, 16132N, 16133N, 16134N, 16135N, 16136N, 16137N, 16138N, 16139N, 16140N, 16141N, 16142N, 16143N, 16144N, 16145N, 16146N, 16147N, 16148N, 16149N, 16150N, 16151N, 16152N, 16153N, 16154N, 16155N, 16156N, 16157N, 16158N, 16159N, 16160N, 16161N, 16162N, 16163N, 16164N, 16165N, 16166N, 16167N, 16168N, 16169N, 16170N, 16171N, 16172N, 16173N, 16174N, 16175N, 16176N, 16177N, 16178N, 16179N, 16180N, 16181N, 16182N, 16183N, 16184N, 16185N, 16186N, 16187N, 16188N, 16189N, 16190N, 16191N, 16192N, 16193N, 16194N, 16195N, 16196N, 16197N, 16198N, 16199N, 16200N, 16201N, 16202N, 16203N, 16204N, 16205N, 16206N, 16207N, 16208N, 16209N, 16210N, 16211N, 16212N, 16213N, 16214N, 16215N, 16216N, 16217N, 16218N, 16219N, 16220N, 16221N, 16222N, 16223N, 16224N, 16225N, 16226N, 16227N, 16228N, 16229N, 16230N, 16231N, 16232N, 16233N, 16234N, 16235N, 16236N, 16237N, 16238N, 16239N, 16240N, 16241N, 16242N, 16243N, 16244N, 16245N, 16246N, 16247N, 16248N, 16249N, 16250N, 16251N, 16252N, 16253N, 16254N, 16255N, 16256N, 16257N, 16258N, 16259N, 16260N, 16261N, 16262N, 16263N, 16264N, 16265N, 16266N, 16267N, 16268N, 16269N, 16270N, 16271N, 16272N, 16273N, 16274N, 16275N, 16276N, 16277N, 16278N, 16279N, 16280N, 16281N, 16282N, 16283N, 16284N, 16285N, 16286N, 16287N, 16288N, 16289N, 16290N, 16291N, 16292N, 16293N, 16294N, 16295N, 16296N, 16297N, 16298N, 16299N, 16300N, 16301N, 16302N, 16303N, 16304N, 16305N, 16306N, 16307N, 16308N, 16309N, 16310N, 16311N, 16312N, 16313N, 16314N, 16315N, 16316N, 16317N, 16318N, 16319N, 16320N, 16321N, 16322N, 16323N, 16324N, 16325N, 16326N, 16327N, 16328N, 16329N, 16330N, 16331N, 16332N, 16333N, 16334N, 16335N, 16336N, 16337N, 16338N, 16339N, 16340N, 16341N, 16342N, 16343N, 16344N, 16345N, 16346N, 16347N, 16348N, 16349N, 16350N, 16351N, 16352N, 16353N, 16354N, 16355N, 16356N, 16357N, 16368N, 16554, 16626, 16651-16657d | L3a2_HQ439473:356 | L3a2_HQ439473:356d, 2226+C, 5239+A, 5277+A, 15381+A, 16126N, 16127N, 16128N, 16129N, 16130N, 16131N, 16132N, 16133N, 16134N, 16135N, 16136N, 16137N, 16138N, 16139N, 16140N, 16141N, 16142N, 16143N, 16144N, 16145N, 16146N, 16147N, 16148N, 16149N, 16150N, 16151N, 16152N, 16153N, 16154N, 16155N, 16156N, 16157N, 16158N, 16159N, 16160N, 16161N, 16162N, 16163N, 16164N, 16165N, 16166N, 16167N, 16168N, 16169N, 16170N, 16171N, 16172N, 16173N, 16174N, 16175N, 16176N, 16177N, 16178N, 16179N, 16180N, 16181N, 16182N, 16183N, 16184N, 16185N, 16186N, 16187N, 16188N, 16189N, 16190N, 16191N, 16192N, 16193N, 16194N, 16195N, 16196N, 16197N, 16198N, 16199N, 16200N, 16201N, 16202N, 16203N, 16204N, 16205N, 16206N, 16207N, 16208N, 16209N, 16210N, 16211N, 16212N, 16213N, 16214N, 16215N, 16216N, 16217N, 16218N, 16219N, 16220N, 16221N, 16222N, 16223N, 16224N, 16225N, 16226N, 16227N, 16228N, 16229N, 16230N, 16231N, 16232N, 16233N, 16234N, 16235N, 16236N, 16237N, 16238N, 16239N, 16240N, 16241N, 16242N, 16243N, 16244N, 16245N, 16246N, 16247N, 16248N, 16249N, 16250N, 16251N, 16252N, 16253N, 16254N, 16255N, 16256N, 16257N, 16258N, 16259N, 16260N, 16261N, 16262N, 16263N, 16264N, 16265N, 16266N, 16267N, 16268N, 16269N, 16270N, 16271N, 16272N, 16273N, 16274N, 16275N, 16276N, 16277N, 16278N, 16279N, 16280N, 16281N, 16282N, 16283N, 16284N, 16285N, 16286N, 16287N, 16288N, 16289N, 16290N, 16291N, 16292N, 16293N, 16294N, 16295N, 16296N, 16297N, 16298N, 16299N, 16300N, 16301N, 16302N, 16303N, 16304N, 16305N, 16306N, 16307N, 16308N, 16309N, 16310N, 16311N, 16312N, 16313N, 16314N, 16315N, 16316N, 16317N, 16318N, 16319N, 16320N, 16321N, 16322N, 16323N, 16324N, 16325N, 16326N, 16327N, 16328N, 16329N, 16330N, 16331N, 16332N, 16333N, 16334N, 16335N, 16336N, 16337N, 16338N, 16339N, 16340N, 16341N, 16342N, 16343N, 16344N, 16345N, 16346N, 16347N, 16348N, 16349N, 16350N, 16351N, 16352N, 16353N, 16354N, 16355N, 16356N, 16357N, 16368N, 16651-16657d | Lippold,S., Matzke,N., Reissman,M., Burbano,H. and Hofreiter,M. BMC Evolutionary Biology. 328 (11), 1471-2148 (2011) |
| HQ439474 (Horse and Przewalski's horse) | Illumina/Solexa Genome Analyzer II | Phantom mutaions: 358d, 2226+C, 5239+A, 5277+A, 15385+T. | B1c1_HQ439474 | 158, 356d, 859, 1896, 2226+C, 4062, 4743, 5239+A, 5277+A, 5929, 6784, 7627, 9961, 10764, 11240, 11375, 14689, 15385+T, 15492, 15647, 15663, 15717, 15823, 16052, 16108, 16110, 16126N, 16127N, 16128N, 16129N, 16130N, 16131N, 16132N, 16133N, 16134N, 16135N, 16136N, 16137N, 16138N, 16139N, 16140N, 16141N, 16142N, 16143N, 16144N, 16145N, 16146N, 16147N, 16148N, 16149N, 16150N, 16151N, 16152N, 16153N, 16154N, 16155N, 16156N, 16157N, 16158N, 16159N, 16160N, 16161N, 16162N, 16163N, 16164N, 16165N, 16166N, 16167N, 16168N, 16169N, 16170N, 16171N, 16172N, 16173N, 16174N, 16175N, 16176N, 16177N, 16178N, 16179N, 16180N, 16181N, 16182N, 16183N, 16184N, 16185N, 16186N, 16187N, 16188N, 16189N, 16190N, 16191N, 16192N, 16193N, 16194N, 16195N, 16196N, 16197N, 16198N, 16199N, 16200N, 16201N, 16202N, 16203N, 16204N, 16205N, 16206N, 16207N, 16208N, 16209N, 16210N, 16211N, 16212N, 16213N, 16214N, 16215N, 16216N, 16217N, 16218N, 16219N, 16220N, 16221N, 16222N, 16223N, 16224N, 16225N, 16226N, 16227N, 16228N, 16229N, 16230N, 16231N, 16232N, 16233N, 16234N, 16235N, 16236N, 16237N, 16238N, 16239N, | B1c1_HQ439474:356 | B1c1_HQ439474:356d, 2226+C, 5239+A, 5277+A, 15385+T, 16126N, 16127N, 16128N, 16129N, 16130N, 16131N, 16132N, 16133N, 16134N, 16135N, 16136N, 16137N, 16138N, 16139N, 16140N, 16141N, 16142N, 16143N, 16144N, 16145N, 16146N, 16147N, 16148N, 16149N, 16150N, 16151N, 16152N, 16153N, 16154N, 16155N, 16156N, 16157N, 16158N, 16159N, 16160N, 16161N, 16162N, 16163N, 16164N, 16165N, 16166N, 16167N, 16168N, 16169N, 16170N, 16171N, 16172N, 16173N, 16174N, 16175N, 16176N, 16177N, 16178N, 16179N, 16180N, 16181N, 16182N, 16183N, 16184N, 16185N, 16186N, 16187N, 16188N, 16189N, 16190N, 16191N, 16192N, 16193N, 16194N, 16195N, 16196N, 16197N, 16198N, 16199N, 16200N, 16201N, 16202N, 16203N, 16204N, 16205N, 16206N, 16207N, 16208N, 16209N, 16210N, 16211N, 16212N, 16213N, 16214N, 16215N, 16216N, 16217N, | Lippold,S., Matzke,N., Reissman,M., Burbano,H. and Hofreiter,M. BMC Evolutionary Biology. 328 (11), 1471-2148 (2011) |
| | | | | 16240N, 16241N, 16242N, 16243N, 16244N, 16245N, 16246N, 16247N, 16248N, 16249N, 16250N, 16251N, 16252N, 16253N, 16254N, 16255N, 16256N, 16257N, 16258N, 16259N, 16260N, 16261N, 16262N, 16263N, 16264N, 16265N, 16266N, 16267N, 16268N, 16269N, 16270N, 16271N, 16272N, 16273N, 16274N, 16275N, 16276N, 16277N, 16278N, 16279N, 16280N, 16281N, 16282N, 16283N, 16284N, 16285N, 16286N, 16287N, 16288N, 16289N, 16290N, 16291N, 16292N, 16293N, 16294N, 16295N, 16296N, 16297N, 16298N, 16299N, 16300N, 16301N, 16302N, 16303N, 16304N, 16305N, 16306N, 16307N, 16308N, 16309N, 16310N, 16311N, 16312N, 16313N, 16314N, 16315N, 16316N, 16317N, 16318N, 16319N, 16320N, 16321N, 16322N, 16323N, 16324N, 16325N, 16326N, 16327N, 16328N, 16329N, 16330N, 16331N, 16332N, 16333N, 16334N, 16335N, 16336N, 16337N, 16338N, 16339N, 16340N, 16341N, 16342N, 16343N, 16344N, 16345N, 16346N, 16347N, 16348N, 16349N, 16350N, 16351N, 16352N, 16353N, 16354N, 16355N, 16356N, 16357N, 16368, 16398A | | 16218N, 16219N, 16220N, 16221N, 16222N, 16223N, 16224N, 16225N, 16226N, 16227N, 16228N, 16229N, 16230N, 16231N, 16232N, 16233N, 16234N, 16235N, 16236N, 16237N, 16238N, 16239N, 16240N, 16241N, 16242N, 16243N, 16244N, 16245N, 16246N, 16247N, 16248N, 16249N, 16250N, 16251N, 16252N, 16253N, 16254N, 16255N, 16256N, 16257N, 16258N, 16259N, 16260N, 16261N, 16262N, 16263N, 16264N, 16265N, 16266N, 16267N, 16268N, 16269N, 16270N, 16271N, 16272N, 16273N, 16274N, 16275N, 16276N, 16277N, 16278N, 16279N, 16280N, 16281N, 16282N, 16283N, 16284N, 16285N, 16286N, 16287N, 16288N, 16289N, 16290N, 16291N, 16292N, 16293N, 16294N, 16295N, 16296N, 16297N, 16298N, 16299N, 16300N, 16301N, 16302N, 16303N, 16304N, 16305N, 16306N, 16307N, 16308N, 16309N, 16310N, 16311N, 16312N, 16313N, 16314N, 16315N, 16316N, 16317N, 16318N, 16319N, 16320N, 16321N, 16322N, 16323N, 16324N, 16325N, 16326N, 16327N, 16328N, 16329N, 16330N, 16331N, 16332N, 16333N, 16334N, 16335N, 16336N, 16337N, 16338N, 16339N, 16340N, 16341N, 16342N, 16343N, 16344N, 16345N, 16346N, 16347N, 16348N, 16349N, 16350N, 16351N, 16352N, 16353N, 16354N, 16355N, 16356N, 16357N | |
| HQ439475 (Horse and Przewalski's horse) | Illumina/Solexa Genome Analyzer II | Phantom mutaions: 358d, 2226+G, 5237+G, 5274+G, 15385+T. | A1b_HQ439475 | 358d, 2226+C, 4520, 5237+G, 5274+G, 6784, 15385+T, 15492, 15823, 16110, 16126N, 16127N, 16128N, 16129N, 16130N, 16131N, 16132N, 16133N, 16134N, 16135N, 16136N, 16137N, 16138N, 16139N, 16140N, 16141N, 16142N, 16143N, 16144N, 16145N, 16146N, 16147N, 16148N, 16149N, 16150N, 16151N, 16152N, 16153N, 16154N, 16155N, 16156N, 16157N, 16158N, 16159N, 16160N, 16161N, 16162N, 16163N, 16164N, 16165N, 16166N, 16167N, 16168N, 16169N, 16170N, 16171N, 16172N, 16173N, 16174N, 16175N, 16176N, 16177N, 16178N, 16179N, 16180N, 16181N, 16182N, 16183N, 16184N, 16185N, 16186N, 16187N, 16188N, 16189N, 16190N, 16191N, 16192N, 16193N, 16194N, 16195N, 16196N, 16197N, 16198N, 16199N, 16200N, 16201N, 16202N, 16203N, 16204N, 16205N, 16206N, 16207N, 16208N, 16209N, 16210N, 16211N, 16212N, 16213N, 16214N, 16215N, 16216N, 16217N, 16218N, 16219N, 16220N, 16221N, 16222N, 16223N, 16224N, 16225N, 16226N, 16227N, 16228N, 16229N, 16230N, 16231N, 16232N, 16233N, 16234N, 16235N, 16236N, 16237N, 16238N, 16239N, 16240N, 16241N, 16242N, 16243N, 16244N, 16245N, 16246N, 16247N, 16248N, 16249N, 16250N, 16251N, 16252N, 16253N, 16254N, 16255N, 16256N, 16257N, 16258N, 16259N, 16260N, 16261N, 16262N, 16263N, 16264N, 16265N, 16266N, 16267N, 16268N, 16269N, 16270N, 16271N, 16272N, 16273N, 16274N, 16275N, 16276N, 16277N, 16278N, 16279N, 16280N, 16281N, 16282N, 16283N, 16284N, 16285N, 16286N, 16287N, 16288N, 16289N, 16290N, 16291N, 16292N, 16293N, 16294N, 16295N, 16296N, 16297N, 16298N, 16299N, 16300N, 16301N, 16302N, 16303N, 16304N, 16305N, 16306N, 16307N, 16308N, 16309N, 16310N, 16311N, 16312N, 16313N, 16314N, 16315N, 16316N, 16317N, 16318N, 16319N, 16320N, 16321N, 16322N, 16323N, 16324N, 16325N, 16326N, 16327N, 16328N, 16329N, 16330N, 16331N, 16332N, 16333N, 16334N, 16335N, 16336N, 16337N, 16338N, 16339N, 16340N, 16341N, 16342N, 16343N, 16344N, 16345N, 16346N, 16347N, 16348N, 16349N, 16350N, 16351N, 16352N, 16353N, 16354N, 16355N, 16356N, 16357N | | A1b_HQ439475:358d, 2226+C, 5237+G, 5274+G, 15385+T, 16126N, 16127N, 16128N, 16129N, 16130N, 16131N, 16132N, 16133N, 16134N, 16135N, 16136N, 16137N, 16138N, 16139N, 16140N, 16141N, 16142N, 16143N, 16144N, 16145N, 16146N, 16147N, 16148N, 16149N, 16150N, 16151N, 16152N, 16153N, 16154N, 16155N, 16156N, 16157N, 16158N, 16159N, 16160N, 16161N, 16162N, 16163N, 16164N, 16165N, 16166N, 16167N, 16168N, 16169N, 16170N, 16171N, 16172N, 16173N, 16174N, 16175N, 16176N, 16177N, 16178N, 16179N, 16180N, 16181N, 16182N, 16183N, 16184N, 16185N, 16186N, 16187N, 16188N, 16189N, 16190N, 16191N, 16192N, 16193N, 16194N, 16195N, 16196N, 16197N, 16198N, 16199N, 16200N, 16201N, 16202N, 16203N, 16204N, 16205N, 16206N, 16207N, 16208N, 16209N, 16210N, 16211N, 16212N, 16213N, 16214N, 16215N, 16216N, 16217N, 16218N, 16219N, 16220N, 16221N, 16222N, 16223N, 16224N, 16225N, 16226N, 16227N, 16228N, 16229N, 16230N, 16231N, 16232N, 16233N, 16234N, 16235N, 16236N, 16237N, 16238N, 16239N, 16240N, 16241N, 16242N, 16243N, 16244N, 16245N, 16246N, 16247N, 16248N, 16249N, 16250N, 16251N, 16252N, 16253N, 16254N, 16255N, 16256N, 16257N, 16258N, 16259N, 16260N, 16261N, 16262N, 16263N, 16264N, 16265N, 16266N, 16267N, 16268N, 16269N, 16270N, 16271N, 16272N, 16273N, 16274N, 16275N, 16276N, 16277N, 16278N, 16279N, 16280N, 16281N, 16282N, 16283N, 16284N, 16285N, 16286N, 16287N, 16288N, 16289N, 16290N, 16291N, 16292N, 16293N, 16294N, 16295N, 16296N, 16297N, 16298N, 16299N, 16300N, 16301N, 16302N, 16303N, 16304N, 16305N, 16306N, 16307N, 16308N, 16309N, 16310N, 16311N, 16312N, 16313N, 16314N, 16315N, 16316N, 16317N, 16318N, 16319N, 16320N, 16321N, 16322N, 16323N, 16324N, 16325N, 16326N, 16327N, 16328N, 16329N, 16330N, 16331N, 16332N, 16333N, 16334N, 16335N, 16336N, 16337N, 16338N, 16339N, | Lippold,S., Matzke,N., Reissman,M., Burbano,H. and Hofreiter,M. BMC Evolutionary Biology. 328 (11), 1471-2148 (2011) |
| | | | | | | 16340N, 16341N, 16342N, 16343N, 16344N, 16345N, 16346N, 16347N, 16348N, 16349N, 16350N, 16351N, 16352N, 16353N, 16354N, 16355N, 16356N, 16357N | |
| HQ439476 (Horse and Przewalski's horse) | Illumina/Solexa Genome Analyzer II | Phantom mutaions: 358d, 1387G, 2227+T, 5239+A, 5274+G, 15385+T. | I1_HQ439476 | 158, 356d, 1387G, 1587, 1791, 2227+T, 2614, 2770, 2788, 4062, 4063, 4392, 4646, 4669, 4830, 5061, 5210, 5239+A, 5274+G, 5884, 6004, 6175, 6247, 6307, 6784, 7001, 8005, 8358, 8379, 8792, 9239, 9948, 10083, 10214, 10238, 10376, 10699, 10793, 11240, 11543, 11827, 11842, 12443, 12683, 12767, 13049, 13333, 13466, 13502, 13761, 14038, 14554, 15385+T, 15492, 15535, 15582, 15594, 15599, 15647, 15706, 15717, 15768, 15823, 16110, 16119, 16126N, 16127N, 16128N, 16129N, 16130N, 16131N, 16132N, 16133N, 16134N, 16135N, 16136N, 16137N, 16138N, 16139N, 16140N, 16141N, 16142N, 16143N, 16144N, 16145N, 16146N, 16147N, 16148N, 16149N, 16150N, 16151N, 16152N, 16153N, 16154N, 16155N, 16156N, 16157N, 16158N, 16159N, 16160N, 16161N, 16162N, 16163N, 16164N, 16165N, 16166N, 16167N, 16168N, 16169N, 16170N, 16171N, 16172N, 16173N, 16174N, 16175N, 16176N, 16177N, 16178N, 16179N, 16180N, 16181N, 16182N, 16183N, 16184N, 16185N, 16186N, 16187N, 16188N, 16189N, 16190N, 16191N, 16192N, 16193N, 16194N, 16195N, 16196N, 16197N, 16198N, 16199N, 16200N, 16201N, 16202N, 16203N, 16204N, 16205N, 16206N, 16207N, 16208N, 16209N, 16210N, 16211N, 16212N, 16213N, 16214N, 16215N, 16216N, 16217N, 16218N, 16219N, 16220N, 16221N, 16222N, 16223N, 16224N, 16225N, 16226N, 16227N, 16228N, 16229N, 16230N, 16231N, 16232N, 16233N, 16234N, 16235N, 16236N, 16237N, 16238N, 16239N, 16240N, 16241N, 16242N, 16243N, 16244N, 16245N, 16246N, 16247N, 16248N, 16249N, 16250N, 16251N, 16252N, 16253N, 16254N, 16255N, 16256N, 16257N, 16258N, 16259N, 16260N, 16261N, 16262N, 16263N, 16264N, 16265N, 16266N, 16267N, 16268N, 16269N, 16270N, 16271N, 16272N, 16273N, 16274N, 16275N, 16276N, 16277N, 16278N, 16279N, 16280N, 16281N, 16282N, 16283N, 16284N, 16285N, 16286N, 16287N, 16288N, 16289N, 16290N, 16291N, 16292N, 16293N, 16294N, 16295N, 16296N, 16297N, 16298N, 16299N, 16300N, 16301N, 16302N, 16303N, 16304N, 16305N, 16306N, 16307N, 16308N, 16309N, 16310N, 16311N, 16312N, 16313N, 16314N, 16315N, 16316N, 16317N, 16318N, 16319N, 16320N, 16321N, 16322N, 16323N, 16324N, 16325N, 16326N, 16327N, 16328N, 16329N, 16330N, 16331N, 16332N, 16333N, 16334N, 16335N, 16336N, 16337N, 16338N, 16339N, 16340N, 16341N, 16342N, 16343N, 16344N, 16345N, 16346N, 16347N, 16348N, 16349N, 16350N, 16351N, 16352N, 16353N, 16354N, 16355N, 16356N, 16357N, 16368 | I1_HQ439476:356 | I1_HQ439476:356d, 2227+T, 5239+A, 5274+G, 15385+T, 16126N, 16127N, 16128N, 16129N, 16130N, 16131N, 16132N, 16133N, 16134N, 16135N, 16136N, 16137N, 16138N, 16139N, 16140N, 16141N, 16142N, 16143N, 16144N, 16145N, 16146N, 16147N, 16148N, 16149N, 16150N, 16151N, 16152N, 16153N, 16154N, 16155N, 16156N, 16157N, 16158N, 16159N, 16160N, 16161N, 16162N, 16163N, 16164N, 16165N, 16166N, 16167N, 16168N, 16169N, 16170N, 16171N, 16172N, 16173N, 16174N, 16175N, 16176N, 16177N, 16178N, 16179N, 16180N, 16181N, 16182N, 16183N, 16184N, 16185N, 16186N, 16187N, 16188N, 16189N, 16190N, 16191N, 16192N, 16193N, 16194N, 16195N, 16196N, 16197N, 16198N, 16199N, 16200N, 16201N, 16202N, 16203N, 16204N, 16205N, 16206N, 16207N, 16208N, 16209N, 16210N, 16211N, 16212N, 16213N, 16214N, 16215N, 16216N, 16217N, 16218N, 16219N, 16220N, 16221N, 16222N, 16223N, 16224N, 16225N, 16226N, 16227N, 16228N, 16229N, 16230N, 16231N, 16232N, 16233N, 16234N, 16235N, 16236N, 16237N, 16238N, 16239N, 16240N, 16241N, 16242N, 16243N, 16244N, 16245N, 16246N, 16247N, 16248N, 16249N, 16250N, 16251N, 16252N, 16253N, 16254N, 16255N, 16256N, 16257N, 16258N, 16259N, 16260N, 16261N, 16262N, 16263N, 16264N, 16265N, 16266N, 16267N, 16268N, 16269N, 16270N, 16271N, 16272N, 16273N, 16274N, 16275N, 16276N, 16277N, 16278N, 16279N, 16280N, 16281N, 16282N, 16283N, 16284N, 16285N, 16286N, 16287N, 16288N, 16289N, 16290N, 16291N, 16292N, 16293N, 16294N, 16295N, 16296N, 16297N, 16298N, 16299N, 16300N, 16301N, 16302N, 16303N, 16304N, 16305N, 16306N, 16307N, 16308N, 16309N, 16310N, 16311N, 16312N, 16313N, 16314N, 16315N, 16316N, 16317N, 16318N, 16319N, 16320N, 16321N, 16322N, 16323N, 16324N, 16325N, 16326N, 16327N, 16328N, 16329N, 16330N, 16331N, 16332N, 16333N, 16334N, 16335N, 16336N, 16337N, 16338N, 16339N, 16340N, 16341N, 16342N, 16343N, 16344N, 16345N, 16346N, 16347N, 16348N, 16349N, 16350N, 16351N, 16352N, 16353N, 16354N, 16355N, 16356N, 16357N | Lippold,S., Matzke,N., Reissman,M., Burbano,H. and Hofreiter,M. BMC Evolutionary Biology. 328 (11), 1471-2148 (2011) |
| HQ439477 (Horse and Przewalski's horse) | Illumina/Solexa Genome Analyzer II | Phantom mutaions: 358d, 1387G, 2227+T, 5239+A, 5274+G, 15385+T. | Q1b1a | 158, 302, 341, 356d, 739, 860, 961, 1387G, 2227+T, 2788, 3070, 3259T, 3271, 3616, 3800, 3942, 4062, 4201, 4536, 4605, 4646, 4669, 5239+A, 5274+G, 5527, 5827, 5884, 6004, 6307, 6529, 6688, 6712, 6784, 7001, 7243, 7294, 7612T, 7666, 7898, 7900, 8005, 8076, 8150, 8238, 8358, 8361, 8556T, 8565, 8855, 9086, 9203, 9239, 9775, 10110, 10173, 10214, 10292, 10376, 10448, 10859, 11240, 11378, 11394, 11424, 11492, 11543, 11842, 11879, 11966, 12167, 12230, 12332, 12404, 12767, 13049, 13307, 13333, 13463, 13629, 13920, 13933, 14626, 14803, 15202, 15342, 15385+T, 15492, 15599, 15601, 15700, 15717, 15723, 15737, 15768, 15808, 15953, 16035, 16054, 16118, 16126N, 16127N, 16128N, 16129N, 16130N, 16131N, 16132N, 16133N, 16134N, 16135N, 16136N, 16137N, 16138N, 16139N, 16140N, 16141N, 16142N, 16143N, 16144N, 16145N, 16146N, 16147N, 16148N, 16149N, 16150N, 16151N, 16152N, 16153N, 16154N, 16155N, 16156N, 16157N, 16158N, 16159N, 16160N, 16161N, 16162N, 16163N, 16164N, 16165N, 16166N, 16167N, 16168N, 16169N, 16170N, 16171N, 16172N, 16173N, 16174N, 16175N, 16176N, 16177N, 16178N, 16179N, 16180N, 16181N, 16182N, 16183N, 16184N, | Q1b1a:356 | Q1b1a:356d, 2227+T, 5239+A, 5274+G, 15385+T, 16126N, 16127N, 16128N, 16129N, 16130N, 16131N, 16132N, 16133N, 16134N, 16135N, 16136N, 16137N, 16138N, 16139N, 16140N, 16141N, 16142N, 16143N, 16144N, 16145N, 16146N, 16147N, 16148N, 16149N, 16150N, 16151N, 16152N, 16153N, 16154N, 16155N, 16156N, 16157N, 16158N, 16159N, 16160N, 16161N, 16162N, 16163N, 16164N, 16165N, 16166N, 16167N, 16168N, 16169N, 16170N, 16171N, 16172N, 16173N, 16174N, 16175N, 16176N, 16177N, 16178N, 16179N, 16180N, 16181N, 16182N, 16183N, 16184N, 16185N, 16186N, 16187N, 16188N, 16189N, 16190N, 16191N, 16192N, 16193N, 16194N, 16195N, 16196N, 16197N, 16198N, 16199N, 16200N, 16201N, 16202N, 16203N, 16204N, 16205N, 16206N, 16207N, 16208N, 16209N, 16210N, 16211N, 16212N, 16213N, 16214N, 16215N, 16216N, 16217N, 16218N, 16219N, | Lippold,S., Matzke,N., Reissman,M., Burbano,H. and Hofreiter,M. BMC Evolutionary Biology. 328 (11), 1471-2148 (2011) |
| | | | | 16185N, 16186N, 16187N, 16188N, 16189N, 16190N, 16191N, 16192N, 16193N, 16194N, 16195N, 16196N, 16197N, 16198N, 16199N, 16200N, 16201N, 16202N, 16203N, 16204N, 16205N, 16206N, 16207N, 16208N, 16209N, 16210N, 16211N, 16212N, 16213N, 16214N, 16215N, 16216N, 16217N, 16218N, 16219N, 16220N, 16221N, 16222N, 16223N, 16224N, 16225N, 16226N, 16227N, 16228N, 16229N, 16230N, 16231N, 16232N, 16233N, 16234N, 16235N, 16236N, 16237N, 16238N, 16239N, 16240N, 16241N, 16242N, 16243N, 16244N, 16245N, 16246N, 16247N, 16248N, 16249N, 16250N, 16251N, 16252N, 16253N, 16254N, 16255N, 16256N, 16257N, 16258N, 16259N, 16260N, 16261N, 16262N, 16263N, 16264N, 16265N, 16266N, 16267N, 16268N, 16269N, 16270N, 16271N, 16272N, 16273N, 16274N, 16275N, 16276N, 16277N, 16278N, 16279N, 16280N, 16281N, 16282N, 16283N, 16284N, 16285N, 16286N, 16287N, 16288N, 16289N, 16290N, 16291N, 16292N, 16293N, 16294N, 16295N, 16296N, 16297N, 16298N, 16299N, 16300N, 16301N, 16302N, 16303N, 16304N, 16305N, 16306N, 16307N, 16308N, 16309N, 16310N, 16311N, 16312N, 16313N, 16314N, 16315N, 16316N, 16317N, 16318N, 16319N, 16320N, 16321N, 16322N, 16323N, 16324N, 16325N, 16326N, 16327N, 16328N, 16329N, 16330N, 16331N, 16332N, 16333N, 16334N, 16335N, 16336N, 16337N, 16338N, 16339N, 16340N, 16341N, 16342N, 16343N, 16344N, 16345N, 16346N, 16347N, 16348N, 16349N, 16350N, 16351N, 16352N, 16353N, 16354N, 16355N, 16356N, 16357N, 16368, 16540A, 16626 | | 16220N, 16221N, 16222N, 16223N, 16224N, 16225N, 16226N, 16227N, 16228N, 16229N, 16230N, 16231N, 16232N, 16233N, 16234N, 16235N, 16236N, 16237N, 16238N, 16239N, 16240N, 16241N, 16242N, 16243N, 16244N, 16245N, 16246N, 16247N, 16248N, 16249N, 16250N, 16251N, 16252N, 16253N, 16254N, 16255N, 16256N, 16257N, 16258N, 16259N, 16260N, 16261N, 16262N, 16263N, 16264N, 16265N, 16266N, 16267N, 16268N, 16269N, 16270N, 16271N, 16272N, 16273N, 16274N, 16275N, 16276N, 16277N, 16278N, 16279N, 16280N, 16281N, 16282N, 16283N, 16284N, 16285N, 16286N, 16287N, 16288N, 16289N, 16290N, 16291N, 16292N, 16293N, 16294N, 16295N, 16296N, 16297N, 16298N, 16299N, 16300N, 16301N, 16302N, 16303N, 16304N, 16305N, 16306N, 16307N, 16308N, 16309N, 16310N, 16311N, 16312N, 16313N, 16314N, 16315N, 16316N, 16317N, 16318N, 16319N, 16320N, 16321N, 16322N, 16323N, 16324N, 16325N, 16326N, 16327N, 16328N, 16329N, 16330N, 16331N, 16332N, 16333N, 16334N, 16335N, 16336N, 16337N, 16338N, 16339N, 16340N, 16341N, 16342N, 16343N, 16344N, 16345N, 16346N, 16347N, 16348N, 16349N, 16350N, 16351N, 16352N, 16353N, 16354N, 16355N, 16356N, 16357N | |
| HQ439478 (Horse and Przewalski's horse) | Illumina/Solexa Genome Analyzer II | Phantom mutaions: 358d, 1387G, 2227+T, 5236+A, 5277+A, 15385+T. | L4a_HQ439478 | 158, 356d, 961, 1375, 1387G, 1459, 2227+T, 2788, 2899, 3517, 3942, 4062, 4536, 4646, 4669, 5236+A, 5277+A, 5527, 5815, 5884, 6004, 6307, 6784, 7001, 7516, 7666, 7900, 8005, 8058, 8301, 8307, 8319, 8358, 8565, 9239, 9951, 10110, 10214, 10292, 10376, 10421, 10613, 11240, 11543, 11693, 11842, 11879, 12119, 12200, 12767, 12896, 12950, 13049, 13333, 13520, 14659, 14803, 14995, 15313, 15385+T, 15491, 15492, 15493, 15531, 15582, 15599, 15600, 15601, 15646, 15717, 15768, 15867, 15868, 15953, 15971, 16065, 16100, 16110, 16126N, 16127N, 16128N, 16129N, 16130N, 16131N, 16132N, 16133N, 16134N, 16135N, 16136N, 16137N, 16138N, 16139N, 16140N, 16141N, 16142N, 16143N, 16144N, 16145N, 16146N, 16147N, 16148N, 16149N, 16150N, 16151N, 16152N, 16153N, 16154N, 16155N, 16156N, 16157N, 16158N, 16159N, 16160N, 16161N, 16162N, 16163N, 16164N, 16165N, 16166N, 16167N, 16168N, 16169N, 16170N, 16171N, 16172N, 16173N, 16174N, 16175N, 16176N, 16177N, 16178N, 16179N, 16180N, 16181N, 16182N, 16183N, 16184N, 16185N, 16186N, 16187N, 16188N, 16189N, 16190N, 16191N, 16192N, 16193N, 16194N, 16195N, 16196N, 16197N, 16198N, 16199N, 16200N, 16201N, 16202N, 16203N, 16204N, 16205N, 16206N, 16207N, 16208N, 16209N, 16210N, 16211N, 16212N, 16213N, 16214N, 16215N, 16216N, 16217N, 16218N, 16219N, 16220N, 16221N, 16222N, 16223N, 16224N, 16225N, 16226N, 16227N, 16228N, 16229N, 16230N, 16231N, 16232N, 16233N, 16234N, 16235N, 16236N, 16237N, 16238N, 16239N, 16240N, 16241N, 16242N, 16243N, 16244N, 16245N, 16246N, 16247N, 16248N, 16249N, 16250N, 16251N, 16252N, 16253N, 16254N, 16255N, 16256N, 16257N, 16258N, 16259N, 16260N, 16261N, 16262N, 16263N, 16264N, 16265N, 16266N, 16267N, 16268N, 16269N, 16270N, 16271N, 16272N, 16273N, 16274N, 16275N, 16276N, 16277N, 16278N, 16279N, 16280N, 16281N, 16282N, 16283N, 16284N, 16285N, 16286N, 16287N, 16288N, 16289N, 16290N, 16291N, 16292N, 16293N, 16294N, 16295N, 16296N, 16297N, 16298N, 16299N, 16300N, 16301N, 16302N, 16303N, 16304N, 16305N, 16306N, 16307N, 16308N, 16309N, 16310N, 16311N, 16312N, 16313N, 16314N, 16315N, 16316N, 16317N, 16318N, 16319N, 16320N, 16321N, 16322N, 16323N, 16324N, 16325N, 16326N, 16327N, 16328N, 16329N, 16330N, 16331N, 16332N, 16333N, 16334N, 16335N, 16336N, 16337N, 16338N, 16339N, 16340N, 16341N, 16342N, 16343N, 16344N, 16345N, 16346N, 16347N, 16348N, 16349N, 16350N, 16351N, 16352N, 16353N, 16354N, 16355N, 16356N, 16357N, 16368, 16398A, 16626 | L4a_HQ439478:356 | L4a_HQ439478:356d, 2227+T, 5236+A, 5277+A, 15385+T, 16126N, 16127N, 16128N, 16129N, 16130N, 16131N, 16132N, 16133N, 16134N, 16135N, 16136N, 16137N, 16138N, 16139N, 16140N, 16141N, 16142N, 16143N, 16144N, 16145N, 16146N, 16147N, 16148N, 16149N, 16150N, 16151N, 16152N, 16153N, 16154N, 16155N, 16156N, 16157N, 16158N, 16159N, 16160N, 16161N, 16162N, 16163N, 16164N, 16165N, 16166N, 16167N, 16168N, 16169N, 16170N, 16171N, 16172N, 16173N, 16174N, 16175N, 16176N, 16177N, 16178N, 16179N, 16180N, 16181N, 16182N, 16183N, 16184N, 16185N, 16186N, 16187N, 16188N, 16189N, 16190N, 16191N, 16192N, 16193N, 16194N, 16195N, 16196N, 16197N, 16198N, 16199N, 16200N, 16201N, 16202N, 16203N, 16204N, 16205N, 16206N, 16207N, 16208N, 16209N, 16210N, 16211N, 16212N, 16213N, 16214N, 16215N, 16216N, 16217N, 16218N, 16219N, 16220N, 16221N, 16222N, 16223N, 16224N, 16225N, 16226N, 16227N, 16228N, 16229N, 16230N, 16231N, 16232N, 16233N, 16234N, 16235N, 16236N, 16237N, 16238N, 16239N, 16240N, 16241N, 16242N, 16243N, 16244N, 16245N, 16246N, 16247N, 16248N, 16249N, 16250N, 16251N, 16252N, 16253N, 16254N, 16255N, 16256N, 16257N, 16258N, 16259N, 16260N, 16261N, 16262N, 16263N, 16264N, 16265N, 16266N, 16267N, 16268N, 16269N, 16270N, 16271N, 16272N, 16273N, 16274N, 16275N, 16276N, 16277N, 16278N, 16279N, 16280N, 16281N, 16282N, 16283N, 16284N, 16285N, 16286N, 16287N, 16288N, 16289N, 16290N, 16291N, 16292N, 16293N, 16294N, 16295N, 16296N, 16297N, 16298N, 16299N, 16300N, 16301N, 16302N, 16303N, 16304N, 16305N, 16306N, 16307N, 16308N, 16309N, 16310N, 16311N, 16312N, 16313N, 16314N, 16315N, 16316N, 16317N, 16318N, 16319N, 16320N, 16321N, 16322N, 16323N, 16324N, 16325N, 16326N, 16327N, 16328N, 16329N, 16330N, 16331N, 16332N, 16333N, 16334N, 16335N, 16336N, 16337N, 16338N, 16339N, | Lippold,S., Matzke,N., Reissman,M., Burbano,H. and Hofreiter,M. BMC Evolutionary Biology. 328 (11), 1471-2148 (2011) |
| | | | | | | 16340N, 16341N, 16342N, 16343N, 16344N, 16345N, 16346N, 16347N, 16348N, 16349N, 16350N, 16351N, 16352N, 16353N, 16354N, 16355N, 16356N, 16357N | |
| HQ439479 (Horse and Przewalski's horse) | Illumina/Solexa Genome Analyzer II | Phantom mutaions: 358d, 1387A, 2226+C, 5237+G, 5277+A, 15385+T. | G1a_HQ439479 | 1-14d, 158, 222, 356d, 382, 387, 416, 1387A, 2226+C, 2788, 2940, 3053, 3576, 4062, 4646, 4669, 4830, 5237+G, 5277+A, 5498, 5669, 5830, 5881, 5884, 6004, 6307, 6688, 6784, 7001, 8005, 8037, 9218, 9239, 9402, 9669, 9741A, 10214, 10376, 10471, 11165, 11240, 11543, 11552, 11842, 12767, 12860, 13049, 13223, 13333, 13502, 14350, 14626, 14651, 14734, 15385+T, 15492, 15539, 15582, 15594, 15599, 15632, 15647, 15663, 15700, 15717, 15867, 16004, 16028, 16110, 16126N, 16127N, 16128N, 16129N, 16130N, 16131N, 16132N, 16133N, 16134N, 16135N, 16136N, 16137N, 16138N, 16139N, 16140N, 16141N, 16142N, 16143N, 16144N, 16145N, 16146N, 16147N, 16148N, 16149N, 16150N, 16151N, 16152N, 16153N, 16154N, 16155N, 16156N, 16157N, 16158N, 16159N, 16160N, 16161N, 16162N, 16163N, 16164N, 16165N, 16166N, 16167N, 16168N, 16169N, 16170N, 16171N, 16172N, 16173N, 16174N, 16175N, 16176N, 16177N, 16178N, 16179N, 16180N, 16181N, 16182N, 16183N, 16184N, 16185N, 16186N, 16187N, 16188N, 16189N, 16190N, 16191N, 16192N, 16193N, 16194N, 16195N, 16196N, 16197N, 16198N, 16199N, 16200N, 16201N, 16202N, 16203N, 16204N, 16205N, 16206N, 16207N, 16208N, 16209N, 16210N, 16211N, 16212N, 16213N, 16214N, 16215N, 16216N, 16217N, 16218N, 16219N, 16220N, 16221N, 16222N, 16223N, 16224N, 16225N, 16226N, 16227N, 16228N, 16229N, 16230N, 16231N, 16232N, 16233N, 16234N, 16235N, 16236N, 16237N, 16238N, 16239N, 16240N, 16241N, 16242N, 16243N, 16244N, 16245N, 16246N, 16247N, 16248N, 16249N, 16250N, 16251N, 16252N, 16253N, 16254N, 16255N, 16256N, 16257N, 16258N, 16259N, 16260N, 16261N, 16262N, 16263N, 16264N, 16265N, 16266N, 16267N, 16268N, 16269N, 16270N, 16271N, 16272N, 16273N, 16274N, 16275N, 16276N, 16277N, 16278N, 16279N, 16280N, 16281N, 16282N, 16283N, 16284N, 16285N, 16286N, 16287N, 16288N, 16289N, 16290N, 16291N, 16292N, 16293N, 16294N, 16295N, 16296N, 16297N, 16298N, 16299N, 16300N, 16301N, 16302N, 16303N, 16304N, 16305N, 16306N, 16307N, 16308N, 16309N, 16310N, 16311N, 16312N, 16313N, 16314N, 16315N, 16316N, 16317N, 16318N, 16319N, 16320N, 16321N, 16322N, 16323N, 16324N, 16325N, 16326N, 16327N, 16328N, 16329N, 16330N, 16331N, 16332N, 16333N, 16334N, 16335N, 16336N, 16337N, 16338N, 16339N, 16340N, 16341N, 16342N, 16343N, 16344N, 16345N, 16346N, 16347N, 16348N, 16349N, 16350N, 16351N, 16352N, 16353N, 16354N, 16355N, 16356N, 16357N, 16368 | G1a_HQ439479:356 | G1a_HQ439479:1-14d, 356d, 2226+C, 5237+G, 5277+A, 15385+T, 16126N, 16127N, 16128N, 16129N, 16130N, 16131N, 16132N, 16133N, 16134N, 16135N, 16136N, 16137N, 16138N, 16139N, 16140N, 16141N, 16142N, 16143N, 16144N, 16145N, 16146N, 16147N, 16148N, 16149N, 16150N, 16151N, 16152N, 16153N, 16154N, 16155N, 16156N, 16157N, 16158N, 16159N, 16160N, 16161N, 16162N, 16163N, 16164N, 16165N, 16166N, 16167N, 16168N, 16169N, 16170N, 16171N, 16172N, 16173N, 16174N, 16175N, 16176N, 16177N, 16178N, 16179N, 16180N, 16181N, 16182N, 16183N, 16184N, 16185N, 16186N, 16187N, 16188N, 16189N, 16190N, 16191N, 16192N, 16193N, 16194N, 16195N, 16196N, 16197N, 16198N, 16199N, 16200N, 16201N, 16202N, 16203N, 16204N, 16205N, 16206N, 16207N, 16208N, 16209N, 16210N, 16211N, 16212N, 16213N, 16214N, 16215N, 16216N, 16217N, 16218N, 16219N, 16220N, 16221N, 16222N, 16223N, 16224N, 16225N, 16226N, 16227N, 16228N, 16229N, 16230N, 16231N, 16232N, 16233N, 16234N, 16235N, 16236N, 16237N, 16238N, 16239N, 16240N, 16241N, 16242N, 16243N, 16244N, 16245N, 16246N, 16247N, 16248N, 16249N, 16250N, 16251N, 16252N, 16253N, 16254N, 16255N, 16256N, 16257N, 16258N, 16259N, 16260N, 16261N, 16262N, 16263N, 16264N, 16265N, 16266N, 16267N, 16268N, 16269N, 16270N, 16271N, 16272N, 16273N, 16274N, 16275N, 16276N, 16277N, 16278N, 16279N, 16280N, 16281N, 16282N, 16283N, 16284N, 16285N, 16286N, 16287N, 16288N, 16289N, 16290N, 16291N, 16292N, 16293N, 16294N, 16295N, 16296N, 16297N, 16298N, 16299N, 16300N, 16301N, 16302N, 16303N, 16304N, 16305N, 16306N, 16307N, 16308N, 16309N, 16310N, 16311N, 16312N, 16313N, 16314N, 16315N, 16316N, 16317N, 16318N, 16319N, 16320N, 16321N, 16322N, 16323N, 16324N, 16325N, 16326N, 16327N, 16328N, 16329N, 16330N, 16331N, 16332N, 16333N, 16334N, 16335N, 16336N, 16337N, 16338N, 16339N, 16340N, 16341N, 16342N, 16343N, 16344N, 16345N, 16346N, 16347N, 16348N, 16349N, 16350N, 16351N, 16352N, 16353N, 16354N, 16355N, 16356N, 16357N | Lippold,S., Matzke,N., Reissman,M., Burbano,H. and Hofreiter,M. BMC Evolutionary Biology. 328 (11), 1471-2148 (2011) |
| HQ439480 (Horse and Przewalski's horse) | Illumina/Solexa Genome Analyzer II | Phantom mutaions: 358d, 1387A, 2227+T, 5239+A, 5277+A, 15385+T. | I2b | 158, 356d, 1387A, 1587, 1791, 2227+T, 2614, 2770, 2788, 4062, 4063, 4290, 4392, 4646, 4669, 4830, 5061, 5210, 5239+A, 5277+A, 5884, 6004, 6175, 6247, 6307, 6784, 7001, 8005, 8074, 8358, 8379, 8792, 9239, 9694, 9948, 10083, 10214, 10238, 10376, 11240, 11543, 11827, 11842, 12404, 12443, 12683, 12725, 12767, 13049, 13333, 13502, 13761, 14554, 15100, 15385+T, 15492, 15535, 15582, 15599, 15647, 15706, 15717, 15768, 15823, 15867, 15971, 16110, 16119, 16126N, 16127N, 16128N, 16129N, 16130N, 16131N, 16132N, 16133N, 16134N, 16135N, 16136N, 16137N, 16138N, 16139N, 16140N, 16141N, 16142N, 16143N, 16144N, 16145N, 16146N, 16147N, 16148N, 16149N, 16150N, 16151N, 16152N, 16153N, 16154N, 16155N, 16156N, 16157N, 16158N, 16159N, 16160N, 16161N, 16162N, 16163N, 16164N, 16165N, 16166N, 16167N, 16168N, 16169N, 16170N, 16171N, 16172N, 16173N, 16174N, 16175N, 16176N, 16177N, 16178N, 16179N, 16180N, 16181N, 16182N, 16183N, 16184N, 16185N, 16186N, 16187N, 16188N, 16189N, 16190N, 16191N, 16192N, 16193N, 16194N, 16195N, 16196N, 16197N, 16198N, 16199N, 16200N, 16201N, 16202N, 16203N, 16204N, 16205N, 16206N, | I2b:356 | I2b:356d, 2227+T, 5239+A, 5277+A, 15385+T, 16126N, 16127N, 16128N, 16129N, 16130N, 16131N, 16132N, 16133N, 16134N, 16135N, 16136N, 16137N, 16138N, 16139N, 16140N, 16141N, 16142N, 16143N, 16144N, 16145N, 16146N, 16147N, 16148N, 16149N, 16150N, 16151N, 16152N, 16153N, 16154N, 16155N, 16156N, 16157N, 16158N, 16159N, 16160N, 16161N, 16162N, 16163N, 16164N, 16165N, 16166N, 16167N, 16168N, 16169N, 16170N, 16171N, 16172N, 16173N, 16174N, 16175N, 16176N, 16177N, 16178N, 16179N, 16180N, 16181N, 16182N, 16183N, 16184N, 16185N, 16186N, 16187N, 16188N, 16189N, 16190N, 16191N, 16192N, 16193N, 16194N, 16195N, 16196N, 16197N, 16198N, 16199N, 16200N, 16201N, 16202N, 16203N, 16204N, 16205N, 16206N, 16207N, 16208N, 16209N, 16210N, 16211N, 16212N, 16213N, 16214N, 16215N, 16216N, 16217N, 16218N, 16219N, | Lippold,S., Matzke,N., Reissman,M., Burbano,H. and Hofreiter,M. BMC Evolutionary Biology. 328 (11), 1471-2148 (2011) |
| | | | | 16207N, 16208N, 16209N, 16210N, 16211N, 16212N, 16213N, 16214N, 16215N, 16216N, 16217N, 16218N, 16219N, 16220N, 16221N, 16222N, 16223N, 16224N, 16225N, 16226N, 16227N, 16228N, 16229N, 16230N, 16231N, 16232N, 16233N, 16234N, 16235N, 16236N, 16237N, 16238N, 16239N, 16240N, 16241N, 16242N, 16243N, 16244N, 16245N, 16246N, 16247N, 16248N, 16249N, 16250N, 16251N, 16252N, 16253N, 16254N, 16255N, 16256N, 16257N, 16258N, 16259N, 16260N, 16261N, 16262N, 16263N, 16264N, 16265N, 16266N, 16267N, 16268N, 16269N, 16270N, 16271N, 16272N, 16273N, 16274N, 16275N, 16276N, 16277N, 16278N, 16279N, 16280N, 16281N, 16282N, 16283N, 16284N, 16285N, 16286N, 16287N, 16288N, 16289N, 16290N, 16291N, 16292N, 16293N, 16294N, 16295N, 16296N, 16297N, 16298N, 16299N, 16300N, 16301N, 16302N, 16303N, 16304N, 16305N, 16306N, 16307N, 16308N, 16309N, 16310N, 16311N, 16312N, 16313N, 16314N, 16315N, 16316N, 16317N, 16318N, 16319N, 16320N, 16321N, 16322N, 16323N, 16324N, 16325N, 16326N, 16327N, 16328N, 16329N, 16330N, 16331N, 16332N, 16333N, 16334N, 16335N, 16336N, 16337N, 16338N, 16339N, 16340N, 16341N, 16342N, 16343N, 16344N, 16345N, 16346N, 16347N, 16348N, 16349N, 16350N, 16351N, 16352N, 16353N, 16354N, 16355N, 16356N, 16357N, 16368 | | 16220N, 16221N, 16222N, 16223N, 16224N, 16225N, 16226N, 16227N, 16228N, 16229N, 16230N, 16231N, 16232N, 16233N, 16234N, 16235N, 16236N, 16237N, 16238N, 16239N, 16240N, 16241N, 16242N, 16243N, 16244N, 16245N, 16246N, 16247N, 16248N, 16249N, 16250N, 16251N, 16252N, 16253N, 16254N, 16255N, 16256N, 16257N, 16258N, 16259N, 16260N, 16261N, 16262N, 16263N, 16264N, 16265N, 16266N, 16267N, 16268N, 16269N, 16270N, 16271N, 16272N, 16273N, 16274N, 16275N, 16276N, 16277N, 16278N, 16279N, 16280N, 16281N, 16282N, 16283N, 16284N, 16285N, 16286N, 16287N, 16288N, 16289N, 16290N, 16291N, 16292N, 16293N, 16294N, 16295N, 16296N, 16297N, 16298N, 16299N, 16300N, 16301N, 16302N, 16303N, 16304N, 16305N, 16306N, 16307N, 16308N, 16309N, 16310N, 16311N, 16312N, 16313N, 16314N, 16315N, 16316N, 16317N, 16318N, 16319N, 16320N, 16321N, 16322N, 16323N, 16324N, 16325N, 16326N, 16327N, 16328N, 16329N, 16330N, 16331N, 16332N, 16333N, 16334N, 16335N, 16336N, 16337N, 16338N, 16339N, 16340N, 16341N, 16342N, 16343N, 16344N, 16345N, 16346N, 16347N, 16348N, 16349N, 16350N, 16351N, 16352N, 16353N, 16354N, 16355N, 16356N, 16357N | |
| HQ439481 (Horse and Przewalski's horse) | Illumina/Solexa Genome Analyzer II | Phantom mutaions: 358d, 1387G, 2226+C, 5239+A, 5277+A, 15383+A. | G1b_HQ439481 | 158, 222, 356d, 382, 387, 416, 1387G, 2226+C, 2788, 2940, 3053, 3513, 3576, 4062, 4646, 4669, 4830, 5239+A, 5277+A, 5498, 5669, 5830, 5881, 5884, 6004, 6097, 6307, 6688, 6784, 7001, 8005, 8037, 9239, 9402, 9669, 9741A, 10214, 10376, 10471, 11165, 11240, 11543, 11552, 11842, 12767, 12860, 13049, 13223, 13333, 13502, 14350, 14473, 14476, 14626, 14734, 14825, 15383+A, 15492, 15539, 15582, 15594, 15599, 15632, 15647, 15663, 15700, 15717, 15867, 16028, 16110, 16126N, 16127N, 16128N, 16129N, 16130N, 16131N, 16132N, 16133N, 16134N, 16135N, 16136N, 16137N, 16138N, 16139N, 16140N, 16141N, 16142N, 16143N, 16144N, 16145N, 16146N, 16147N, 16148N, 16149N, 16150N, 16151N, 16152N, 16153N, 16154N, 16155N, 16156N, 16157N, 16158N, 16159N, 16160N, 16161N, 16162N, 16163N, 16164N, 16165N, 16166N, 16167N, 16168N, 16169N, 16170N, 16171N, 16172N, 16173N, 16174N, 16175N, 16176N, 16177N, 16178N, 16179N, 16180N, 16181N, 16182N, 16183N, 16184N, 16185N, 16186N, 16187N, 16188N, 16189N, 16190N, 16191N, 16192N, 16193N, 16194N, 16195N, 16196N, 16197N, 16198N, 16199N, 16200N, 16201N, 16202N, 16203N, 16204N, 16205N, 16206N, 16207N, 16208N, 16209N, 16210N, 16211N, 16212N, 16213N, 16214N, 16215N, 16216N, 16217N, 16218N, 16219N, 16220N, 16221N, 16222N, 16223N, 16224N, 16225N, 16226N, 16227N, 16228N, 16229N, 16230N, 16231N, 16232N, 16233N, 16234N, 16235N, 16236N, 16237N, 16238N, 16239N, 16240N, 16241N, 16242N, 16243N, 16244N, 16245N, 16246N, 16247N, 16248N, 16249N, 16250N, 16251N, 16252N, 16253N, 16254N, 16255N, 16256N, 16257N, 16258N, 16259N, 16260N, 16261N, 16262N, 16263N, 16264N, 16265N, 16266N, 16267N, 16268N, 16269N, 16270N, 16271N, 16272N, 16273N, 16274N, 16275N, 16276N, 16277N, 16278N, 16279N, 16280N, 16281N, 16282N, 16283N, 16284N, 16285N, 16286N, 16287N, 16288N, 16289N, 16290N, 16291N, 16292N, 16293N, 16294N, 16295N, 16296N, 16297N, 16298N, 16299N, 16300N, 16301N, 16302N, 16303N, 16304N, 16305N, 16306N, 16307N, 16308N, 16309N, 16310N, 16311N, 16312N, 16313N, 16314N, 16315N, 16316N, 16317N, 16318N, 16319N, 16320N, 16321N, 16322N, 16323N, 16324N, 16325N, 16326N, 16327N, 16328N, 16329N, 16330N, 16331N, 16332N, 16333N, 16334N, 16335N, 16336N, 16337N, 16338N, 16339N, 16340N, 16341N, 16342N, 16343N, 16344N, 16345N, 16346N, 16347N, 16348N, 16349N, 16350N, 16351N, 16352N, 16353N, 16354N, 16355N, 16356N, 16357N, 16368N | G1b_HQ439481:356, 16368 | G1b_HQ439481:356d, 2226+C, 5239+A, 5277+A, 15383+A, 16126N, 16127N, 16128N, 16129N, 16130N, 16131N, 16132N, 16133N, 16134N, 16135N, 16136N, 16137N, 16138N, 16139N, 16140N, 16141N, 16142N, 16143N, 16144N, 16145N, 16146N, 16147N, 16148N, 16149N, 16150N, 16151N, 16152N, 16153N, 16154N, 16155N, 16156N, 16157N, 16158N, 16159N, 16160N, 16161N, 16162N, 16163N, 16164N, 16165N, 16166N, 16167N, 16168N, 16169N, 16170N, 16171N, 16172N, 16173N, 16174N, 16175N, 16176N, 16177N, 16178N, 16179N, 16180N, 16181N, 16182N, 16183N, 16184N, 16185N, 16186N, 16187N, 16188N, 16189N, 16190N, 16191N, 16192N, 16193N, 16194N, 16195N, 16196N, 16197N, 16198N, 16199N, 16200N, 16201N, 16202N, 16203N, 16204N, 16205N, 16206N, 16207N, 16208N, 16209N, 16210N, 16211N, 16212N, 16213N, 16214N, 16215N, 16216N, 16217N, 16218N, 16219N, 16220N, 16221N, 16222N, 16223N, 16224N, 16225N, 16226N, 16227N, 16228N, 16229N, 16230N, 16231N, 16232N, 16233N, 16234N, 16235N, 16236N, 16237N, 16238N, 16239N, 16240N, 16241N, 16242N, 16243N, 16244N, 16245N, 16246N, 16247N, 16248N, 16249N, 16250N, 16251N, 16252N, 16253N, 16254N, 16255N, 16256N, 16257N, 16258N, 16259N, 16260N, 16261N, 16262N, 16263N, 16264N, 16265N, 16266N, 16267N, 16268N, 16269N, 16270N, 16271N, 16272N, 16273N, 16274N, 16275N, 16276N, 16277N, 16278N, 16279N, 16280N, 16281N, 16282N, 16283N, 16284N, 16285N, 16286N, 16287N, 16288N, 16289N, 16290N, 16291N, 16292N, 16293N, 16294N, 16295N, 16296N, 16297N, 16298N, 16299N, 16300N, 16301N, 16302N, 16303N, 16304N, 16305N, 16306N, 16307N, 16308N, 16309N, 16310N, 16311N, 16312N, 16313N, 16314N, 16315N, 16316N, 16317N, 16318N, 16319N, 16320N, 16321N, 16322N, 16323N, 16324N, 16325N, 16326N, 16327N, 16328N, 16329N, 16330N, 16331N, 16332N, 16333N, 16334N, 16335N, 16336N, 16337N, 16338N, | Lippold,S., Matzke,N., Reissman,M., Burbano,H. and Hofreiter,M. BMC Evolutionary Biology. 328 (11), 1471-2148 (2011) |
| | | | | | | 16339N, 16340N, 16341N, 16342N, 16343N, 16344N, 16345N, 16346N, 16347N, 16348N, 16349N, 16350N, 16351N, 16352N, 16353N, 16354N, 16355N, 16356N, 16357N, 16368N | |
| HQ439482 (Horse and Przewalski's horse) | Illumina/Solexa Genome Analyzer II | Phantom mutaions: 358d, 1387A, 2226+C, 5239+A, 5277+A, 15385+T. | L1a1a_HQ439482 | 158, 356d, 961, 1375, 1387A, 2226+C, 2788, 2899, 3517, 3942, 4062, 4536, 4646, 4669, 5239+A, 5277+A, 5527, 5815, 5884, 6004, 6307, 6784, 6975N, 6976N, 6977N, 6978N, 6979N, 6980N, 6981N, 7001N, 7516, 7666, 7900, 8005, 8058, 8301, 8319, 8358, 8565, 9239, 9684, 9951, 10110, 10214, 10292, 10376, 10421, 10613, 11240, 11543, 11682, 11693, 11842, 11879, 12119, 12200, 12767, 12896, 12950, 13049, 13333, 13520, 14803, 14995, 15313, 15385+T, 15491, 15492, 15493, 15531, 15599, 15600, 15646, 15717, 15768, 15867, 15868, 15953, 15971, 16065, 16084, 16100, 16126N, 16127N, 16128N, 16129N, 16130N, 16131N, 16132N, 16133N, 16134N, 16135N, 16136N, 16137N, 16138N, 16139N, 16140N, 16141N, 16142N, 16143N, 16144N, 16145N, 16146N, 16147N, 16148N, 16149N, 16150N, 16151N, 16152N, 16153N, 16154N, 16155N, 16156N, 16157N, 16158N, 16159N, 16160N, 16161N, 16162N, 16163N, 16164N, 16165N, 16166N, 16167N, 16168N, 16169N, 16170N, 16171N, 16172N, 16173N, 16174N, 16175N, 16176N, 16177N, 16178N, 16179N, 16180N, 16181N, 16182N, 16183N, 16184N, 16185N, 16186N, 16187N, 16188N, 16189N, 16190N, 16191N, 16192N, 16193N, 16194N, 16195N, 16196N, 16197N, 16198N, 16199N, 16200N, 16201N, 16202N, 16203N, 16204N, 16205N, 16206N, 16207N, 16208N, 16209N, 16210N, 16211N, 16212N, 16213N, 16214N, 16215N, 16216N, 16217N, 16218N, 16219N, 16220N, 16221N, 16222N, 16223N, 16224N, 16225N, 16226N, 16227N, 16228N, 16229N, 16230N, 16231N, 16232N, 16233N, 16234N, 16235N, 16236N, 16237N, 16238N, 16239N, 16240N, 16241N, 16242N, 16243N, 16244N, 16245N, 16246N, 16247N, 16248N, 16249N, 16250N, 16251N, 16252N, 16253N, 16254N, 16255N, 16256N, 16257N, 16258N, 16259N, 16260N, 16261N, 16262N, 16263N, 16264N, 16265N, 16266N, 16267N, 16268N, 16269N, 16270N, 16271N, 16272N, 16273N, 16274N, 16275N, 16276N, 16277N, 16278N, 16279N, 16280N, 16281N, 16282N, 16283N, 16284N, 16285N, 16286N, 16287N, 16288N, 16289N, 16290N, 16291N, 16292N, 16293N, 16294N, 16295N, 16296N, 16297N, 16298N, 16299N, 16300N, 16301N, 16302N, 16303N, 16304N, 16305N, 16306N, 16307N, 16308N, 16309N, 16310N, 16311N, 16312N, 16313N, 16314N, 16315N, 16316N, 16317N, 16318N, 16319N, 16320N, 16321N, 16322N, 16323N, 16324N, 16325N, 16326N, 16327N, 16328N, 16329N, 16330N, 16331N, 16332N, 16333N, 16334N, 16335N, 16336N, 16337N, 16338N, 16339N, 16340N, 16341N, 16342N, 16343N, 16344N, 16345N, 16346N, 16347N, 16348N, 16349N, 16350N, 16351N, 16352N, 16353N, 16354N, 16355N, 16356N, 16357N, 16368, 16398A, 16626 | L1a1a_HQ439482:356, 7001 | L1a1a_HQ439482:356d, 2226+C, 5239+A, 5277+A, 6975N, 6976N, 6977N, 6978N, 6979N, 6980N, 6981N, 7001N, 15385+T, 16126N, 16127N, 16128N, 16129N, 16130N, 16131N, 16132N, 16133N, 16134N, 16135N, 16136N, 16137N, 16138N, 16139N, 16140N, 16141N, 16142N, 16143N, 16144N, 16145N, 16146N, 16147N, 16148N, 16149N, 16150N, 16151N, 16152N, 16153N, 16154N, 16155N, 16156N, 16157N, 16158N, 16159N, 16160N, 16161N, 16162N, 16163N, 16164N, 16165N, 16166N, 16167N, 16168N, 16169N, 16170N, 16171N, 16172N, 16173N, 16174N, 16175N, 16176N, 16177N, 16178N, 16179N, 16180N, 16181N, 16182N, 16183N, 16184N, 16185N, 16186N, 16187N, 16188N, 16189N, 16190N, 16191N, 16192N, 16193N, 16194N, 16195N, 16196N, 16197N, 16198N, 16199N, 16200N, 16201N, 16202N, 16203N, 16204N, 16205N, 16206N, 16207N, 16208N, 16209N, 16210N, 16211N, 16212N, 16213N, 16214N, 16215N, 16216N, 16217N, 16218N, 16219N, 16220N, 16221N, 16222N, 16223N, 16224N, 16225N, 16226N, 16227N, 16228N, 16229N, 16230N, 16231N, 16232N, 16233N, 16234N, 16235N, 16236N, 16237N, 16238N, 16239N, 16240N, 16241N, 16242N, 16243N, 16244N, 16245N, 16246N, 16247N, 16248N, 16249N, 16250N, 16251N, 16252N, 16253N, 16254N, 16255N, 16256N, 16257N, 16258N, 16259N, 16260N, 16261N, 16262N, 16263N, 16264N, 16265N, 16266N, 16267N, 16268N, 16269N, 16270N, 16271N, 16272N, 16273N, 16274N, 16275N, 16276N, 16277N, 16278N, 16279N, 16280N, 16281N, 16282N, 16283N, 16284N, 16285N, 16286N, 16287N, 16288N, 16289N, 16290N, 16291N, 16292N, 16293N, 16294N, 16295N, 16296N, 16297N, 16298N, 16299N, 16300N, 16301N, 16302N, 16303N, 16304N, 16305N, 16306N, 16307N, 16308N, 16309N, 16310N, 16311N, 16312N, 16313N, 16314N, 16315N, 16316N, 16317N, 16318N, 16319N, 16320N, 16321N, 16322N, 16323N, 16324N, 16325N, 16326N, 16327N, 16328N, 16329N, 16330N, 16331N, 16332N, 16333N, 16334N, 16335N, 16336N, 16337N, 16338N, 16339N, 16340N, 16341N, 16342N, 16343N, 16344N, 16345N, 16346N, 16347N, 16348N, 16349N, 16350N, 16351N, 16352N, 16353N, 16354N, 16355N, 16356N, 16357N | Lippold,S., Matzke,N., Reissman,M., Burbano,H. and Hofreiter,M. BMC Evolutionary Biology. 328 (11), 1471-2148 (2011) |
| HQ439483 (Horse and Przewalski's horse) | Illumina/Solexa Genome Analyzer II | Phantom mutaions: 358d, 1387A, 2226+C, 5237+G, 5277+A, 15385+T. | L1a1a_HQ439483 | 158, 356d, 961, 1375, 1387A, 2226+C, 2788, 2899, 3517, 3942, 4062, 4536, 4646, 4669, 5237+G, 5277+A, 5527, 5815, 5884, 6004, 6307, 6784, 7001, 7516, 7666, 7900, 8005, 8058, 8301, 8319, 8358, 8565, 9239, 9684, 9951, 10110, 10214, 10292, 10376, 10421, 10613, 10956, 11240, 11543, 11682, 11693, 11842, 11879, 12119, 12200, 12767, 12896, 12950, 13049, 13333, 13520, 14803, 14995, 15313, 15385+T, 15491, 15492, 15493, 15531, 15599, 15600, 15646, 15717, 15768, 15867, 15868, 15953, 15971, 16065, 16091, 16100, 16126N, 16127N, 16128N, 16129N, 16130N, 16131N, 16132N, 16133N, 16134N, 16135N, 16136N, 16137N, 16138N, 16139N, 16140N, 16141N, 16142N, 16143N, 16144N, 16145N, 16146N, 16147N, 16148N, 16149N, 16150N, 16151N, 16152N, 16153N, 16154N, 16155N, 16156N, 16157N, 16158N, 16159N, 16160N, 16161N, 16162N, 16163N, 16164N, 16165N, 16166N, 16167N, 16168N, 16169N, 16170N, 16171N, 16172N, 16173N, 16174N, 16175N, 16176N, 16177N, 16178N, 16179N, 16180N, 16181N, 16182N, 16183N, 16184N, | L1a1a_HQ439483:356 | L1a1a_HQ439483:356d, 2226+C, 5237+G, 5277+A, 15385+T, 16126N, 16127N, 16128N, 16129N, 16130N, 16131N, 16132N, 16133N, 16134N, 16135N, 16136N, 16137N, 16138N, 16139N, 16140N, 16141N, 16142N, 16143N, 16144N, 16145N, 16146N, 16147N, 16148N, 16149N, 16150N, 16151N, 16152N, 16153N, 16154N, 16155N, 16156N, 16157N, 16158N, 16159N, 16160N, 16161N, 16162N, 16163N, 16164N, 16165N, 16166N, 16167N, 16168N, 16169N, 16170N, 16171N, 16172N, 16173N, 16174N, 16175N, 16176N, 16177N, 16178N, 16179N, 16180N, 16181N, 16182N, 16183N, 16184N, 16185N, 16186N, 16187N, 16188N, 16189N, 16190N, 16191N, 16192N, 16193N, 16194N, 16195N, 16196N, 16197N, 16198N, 16199N, 16200N, 16201N, 16202N, 16203N, 16204N, 16205N, 16206N, | Lippold,S., Matzke,N., Reissman,M., Burbano,H. and Hofreiter,M. BMC Evolutionary Biology. 328 (11), 1471-2148 (2011) |
| | | | | 16185N, 16186N, 16187N, 16188N, 16189N, 16190N, 16191N, 16192N, 16193N, 16194N, 16195N, 16196N, 16197N, 16198N, 16199N, 16200N, 16201N, 16202N, 16203N, 16204N, 16205N, 16206N, 16207N, 16208N, 16209N, 16210N, 16211N, 16212N, 16213N, 16214N, 16215N, 16216N, 16217N, 16218N, 16219N, 16220N, 16221N, 16222N, 16223N, 16224N, 16225N, 16226N, 16227N, 16228N, 16229N, 16230N, 16231N, 16232N, 16233N, 16234N, 16235N, 16236N, 16237N, 16238N, 16239N, 16240N, 16241N, 16242N, 16243N, 16244N, 16245N, 16246N, 16247N, 16248N, 16249N, 16250N, 16251N, 16252N, 16253N, 16254N, 16255N, 16256N, 16257N, 16258N, 16259N, 16260N, 16261N, 16262N, 16263N, 16264N, 16265N, 16266N, 16267N, 16268N, 16269N, 16270N, 16271N, 16272N, 16273N, 16274N, 16275N, 16276N, 16277N, 16278N, 16279N, 16280N, 16281N, 16282N, 16283N, 16284N, 16285N, 16286N, 16287N, 16288N, 16289N, 16290N, 16291N, 16292N, 16293N, 16294N, 16295N, 16296N, 16297N, 16298N, 16299N, 16300N, 16301N, 16302N, 16303N, 16304N, 16305N, 16306N, 16307N, 16308N, 16309N, 16310N, 16311N, 16312N, 16313N, 16314N, 16315N, 16316N, 16317N, 16318N, 16319N, 16320N, 16321N, 16322N, 16323N, 16324N, 16325N, 16326N, 16327N, 16328N, 16329N, 16330N, 16331N, 16332N, 16333N, 16334N, 16335N, 16336N, 16337N, 16338N, 16339N, 16340N, 16341N, 16342N, 16343N, 16344N, 16345N, 16346N, 16347N, 16348N, 16349N, 16350N, 16351N, 16352N, 16353N, 16354N, 16355N, 16356N, 16357N, 16368, 16398A, 16626 | | 16207N, 16208N, 16209N, 16210N, 16211N, 16212N, 16213N, 16214N, 16215N, 16216N, 16217N, 16218N, 16219N, 16220N, 16221N, 16222N, 16223N, 16224N, 16225N, 16226N, 16227N, 16228N, 16229N, 16230N, 16231N, 16232N, 16233N, 16234N, 16235N, 16236N, 16237N, 16238N, 16239N, 16240N, 16241N, 16242N, 16243N, 16244N, 16245N, 16246N, 16247N, 16248N, 16249N, 16250N, 16251N, 16252N, 16253N, 16254N, 16255N, 16256N, 16257N, 16258N, 16259N, 16260N, 16261N, 16262N, 16263N, 16264N, 16265N, 16266N, 16267N, 16268N, 16269N, 16270N, 16271N, 16272N, 16273N, 16274N, 16275N, 16276N, 16277N, 16278N, 16279N, 16280N, 16281N, 16282N, 16283N, 16284N, 16285N, 16286N, 16287N, 16288N, 16289N, 16290N, 16291N, 16292N, 16293N, 16294N, 16295N, 16296N, 16297N, 16298N, 16299N, 16300N, 16301N, 16302N, 16303N, 16304N, 16305N, 16306N, 16307N, 16308N, 16309N, 16310N, 16311N, 16312N, 16313N, 16314N, 16315N, 16316N, 16317N, 16318N, 16319N, 16320N, 16321N, 16322N, 16323N, 16324N, 16325N, 16326N, 16327N, 16328N, 16329N, 16330N, 16331N, 16332N, 16333N, 16334N, 16335N, 16336N, 16337N, 16338N, 16339N, 16340N, 16341N, 16342N, 16343N, 16344N, 16345N, 16346N, 16347N, 16348N, 16349N, 16350N, 16351N, 16352N, 16353N, 16354N, 16355N, 16356N, 16357N | |
| HQ439484 (Horse and Przewalski's horse) | Illumina/Solexa Genome Analyzer II | Phantom mutaions: 358d, 1387G, 2226+C, 5239+A, 5277+A, 15385+T. | F | 158, 356d, 382, 387, 416, 1387G, 2226+C, 2788, 2940, 3371, 3562, 3576, 4062, 4646, 4669, 4830, 5239+A, 5277+A, 5830, 5884, 6004, 6307, 6784, 7001, 7504, 7746, 8005, 8037, 8789, 9239, 9669, 10214, 10376, 10471, 11165, 11240, 11543, 11820, 11842, 12767, 12860, 13042, 13049, 13333, 13502, 14350, 14626, 14734, 15385+T, 15492, 15539, 15592, 15599, 15647, 15663, 15717, 15865, 15867, 16125N, 16126N, 16127N, 16128N, 16129N, 16130N, 16131N, 16132N, 16133N, 16134N, 16135N, 16136N, 16137N, 16138N, 16139N, 16140N, 16141N, 16142N, 16143N, 16144N, 16145N, 16146N, 16147N, 16148N, 16149N, 16150N, 16151N, 16152N, 16153N, 16154N, 16155N, 16156N, 16157N, 16158N, 16159N, 16160N, 16161N, 16162N, 16163N, 16164N, 16165N, 16166N, 16167N, 16168N, 16169N, 16170N, 16171N, 16172N, 16173N, 16174N, 16175N, 16176N, 16177N, 16178N, 16179N, 16180N, 16181N, 16182N, 16183N, 16184N, 16185N, 16186N, 16187N, 16188N, 16189N, 16190N, 16191N, 16192N, 16193N, 16194N, 16195N, 16196N, 16197N, 16198N, 16199N, 16200N, 16201N, 16202N, 16203N, 16204N, 16205N, 16206N, 16207N, 16208N, 16209N, 16210N, 16211N, 16212N, 16213N, 16214N, 16215N, 16216N, 16217N, 16218N, 16219N, 16220N, 16221N, 16222N, 16223N, 16224N, 16225N, 16226N, 16227N, 16228N, 16229N, 16230N, 16231N, 16232N, 16233N, 16234N, 16235N, 16236N, 16237N, 16238N, 16239N, 16240N, 16241N, 16242N, 16243N, 16244N, 16245N, 16246N, 16247N, 16248N, 16249N, 16250N, 16251N, 16252N, 16253N, 16254N, 16255N, 16256N, 16257N, 16258N, 16259N, 16260N, 16261N, 16262N, 16263N, 16264N, 16265N, 16266N, 16267N, 16268N, 16269N, 16270N, 16271N, 16272N, 16273N, 16274N, 16275N, 16276N, 16277N, 16278N, 16279N, 16280N, 16281N, 16282N, 16283N, 16284N, 16285N, 16286N, 16287N, 16288N, 16289N, 16290N, 16291N, 16292N, 16293N, 16294N, 16295N, 16296N, 16297N, 16298N, 16299N, 16300N, 16301N, 16302N, 16303N, 16304N, 16305N, 16306N, 16307N, 16308N, 16309N, 16310N, 16311N, 16312N, 16313N, 16314N, 16315N, 16316N, 16317N, 16318N, 16319N, 16320N, 16321N, 16322N, 16323N, 16324N, 16325N, 16326N, 16327N, 16328N, 16329N, 16330N, 16331N, 16332N, 16333N, 16334N, 16335N, 16336N, 16337N, 16338N, 16339N, 16340N, 16341N, 16342N, 16343N, 16344N, 16345N, 16346N, 16347N, 16348N, 16349N, 16350N, 16351N, 16352N, 16353N, 16354N, 16355N, 16356N, 16357N, 16361N, 16368, 16511 | F:356, 1387d | F:356d, 1387G, 2226+C, 5239+A, 5277+A, 15385+T, 16125N, 16126N, 16127N, 16128N, 16129N, 16130N, 16131N, 16132N, 16133N, 16134N, 16135N, 16136N, 16137N, 16138N, 16139N, 16140N, 16141N, 16142N, 16143N, 16144N, 16145N, 16146N, 16147N, 16148N, 16149N, 16150N, 16151N, 16152N, 16153N, 16154N, 16155N, 16156N, 16157N, 16158N, 16159N, 16160N, 16161N, 16162N, 16163N, 16164N, 16165N, 16166N, 16167N, 16168N, 16169N, 16170N, 16171N, 16172N, 16173N, 16174N, 16175N, 16176N, 16177N, 16178N, 16179N, 16180N, 16181N, 16182N, 16183N, 16184N, 16185N, 16186N, 16187N, 16188N, 16189N, 16190N, 16191N, 16192N, 16193N, 16194N, 16195N, 16196N, 16197N, 16198N, 16199N, 16200N, 16201N, 16202N, 16203N, 16204N, 16205N, 16206N, 16207N, 16208N, 16209N, 16210N, 16211N, 16212N, 16213N, 16214N, 16215N, 16216N, 16217N, 16218N, 16219N, 16220N, 16221N, 16222N, 16223N, 16224N, 16225N, 16226N, 16227N, 16228N, 16229N, 16230N, 16231N, 16232N, 16233N, 16234N, 16235N, 16236N, 16237N, 16238N, 16239N, 16240N, 16241N, 16242N, 16243N, 16244N, 16245N, 16246N, 16247N, 16248N, 16249N, 16250N, 16251N, 16252N, 16253N, 16254N, 16255N, 16256N, 16257N, 16258N, 16259N, 16260N, 16261N, 16262N, 16263N, 16264N, 16265N, 16266N, 16267N, 16268N, 16269N, 16270N, 16271N, 16272N, 16273N, 16274N, 16275N, 16276N, 16277N, 16278N, 16279N, 16280N, 16281N, 16282N, 16283N, 16284N, 16285N, 16286N, 16287N, 16288N, 16289N, 16290N, 16291N, 16292N, 16293N, 16294N, 16295N, 16296N, 16297N, 16298N, 16299N, 16300N, 16301N, 16302N, 16303N, 16304N, 16305N, 16306N, 16307N, 16308N, 16309N, 16310N, 16311N, 16312N, 16313N, 16314N, 16315N, 16316N, 16317N, 16318N, 16319N, 16320N, 16321N, 16322N, 16323N, 16324N, 16325N, 16326N, 16327N, | Lippold,S., Matzke,N., Reissman,M., Burbano,H. and Hofreiter,M. BMC Evolutionary Biology. 328 (11), 1471-2148 (2011) |
| | | | | | | 16328N, 16329N, 16330N, 16331N, 16332N, 16333N, 16334N, 16335N, 16336N, 16337N, 16338N, 16339N, 16340N, 16341N, 16342N, 16343N, 16344N, 16345N, 16346N, 16347N, 16348N, 16349N, 16350N, 16351N, 16352N, 16353N, 16354N, 16355N, 16356N, 16357N, 16361N | |
| HQ439485 (Horse and Przewalski's horse) | Illumina/Solexa Genome Analyzer II | Phantom mutaions: 358d, 1387A, 2226+A, 5239+A, 5277+A, 15385+T. | M1a_HQ439485 | 158, 356d, 874, 961, 1387A, 1609T, 2226+A, 2339A, 2788, 3070, 3100, 3475, 3800, 4062N, 4494N, 4526N, 4536N, 4599G, 4605, 4646, 4669, 4898, 5103, 5239+A, 5277+A, 5527, 5827, 5884, 6004, 6076, 6307, 6712, 6784, 7001N, 7432, 7666, 7900, 8005, 8043, 8076, 8150, 8175, 8238, 8358, 8556T, 8565, 8798, 9002, 9086, 9332, 9540, 10110, 10173, 10214, 10292, 10376, 10448, 10460, 10646, 10859, 11240, 11394, 11492, 11543, 11842, 11879, 11966, 12029, 12095, 12332, 12767, 13049, 13100, 13333, 13356, 13502, 13615, 13629, 13720, 13920, 13933, 14012N, 14013N, 14014N, 14015N, 14016N, 14017N, 14018N, 14019N, 14020N, 14021N, 14022N, 14023N, 14024N, 14025N, 14026N, 14027N, 14028N, 14029N, 14030N, 14031N, 14032N, 14033N, 14034N, 14035N, 14036N, 14037N, 14038N, 14039N, 14040N, 14183, 14422, 14626, 14671, 14803, 14815, 15052A, 15133, 15342, 15385+T, 15492, 15599, 15614, 15656, 15717, 15768, 15803, 15824, 15866, 15953, 16065, 16077, 16114N, 16115N, 16116N, 16117N, 16118N, 16119N, 16120N, 16121N, 16122N, 16123N, 16124N, 16125N, 16126N, 16127N, 16128N, 16129N, 16130N, 16131N, 16132N, 16133N, 16134N, 16135N, 16136N, 16137N, 16138N, 16139N, 16140N, 16141N, 16142N, 16143N, 16144N, 16145N, 16146N, 16147N, 16148N, 16149N, 16150N, 16151N, 16152N, 16153N, 16154N, 16155N, 16156N, 16157N, 16158N, 16159N, 16160N, 16161N, 16162N, 16163N, 16164N, 16165N, 16166N, 16167N, 16168N, 16169N, 16170N, 16171N, 16172N, 16173N, 16174N, 16175N, 16176N, 16177N, 16178N, 16179N, 16180N, 16181N, 16182N, 16183N, 16184N, 16185N, 16186N, 16187N, 16188N, 16189N, 16190N, 16191N, 16192N, 16193N, 16194N, 16195N, 16196N, 16197N, 16198N, 16199N, 16200N, 16201N, 16202N, 16203N, 16204N, 16205N, 16206N, 16207N, 16208N, 16209N, 16210N, 16211N, 16212N, 16213N, 16214N, 16215N, 16216N, 16217N, 16218N, 16219N, 16220N, 16221N, 16222N, 16223N, 16224N, 16225N, 16226N, 16227N, 16228N, 16229N, 16230N, 16231N, 16232N, 16233N, 16234N, 16235N, 16236N, 16237N, 16238N, 16239N, 16240N, 16241N, 16242N, 16243N, 16244N, 16245N, 16246N, 16247N, 16248N, 16249N, 16250N, 16251N, 16252N, 16253N, 16254N, 16255N, 16256N, 16257N, 16258N, 16259N, 16260N, 16261N, 16262N, 16263N, 16264N, 16265N, 16266N, 16267N, 16268N, 16269N, 16270N, 16271N, 16272N, 16273N, 16274N, 16275N, 16276N, 16277N, 16278N, 16279N, 16280N, 16281N, 16282N, 16283N, 16284N, 16285N, 16286N, 16287N, 16288N, 16289N, 16290N, 16291N, 16292N, 16293N, 16294N, 16295N, 16296N, 16297N, 16298N, 16299N, 16300N, 16301N, 16302N, 16303N, 16304N, 16305N, 16306N, 16307N, 16308N, 16309N, 16310N, 16311N, 16312N, 16313N, 16314N, 16315N, 16316N, 16317N, 16318N, 16319N, 16320N, 16321N, 16322N, 16323N, 16324N, 16325N, 16326N, 16327N, 16328N, 16329N, 16330N, 16331N, 16332N, 16333N, 16334N, 16335N, 16336N, 16337N, 16338N, 16339N, 16340N, 16341N, 16342N, 16343N, 16344N, 16345N, 16346N, 16347N, 16348N, 16349N, 16350N, 16351N, 16352N, 16353N, 16354N, 16355N, 16356N, 16357N, 16358N, 16359N, 16360N, 16361N, 16362N, 16368N, 16540A, 16543, 16556, 16626, 16657d | M1a_HQ439485:356, 4062, 4526, 4536, 7001, 16118, 16368 | M1a_HQ439485:356d, 1387A, 2226+A, 4062N, 4494N, 4526N, 4536N, 5239+A, 5277+A, 7001N, 14012N, 14013N, 14014N, 14015N, 14016N, 14017N, 14018N, 14019N, 14020N, 14021N, 14022N, 14023N, 14024N, 14025N, 14026N, 14027N, 14028N, 14029N, 14030N, 14031N, 14032N, 14033N, 14034N, 14035N, 14036N, 14037N, 14038N, 14039N, 14040N, 15385+T, 16114N, 16115N, 16116N, 16117N, 16118N, 16119N, 16120N, 16121N, 16122N, 16123N, 16124N, 16125N, 16126N, 16127N, 16128N, 16129N, 16130N, 16131N, 16132N, 16133N, 16134N, 16135N, 16136N, 16137N, 16138N, 16139N, 16140N, 16141N, 16142N, 16143N, 16144N, 16145N, 16146N, 16147N, 16148N, 16149N, 16150N, 16151N, 16152N, 16153N, 16154N, 16155N, 16156N, 16157N, 16158N, 16159N, 16160N, 16161N, 16162N, 16163N, 16164N, 16165N, 16166N, 16167N, 16168N, 16169N, 16170N, 16171N, 16172N, 16173N, 16174N, 16175N, 16176N, 16177N, 16178N, 16179N, 16180N, 16181N, 16182N, 16183N, 16184N, 16185N, 16186N, 16187N, 16188N, 16189N, 16190N, 16191N, 16192N, 16193N, 16194N, 16195N, 16196N, 16197N, 16198N, 16199N, 16200N, 16201N, 16202N, 16203N, 16204N, 16205N, 16206N, 16207N, 16208N, 16209N, 16210N, 16211N, 16212N, 16213N, 16214N, 16215N, 16216N, 16217N, 16218N, 16219N, 16220N, 16221N, 16222N, 16223N, 16224N, 16225N, 16226N, 16227N, 16228N, 16229N, 16230N, 16231N, 16232N, 16233N, 16234N, 16235N, 16236N, 16237N, 16238N, 16239N, 16240N, 16241N, 16242N, 16243N, 16244N, 16245N, 16246N, 16247N, 16248N, 16249N, 16250N, 16251N, 16252N, 16253N, 16254N, 16255N, 16256N, 16257N, 16258N, 16259N, 16260N, 16261N, 16262N, 16263N, 16264N, 16265N, 16266N, 16267N, 16268N, 16269N, 16270N, 16271N, 16272N, 16273N, 16274N, 16275N, 16276N, 16277N, 16278N, 16279N, 16280N, 16281N, 16282N, 16283N, 16284N, 16285N, 16286N, 16287N, 16288N, 16289N, 16290N, 16291N, 16292N, 16293N, 16294N, 16295N, 16296N, 16297N, 16298N, 16299N, 16300N, 16301N, 16302N, 16303N, 16304N, 16305N, 16306N, 16307N, 16308N, 16309N, 16310N, 16311N, 16312N, 16313N, 16314N, 16315N, 16316N, 16317N, 16318N, 16319N, 16320N, 16321N, 16322N, 16323N, 16324N, 16325N, 16326N, 16327N, 16328N, 16329N, 16330N, 16331N, 16332N, 16333N, 16334N, 16335N, 16336N, 16337N, 16338N, 16339N, 16340N, 16341N, 16342N, 16343N, 16344N, 16345N, 16346N, 16347N, 16348N, 16349N, 16350N, 16351N, 16352N, 16353N, 16354N, 16355N, 16356N, 16357N, 16358N, 16359N, 16360N, 16361N, 16362N, 16368N, 16657d | Lippold,S., Matzke,N., Reissman,M., Burbano,H. and Hofreiter,M. BMC Evolutionary Biology. 328 (11), 1471-2148 (2011) |
| HQ439486 (Horse and Przewalski's | Illumina/Solexa Genome Analyzer II | Phantom mutaions: 358d,, 2226+A, 5237+G, | B1c1, B1c_HQ439486 | 158, 356d, 859, 2226+C, 4062, 4743, 5237+G, 5274+G, 5929, 6784, 7627, 9961, 10764, 11240, 13741, 14689, 15385+T, 15492, 15647, 15663, 15717, 15823, 16052, 16108, 16110, 16126N, 16127N, 16128N, 16129N, 16130N, 16131N, 16132N, 16133N, 16134N, 16135N, 16136N, 16137N, 16138N, 16139N, 16140N, 16141N, 16142N, 16143N, 16144N, 16145N, 16146N, 16147N, 16148N, 16149N, 16150N, | B1c1:356; B1c_HQ439486:356 | B1c1:356d, 2226+C, 5237+G, 5274+G, 13741, 15385+T, 16126N, 16127N, 16128N, 16129N, 16130N, 16131N, 16132N, 16133N, 16134N, 16135N, 16136N, 16137N, 16138N, 16139N, 16140N, 16141N, 16142N, 16143N, 16144N, 16145N, 16146N, 16147N, 16148N, 16149N, 16150N, 16151N, 16152N, | Lippold,S., Matzke,N., Reissman,M., Burbano,H. and Hofreiter,M. BMC Evolutionary |
| horse) | | 5274+G, 15385+T. | | 16151N, 16152N, 16153N, 16154N, 16155N, 16156N, 16157N, 16158N, 16159N, 16160N, 16161N, 16162N, 16163N, 16164N, 16165N, 16166N, 16167N, 16168N, 16169N, 16170N, 16171N, 16172N, 16173N, 16174N, 16175N, 16176N, 16177N, 16178N, 16179N, 16180N, 16181N, 16182N, 16183N, 16184N, 16185N, 16186N, 16187N, 16188N, 16189N, 16190N, 16191N, 16192N, 16193N, 16194N, 16195N, 16196N, 16197N, 16198N, 16199N, 16200N, 16201N, 16202N, 16203N, 16204N, 16205N, 16206N, 16207N, 16208N, 16209N, 16210N, 16211N, 16212N, 16213N, 16214N, 16215N, 16216N, 16217N, 16218N, 16219N, 16220N, 16221N, 16222N, 16223N, 16224N, 16225N, 16226N, 16227N, 16228N, 16229N, 16230N, 16231N, 16232N, 16233N, 16234N, 16235N, 16236N, 16237N, 16238N, 16239N, 16240N, 16241N, 16242N, 16243N, 16244N, 16245N, 16246N, 16247N, 16248N, 16249N, 16250N, 16251N, 16252N, 16253N, 16254N, 16255N, 16256N, 16257N, 16258N, 16259N, 16260N, 16261N, 16262N, 16263N, 16264N, 16265N, 16266N, 16267N, 16268N, 16269N, 16270N, 16271N, 16272N, 16273N, 16274N, 16275N, 16276N, 16277N, 16278N, 16279N, 16280N, 16281N, 16282N, 16283N, 16284N, 16285N, 16286N, 16287N, 16288N, 16289N, 16290N, 16291N, 16292N, 16293N, 16294N, 16295N, 16296N, 16297N, 16298N, 16299N, 16300N, 16301N, 16302N, 16303N, 16304N, 16305N, 16306N, 16307N, 16308N, 16309N, 16310N, 16311N, 16312N, 16313N, 16314N, 16315N, 16316N, 16317N, 16318N, 16319N, 16320N, 16321N, 16322N, 16323N, 16324N, 16325N, 16326N, 16327N, 16328N, 16329N, 16330N, 16331N, 16332N, 16333N, 16334N, 16335N, 16336N, 16337N, 16338N, 16339N, 16340N, 16341N, 16342N, 16343N, 16344N, 16345N, 16346N, 16347N, 16348N, 16349N, 16350N, 16351N, 16352N, 16353N, 16354N, 16355N, 16356N, 16357N, 16368 | | 16153N, 16154N, 16155N, 16156N, 16157N, 16158N, 16159N, 16160N, 16161N, 16162N, 16163N, 16164N, 16165N, 16166N, 16167N, 16168N, 16169N, 16170N, 16171N, 16172N, 16173N, 16174N, 16175N, 16176N, 16177N, 16178N, 16179N, 16180N, 16181N, 16182N, 16183N, 16184N, 16185N, 16186N, 16187N, 16188N, 16189N, 16190N, 16191N, 16192N, 16193N, 16194N, 16195N, 16196N, 16197N, 16198N, 16199N, 16200N, 16201N, 16202N, 16203N, 16204N, 16205N, 16206N, 16207N, 16208N, 16209N, 16210N, 16211N, 16212N, 16213N, 16214N, 16215N, 16216N, 16217N, 16218N, 16219N, 16220N, 16221N, 16222N, 16223N, 16224N, 16225N, 16226N, 16227N, 16228N, 16229N, 16230N, 16231N, 16232N, 16233N, 16234N, 16235N, 16236N, 16237N, 16238N, 16239N, 16240N, 16241N, 16242N, 16243N, 16244N, 16245N, 16246N, 16247N, 16248N, 16249N, 16250N, 16251N, 16252N, 16253N, 16254N, 16255N, 16256N, 16257N, 16258N, 16259N, 16260N, 16261N, 16262N, 16263N, 16264N, 16265N, 16266N, 16267N, 16268N, 16269N, 16270N, 16271N, 16272N, 16273N, 16274N, 16275N, 16276N, 16277N, 16278N, 16279N, 16280N, 16281N, 16282N, 16283N, 16284N, 16285N, 16286N, 16287N, 16288N, 16289N, 16290N, 16291N, 16292N, 16293N, 16294N, 16295N, 16296N, 16297N, 16298N, 16299N, 16300N, 16301N, 16302N, 16303N, 16304N, 16305N, 16306N, 16307N, 16308N, 16309N, 16310N, 16311N, 16312N, 16313N, 16314N, 16315N, 16316N, 16317N, 16318N, 16319N, 16320N, 16321N, 16322N, 16323N, 16324N, 16325N, 16326N, 16327N, 16328N, 16329N, 16330N, 16331N, 16332N, 16333N, 16334N, 16335N, 16336N, 16337N, 16338N, 16339N, 16340N, 16341N, 16342N, 16343N, 16344N, 16345N, 16346N, 16347N, 16348N, 16349N, 16350N, 16351N, 16352N, 16353N, 16354N, 16355N, 16356N, 16357N; B1c_HQ439486:356d, 2226+C, 4743, 5237+G, 5274+G, 15385+T, 16126N, 16127N, 16128N, 16129N, 16130N, 16131N, 16132N, 16133N, 16134N, 16135N, 16136N, 16137N, 16138N, 16139N, 16140N, 16141N, 16142N, 16143N, 16144N, 16145N, 16146N, 16147N, 16148N, 16149N, 16150N, 16151N, 16152N, 16153N, 16154N, 16155N, 16156N, 16157N, 16158N, 16159N, 16160N, 16161N, 16162N, 16163N, 16164N, 16165N, 16166N, 16167N, 16168N, 16169N, 16170N, 16171N, 16172N, 16173N, 16174N, 16175N, 16176N, 16177N, 16178N, 16179N, 16180N, 16181N, 16182N, 16183N, 16184N, 16185N, 16186N, 16187N, 16188N, 16189N, 16190N, 16191N, 16192N, 16193N, 16194N, 16195N, 16196N, 16197N, 16198N, 16199N, 16200N, 16201N, 16202N, 16203N, 16204N, 16205N, 16206N, 16207N, 16208N, 16209N, 16210N, 16211N, 16212N, 16213N, 16214N, 16215N, 16216N, 16217N, 16218N, 16219N, 16220N, 16221N, 16222N, 16223N, 16224N, 16225N, 16226N, 16227N, 16228N, 16229N, 16230N, 16231N, 16232N, 16233N, 16234N, 16235N, 16236N, 16237N, 16238N, 16239N, 16240N, 16241N, 16242N, 16243N, 16244N, 16245N, 16246N, 16247N, 16248N, 16249N, 16250N, 16251N, 16252N, 16253N, 16254N, 16255N, 16256N, 16257N, 16258N, 16259N, 16260N, 16261N, 16262N, 16263N, 16264N, 16265N, 16266N, 16267N, 16268N, 16269N, 16270N, 16271N, 16272N, 16273N, 16274N, 16275N, 16276N, | Biology. 328 (11), 1471-2148 (2011) |
| | | | | | | 16277N, 16278N, 16279N, 16280N, 16281N, 16282N, 16283N, 16284N, 16285N, 16286N, 16287N, 16288N, 16289N, 16290N, 16291N, 16292N, 16293N, 16294N, 16295N, 16296N, 16297N, 16298N, 16299N, 16300N, 16301N, 16302N, 16303N, 16304N, 16305N, 16306N, 16307N, 16308N, 16309N, 16310N, 16311N, 16312N, 16313N, 16314N, 16315N, 16316N, 16317N, 16318N, 16319N, 16320N, 16321N, 16322N, 16323N, 16324N, 16325N, 16326N, 16327N, 16328N, 16329N, 16330N, 16331N, 16332N, 16333N, 16334N, 16335N, 16336N, 16337N, 16338N, 16339N, 16340N, 16341N, 16342N, 16343N, 16344N, 16345N, 16346N, 16347N, 16348N, 16349N, 16350N, 16351N, 16352N, 16353N, 16354N, 16355N, 16356N, 16357N | |
| HQ439487 (Horse and Przewalski's horse) | Illumina/Solexa Genome Analyzer II | Phantom mutaions: 358d, 1387A, 2227+T, 5237+G, 5277+A, 15383+G. | C2a_HQ439487 | 158, 267, 356d, 957, 1387A, 2227+T, 2238, 2788, 3800, 4062, 4599, 4884, 4993, 5237+G, 5277+A, 5500, 6076, 6784, 7001, 7811, 7829, 7942, 8161, 9239, 9664, 10214, 10217, 11046, 11129, 11240, 11543, 12352, 12806, 13079, 15383+G, 15492, 15594, 15599, 15647, 15717, 15823, 15867, 15876, 15953, 15971, 16110, 16126N, 16127N, 16128N, 16129N, 16130N, 16131N, 16132N, 16133N, 16134N, 16135N, 16136N, 16137N, 16138N, 16139N, 16140N, 16141N, 16142N, 16143N, 16144N, 16145N, 16146N, 16147N, 16148N, 16149N, 16150N, 16151N, 16152N, 16153N, 16154N, 16155N, 16156N, 16157N, 16158N, 16159N, 16160N, 16161N, 16162N, 16163N, 16164N, 16165N, 16166N, 16167N, 16168N, 16169N, 16170N, 16171N, 16172N, 16173N, 16174N, 16175N, 16176N, 16177N, 16178N, 16179N, 16180N, 16181N, 16182N, 16183N, 16184N, 16185N, 16186N, 16187N, 16188N, 16189N, 16190N, 16191N, 16192N, 16193N, 16194N, 16195N, 16196N, 16197N, 16198N, 16199N, 16200N, 16201N, 16202N, 16203N, 16204N, 16205N, 16206N, 16207N, 16208N, 16209N, 16210N, 16211N, 16212N, 16213N, 16214N, 16215N, 16216N, 16217N, 16218N, 16219N, 16220N, 16221N, 16222N, 16223N, 16224N, 16225N, 16226N, 16227N, 16228N, 16229N, 16230N, 16231N, 16232N, 16233N, 16234N, 16235N, 16236N, 16237N, 16238N, 16239N, 16240N, 16241N, 16242N, 16243N, 16244N, 16245N, 16246N, 16247N, 16248N, 16249N, 16250N, 16251N, 16252N, 16253N, 16254N, 16255N, 16256N, 16257N, 16258N, 16259N, 16260N, 16261N, 16262N, 16263N, 16264N, 16265N, 16266N, 16267N, 16268N, 16269N, 16270N, 16271N, 16272N, 16273N, 16274N, 16275N, 16276N, 16277N, 16278N, 16279N, 16280N, 16281N, 16282N, 16283N, 16284N, 16285N, 16286N, 16287N, 16288N, 16289N, 16290N, 16291N, 16292N, 16293N, 16294N, 16295N, 16296N, 16297N, 16298N, 16299N, 16300N, 16301N, 16302N, 16303N, 16304N, 16305N, 16306N, 16307N, 16308N, 16309N, 16310N, 16311N, 16312N, 16313N, 16314N, 16315N, 16316N, 16317N, 16318N, 16319N, 16320N, 16321N, 16322N, 16323N, 16324N, 16325N, 16326N, 16327N, 16328N, 16329N, 16330N, 16331N, 16332N, 16333N, 16334N, 16335N, 16336N, 16337N, 16338N, 16339N, 16340N, 16341N, 16342N, 16343N, 16344N, 16345N, 16346N, 16347N, 16348N, 16349N, 16350N, 16351N, 16352N, 16353N, 16354N, 16355N, 16356N, 16357N, 16368 | C2a_HQ439487:356 | C2a_HQ439487:356d, 2227+T, 5237+G, 5277+A, 15383+G, 16126N, 16127N, 16128N, 16129N, 16130N, 16131N, 16132N, 16133N, 16134N, 16135N, 16136N, 16137N, 16138N, 16139N, 16140N, 16141N, 16142N, 16143N, 16144N, 16145N, 16146N, 16147N, 16148N, 16149N, 16150N, 16151N, 16152N, 16153N, 16154N, 16155N, 16156N, 16157N, 16158N, 16159N, 16160N, 16161N, 16162N, 16163N, 16164N, 16165N, 16166N, 16167N, 16168N, 16169N, 16170N, 16171N, 16172N, 16173N, 16174N, 16175N, 16176N, 16177N, 16178N, 16179N, 16180N, 16181N, 16182N, 16183N, 16184N, 16185N, 16186N, 16187N, 16188N, 16189N, 16190N, 16191N, 16192N, 16193N, 16194N, 16195N, 16196N, 16197N, 16198N, 16199N, 16200N, 16201N, 16202N, 16203N, 16204N, 16205N, 16206N, 16207N, 16208N, 16209N, 16210N, 16211N, 16212N, 16213N, 16214N, 16215N, 16216N, 16217N, 16218N, 16219N, 16220N, 16221N, 16222N, 16223N, 16224N, 16225N, 16226N, 16227N, 16228N, 16229N, 16230N, 16231N, 16232N, 16233N, 16234N, 16235N, 16236N, 16237N, 16238N, 16239N, 16240N, 16241N, 16242N, 16243N, 16244N, 16245N, 16246N, 16247N, 16248N, 16249N, 16250N, 16251N, 16252N, 16253N, 16254N, 16255N, 16256N, 16257N, 16258N, 16259N, 16260N, 16261N, 16262N, 16263N, 16264N, 16265N, 16266N, 16267N, 16268N, 16269N, 16270N, 16271N, 16272N, 16273N, 16274N, 16275N, 16276N, 16277N, 16278N, 16279N, 16280N, 16281N, 16282N, 16283N, 16284N, 16285N, 16286N, 16287N, 16288N, 16289N, 16290N, 16291N, 16292N, 16293N, 16294N, 16295N, 16296N, 16297N, 16298N, 16299N, 16300N, 16301N, 16302N, 16303N, 16304N, 16305N, 16306N, 16307N, 16308N, 16309N, 16310N, 16311N, 16312N, 16313N, 16314N, 16315N, 16316N, 16317N, 16318N, 16319N, 16320N, 16321N, 16322N, 16323N, 16324N, 16325N, 16326N, 16327N, 16328N, 16329N, 16330N, 16331N, 16332N, 16333N, 16334N, 16335N, 16336N, 16337N, 16338N, 16339N, 16340N, 16341N, 16342N, 16343N, 16344N, 16345N, 16346N, 16347N, 16348N, 16349N, 16350N, 16351N, 16352N, 16353N, 16354N, 16355N, 16356N, 16357N | Lippold,S., Matzke,N., Reissman,M., Burbano,H. and Hofreiter,M. BMC Evolutionary Biology. 328 (11), 1471-2148 (2011) |
| HQ439488 (Horse and Przewalski's | Illumina/Solexa Genome Analyzer II | Phantom mutaions: 358d, 2227+T, 5239+A, | B1a1_HQ439488 | 1-3d, 158, 356d, 859, 2227+T, 4062, 4230, 5239+A, 5277+A, 6784, 7627, 9961, 10764, 11240, 13816, 15383+G, 15492, 15582, 15647, 15663, 15717, 15807, 15823, 16052, 16110, 16126N, 16127N, 16128N, 16129N, 16130N, 16131N, 16132N, 16133N, 16134N, 16135N, 16136N, 16137N, 16138N, 16139N, 16140N, 16141N, 16142N, 16143N, 16144N, 16145N, 16146N, 16147N, 16148N, 16149N, 16150N, | B1a1_HQ439488:356 | B1a1_HQ439488:1-3d, 356d, 2227+T, 5239+A, 5277+A, 15383+G, 16126N, 16127N, 16128N, 16129N, 16130N, 16131N, 16132N, 16133N, 16134N, 16135N, 16136N, 16137N, 16138N, 16139N, 16140N, 16141N, 16142N, 16143N, 16144N, 16145N, 16146N, 16147N, 16148N, 16149N, 16150N, 16151N, | Lippold,S., Matzke,N., Reissman,M., Burbano,H. and Hofreiter,M. BMC Evolutionary |
| horse) | | 5277+A, 15383+G. | | 16151N, 16152N, 16153N, 16154N, 16155N, 16156N, 16157N, 16158N, 16159N, 16160N, 16161N, 16162N, 16163N, 16164N, 16165N, 16166N, 16167N, 16168N, 16169N, 16170N, 16171N, 16172N, 16173N, 16174N, 16175N, 16176N, 16177N, 16178N, 16179N, 16180N, 16181N, 16182N, 16183N, 16184N, 16185N, 16186N, 16187N, 16188N, 16189N, 16190N, 16191N, 16192N, 16193N, 16194N, 16195N, 16196N, 16197N, 16198N, 16199N, 16200N, 16201N, 16202N, 16203N, 16204N, 16205N, 16206N, 16207N, 16208N, 16209N, 16210N, 16211N, 16212N, 16213N, 16214N, 16215N, 16216N, 16217N, 16218N, 16219N, 16220N, 16221N, 16222N, 16223N, 16224N, 16225N, 16226N, 16227N, 16228N, 16229N, 16230N, 16231N, 16232N, 16233N, 16234N, 16235N, 16236N, 16237N, 16238N, 16239N, 16240N, 16241N, 16242N, 16243N, 16244N, 16245N, 16246N, 16247N, 16248N, 16249N, 16250N, 16251N, 16252N, 16253N, 16254N, 16255N, 16256N, 16257N, 16258N, 16259N, 16260N, 16261N, 16262N, 16263N, 16264N, 16265N, 16266N, 16267N, 16268N, 16269N, 16270N, 16271N, 16272N, 16273N, 16274N, 16275N, 16276N, 16277N, 16278N, 16279N, 16280N, 16281N, 16282N, 16283N, 16284N, 16285N, 16286N, 16287N, 16288N, 16289N, 16290N, 16291N, 16292N, 16293N, 16294N, 16295N, 16296N, 16297N, 16298N, 16299N, 16300N, 16301N, 16302N, 16303N, 16304N, 16305N, 16306N, 16307N, 16308N, 16309N, 16310N, 16311N, 16312N, 16313N, 16314N, 16315N, 16316N, 16317N, 16318N, 16319N, 16320N, 16321N, 16322N, 16323N, 16324N, 16325N, 16326N, 16327N, 16328N, 16329N, 16330N, 16331N, 16332N, 16333N, 16334N, 16335N, 16336N, 16337N, 16338N, 16339N, 16340N, 16341N, 16342N, 16343N, 16344N, 16345N, 16346N, 16347N, 16348N, 16349N, 16350N, 16351N, 16352N, 16353N, 16354N, 16355N, 16356N, 16357N, 16368 | | 16152N, 16153N, 16154N, 16155N, 16156N, 16157N, 16158N, 16159N, 16160N, 16161N, 16162N, 16163N, 16164N, 16165N, 16166N, 16167N, 16168N, 16169N, 16170N, 16171N, 16172N, 16173N, 16174N, 16175N, 16176N, 16177N, 16178N, 16179N, 16180N, 16181N, 16182N, 16183N, 16184N, 16185N, 16186N, 16187N, 16188N, 16189N, 16190N, 16191N, 16192N, 16193N, 16194N, 16195N, 16196N, 16197N, 16198N, 16199N, 16200N, 16201N, 16202N, 16203N, 16204N, 16205N, 16206N, 16207N, 16208N, 16209N, 16210N, 16211N, 16212N, 16213N, 16214N, 16215N, 16216N, 16217N, 16218N, 16219N, 16220N, 16221N, 16222N, 16223N, 16224N, 16225N, 16226N, 16227N, 16228N, 16229N, 16230N, 16231N, 16232N, 16233N, 16234N, 16235N, 16236N, 16237N, 16238N, 16239N, 16240N, 16241N, 16242N, 16243N, 16244N, 16245N, 16246N, 16247N, 16248N, 16249N, 16250N, 16251N, 16252N, 16253N, 16254N, 16255N, 16256N, 16257N, 16258N, 16259N, 16260N, 16261N, 16262N, 16263N, 16264N, 16265N, 16266N, 16267N, 16268N, 16269N, 16270N, 16271N, 16272N, 16273N, 16274N, 16275N, 16276N, 16277N, 16278N, 16279N, 16280N, 16281N, 16282N, 16283N, 16284N, 16285N, 16286N, 16287N, 16288N, 16289N, 16290N, 16291N, 16292N, 16293N, 16294N, 16295N, 16296N, 16297N, 16298N, 16299N, 16300N, 16301N, 16302N, 16303N, 16304N, 16305N, 16306N, 16307N, 16308N, 16309N, 16310N, 16311N, 16312N, 16313N, 16314N, 16315N, 16316N, 16317N, 16318N, 16319N, 16320N, 16321N, 16322N, 16323N, 16324N, 16325N, 16326N, 16327N, 16328N, 16329N, 16330N, 16331N, 16332N, 16333N, 16334N, 16335N, 16336N, 16337N, 16338N, 16339N, 16340N, 16341N, 16342N, 16343N, 16344N, 16345N, 16346N, 16347N, 16348N, 16349N, 16350N, 16351N, 16352N, 16353N, 16354N, 16355N, 16356N, 16357N | Biology. 328 (11), 1471-2148 (2011) |
| HQ439489 (Horse and Przewalski's horse) | Illumina/Solexa Genome Analyzer II | Phantom mutaions: 358d, 1387G, 2227+T, 5239+A, 5274+G, 15385+T. | D1a_HQ439489 | 158, 356d, 1387G, 1694, 2217, 2227+T, 2788, 3557, 4062, 4646, 5239+A, 5274+G, 5497, 5884, 6004, 6532, 6784, 7001, 8867, 9239, 9402, 10214, 10376, 11240, 11417, 11457, 11543, 12191, 12218, 12791, 13049, 13333, 13589, 13948, 14200, 15385+T, 15492, 15518, 15599, 15717, 15734, 15767, 15807, 15867, 15981, 16084, 16108, 16110, 16126N, 16127N, 16128N, 16129N, 16130N, 16131N, 16132N, 16133N, 16134N, 16135N, 16136N, 16137N, 16138N, 16139N, 16140N, 16141N, 16142N, 16143N, 16144N, 16145N, 16146N, 16147N, 16148N, 16149N, 16150N, 16151N, 16152N, 16153N, 16154N, 16155N, 16156N, 16157N, 16158N, 16159N, 16160N, 16161N, 16162N, 16163N, 16164N, 16165N, 16166N, 16167N, 16168N, 16169N, 16170N, 16171N, 16172N, 16173N, 16174N, 16175N, 16176N, 16177N, 16178N, 16179N, 16180N, 16181N, 16182N, 16183N, 16184N, 16185N, 16186N, 16187N, 16188N, 16189N, 16190N, 16191N, 16192N, 16193N, 16194N, 16195N, 16196N, 16197N, 16198N, 16199N, 16200N, 16201N, 16202N, 16203N, 16204N, 16205N, 16206N, 16207N, 16208N, 16209N, 16210N, 16211N, 16212N, 16213N, 16214N, 16215N, 16216N, 16217N, 16218N, 16219N, 16220N, 16221N, 16222N, 16223N, 16224N, 16225N, 16226N, 16227N, 16228N, 16229N, 16230N, 16231N, 16232N, 16233N, 16234N, 16235N, 16236N, 16237N, 16238N, 16239N, 16240N, 16241N, 16242N, 16243N, 16244N, 16245N, 16246N, 16247N, 16248N, 16249N, 16250N, 16251N, 16252N, 16253N, 16254N, 16255N, 16256N, 16257N, 16258N, 16259N, 16260N, 16261N, 16262N, 16263N, 16264N, 16265N, 16266N, 16267N, 16268N, 16269N, 16270N, 16271N, 16272N, 16273N, 16274N, 16275N, 16276N, 16277N, 16278N, 16279N, 16280N, 16281N, 16282N, 16283N, 16284N, 16285N, 16286N, 16287N, 16288N, 16289N, 16290N, 16291N, 16292N, 16293N, 16294N, 16295N, 16296N, | D1a_HQ439489:356, 1387d | D1a_HQ439489:356d, 1387G, 2227+T, 5239+A, 5274+G, 15385+T, 16126N, 16127N, 16128N, 16129N, 16130N, 16131N, 16132N, 16133N, 16134N, 16135N, 16136N, 16137N, 16138N, 16139N, 16140N, 16141N, 16142N, 16143N, 16144N, 16145N, 16146N, 16147N, 16148N, 16149N, 16150N, 16151N, 16152N, 16153N, 16154N, 16155N, 16156N, 16157N, 16158N, 16159N, 16160N, 16161N, 16162N, 16163N, 16164N, 16165N, 16166N, 16167N, 16168N, 16169N, 16170N, 16171N, 16172N, 16173N, 16174N, 16175N, 16176N, 16177N, 16178N, 16179N, 16180N, 16181N, 16182N, 16183N, 16184N, 16185N, 16186N, 16187N, 16188N, 16189N, 16190N, 16191N, 16192N, 16193N, 16194N, 16195N, 16196N, 16197N, 16198N, 16199N, 16200N, 16201N, 16202N, 16203N, 16204N, 16205N, 16206N, 16207N, 16208N, 16209N, 16210N, 16211N, 16212N, 16213N, 16214N, 16215N, 16216N, 16217N, 16218N, 16219N, 16220N, 16221N, 16222N, 16223N, 16224N, 16225N, 16226N, 16227N, 16228N, 16229N, 16230N, 16231N, 16232N, 16233N, 16234N, 16235N, 16236N, 16237N, 16238N, 16239N, 16240N, 16241N, 16242N, 16243N, 16244N, 16245N, 16246N, 16247N, 16248N, 16249N, 16250N, 16251N, 16252N, 16253N, 16254N, 16255N, 16256N, 16257N, 16258N, 16259N, 16260N, 16261N, 16262N, 16263N, 16264N, 16265N, 16266N, 16267N, 16268N, 16269N, 16270N, 16271N, 16272N, | Lippold,S., Matzke,N., Reissman,M., Burbano,H. and Hofreiter,M. BMC Evolutionary Biology. 328 (11), 1471-2148 (2011) |
| | | | | 16297N, 16298N, 16299N, 16300N, 16301N, 16302N, 16303N, 16304N, 16305N, 16306N, 16307N, 16308N, 16309N, 16310N, 16311N, 16312N, 16313N, 16314N, 16315N, 16316N, 16317N, 16318N, 16319N, 16320N, 16321N, 16322N, 16323N, 16324N, 16325N, 16326N, 16327N, 16328N, 16329N, 16330N, 16331N, 16332N, 16333N, 16334N, 16335N, 16336N, 16337N, 16338N, 16339N, 16340N, 16341N, 16342N, 16343N, 16344N, 16345N, 16346N, 16347N, 16348N, 16349N, 16350N, 16351N, 16352N, 16353N, 16354N, 16355N, 16356N, 16357N, 16368, 16391 | | 16273N, 16274N, 16275N, 16276N, 16277N, 16278N, 16279N, 16280N, 16281N, 16282N, 16283N, 16284N, 16285N, 16286N, 16287N, 16288N, 16289N, 16290N, 16291N, 16292N, 16293N, 16294N, 16295N, 16296N, 16297N, 16298N, 16299N, 16300N, 16301N, 16302N, 16303N, 16304N, 16305N, 16306N, 16307N, 16308N, 16309N, 16310N, 16311N, 16312N, 16313N, 16314N, 16315N, 16316N, 16317N, 16318N, 16319N, 16320N, 16321N, 16322N, 16323N, 16324N, 16325N, 16326N, 16327N, 16328N, 16329N, 16330N, 16331N, 16332N, 16333N, 16334N, 16335N, 16336N, 16337N, 16338N, 16339N, 16340N, 16341N, 16342N, 16343N, 16344N, 16345N, 16346N, 16347N, 16348N, 16349N, 16350N, 16351N, 16352N, 16353N, 16354N, 16355N, 16356N, 16357N | |
| HQ439490 (Horse and Przewalski's horse) | Illumina/Solexa Genome Analyzer II | Phantom mutaions: 358d, 1387A, 2226+C, 5237+G, 5277+A, 15385+T. | H_HQ439490 | 158, 356d, 1387A, 1546N, 1551N, 1552N, 1553N, 1554N, 1668, 1735N, 1736N, 1737N, 1738N, 1739N, 1740N, 1741N, 1742N, 1743N, 1744N, 1745N, 1746N, 1747N, 1748N, 1749N, 1750N, 1751N, 1752N, 1753N, 1754N, 1755N, 1756N, 1757N, 1758N, 1759N, 2226+C, 2788, 3064, 4062, 4182, 4646, 4669, 4830, 5237+G, 5272, 5277+A, 5884, 6004, 6076, 6127, 6307, 6565, 6784, 6835, 7001, 7093, 7111, 7258, 8005, 8481, 8741, 8763, 8776, 9209, 9239, 9856, 10083, 10214, 10376, 10914, 11240, 11543, 11842, 12767, 13049, 13333, 13502, 13761, 14335, 15004, 15385+T, 15492, 15523, 15537, 15582, 15599, 15646, 15715, 15717, 15768, 15867, 15971, 16078, 16108, 16110, 16126N, 16127N, 16128N, 16129N, 16130N, 16131N, 16132N, 16133N, 16134N, 16135N, 16136N, 16137N, 16138N, 16139N, 16140N, 16141N, 16142N, 16143N, 16144N, 16145N, 16146N, 16147N, 16148N, 16149N, 16150N, 16151N, 16152N, 16153N, 16154N, 16155N, 16156N, 16157N, 16158N, 16159N, 16160N, 16161N, 16162N, 16163N, 16164N, 16165N, 16166N, 16167N, 16168N, 16169N, 16170N, 16171N, 16172N, 16173N, 16174N, 16175N, 16176N, 16177N, 16178N, 16179N, 16180N, 16181N, 16182N, 16183N, 16184N, 16185N, 16186N, 16187N, 16188N, 16189N, 16190N, 16191N, 16192N, 16193N, 16194N, 16195N, 16196N, 16197N, 16198N, 16199N, 16200N, 16201N, 16202N, 16203N, 16204N, 16205N, 16206N, 16207N, 16208N, 16209N, 16210N, 16211N, 16212N, 16213N, 16214N, 16215N, 16216N, 16217N, 16218N, 16219N, 16220N, 16221N, 16222N, 16223N, 16224N, 16225N, 16226N, 16227N, 16228N, 16229N, 16230N, 16231N, 16232N, 16233N, 16234N, 16235N, 16236N, 16237N, 16238N, 16239N, 16240N, 16241N, 16242N, 16243N, 16244N, 16245N, 16246N, 16247N, 16248N, 16249N, 16250N, 16251N, 16252N, 16253N, 16254N, 16255N, 16256N, 16257N, 16258N, 16259N, 16260N, 16261N, 16262N, 16263N, 16264N, 16265N, 16266N, 16267N, 16268N, 16269N, 16270N, 16271N, 16272N, 16273N, 16274N, 16275N, 16276N, 16277N, 16278N, 16279N, 16280N, 16281N, 16282N, 16283N, 16284N, 16285N, 16286N, 16287N, 16288N, 16289N, 16290N, 16291N, 16292N, 16293N, 16294N, 16295N, 16296N, 16297N, 16298N, 16299N, 16300N, 16301N, 16302N, 16303N, 16304N, 16305N, 16306N, 16307N, 16308N, 16309N, 16310N, 16311N, 16312N, 16313N, 16314N, 16315N, 16316N, 16317N, 16318N, 16319N, 16320N, 16321N, 16322N, 16323N, 16324N, 16325N, 16326N, 16327N, 16328N, 16329N, 16330N, 16331N, 16332N, 16333N, 16334N, 16335N, 16336N, 16337N, 16338N, 16339N, 16340N, 16341N, 16342N, 16343N, 16344N, 16345N, 16346N, 16347N, 16348N, 16349N, 16350N, 16351N, 16352N, 16353N, 16354N, 16355N, 16356N, 16357N, 16368, 16657d | H_HQ439490:356, 1387d, 1546 | H_HQ439490:356d, 1387A, 1546N, 1551N, 1552N, 1553N, 1554N, 1735N, 1736N, 1737N, 1738N, 1739N, 1740N, 1741N, 1742N, 1743N, 1744N, 1745N, 1746N, 1747N, 1748N, 1749N, 1750N, 1751N, 1752N, 1753N, 1754N, 1755N, 1756N, 1757N, 1758N, 1759N, 2226+C, 5237+G, 5277+A, 15385+T, 16126N, 16127N, 16128N, 16129N, 16130N, 16131N, 16132N, 16133N, 16134N, 16135N, 16136N, 16137N, 16138N, 16139N, 16140N, 16141N, 16142N, 16143N, 16144N, 16145N, 16146N, 16147N, 16148N, 16149N, 16150N, 16151N, 16152N, 16153N, 16154N, 16155N, 16156N, 16157N, 16158N, 16159N, 16160N, 16161N, 16162N, 16163N, 16164N, 16165N, 16166N, 16167N, 16168N, 16169N, 16170N, 16171N, 16172N, 16173N, 16174N, 16175N, 16176N, 16177N, 16178N, 16179N, 16180N, 16181N, 16182N, 16183N, 16184N, 16185N, 16186N, 16187N, 16188N, 16189N, 16190N, 16191N, 16192N, 16193N, 16194N, 16195N, 16196N, 16197N, 16198N, 16199N, 16200N, 16201N, 16202N, 16203N, 16204N, 16205N, 16206N, 16207N, 16208N, 16209N, 16210N, 16211N, 16212N, 16213N, 16214N, 16215N, 16216N, 16217N, 16218N, 16219N, 16220N, 16221N, 16222N, 16223N, 16224N, 16225N, 16226N, 16227N, 16228N, 16229N, 16230N, 16231N, 16232N, 16233N, 16234N, 16235N, 16236N, 16237N, 16238N, 16239N, 16240N, 16241N, 16242N, 16243N, 16244N, 16245N, 16246N, 16247N, 16248N, 16249N, 16250N, 16251N, 16252N, 16253N, 16254N, 16255N, 16256N, 16257N, 16258N, 16259N, 16260N, 16261N, 16262N, 16263N, 16264N, 16265N, 16266N, 16267N, 16268N, 16269N, 16270N, 16271N, 16272N, 16273N, 16274N, 16275N, 16276N, 16277N, 16278N, 16279N, 16280N, 16281N, 16282N, 16283N, 16284N, 16285N, 16286N, 16287N, 16288N, 16289N, 16290N, 16291N, 16292N, 16293N, 16294N, 16295N, 16296N, 16297N, 16298N, 16299N, 16300N, 16301N, 16302N, 16303N, 16304N, 16305N, 16306N, 16307N, 16308N, 16309N, 16310N, 16311N, 16312N, 16313N, 16314N, 16315N, 16316N, 16317N, 16318N, 16319N, 16320N, 16321N, 16322N, 16323N, 16324N, 16325N, 16326N, 16327N, 16328N, 16329N, 16330N, 16331N, 16332N, 16333N, 16334N, 16335N, 16336N, 16337N, 16338N, 16339N, 16340N, 16341N, 16342N, 16343N, 16344N, 16345N, 16346N, 16347N, 16348N, 16349N, 16350N, 16351N, 16352N, 16353N, 16354N, 16355N, 16356N, 16357N | Lippold,S., Matzke,N., Reissman,M., Burbano,H. and Hofreiter,M. BMC Evolutionary Biology. 328 (11), 1471-2148 (2011) |
| HQ439491 (Horse and Przewalski's horse) | Illumina/Solexa Genome Analyzer II | Phantom mutaions: 358d,, 2227+T, 5239+A, 5277+A, 15385+T. | A1a | 358d, 2226+C, 3557, 5239+A, 5277+A, 6784, 12653, 15385+T, 15629, 16126N, 16127N, 16128N, 16129N, 16130N, 16131N, 16132N, 16133N, 16134N, 16135N, 16136N, 16137N, 16138N, 16139N, 16140N, 16141N, 16142N, 16143N, 16144N, 16145N, 16146N, 16147N, 16148N, 16149N, 16150N, 16151N, 16152N, 16153N, 16154N, 16155N, 16156N, 16157N, 16158N, 16159N, 16160N, 16161N, 16162N, 16163N, 16164N, 16165N, 16166N, 16167N, 16168N, 16169N, 16170N, 16171N, 16172N, 16173N, 16174N, 16175N, 16176N, 16177N, 16178N, 16179N, 16180N, 16181N, 16182N, 16183N, 16184N, 16185N, 16186N, 16187N, 16188N, 16189N, 16190N, 16191N, 16192N, 16193N, 16194N, 16195N, 16196N, 16197N, 16198N, 16199N, 16200N, 16201N, 16202N, 16203N, 16204N, 16205N, 16206N, 16207N, 16208N, 16209N, 16210N, 16211N, 16212N, 16213N, 16214N, 16215N, 16216N, 16217N, 16218N, 16219N, 16220N, 16221N, 16222N, 16223N, 16224N, 16225N, 16226N, 16227N, 16228N, 16229N, 16230N, 16231N, 16232N, 16233N, 16234N, 16235N, 16236N, 16237N, 16238N, 16239N, 16240N, 16241N, 16242N, 16243N, 16244N, 16245N, 16246N, 16247N, 16248N, 16249N, 16250N, 16251N, 16252N, 16253N, 16254N, 16255N, 16256N, 16257N, 16258N, 16259N, 16260N, 16261N, 16262N, 16263N, 16264N, 16265N, 16266N, 16267N, 16268N, 16269N, 16270N, 16271N, 16272N, 16273N, 16274N, 16275N, 16276N, 16277N, 16278N, 16279N, 16280N, 16281N, 16282N, 16283N, 16284N, 16285N, 16286N, 16287N, 16288N, 16289N, 16290N, 16291N, 16292N, 16293N, 16294N, 16295N, 16296N, 16297N, 16298N, 16299N, 16300N, 16301N, 16302N, 16303N, 16304N, 16305N, 16306N, 16307N, 16308N, 16309N, 16310N, 16311N, 16312N, 16313N, 16314N, 16315N, 16316N, 16317N, 16318N, 16319N, 16320N, 16321N, 16322N, 16323N, 16324N, 16325N, 16326N, 16327N, 16328N, 16329N, 16330N, 16331N, 16332N, 16333N, 16334N, 16335N, 16336N, 16337N, 16338N, 16339N, 16340N, 16341N, 16342N, 16343N, 16344N, 16345N, 16346N, 16347N, 16348N, 16349N, 16350N, 16351N, 16352N, 16353N, 16354N, 16355N, 16356N, 16357N | | A1a:358d, 2226+C, 3557, 5239+A, 5277+A, 6784, 12653, 15385+T, 15629, 16126N, 16127N, 16128N, 16129N, 16130N, 16131N, 16132N, 16133N, 16134N, 16135N, 16136N, 16137N, 16138N, 16139N, 16140N, 16141N, 16142N, 16143N, 16144N, 16145N, 16146N, 16147N, 16148N, 16149N, 16150N, 16151N, 16152N, 16153N, 16154N, 16155N, 16156N, 16157N, 16158N, 16159N, 16160N, 16161N, 16162N, 16163N, 16164N, 16165N, 16166N, 16167N, 16168N, 16169N, 16170N, 16171N, 16172N, 16173N, 16174N, 16175N, 16176N, 16177N, 16178N, 16179N, 16180N, 16181N, 16182N, 16183N, 16184N, 16185N, 16186N, 16187N, 16188N, 16189N, 16190N, 16191N, 16192N, 16193N, 16194N, 16195N, 16196N, 16197N, 16198N, 16199N, 16200N, 16201N, 16202N, 16203N, 16204N, 16205N, 16206N, 16207N, 16208N, 16209N, 16210N, 16211N, 16212N, 16213N, 16214N, 16215N, 16216N, 16217N, 16218N, 16219N, 16220N, 16221N, 16222N, 16223N, 16224N, 16225N, 16226N, 16227N, 16228N, 16229N, 16230N, 16231N, 16232N, 16233N, 16234N, 16235N, 16236N, 16237N, 16238N, 16239N, 16240N, 16241N, 16242N, 16243N, 16244N, 16245N, 16246N, 16247N, 16248N, 16249N, 16250N, 16251N, 16252N, 16253N, 16254N, 16255N, 16256N, 16257N, 16258N, 16259N, 16260N, 16261N, 16262N, 16263N, 16264N, 16265N, 16266N, 16267N, 16268N, 16269N, 16270N, 16271N, 16272N, 16273N, 16274N, 16275N, 16276N, 16277N, 16278N, 16279N, 16280N, 16281N, 16282N, 16283N, 16284N, 16285N, 16286N, 16287N, 16288N, 16289N, 16290N, 16291N, 16292N, 16293N, 16294N, 16295N, 16296N, 16297N, 16298N, 16299N, 16300N, 16301N, 16302N, 16303N, 16304N, 16305N, 16306N, 16307N, 16308N, 16309N, 16310N, 16311N, 16312N, 16313N, 16314N, 16315N, 16316N, 16317N, 16318N, 16319N, 16320N, 16321N, 16322N, 16323N, 16324N, 16325N, 16326N, 16327N, 16328N, 16329N, 16330N, 16331N, 16332N, 16333N, 16334N, 16335N, 16336N, 16337N, 16338N, 16339N, 16340N, 16341N, 16342N, 16343N, 16344N, 16345N, 16346N, 16347N, 16348N, 16349N, 16350N, 16351N, 16352N, 16353N, 16354N, 16355N, 16356N, 16357N | Lippold,S., Matzke,N., Reissman,M., Burbano,H. and Hofreiter,M. BMC Evolutionary Biology. 328 (11), 1471-2148 (2011) |
| HQ439492 (Horse and Przewalski's horse) | Illumina/Solexa Genome Analyzer II | Phantom mutaions: 358d, 1387G, 2226+C, 5239+A, 5274+G, 15385+T. | L2a2a_HQ439492 | 158, 356d, 961, 1375, 1387G, 2226+C, 2607, 2788, 2899, 3517, 3942, 4062, 4536, 4646, 4669, 5239+A, 5274+G, 5527, 5815, 5884, 6004, 6307, 6619, 6784, 7001, 7516, 7666, 7900, 8005, 8058, 8199, 8301, 8319, 8358, 8403, 8565, 9239, 9951, 10110, 10214, 10292, 10376, 10421, 10613, 11240, 11543, 11693, 11842, 11879, 12119, 12200, 12767, 12896, 12950, 13049, 13333, 13520, 14803, 14825, 14995, 15313, 15385+T, 15491, 15492, 15493, 15531, 15582, 15600, 15601, 15646, 15717, 15768, 15867, 15868, 15953, 15971, 16065, 16100, 16126N, 16127N, 16128N, 16129N, 16130N, 16131N, 16132N, 16133N, 16134N, 16135N, 16136N, 16137N, 16138N, 16139N, 16140N, 16141N, 16142N, 16143N, 16144N, 16145N, 16146N, 16147N, 16148N, 16149N, 16150N, 16151N, 16152N, 16153N, 16154N, 16155N, 16156N, 16157N, 16158N, 16159N, 16160N, 16161N, 16162N, 16163N, 16164N, 16165N, 16166N, 16167N, 16168N, 16169N, 16170N, 16171N, 16172N, 16173N, 16174N, 16175N, 16176N, 16177N, 16178N, 16179N, 16180N, 16181N, 16182N, 16183N, 16184N, 16185N, 16186N, 16187N, 16188N, 16189N, 16190N, 16191N, 16192N, 16193N, 16194N, 16195N, 16196N, 16197N, 16198N, 16199N, 16200N, 16201N, 16202N, 16203N, 16204N, 16205N, 16206N, 16207N, 16208N, 16209N, 16210N, 16211N, 16212N, 16213N, 16214N, 16215N, 16216N, 16217N, 16218N, 16219N, 16220N, 16221N, 16222N, 16223N, 16224N, 16225N, 16226N, 16227N, 16228N, 16229N, | L2a2a_HQ439492:356 | L2a2a_HQ439492:356d, 2226+C, 5239+A, 5274+G, 15385+T, 16126N, 16127N, 16128N, 16129N, 16130N, 16131N, 16132N, 16133N, 16134N, 16135N, 16136N, 16137N, 16138N, 16139N, 16140N, 16141N, 16142N, 16143N, 16144N, 16145N, 16146N, 16147N, 16148N, 16149N, 16150N, 16151N, 16152N, 16153N, 16154N, 16155N, 16156N, 16157N, 16158N, 16159N, 16160N, 16161N, 16162N, 16163N, 16164N, 16165N, 16166N, 16167N, 16168N, 16169N, 16170N, 16171N, 16172N, 16173N, 16174N, 16175N, 16176N, 16177N, 16178N, 16179N, 16180N, 16181N, 16182N, 16183N, 16184N, 16185N, 16186N, 16187N, 16188N, 16189N, 16190N, 16191N, 16192N, 16193N, 16194N, 16195N, 16196N, 16197N, 16198N, 16199N, 16200N, 16201N, 16202N, 16203N, 16204N, 16205N, 16206N, 16207N, 16208N, 16209N, 16210N, 16211N, 16212N, 16213N, 16214N, 16215N, 16216N, 16217N, 16218N, 16219N, 16220N, 16221N, 16222N, 16223N, 16224N, 16225N, 16226N, 16227N, 16228N, 16229N, 16230N, 16231N, 16232N, 16233N, 16234N, 16235N, 16236N, 16237N, 16238N, 16239N, | Lippold,S., Matzke,N., Reissman,M., Burbano,H. and Hofreiter,M. BMC Evolutionary Biology. 328 (11), 1471-2148 (2011) |
| | | | | 16230N, 16231N, 16232N, 16233N, 16234N, 16235N, 16236N, 16237N, 16238N, 16239N, 16240N, 16241N, 16242N, 16243N, 16244N, 16245N, 16246N, 16247N, 16248N, 16249N, 16250N, 16251N, 16252N, 16253N, 16254N, 16255N, 16256N, 16257N, 16258N, 16259N, 16260N, 16261N, 16262N, 16263N, 16264N, 16265N, 16266N, 16267N, 16268N, 16269N, 16270N, 16271N, 16272N, 16273N, 16274N, 16275N, 16276N, 16277N, 16278N, 16279N, 16280N, 16281N, 16282N, 16283N, 16284N, 16285N, 16286N, 16287N, 16288N, 16289N, 16290N, 16291N, 16292N, 16293N, 16294N, 16295N, 16296N, 16297N, 16298N, 16299N, 16300N, 16301N, 16302N, 16303N, 16304N, 16305N, 16306N, 16307N, 16308N, 16309N, 16310N, 16311N, 16312N, 16313N, 16314N, 16315N, 16316N, 16317N, 16318N, 16319N, 16320N, 16321N, 16322N, 16323N, 16324N, 16325N, 16326N, 16327N, 16328N, 16329N, 16330N, 16331N, 16332N, 16333N, 16334N, 16335N, 16336N, 16337N, 16338N, 16339N, 16340N, 16341N, 16342N, 16343N, 16344N, 16345N, 16346N, 16347N, 16348N, 16349N, 16350N, 16351N, 16352N, 16353N, 16354N, 16355N, 16356N, 16357N, 16368, 16398A, 16626, 16649 | | 16240N, 16241N, 16242N, 16243N, 16244N, 16245N, 16246N, 16247N, 16248N, 16249N, 16250N, 16251N, 16252N, 16253N, 16254N, 16255N, 16256N, 16257N, 16258N, 16259N, 16260N, 16261N, 16262N, 16263N, 16264N, 16265N, 16266N, 16267N, 16268N, 16269N, 16270N, 16271N, 16272N, 16273N, 16274N, 16275N, 16276N, 16277N, 16278N, 16279N, 16280N, 16281N, 16282N, 16283N, 16284N, 16285N, 16286N, 16287N, 16288N, 16289N, 16290N, 16291N, 16292N, 16293N, 16294N, 16295N, 16296N, 16297N, 16298N, 16299N, 16300N, 16301N, 16302N, 16303N, 16304N, 16305N, 16306N, 16307N, 16308N, 16309N, 16310N, 16311N, 16312N, 16313N, 16314N, 16315N, 16316N, 16317N, 16318N, 16319N, 16320N, 16321N, 16322N, 16323N, 16324N, 16325N, 16326N, 16327N, 16328N, 16329N, 16330N, 16331N, 16332N, 16333N, 16334N, 16335N, 16336N, 16337N, 16338N, 16339N, 16340N, 16341N, 16342N, 16343N, 16344N, 16345N, 16346N, 16347N, 16348N, 16349N, 16350N, 16351N, 16352N, 16353N, 16354N, 16355N, 16356N, 16357N | |
| HQ439493 (Horse and Przewalski's horse) | Illumina/Solexa Genome Analyzer II | Phantom mutaions: 358d, 1387A, 2226+C, 5237+G, 5274+G, 15385+T. | L2a2_HQ439493 | 158, 180, 356d, 961, 1375, 1387A, 2226+C, 2607, 2788, 2899, 3517, 3942, 4062, 4536, 4646, 4669, 5237+G, 5274+G, 5527, 5815, 5884, 6004, 6307, 6784, 7001, 7516, 7666, 7900, 8005, 8058, 8199, 8301, 8319, 8358, 8403, 8565, 9239, 9951, 10110, 10214, 10292, 10376, 10421, 10613, 11240, 11543, 11693, 11842, 11879, 12119, 12200, 12767, 12896, 12950, 13049, 13333, 13520, 14803, 14995, 15313, 15385+T, 15491, 15492, 15493, 15531, 15600, 15646, 15717, 15768, 15867, 15868, 15953, 15971, 16065, 16100, 16126N, 16127N, 16128N, 16129N, 16130N, 16131N, 16132N, 16133N, 16134N, 16135N, 16136N, 16137N, 16138N, 16139N, 16140N, 16141N, 16142N, 16143N, 16144N, 16145N, 16146N, 16147N, 16148N, 16149N, 16150N, 16151N, 16152N, 16153N, 16154N, 16155N, 16156N, 16157N, 16158N, 16159N, 16160N, 16161N, 16162N, 16163N, 16164N, 16165N, 16166N, 16167N, 16168N, 16169N, 16170N, 16171N, 16172N, 16173N, 16174N, 16175N, 16176N, 16177N, 16178N, 16179N, 16180N, 16181N, 16182N, 16183N, 16184N, 16185N, 16186N, 16187N, 16188N, 16189N, 16190N, 16191N, 16192N, 16193N, 16194N, 16195N, 16196N, 16197N, 16198N, 16199N, 16200N, 16201N, 16202N, 16203N, 16204N, 16205N, 16206N, 16207N, 16208N, 16209N, 16210N, 16211N, 16212N, 16213N, 16214N, 16215N, 16216N, 16217N, 16218N, 16219N, 16220N, 16221N, 16222N, 16223N, 16224N, 16225N, 16226N, 16227N, 16228N, 16229N, 16230N, 16231N, 16232N, 16233N, 16234N, 16235N, 16236N, 16237N, 16238N, 16239N, 16240N, 16241N, 16242N, 16243N, 16244N, 16245N, 16246N, 16247N, 16248N, 16249N, 16250N, 16251N, 16252N, 16253N, 16254N, 16255N, 16256N, 16257N, 16258N, 16259N, 16260N, 16261N, 16262N, 16263N, 16264N, 16265N, 16266N, 16267N, 16268N, 16269N, 16270N, 16271N, 16272N, 16273N, 16274N, 16275N, 16276N, 16277N, 16278N, 16279N, 16280N, 16281N, 16282N, 16283N, 16284N, 16285N, 16286N, 16287N, 16288N, 16289N, 16290N, 16291N, 16292N, 16293N, 16294N, 16295N, 16296N, 16297N, 16298N, 16299N, 16300N, 16301N, 16302N, 16303N, 16304N, 16305N, 16306N, 16307N, 16308N, 16309N, 16310N, 16311N, 16312N, 16313N, 16314N, 16315N, 16316N, 16317N, 16318N, 16319N, 16320N, 16321N, 16322N, 16323N, 16324N, 16325N, 16326N, 16327N, 16328N, 16329N, 16330N, 16331N, 16332N, 16333N, 16334N, 16335N, 16336N, 16337N, 16338N, 16339N, 16340N, 16341N, 16342N, 16343N, 16344N, 16345N, 16346N, 16347N, 16348N, 16349N, 16350N, 16351N, 16352N, 16353N, 16354N, 16355N, 16356N, 16357N, 16368, 16398A, 16626, 16657d | L2a2_HQ439493:356 | L2a2_HQ439493:356d, 2226+C, 5237+G, 5274+G, 15385+T, 16126N, 16127N, 16128N, 16129N, 16130N, 16131N, 16132N, 16133N, 16134N, 16135N, 16136N, 16137N, 16138N, 16139N, 16140N, 16141N, 16142N, 16143N, 16144N, 16145N, 16146N, 16147N, 16148N, 16149N, 16150N, 16151N, 16152N, 16153N, 16154N, 16155N, 16156N, 16157N, 16158N, 16159N, 16160N, 16161N, 16162N, 16163N, 16164N, 16165N, 16166N, 16167N, 16168N, 16169N, 16170N, 16171N, 16172N, 16173N, 16174N, 16175N, 16176N, 16177N, 16178N, 16179N, 16180N, 16181N, 16182N, 16183N, 16184N, 16185N, 16186N, 16187N, 16188N, 16189N, 16190N, 16191N, 16192N, 16193N, 16194N, 16195N, 16196N, 16197N, 16198N, 16199N, 16200N, 16201N, 16202N, 16203N, 16204N, 16205N, 16206N, 16207N, 16208N, 16209N, 16210N, 16211N, 16212N, 16213N, 16214N, 16215N, 16216N, 16217N, 16218N, 16219N, 16220N, 16221N, 16222N, 16223N, 16224N, 16225N, 16226N, 16227N, 16228N, 16229N, 16230N, 16231N, 16232N, 16233N, 16234N, 16235N, 16236N, 16237N, 16238N, 16239N, 16240N, 16241N, 16242N, 16243N, 16244N, 16245N, 16246N, 16247N, 16248N, 16249N, 16250N, 16251N, 16252N, 16253N, 16254N, 16255N, 16256N, 16257N, 16258N, 16259N, 16260N, 16261N, 16262N, 16263N, 16264N, 16265N, 16266N, 16267N, 16268N, 16269N, 16270N, 16271N, 16272N, 16273N, 16274N, 16275N, 16276N, 16277N, 16278N, 16279N, 16280N, 16281N, 16282N, 16283N, 16284N, 16285N, 16286N, 16287N, 16288N, 16289N, 16290N, 16291N, 16292N, 16293N, 16294N, 16295N, 16296N, 16297N, 16298N, 16299N, 16300N, 16301N, 16302N, 16303N, 16304N, 16305N, 16306N, 16307N, 16308N, 16309N, 16310N, 16311N, 16312N, 16313N, 16314N, 16315N, 16316N, 16317N, 16318N, 16319N, 16320N, 16321N, 16322N, 16323N, 16324N, 16325N, 16326N, 16327N, 16328N, 16329N, 16330N, 16331N, 16332N, 16333N, 16334N, 16335N, 16336N, 16337N, 16338N, 16339N, 16340N, 16341N, 16342N, 16343N, 16344N, 16345N, 16346N, 16347N, 16348N, 16349N, 16350N, 16351N, 16352N, 16353N, 16354N, 16355N, 16356N, 16357N | Lippold,S., Matzke,N., Reissman,M., Burbano,H. and Hofreiter,M. BMC Evolutionary Biology. 328 (11), 1471-2148 (2011) |
| HQ439494 (Horse and Przewalski's horse) | Illumina/Solexa Genome Analyzer II | Phantom mutaions: 358d, 2227+T, 5237+G, 5277+A, 15385+T. | L1a1a_HQ439494 | 158, 356d, 961, 1375, 2227+T, 2607, 2788, 2899, 3517, 3942, 4062, 4536, 4646N, 4669, 5237+G, 5277+A, 5500, 5527, 5815, 5884, 5932, 6004, 6307, 6784, 7001, 7516, 7666, 7900, 8005, 8058, 8301, 8319, 8358, 8403, 8481, 8565, 9239, 9951, 10110, 10214, 10292, 10376, 10421, 10613, 11240, 11543, 11693, 11842, 11879, 12119, 12200, 12671, 12767, 12896, 12950, 13049, 13333, 13520, 13785, 14029, 14803, 14995, 15313, 15385+T, 15491, 15492, 15493, 15531, 15582, 15600, 15646, 15682, 15717, 15768, 15867, 15868, 15953, 15971, 16100, 16126N, 16127N, 16128N, 16129N, 16130N, 16131N, 16132N, 16133N, 16134N, 16135N, 16136N, 16137N, 16138N, 16139N, 16140N, 16141N, 16142N, 16143N, 16144N, 16145N, 16146N, 16147N, 16148N, 16149N, 16150N, 16151N, 16152N, 16153N, 16154N, 16155N, 16156N, 16157N, 16158N, 16159N, 16160N, 16161N, 16162N, 16163N, 16164N, 16165N, 16166N, 16167N, 16168N, 16169N, 16170N, 16171N, 16172N, 16173N, 16174N, 16175N, 16176N, 16177N, 16178N, 16179N, 16180N, 16181N, 16182N, 16183N, 16184N, 16185N, 16186N, 16187N, 16188N, 16189N, 16190N, 16191N, 16192N, 16193N, 16194N, 16195N, 16196N, 16197N, 16198N, 16199N, 16200N, 16201N, 16202N, 16203N, 16204N, 16205N, 16206N, 16207N, 16208N, 16209N, 16210N, 16211N, 16212N, 16213N, 16214N, 16215N, 16216N, 16217N, 16218N, 16219N, 16220N, 16221N, 16222N, 16223N, 16224N, 16225N, 16226N, 16227N, 16228N, 16229N, 16230N, 16231N, 16232N, 16233N, 16234N, 16235N, 16236N, 16237N, 16238N, 16239N, 16240N, 16241N, 16242N, 16243N, 16244N, 16245N, 16246N, 16247N, 16248N, 16249N, 16250N, 16251N, 16252N, 16253N, 16254N, 16255N, 16256N, 16257N, 16258N, 16259N, 16260N, 16261N, 16262N, 16263N, 16264N, 16265N, 16266N, 16267N, 16268N, 16269N, 16270N, 16271N, 16272N, 16273N, 16274N, 16275N, 16276N, 16277N, 16278N, 16279N, 16280N, 16281N, 16282N, 16283N, 16284N, 16285N, 16286N, 16287N, 16288N, 16289N, 16290N, 16291N, 16292N, 16293N, 16294N, 16295N, 16296N, 16297N, 16298N, 16299N, 16300N, 16301N, 16302N, 16303N, 16304N, 16305N, 16306N, 16307N, 16308N, 16309N, 16310N, 16311N, 16312N, 16313N, 16314N, 16315N, 16316N, 16317N, 16318N, 16319N, 16320N, 16321N, 16322N, 16323N, 16324N, 16325N, 16326N, 16327N, 16328N, 16329N, 16330N, 16331N, 16332N, 16333N, 16334N, 16335N, 16336N, 16337N, 16338N, 16339N, 16340N, 16341N, 16342N, 16343N, 16344N, 16345N, 16346N, 16347N, 16348N, 16349N, 16350N, 16351N, 16352N, 16353N, 16354N, 16355N, 16356N, 16357N, 16368, 16398A, 16626 | L1a1a_HQ439494:356, 4646 | L1a1a_HQ439494:356d, 2227+T, 4646N, 5237+G, 5277+A, 15385+T, 16126N, 16127N, 16128N, 16129N, 16130N, 16131N, 16132N, 16133N, 16134N, 16135N, 16136N, 16137N, 16138N, 16139N, 16140N, 16141N, 16142N, 16143N, 16144N, 16145N, 16146N, 16147N, 16148N, 16149N, 16150N, 16151N, 16152N, 16153N, 16154N, 16155N, 16156N, 16157N, 16158N, 16159N, 16160N, 16161N, 16162N, 16163N, 16164N, 16165N, 16166N, 16167N, 16168N, 16169N, 16170N, 16171N, 16172N, 16173N, 16174N, 16175N, 16176N, 16177N, 16178N, 16179N, 16180N, 16181N, 16182N, 16183N, 16184N, 16185N, 16186N, 16187N, 16188N, 16189N, 16190N, 16191N, 16192N, 16193N, 16194N, 16195N, 16196N, 16197N, 16198N, 16199N, 16200N, 16201N, 16202N, 16203N, 16204N, 16205N, 16206N, 16207N, 16208N, 16209N, 16210N, 16211N, 16212N, 16213N, 16214N, 16215N, 16216N, 16217N, 16218N, 16219N, 16220N, 16221N, 16222N, 16223N, 16224N, 16225N, 16226N, 16227N, 16228N, 16229N, 16230N, 16231N, 16232N, 16233N, 16234N, 16235N, 16236N, 16237N, 16238N, 16239N, 16240N, 16241N, 16242N, 16243N, 16244N, 16245N, 16246N, 16247N, 16248N, 16249N, 16250N, 16251N, 16252N, 16253N, 16254N, 16255N, 16256N, 16257N, 16258N, 16259N, 16260N, 16261N, 16262N, 16263N, 16264N, 16265N, 16266N, 16267N, 16268N, 16269N, 16270N, 16271N, 16272N, 16273N, 16274N, 16275N, 16276N, 16277N, 16278N, 16279N, 16280N, 16281N, 16282N, 16283N, 16284N, 16285N, 16286N, 16287N, 16288N, 16289N, 16290N, 16291N, 16292N, 16293N, 16294N, 16295N, 16296N, 16297N, 16298N, 16299N, 16300N, 16301N, 16302N, 16303N, 16304N, 16305N, 16306N, 16307N, 16308N, 16309N, 16310N, 16311N, 16312N, 16313N, 16314N, 16315N, 16316N, 16317N, 16318N, 16319N, 16320N, 16321N, 16322N, 16323N, 16324N, 16325N, 16326N, 16327N, 16328N, 16329N, 16330N, 16331N, 16332N, 16333N, 16334N, 16335N, 16336N, 16337N, 16338N, 16339N, 16340N, 16341N, 16342N, 16343N, 16344N, 16345N, 16346N, 16347N, 16348N, 16349N, 16350N, 16351N, 16352N, 16353N, 16354N, 16355N, 16356N, 16357N | Lippold,S., Matzke,N., Reissman,M., Burbano,H. and Hofreiter,M. BMC Evolutionary Biology. 328 (11), 1471-2148 (2011) |
| HQ439495 (Horse and Przewalski's horse) | Illumina/Solexa Genome Analyzer II | Phantom mutaions: 358d, 1387A, 2227+T, 5239+A, 5274+G, 15385+T. | M1b, M1b_HQ439495 | 158, 356d, 427, 961, 1387A, 1609T, 2227+T, 2339A, 2788, 3070, 3100, 3211T, 3475, 3800, 4062, 4526, 4536, 4599G, 4605, 4646, 4669, 4898, 5103, 5239+A, 5274+G, 5527, 5827, 5884, 6004, 6076, 6307, 6712, 6784, 7001, 7432, 7666, 7900, 8005, 8043, 8076, 8150, 8175, 8238, 8358, 8556T, 8565, 8798, 9086, 9332, 9540, 10110, 10173, 10214, 10292, 10376, 10448, 10460, 10515, 10646, 10859, 11240, 11394, 11492, 11513, 11543, 11803, 11842, 11879, 11966, 12029, 12095, 12332, 12767, 13049, 13100, 13333, 13356, 13502, 13615, 13629, 13720, 13920, 13933, 14422, 14626, 14671, 14803, 14815, 15052A, 15133, 15342, 15385+T, 15492, 15582, 15599, 15614, 15656, 15717, 15768, 15803, 15824, 15866, 15953, 16065, 16077, 16118, 16126N, 16127N, 16128N, 16129N, 16130N, 16131N, 16132N, 16133N, 16134N, 16135N, 16136N, 16137N, 16138N, 16139N, 16140N, 16141N, 16142N, 16143N, 16144N, 16145N, 16146N, 16147N, 16148N, 16149N, 16150N, 16151N, 16152N, 16153N, 16154N, 16155N, 16156N, 16157N, 16158N, 16159N, 16160N, 16161N, 16162N, 16163N, 16164N, 16165N, 16166N, 16167N, 16168N, 16169N, 16170N, 16171N, 16172N, 16173N, 16174N, 16175N, 16176N, 16177N, 16178N, 16179N, 16180N, 16181N, 16182N, 16183N, 16184N, 16185N, 16186N, 16187N, 16188N, 16189N, 16190N, 16191N, 16192N, 16193N, 16194N, 16195N, 16196N, 16197N, 16198N, 16199N, 16200N, 16201N, 16202N, 16203N, 16204N, 16205N, 16206N, | M1b:356;<br>M1b_HQ439495:356 | M1b:356d, 2227+T, 5239+A, 5274+G, 11803, 15385+T, 15582, 16126N, 16127N, 16128N, 16129N, 16130N, 16131N, 16132N, 16133N, 16134N, 16135N, 16136N, 16137N, 16138N, 16139N, 16140N, 16141N, 16142N, 16143N, 16144N, 16145N, 16146N, 16147N, 16148N, 16149N, 16150N, 16151N, 16152N, 16153N, 16154N, 16155N, 16156N, 16157N, 16158N, 16159N, 16160N, 16161N, 16162N, 16163N, 16164N, 16165N, 16166N, 16167N, 16168N, 16169N, 16170N, 16171N, 16172N, 16173N, 16174N, 16175N, 16176N, 16177N, 16178N, 16179N, 16180N, 16181N, 16182N, 16183N, 16184N, 16185N, 16186N, 16187N, 16188N, 16189N, 16190N, 16191N, 16192N, 16193N, 16194N, 16195N, 16196N, 16197N, 16198N, 16199N, 16200N, 16201N, 16202N, 16203N, 16204N, 16205N, 16206N, 16207N, 16208N, 16209N, 16210N, 16211N, 16212N, 16213N, 16214N, 16215N, 16216N, 16217N, 16218N, 16219N, 16220N, 16221N, 16222N, 16223N, 16224N, 16225N, 16226N, 16227N, 16228N, 16229N, 16230N, 16231N, 16232N, 16233N, 16234N, 16235N, 16236N, 16237N, 16238N, 16239N, | Lippold,S., Matzke,N., Reissman,M., Burbano,H. and Hofreiter,M. BMC Evolutionary Biology. 328 (11), 1471-2148 (2011) |
| | | | | 16207N, 16208N, 16209N, 16210N, 16211N, 16212N, 16213N, 16214N, 16215N, 16216N, 16217N, 16218N, 16219N, 16220N, 16221N, 16222N, 16223N, 16224N, 16225N, 16226N, 16227N, 16228N, 16229N, 16230N, 16231N, 16232N, 16233N, 16234N, 16235N, 16236N, 16237N, 16238N, 16239N, 16240N, 16241N, 16242N, 16243N, 16244N, 16245N, 16246N, 16247N, 16248N, 16249N, 16250N, 16251N, 16252N, 16253N, 16254N, 16255N, 16256N, 16257N, 16258N, 16259N, 16260N, 16261N, 16262N, 16263N, 16264N, 16265N, 16266N, 16267N, 16268N, 16269N, 16270N, 16271N, 16272N, 16273N, 16274N, 16275N, 16276N, 16277N, 16278N, 16279N, 16280N, 16281N, 16282N, 16283N, 16284N, 16285N, 16286N, 16287N, 16288N, 16289N, 16290N, 16291N, 16292N, 16293N, 16294N, 16295N, 16296N, 16297N, 16298N, 16299N, 16300N, 16301N, 16302N, 16303N, 16304N, 16305N, 16306N, 16307N, 16308N, 16309N, 16310N, 16311N, 16312N, 16313N, 16314N, 16315N, 16316N, 16317N, 16318N, 16319N, 16320N, 16321N, 16322N, 16323N, 16324N, 16325N, 16326N, 16327N, 16328N, 16329N, 16330N, 16331N, 16332N, 16333N, 16334N, 16335N, 16336N, 16337N, 16338N, 16339N, 16340N, 16341N, 16342N, 16343N, 16344N, 16345N, 16346N, 16347N, 16348N, 16349N, 16350N, 16351N, 16352N, 16353N, 16354N, 16355N, 16356N, 16357N, 16368, 16540A, 16543, 16556, 16626, 16653N | | 16240N, 16241N, 16242N, 16243N, 16244N, 16245N, 16246N, 16247N, 16248N, 16249N, 16250N, 16251N, 16252N, 16253N, 16254N, 16255N, 16256N, 16257N, 16258N, 16259N, 16260N, 16261N, 16262N, 16263N, 16264N, 16265N, 16266N, 16267N, 16268N, 16269N, 16270N, 16271N, 16272N, 16273N, 16274N, 16275N, 16276N, 16277N, 16278N, 16279N, 16280N, 16281N, 16282N, 16283N, 16284N, 16285N, 16286N, 16287N, 16288N, 16289N, 16290N, 16291N, 16292N, 16293N, 16294N, 16295N, 16296N, 16297N, 16298N, 16299N, 16300N, 16301N, 16302N, 16303N, 16304N, 16305N, 16306N, 16307N, 16308N, 16309N, 16310N, 16311N, 16312N, 16313N, 16314N, 16315N, 16316N, 16317N, 16318N, 16319N, 16320N, 16321N, 16322N, 16323N, 16324N, 16325N, 16326N, 16327N, 16328N, 16329N, 16330N, 16331N, 16332N, 16333N, 16334N, 16335N, 16336N, 16337N, 16338N, 16339N, 16340N, 16341N, 16342N, 16343N, 16344N, 16345N, 16346N, 16347N, 16348N, 16349N, 16350N, 16351N, 16352N, 16353N, 16354N, 16355N, 16356N, 16357N, 16653N;<br>M1b_HQ439495:356d, 1387A, 2227+T, 5239+A, 5274+G, 10646, 15385+T, 16126N, 16127N, 16128N, 16129N, 16130N, 16131N, 16132N, 16133N, 16134N, 16135N, 16136N, 16137N, 16138N, 16139N, 16140N, 16141N, 16142N, 16143N, 16144N, 16145N, 16146N, 16147N, 16148N, 16149N, 16150N, 16151N, 16152N, 16153N, 16154N, 16155N, 16156N, 16157N, 16158N, 16159N, 16160N, 16161N, 16162N, 16163N, 16164N, 16165N, 16166N, 16167N, 16168N, 16169N, 16170N, 16171N, 16172N, 16173N, 16174N, 16175N, 16176N, 16177N, 16178N, 16179N, 16180N, 16181N, 16182N, 16183N, 16184N, 16185N, 16186N, 16187N, 16188N, 16189N, 16190N, 16191N, 16192N, 16193N, 16194N, 16195N, 16196N, 16197N, 16198N, 16199N, 16200N, 16201N, 16202N, 16203N, 16204N, 16205N, 16206N, 16207N, 16208N, 16209N, 16210N, 16211N, 16212N, 16213N, 16214N, 16215N, 16216N, 16217N, 16218N, 16219N, 16220N, 16221N, 16222N, 16223N, 16224N, 16225N, 16226N, 16227N, 16228N, 16229N, 16230N, 16231N, 16232N, 16233N, 16234N, 16235N, 16236N, 16237N, 16238N, 16239N, 16240N, 16241N, 16242N, 16243N, 16244N, 16245N, 16246N, 16247N, 16248N, 16249N, 16250N, 16251N, 16252N, 16253N, 16254N, 16255N, 16256N, 16257N, 16258N, 16259N, 16260N, 16261N, 16262N, 16263N, 16264N, 16265N, 16266N, 16267N, 16268N, 16269N, 16270N, 16271N, 16272N, 16273N, 16274N, 16275N, 16276N, 16277N, 16278N, 16279N, 16280N, 16281N, 16282N, 16283N, 16284N, 16285N, 16286N, 16287N, 16288N, 16289N, 16290N, 16291N, 16292N, 16293N, 16294N, 16295N, 16296N, 16297N, 16298N, 16299N, 16300N, 16301N, 16302N, 16303N, 16304N, 16305N, 16306N, 16307N, 16308N, 16309N, 16310N, 16311N, 16312N, 16313N, 16314N, 16315N, 16316N, 16317N, 16318N, 16319N, 16320N, 16321N, 16322N, 16323N, 16324N, 16325N, 16326N, 16327N, 16328N, 16329N, 16330N, 16331N, 16332N, 16333N, 16334N, 16335N, 16336N, 16337N, 16338N, 16339N, 16340N, 16341N, 16342N, 16343N, 16344N, 16345N, 16346N, 16347N, 16348N, 16349N, 16350N, 16351N, 16352N, 16353N, 16354N, 16355N, 16356N, 16357N, 16653N | |
| HQ439496 (Horse and Przewalski's horse) | Illumina/Solexa Genome Analyzer II | Phantom mutaions: 358d, 1387G, 2227+T, 5237+G, 5274+G, 15385+T. | I2a4_HQ439496 | 158, 356d, 1387G, 1587, 1791, 2227+T, 2614, 2770, 2788, 4062, 4063, 4392, 4646, 4669, 4830, 5061, 5210, 5237+G, 5274+G, 5884, 6004, 6175, 6247, 6307, 6784, 7001, 8005, 8358, 8379, 8792, 9239, 9694, 9948, 10083, 10214, 10238, 10376, 11240, 11543, 11827, 11842, 12404, 12443, 12683, 12767, 13049, 13333, 13502, 13761, 14554, 15385+T, 15492, 15535, 15582, 15599, 15647, 15648, 15706, 15717, 15768, 15823, 15867, 15971, 16110, 16119, 16126N, 16127N, 16128N, 16129N, 16130N, 16131N, 16132N, 16133N, 16134N, 16135N, 16136N, 16137N, 16138N, 16139N, 16140N, 16141N, 16142N, 16143N, 16144N, 16145N, 16146N, 16147N, 16148N, 16149N, 16150N, 16151N, 16152N, 16153N, 16154N, 16155N, 16156N, 16157N, 16158N, 16159N, 16160N, 16161N, 16162N, 16163N, 16164N, 16165N, 16166N, 16167N, 16168N, 16169N, 16170N, 16171N, 16172N, 16173N, 16174N, 16175N, 16176N, 16177N, 16178N, 16179N, 16180N, 16181N, 16182N, 16183N, 16184N, 16185N, 16186N, 16187N, 16188N, 16189N, 16190N, 16191N, 16192N, 16193N, 16194N, 16195N, 16196N, 16197N, 16198N, 16199N, 16200N, 16201N, 16202N, 16203N, 16204N, 16205N, 16206N, 16207N, 16208N, 16209N, 16210N, 16211N, 16212N, 16213N, 16214N, 16215N, 16216N, 16217N, 16218N, 16219N, 16220N, 16221N, 16222N, 16223N, 16224N, 16225N, 16226N, 16227N, 16228N, 16229N, 16230N, 16231N, 16232N, 16233N, 16234N, 16235N, 16236N, 16237N, 16238N, 16239N, 16240N, 16241N, 16242N, 16243N, 16244N, 16245N, 16246N, 16247N, 16248N, 16249N, 16250N, 16251N, 16252N, 16253N, 16254N, 16255N, 16256N, 16257N, 16258N, 16259N, 16260N, 16261N, 16262N, 16263N, 16264N, 16265N, 16266N, 16267N, 16268N, 16269N, 16270N, 16271N, 16272N, 16273N, 16274N, 16275N, 16276N, 16277N, 16278N, 16279N, 16280N, 16281N, 16282N, 16283N, 16284N, 16285N, 16286N, 16287N, 16288N, 16289N, 16290N, 16291N, 16292N, 16293N, 16294N, 16295N, 16296N, 16297N, 16298N, 16299N, 16300N, 16301N, 16302N, 16303N, 16304N, 16305N, 16306N, 16307N, 16308N, 16309N, 16310N, 16311N, 16312N, 16313N, 16314N, 16315N, 16316N, 16317N, 16318N, 16319N, 16320N, 16321N, 16322N, 16323N, 16324N, 16325N, 16326N, 16327N, 16328N, 16329N, 16330N, 16331N, 16332N, 16333N, 16334N, 16335N, 16336N, 16337N, 16338N, 16339N, 16340N, 16341N, 16342N, 16343N, 16344N, 16345N, 16346N, 16347N, 16348N, 16349N, 16350N, 16351N, 16352N, 16353N, 16354N, 16355N, 16356N, 16357N, 16368, 16436 | I2a4_HQ439496:356 | I2a4_HQ439496:356d, 2227+T, 5237+G, 5274+G, 15385+T, 16126N, 16127N, 16128N, 16129N, 16130N, 16131N, 16132N, 16133N, 16134N, 16135N, 16136N, 16137N, 16138N, 16139N, 16140N, 16141N, 16142N, 16143N, 16144N, 16145N, 16146N, 16147N, 16148N, 16149N, 16150N, 16151N, 16152N, 16153N, 16154N, 16155N, 16156N, 16157N, 16158N, 16159N, 16160N, 16161N, 16162N, 16163N, 16164N, 16165N, 16166N, 16167N, 16168N, 16169N, 16170N, 16171N, 16172N, 16173N, 16174N, 16175N, 16176N, 16177N, 16178N, 16179N, 16180N, 16181N, 16182N, 16183N, 16184N, 16185N, 16186N, 16187N, 16188N, 16189N, 16190N, 16191N, 16192N, 16193N, 16194N, 16195N, 16196N, 16197N, 16198N, 16199N, 16200N, 16201N, 16202N, 16203N, 16204N, 16205N, 16206N, 16207N, 16208N, 16209N, 16210N, 16211N, 16212N, 16213N, 16214N, 16215N, 16216N, 16217N, 16218N, 16219N, 16220N, 16221N, 16222N, 16223N, 16224N, 16225N, 16226N, 16227N, 16228N, 16229N, 16230N, 16231N, 16232N, 16233N, 16234N, 16235N, 16236N, 16237N, 16238N, 16239N, 16240N, 16241N, 16242N, 16243N, 16244N, 16245N, 16246N, 16247N, 16248N, 16249N, 16250N, 16251N, 16252N, 16253N, 16254N, 16255N, 16256N, 16257N, 16258N, 16259N, 16260N, 16261N, 16262N, 16263N, 16264N, 16265N, 16266N, 16267N, 16268N, 16269N, 16270N, 16271N, 16272N, 16273N, 16274N, 16275N, 16276N, 16277N, 16278N, 16279N, 16280N, 16281N, 16282N, 16283N, 16284N, 16285N, 16286N, 16287N, 16288N, 16289N, 16290N, 16291N, 16292N, 16293N, 16294N, 16295N, 16296N, 16297N, 16298N, 16299N, 16300N, 16301N, 16302N, 16303N, 16304N, 16305N, 16306N, 16307N, 16308N, 16309N, 16310N, 16311N, 16312N, 16313N, 16314N, 16315N, 16316N, 16317N, 16318N, 16319N, 16320N, 16321N, 16322N, 16323N, 16324N, 16325N, 16326N, 16327N, 16328N, 16329N, 16330N, 16331N, 16332N, 16333N, 16334N, 16335N, 16336N, 16337N, 16338N, 16339N, 16340N, 16341N, 16342N, 16343N, 16344N, 16345N, 16346N, 16347N, 16348N, 16349N, 16350N, 16351N, 16352N, 16353N, 16354N, 16355N, 16356N, 16357N | Lippold,S., Matzke,N., Reissman,M., Burbano,H. and Hofreiter,M. BMC Evolutionary Biology. 328 (11), 1471-2148 (2011) |
| HQ439497 (Horse and Przewalski's horse) | Illumina/Solexa Genome Analyzer II | Phantom mutaions: 358d, 1387A, 2227+T, 5237+G, 5274+G, 15385+T. | I2b2b_HQ439497 | 158, 356d, 1387A, 1587, 1606, 1791, 2227+T, 2614, 2770, 2788, 4062, 4063, 4392, 4646, 4669, 4830, 5061, 5210, 5237+G, 5274+G, 5884, 6004, 6175, 6247, 6307, 6784, 7001, 8005, 8358, 8379, 8792, 9239, 9694, 9948, 10083, 10214, 10238, 10376, 11240, 11543, 11827, 11842, 12404, 12443, 12683, 12767, 13049, 13333, 13502, 13761, 14554, 15385+T, 15492, 15535, 15582, 15599, 15647, 15706, 15717, 15768, 15823, 15867, 15971, 16108, 16110, 16119, 16126N, 16127N, 16128N, 16129N, 16130N, 16131N, 16132N, 16133N, 16134N, 16135N, 16136N, 16137N, 16138N, 16139N, 16140N, 16141N, 16142N, 16143N, 16144N, 16145N, 16146N, 16147N, 16148N, 16149N, 16150N, 16151N, 16152N, 16153N, 16154N, 16155N, 16156N, 16157N, 16158N, 16159N, 16160N, 16161N, 16162N, 16163N, 16164N, 16165N, 16166N, 16167N, 16168N, 16169N, 16170N, 16171N, 16172N, 16173N, 16174N, 16175N, 16176N, 16177N, 16178N, 16179N, 16180N, 16181N, 16182N, 16183N, 16184N, 16185N, 16186N, 16187N, 16188N, 16189N, 16190N, 16191N, 16192N, 16193N, 16194N, 16195N, 16196N, 16197N, 16198N, 16199N, 16200N, 16201N, 16202N, 16203N, 16204N, 16205N, 16206N, 16207N, 16208N, 16209N, 16210N, 16211N, 16212N, 16213N, 16214N, 16215N, 16216N, 16217N, 16218N, 16219N, 16220N, 16221N, 16222N, 16223N, 16224N, 16225N, 16226N, 16227N, 16228N, 16229N, 16230N, 16231N, 16232N, 16233N, 16234N, 16235N, 16236N, 16237N, 16238N, | I2b2b_HQ439497:356 | I2b2b_HQ439497:356d, 2227+T, 5237+G, 5274+G, 15385+T, 16126N, 16127N, 16128N, 16129N, 16130N, 16131N, 16132N, 16133N, 16134N, 16135N, 16136N, 16137N, 16138N, 16139N, 16140N, 16141N, 16142N, 16143N, 16144N, 16145N, 16146N, 16147N, 16148N, 16149N, 16150N, 16151N, 16152N, 16153N, 16154N, 16155N, 16156N, 16157N, 16158N, 16159N, 16160N, 16161N, 16162N, 16163N, 16164N, 16165N, 16166N, 16167N, 16168N, 16169N, 16170N, 16171N, 16172N, 16173N, 16174N, 16175N, 16176N, 16177N, 16178N, 16179N, 16180N, 16181N, 16182N, 16183N, 16184N, 16185N, 16186N, 16187N, 16188N, 16189N, 16190N, 16191N, 16192N, 16193N, 16194N, 16195N, 16196N, 16197N, 16198N, 16199N, 16200N, 16201N, 16202N, 16203N, 16204N, 16205N, 16206N, 16207N, 16208N, 16209N, 16210N, 16211N, 16212N, 16213N, 16214N, 16215N, 16216N, 16217N, 16218N, 16219N, 16220N, 16221N, 16222N, 16223N, 16224N, 16225N, 16226N, 16227N, 16228N, 16229N, 16230N, 16231N, 16232N, 16233N, 16234N, 16235N, 16236N, 16237N, 16238N, 16239N, | Lippold,S., Matzke,N., Reissman,M., Burbano,H. and Hofreiter,M. BMC Evolutionary Biology. 328 (11), 1471-2148 (2011) |
| | | | | 16239N, 16240N, 16241N, 16242N, 16243N, 16244N, 16245N, 16246N, 16247N, 16248N, 16249N, 16250N, 16251N, 16252N, 16253N, 16254N, 16255N, 16256N, 16257N, 16258N, 16259N, 16260N, 16261N, 16262N, 16263N, 16264N, 16265N, 16266N, 16267N, 16268N, 16269N, 16270N, 16271N, 16272N, 16273N, 16274N, 16275N, 16276N, 16277N, 16278N, 16279N, 16280N, 16281N, 16282N, 16283N, 16284N, 16285N, 16286N, 16287N, 16288N, 16289N, 16290N, 16291N, 16292N, 16293N, 16294N, 16295N, 16296N, 16297N, 16298N, 16299N, 16300N, 16301N, 16302N, 16303N, 16304N, 16305N, 16306N, 16307N, 16308N, 16309N, 16310N, 16311N, 16312N, 16313N, 16314N, 16315N, 16316N, 16317N, 16318N, 16319N, 16320N, 16321N, 16322N, 16323N, 16324N, 16325N, 16326N, 16327N, 16328N, 16329N, 16330N, 16331N, 16332N, 16333N, 16334N, 16335N, 16336N, 16337N, 16338N, 16339N, 16340N, 16341N, 16342N, 16343N, 16344N, 16345N, 16346N, 16347N, 16348N, 16349N, 16350N, 16351N, 16352N, 16353N, 16354N, 16355N, 16356N, 16357N, 16368, 16436 | | 16240N, 16241N, 16242N, 16243N, 16244N, 16245N, 16246N, 16247N, 16248N, 16249N, 16250N, 16251N, 16252N, 16253N, 16254N, 16255N, 16256N, 16257N, 16258N, 16259N, 16260N, 16261N, 16262N, 16263N, 16264N, 16265N, 16266N, 16267N, 16268N, 16269N, 16270N, 16271N, 16272N, 16273N, 16274N, 16275N, 16276N, 16277N, 16278N, 16279N, 16280N, 16281N, 16282N, 16283N, 16284N, 16285N, 16286N, 16287N, 16288N, 16289N, 16290N, 16291N, 16292N, 16293N, 16294N, 16295N, 16296N, 16297N, 16298N, 16299N, 16300N, 16301N, 16302N, 16303N, 16304N, 16305N, 16306N, 16307N, 16308N, 16309N, 16310N, 16311N, 16312N, 16313N, 16314N, 16315N, 16316N, 16317N, 16318N, 16319N, 16320N, 16321N, 16322N, 16323N, 16324N, 16325N, 16326N, 16327N, 16328N, 16329N, 16330N, 16331N, 16332N, 16333N, 16334N, 16335N, 16336N, 16337N, 16338N, 16339N, 16340N, 16341N, 16342N, 16343N, 16344N, 16345N, 16346N, 16347N, 16348N, 16349N, 16350N, 16351N, 16352N, 16353N, 16354N, 16355N, 16356N, 16357N | |
| HQ439498 (Horse and Przewalski's horse) | Illumina/Solexa Genome Analyzer II | Phantom mutaions: 358d, 1387A, 2226+C, 5239+A, 5277+A, 15385+T. | L1b1_HQ439498 | 158, 356d, 961, 1375, 1387A, 2226+C, 2788, 2899, 3379, 3517, 3942, 4062, 4536, 4646, 4669, 5239+A, 5277+A, 5527, 5815, 5884, 6004, 6307, 6784, 7001, 7516, 7666, 7900, 8005, 8058, 8172, 8301, 8319, 8358, 8361, 8565, 9239, 9951, 10110, 10214, 10292, 10376, 10421, 10613, 11240, 11543, 11682, 11693, 11842, 11879, 12119, 12200, 12767, 12896, 12950, 13049, 13333, 13520, 14359, 14803, 14995, 15313, 15385+T, 15491, 15492, 15493, 15531, 15599, 15600, 15646, 15717, 15768, 15867, 15868, 15953, 15971, 16065, 16100, 16126N, 16127N, 16128N, 16129N, 16130N, 16131N, 16132N, 16133N, 16134N, 16135N, 16136N, 16137N, 16138N, 16139N, 16140N, 16141N, 16142N, 16143N, 16144N, 16145N, 16146N, 16147N, 16148N, 16149N, 16150N, 16151N, 16152N, 16153N, 16154N, 16155N, 16156N, 16157N, 16158N, 16159N, 16160N, 16161N, 16162N, 16163N, 16164N, 16165N, 16166N, 16167N, 16168N, 16169N, 16170N, 16171N, 16172N, 16173N, 16174N, 16175N, 16176N, 16177N, 16178N, 16179N, 16180N, 16181N, 16182N, 16183N, 16184N, 16185N, 16186N, 16187N, 16188N, 16189N, 16190N, 16191N, 16192N, 16193N, 16194N, 16195N, 16196N, 16197N, 16198N, 16199N, 16200N, 16201N, 16202N, 16203N, 16204N, 16205N, 16206N, 16207N, 16208N, 16209N, 16210N, 16211N, 16212N, 16213N, 16214N, 16215N, 16216N, 16217N, 16218N, 16219N, 16220N, 16221N, 16222N, 16223N, 16224N, 16225N, 16226N, 16227N, 16228N, 16229N, 16230N, 16231N, 16232N, 16233N, 16234N, 16235N, 16236N, 16237N, 16238N, 16239N, 16240N, 16241N, 16242N, 16243N, 16244N, 16245N, 16246N, 16247N, 16248N, 16249N, 16250N, 16251N, 16252N, 16253N, 16254N, 16255N, 16256N, 16257N, 16258N, 16259N, 16260N, 16261N, 16262N, 16263N, 16264N, 16265N, 16266N, 16267N, 16268N, 16269N, 16270N, 16271N, 16272N, 16273N, 16274N, 16275N, 16276N, 16277N, 16278N, 16279N, 16280N, 16281N, 16282N, 16283N, 16284N, 16285N, 16286N, 16287N, 16288N, 16289N, 16290N, 16291N, 16292N, 16293N, 16294N, 16295N, 16296N, 16297N, 16298N, 16299N, 16300N, 16301N, 16302N, 16303N, 16304N, 16305N, 16306N, 16307N, 16308N, 16309N, 16310N, 16311N, 16312N, 16313N, 16314N, 16315N, 16316N, 16317N, 16318N, 16319N, 16320N, 16321N, 16322N, 16323N, 16324N, 16325N, 16326N, 16327N, 16328N, 16329N, 16330N, 16331N, 16332N, 16333N, 16334N, 16335N, 16336N, 16337N, 16338N, 16339N, 16340N, 16341N, 16342N, 16343N, 16344N, 16345N, 16346N, 16347N, 16348N, 16349N, 16350N, 16351N, 16352N, 16353N, 16354N, 16355N, 16356N, 16357N, 16368, 16626 | L1b1_HQ439498:356 | L1b1_HQ439498:356d, 2226+C, 5239+A, 5277+A, 15385+T, 16126N, 16127N, 16128N, 16129N, 16130N, 16131N, 16132N, 16133N, 16134N, 16135N, 16136N, 16137N, 16138N, 16139N, 16140N, 16141N, 16142N, 16143N, 16144N, 16145N, 16146N, 16147N, 16148N, 16149N, 16150N, 16151N, 16152N, 16153N, 16154N, 16155N, 16156N, 16157N, 16158N, 16159N, 16160N, 16161N, 16162N, 16163N, 16164N, 16165N, 16166N, 16167N, 16168N, 16169N, 16170N, 16171N, 16172N, 16173N, 16174N, 16175N, 16176N, 16177N, 16178N, 16179N, 16180N, 16181N, 16182N, 16183N, 16184N, 16185N, 16186N, 16187N, 16188N, 16189N, 16190N, 16191N, 16192N, 16193N, 16194N, 16195N, 16196N, 16197N, 16198N, 16199N, 16200N, 16201N, 16202N, 16203N, 16204N, 16205N, 16206N, 16207N, 16208N, 16209N, 16210N, 16211N, 16212N, 16213N, 16214N, 16215N, 16216N, 16217N, 16218N, 16219N, 16220N, 16221N, 16222N, 16223N, 16224N, 16225N, 16226N, 16227N, 16228N, 16229N, 16230N, 16231N, 16232N, 16233N, 16234N, 16235N, 16236N, 16237N, 16238N, 16239N, 16240N, 16241N, 16242N, 16243N, 16244N, 16245N, 16246N, 16247N, 16248N, 16249N, 16250N, 16251N, 16252N, 16253N, 16254N, 16255N, 16256N, 16257N, 16258N, 16259N, 16260N, 16261N, 16262N, 16263N, 16264N, 16265N, 16266N, 16267N, 16268N, 16269N, 16270N, 16271N, 16272N, 16273N, 16274N, 16275N, 16276N, 16277N, 16278N, 16279N, 16280N, 16281N, 16282N, 16283N, 16284N, 16285N, 16286N, 16287N, 16288N, 16289N, 16290N, 16291N, 16292N, 16293N, 16294N, 16295N, 16296N, 16297N, 16298N, 16299N, 16300N, 16301N, 16302N, 16303N, 16304N, 16305N, 16306N, 16307N, 16308N, 16309N, 16310N, 16311N, 16312N, 16313N, 16314N, 16315N, 16316N, 16317N, 16318N, 16319N, 16320N, 16321N, 16322N, 16323N, 16324N, 16325N, 16326N, 16327N, 16328N, 16329N, 16330N, 16331N, 16332N, 16333N, 16334N, 16335N, 16336N, 16337N, 16338N, 16339N, 16340N, 16341N, 16342N, 16343N, 16344N, 16345N, 16346N, 16347N, 16348N, 16349N, 16350N, 16351N, 16352N, 16353N, 16354N, 16355N, 16356N, 16357N | Lippold,S., Matzke,N., Reissman,M., Burbano,H. and Hofreiter,M. BMC Evolutionary Biology. 328 (11), 1471-2148 (2011) |
| HQ439499 (Horse and Przewalski's horse) | Illumina/Solexa Genome Analyzer II | Phantom mutaions: 358d, 1387G, 2226+C, 5239+A, 5277+A, 15385+T. | I2a3_HQ439499 | 158, 356d, 1387G, 1587, 1606, 1791, 2226+C, 2614, 2770, 2788, 4062, 4063, 4392, 4646, 4669, 4830, 5061, 5210, 5239+A, 5277+A, 5884, 6004, 6175, 6247, 6307, 6784, 7001, 8005, 8358, 8379, 8792, 9239, 9694, 9948, 10083, 10214, 10238, 10376, 11240, 11543, 11827, 11842, 12404, 12443, 12683, 12767, 13049, 13333, 13502, 13761, 14554, 15385+T, 15492, 15535, 15582, 15599, 15647, 15706, 15717, 15768, 15823, 15867, 15971, 16108, 16110, 16119, 16126N, 16127N, 16128N, 16129N, 16130N, 16131N, 16132N, 16133N, 16134N, 16135N, 16136N, 16137N, 16138N, 16139N, 16140N, 16141N, 16142N, 16143N, 16144N, 16145N, 16146N, 16147N, 16148N, 16149N, 16150N, 16151N, 16152N, 16153N, 16154N, 16155N, 16156N, 16157N, 16158N, 16159N, 16160N, 16161N, 16162N, 16163N, 16164N, 16165N, 16166N, 16167N, 16168N, 16169N, 16170N, 16171N, 16172N, 16173N, 16174N, 16175N, 16176N, 16177N, 16178N, 16179N, 16180N, 16181N, 16182N, 16183N, 16184N, 16185N, 16186N, 16187N, 16188N, 16189N, 16190N, 16191N, 16192N, 16193N, 16194N, 16195N, 16196N, 16197N, 16198N, 16199N, 16200N, 16201N, 16202N, 16203N, 16204N, 16205N, 16206N, 16207N, 16208N, 16209N, 16210N, 16211N, 16212N, 16213N, 16214N, 16215N, 16216N, 16217N, 16218N, 16219N, 16220N, 16221N, 16222N, 16223N, 16224N, 16225N, 16226N, 16227N, 16228N, 16229N, 16230N, 16231N, 16232N, 16233N, 16234N, 16235N, 16236N, 16237N, 16238N, 16239N, 16240N, 16241N, 16242N, 16243N, 16244N, 16245N, 16246N, 16247N, 16248N, 16249N, 16250N, 16251N, 16252N, 16253N, 16254N, 16255N, 16256N, 16257N, 16258N, 16259N, 16260N, 16261N, 16262N, 16263N, 16264N, 16265N, 16266N, 16267N, 16268N, 16269N, 16270N, 16271N, 16272N, 16273N, 16274N, 16275N, 16276N, 16277N, 16278N, 16279N, 16280N, 16281N, 16282N, 16283N, 16284N, 16285N, 16286N, 16287N, 16288N, 16289N, 16290N, 16291N, 16292N, 16293N, 16294N, 16295N, 16296N, 16297N, 16298N, 16299N, 16300N, 16301N, 16302N, 16303N, 16304N, 16305N, 16306N, 16307N, 16308N, 16309N, 16310N, 16311N, 16312N, 16313N, 16314N, 16315N, 16316N, 16317N, 16318N, 16319N, 16320N, 16321N, 16322N, 16323N, 16324N, 16325N, 16326N, 16327N, 16328N, 16329N, 16330N, 16331N, 16332N, 16333N, 16334N, 16335N, 16336N, 16337N, 16338N, 16339N, 16340N, 16341N, 16342N, 16343N, 16344N, 16345N, 16346N, 16347N, 16348N, 16349N, 16350N, 16351N, 16352N, 16353N, 16354N, 16355N, 16356N, 16357N, 16368, 16436 | I2a3_HQ439499:356 | I2a3_HQ439499:356d, 2226+C, 5239+A, 5277+A, 15385+T, 16126N, 16127N, 16128N, 16129N, 16130N, 16131N, 16132N, 16133N, 16134N, 16135N, 16136N, 16137N, 16138N, 16139N, 16140N, 16141N, 16142N, 16143N, 16144N, 16145N, 16146N, 16147N, 16148N, 16149N, 16150N, 16151N, 16152N, 16153N, 16154N, 16155N, 16156N, 16157N, 16158N, 16159N, 16160N, 16161N, 16162N, 16163N, 16164N, 16165N, 16166N, 16167N, 16168N, 16169N, 16170N, 16171N, 16172N, 16173N, 16174N, 16175N, 16176N, 16177N, 16178N, 16179N, 16180N, 16181N, 16182N, 16183N, 16184N, 16185N, 16186N, 16187N, 16188N, 16189N, 16190N, 16191N, 16192N, 16193N, 16194N, 16195N, 16196N, 16197N, 16198N, 16199N, 16200N, 16201N, 16202N, 16203N, 16204N, 16205N, 16206N, 16207N, 16208N, 16209N, 16210N, 16211N, 16212N, 16213N, 16214N, 16215N, 16216N, 16217N, 16218N, 16219N, 16220N, 16221N, 16222N, 16223N, 16224N, 16225N, 16226N, 16227N, 16228N, 16229N, 16230N, 16231N, 16232N, 16233N, 16234N, 16235N, 16236N, 16237N, 16238N, 16239N, 16240N, 16241N, 16242N, 16243N, 16244N, 16245N, 16246N, 16247N, 16248N, 16249N, 16250N, 16251N, 16252N, 16253N, 16254N, 16255N, 16256N, 16257N, 16258N, 16259N, 16260N, 16261N, 16262N, 16263N, 16264N, 16265N, 16266N, 16267N, 16268N, 16269N, 16270N, 16271N, 16272N, 16273N, 16274N, 16275N, 16276N, 16277N, 16278N, 16279N, 16280N, 16281N, 16282N, 16283N, 16284N, 16285N, 16286N, 16287N, 16288N, 16289N, 16290N, 16291N, 16292N, 16293N, 16294N, 16295N, 16296N, 16297N, 16298N, 16299N, 16300N, 16301N, 16302N, 16303N, 16304N, 16305N, 16306N, 16307N, 16308N, 16309N, 16310N, 16311N, 16312N, 16313N, 16314N, 16315N, 16316N, 16317N, 16318N, 16319N, 16320N, 16321N, 16322N, 16323N, 16324N, 16325N, 16326N, 16327N, 16328N, 16329N, 16330N, 16331N, 16332N, 16333N, 16334N, 16335N, 16336N, 16337N, 16338N, 16339N, 16340N, 16341N, 16342N, 16343N, 16344N, 16345N, 16346N, 16347N, 16348N, 16349N, 16350N, 16351N, 16352N, 16353N, 16354N, 16355N, 16356N, 16357N | Lippold,S., Matzke,N., Reissman,M., Burbano,H. and Hofreiter,M. BMC Evolutionary Biology. 328 (11), 1471-2148 (2011) |
| HQ439500 (Horse and Przewalski's horse) | Illumina/Solexa Genome Analyzer II | Phantom mutaions: 358d, 2227+T, 5239+A, 5277+A, 15385+T. | P4a_HQ439500 | 158, 356d, 739, 860, 961, 1383, 1387A, 1684C, 2227+T, 2788, 3070, 3259T, 3271, 3557, 3616, 3800, 3942, 4062, 4536, 4605N, 4646, 4669, 5239+A, 5277+A, 5527, 5827, 5884, 6004, 6307, 6505, 6529, 6712, 6784, 7001, 7243, 7427, 7612T, 7666, 7898, 7900, 7981, 8005, 8076, 8150, 8238, 8358, 8361, 8556T, 8565, 8855, 9053, 9086, 9239, 9775, 10110, 10173, 10214, 10292, 10376, 10448, 10859, 11240, 11378, 11394, 11424, 11492, 11543, 11842, 11879, 11966, 12167, 12230, 12287, 12332, 12404, 12503, 12767, 13049, 13333, 13370, 13463, 13466, 13629, 13920, 13933, 14626, 14803, 15202, 15342, 15385+T, 15492, 15599, 15601, 15664, 15700, 15717, 15768, 15774, 15806, 15953, 15992, 16035, 16118, 16126N, 16127N, 16128N, 16129N, 16130N, 16131N, 16132N, 16133N, 16134N, 16135N, 16136N, 16137N, 16138N, 16139N, 16140N, 16141N, 16142N, 16143N, 16144N, 16145N, 16146N, 16147N, 16148N, 16149N, 16150N, 16151N, 16152N, 16153N, 16154N, 16155N, 16156N, 16157N, 16158N, 16159N, 16160N, 16161N, 16162N, 16163N, 16164N, 16165N, 16166N, 16167N, 16168N, 16169N, 16170N, 16171N, 16172N, 16173N, 16174N, 16175N, 16176N, 16177N, 16178N, 16179N, 16180N, 16181N, 16182N, 16183N, 16184N, 16185N, 16186N, 16187N, 16188N, 16189N, 16190N, 16191N, 16192N, 16193N, 16194N, 16195N, 16196N, 16197N, 16198N, 16199N, 16200N, 16201N, 16202N, 16203N, 16204N, 16205N, 16206N, 16207N, 16208N, 16209N, 16210N, | P4a_HQ439500:356, 4605 | P4a_HQ439500:356d, 2227+T, 4605N, 5239+A, 5277+A, 15385+T, 16126N, 16127N, 16128N, 16129N, 16130N, 16131N, 16132N, 16133N, 16134N, 16135N, 16136N, 16137N, 16138N, 16139N, 16140N, 16141N, 16142N, 16143N, 16144N, 16145N, 16146N, 16147N, 16148N, 16149N, 16150N, 16151N, 16152N, 16153N, 16154N, 16155N, 16156N, 16157N, 16158N, 16159N, 16160N, 16161N, 16162N, 16163N, 16164N, 16165N, 16166N, 16167N, 16168N, 16169N, 16170N, 16171N, 16172N, 16173N, 16174N, 16175N, 16176N, 16177N, 16178N, 16179N, 16180N, 16181N, 16182N, 16183N, 16184N, 16185N, 16186N, 16187N, 16188N, 16189N, 16190N, 16191N, 16192N, 16193N, 16194N, 16195N, 16196N, 16197N, 16198N, 16199N, 16200N, 16201N, 16202N, 16203N, 16204N, 16205N, 16206N, 16207N, 16208N, 16209N, 16210N, 16211N, 16212N, 16213N, 16214N, 16215N, 16216N, 16217N, 16218N, 16219N, 16220N, 16221N, 16222N, 16223N, 16224N, 16225N, 16226N, 16227N, 16228N, 16229N, 16230N, 16231N, 16232N, 16233N, 16234N, 16235N, 16236N, 16237N, 16238N, 16239N, | Lippold,S., Matzke,N., Reissman,M., Burbano,H. and Hofreiter,M. BMC Evolutionary Biology. 328 (11), 1471-2148 (2011) |
| | | | | 16211N, 16212N, 16213N, 16214N, 16215N, 16216N, 16217N, 16218N, 16219N, 16220N, 16221N, 16222N, 16223N, 16224N, 16225N, 16226N, 16227N, 16228N, 16229N, 16230N, 16231N, 16232N, 16233N, 16234N, 16235N, 16236N, 16237N, 16238N, 16239N, 16240N, 16241N, 16242N, 16243N, 16244N, 16245N, 16246N, 16247N, 16248N, 16249N, 16250N, 16251N, 16252N, 16253N, 16254N, 16255N, 16256N, 16257N, 16258N, 16259N, 16260N, 16261N, 16262N, 16263N, 16264N, 16265N, 16266N, 16267N, 16268N, 16269N, 16270N, 16271N, 16272N, 16273N, 16274N, 16275N, 16276N, 16277N, 16278N, 16279N, 16280N, 16281N, 16282N, 16283N, 16284N, 16285N, 16286N, 16287N, 16288N, 16289N, 16290N, 16291N, 16292N, 16293N, 16294N, 16295N, 16296N, 16297N, 16298N, 16299N, 16300N, 16301N, 16302N, 16303N, 16304N, 16305N, 16306N, 16307N, 16308N, 16309N, 16310N, 16311N, 16312N, 16313N, 16314N, 16315N, 16316N, 16317N, 16318N, 16319N, 16320N, 16321N, 16322N, 16323N, 16324N, 16325N, 16326N, 16327N, 16328N, 16329N, 16330N, 16331N, 16332N, 16333N, 16334N, 16335N, 16336N, 16337N, 16338N, 16339N, 16340N, 16341N, 16342N, 16343N, 16344N, 16345N, 16346N, 16347N, 16348N, 16349N, 16350N, 16351N, 16352N, 16353N, 16354N, 16355N, 16356N, 16357N, 16368, 16540A, 16626, 16657d | | 16240N, 16241N, 16242N, 16243N, 16244N, 16245N, 16246N, 16247N, 16248N, 16249N, 16250N, 16251N, 16252N, 16253N, 16254N, 16255N, 16256N, 16257N, 16258N, 16259N, 16260N, 16261N, 16262N, 16263N, 16264N, 16265N, 16266N, 16267N, 16268N, 16269N, 16270N, 16271N, 16272N, 16273N, 16274N, 16275N, 16276N, 16277N, 16278N, 16279N, 16280N, 16281N, 16282N, 16283N, 16284N, 16285N, 16286N, 16287N, 16288N, 16289N, 16290N, 16291N, 16292N, 16293N, 16294N, 16295N, 16296N, 16297N, 16298N, 16299N, 16300N, 16301N, 16302N, 16303N, 16304N, 16305N, 16306N, 16307N, 16308N, 16309N, 16310N, 16311N, 16312N, 16313N, 16314N, 16315N, 16316N, 16317N, 16318N, 16319N, 16320N, 16321N, 16322N, 16323N, 16324N, 16325N, 16326N, 16327N, 16328N, 16329N, 16330N, 16331N, 16332N, 16333N, 16334N, 16335N, 16336N, 16337N, 16338N, 16339N, 16340N, 16341N, 16342N, 16343N, 16344N, 16345N, 16346N, 16347N, 16348N, 16349N, 16350N, 16351N, 16352N, 16353N, 16354N, 16355N, 16356N, 16357N | |
| AP012268 (Horse and Przewalski's horse) | Illumina/Solexa Genome Analyzer System II | Phantom mutaions: 356d, 1387A, 2227+T, 5239+A, 5277+A, 15385+T | F | 158, 356d, 382, 387, 416, 1387A, 2227+T, 2788, 2940, 3371, 3562, 3576, 4062, 4646, 4669, 4830, 5239+A, 5277+A, 5830, 5884, 6004, 6307, 6784, 7001, 7504, 7746, 8005, 8037, 8789, 9239, 9669, 10214, 10376, 10471, 11165, 11240, 11543, 11820, 11842, 12767, 13333, 13502, 14350, 14626, 15385+T, 15492, 15539, 15592, 15599, 15647, 15663, 15717, 15865, 15867, 16127, 16368, 16511 | F:356, 1387d, 12860, 13042, 13049, 14734 | F:356d, 1387A, 2227+T, 5239+A, 5277+A, 15385+T, 16127 | Goto,H., Ryder,O.A., Fisher,A.R., Schultz,B., Kosakovsky Pond,S.L., Nekrutenko,A. and Makova,K.D. Genome Biol Evol 3, 1096-1106 (2012) |
| AP012269 (Horse and Przewalski's horse) | Illumina/Solexa Genome Analyzer System II | Phantom mutaions: 356d, 2227+T, 5239+A, 5277+A, 15385+T | JK | 158, 297, 356d, 961, 1155, 1326, 2227+T, 2788, 2863, 3043, 3070, 3523, 3557, 3942, 3963, 4062, 4296, 4332, 4536, 4608, 4646, 4669, 5239+A, 5277+A, 5527, 5884, 6004, 6172, 6307, 6457, 6463, 6784, 6877, 7001, 7057, 7201, 7666, 8005, 8040, 8082, 8106, 8358, 8565, 8840, 9071, 9188, 9239, 9347, 9597, 9694, 10110, 10177, 10376, 10400, 11165, 11426, 11543, 11842, 11879, 12332, 12407, 12422, 12767, 13049, 13333, 13336, 13367, 13948, 14395, 14551, 14572, 14626, 14765, 14803, 15001, 15385+T, 15492, 15566, 15582, 15599, 15717, 15768, 15772, 15867, 15868, 16034, 16110, 16127, 16368 | JK:356, 1387d, 6388, 8442, 9396, 10214, 10217, 11240, 13229, 15647, 15804, 16068 | JK:297, 356d, 1326, 2227+T, 3043, 3523, 3557, 5239+A, 5277+A, 6172, 6457, 6463, 8040, 8106, 8840, 9188, 9347, 9597, 9694, 10177, 10400, 11165, 11426, 12407, 12422, 13367, 14572, 14626, 14765, 15001, 15385+T, 15566, 15582, 15772, 16034, 16110, 16127 | Goto,H., Ryder,O.A., Fisher,A.R., Schultz,B., Kosakovsky Pond,S.L., Nekrutenko,A. and Makova,K.D. Genome Biol Evol 3, 1096-1106 (2013) |
| AP012270 (Horse and Przewalski's horse) | Illumina/Solexa Genome Analyzer System II | Phantom mutaions: 356d, 2229+A, 5239+A, 5277+A, 15385+T | F | 1-12d, 356d, 382, 387, 416, 2229+A, 2788, 3371, 3562, 4062, 4646, 4669, 4830, 5239+A, 5277+A, 5830, 5884, 6004, 6307, 6784, 7001, 7504, 7746, 8005, 8037, 8120A, 8789, 9239, 10214, 10376, 10471, 11543, 12767, 13333, 14350, 15385+T, 15492, 15539, 15592, 15599, 15647, 15663, 15717, 15865, 15867, 16368, 16511, 16657d | F:158, 356, 1387d, 2940, 3576, 9669, 11165, 11240, 11820, 11842, 12860, 13042, 13049, 13502, 14626, 14734 | F:1-12d, 356d, 2229+A, 5239+A, 5277+A, 8120A, 15385+T, 16657d | Goto,H., Ryder,O.A., Fisher,A.R., Schultz,B., Kosakovsky Pond,S.L., Nekrutenko,A. and Makova,K.D. Genome Biol Evol 3, 1096-1106 (2011) |
| KF038159 (Horse and Przewalski's horse) | Illumina | Phantom mutaions: 355+C | Q3_JN398455 | 158, 287, 302, 341, 355+C, 542, 739, 860, 961, 1387d, 1427, 2788, 3070, 3259T, 3271, 3616, 3800, 3942, 4062, 4201, 4536, 4605, 4646, 4669, 5527, 5827, 5884, 6004, 6307, 6529, 6712, 6784, 7001, 7243, 7294, 7612T, 7666, 7898, 7900, 8005, 8076, 8150, 8238, 8358, 8361, 8556T, 8565, 8855, 9086, 9203, 9239, 9775, 10110, 10173, 10214, 10292, 10376, 10448, 10859, 11240, 11378, 11394, 11424, 11492, 11543, 11842, 11879, 11942, 11966, 12167, 12230, 12332, 12404, 12767, 13020, 13049, 13232, 13333, 13463, 13629, 13920, 13933, 14626, 14803, 15202, 15342, 15492, 15582, 15599, 15601, 15700, 15717, 15737, 15768, 15774, 15808, 15953, 16034, 16035, 16054, 16060+C, 16368, 16540A, 16626 | Q3_JN398455:356, 16118 | Q3_JN398455:287, 355+C, 542, 1427, 13020 | Yoon,S.H.Unpublished |
| KF038160 | Illumina | Phantom | P | 142, 158, 287, 355+C, 542, 721, 739, 860, 961, 1383, 1387d, 1684C, 1892, 2739, 2788, 3070, 3259T, 3271, 3557, 3616, 3727, 3800, | P:356, 15594, 15599, 16118 | P:142, 287, 355+C, 542, 721, 1892, 2739, 3727, 5068, 5359, 9696, 10379, 10631, 13304, 13696, 13861, | Yoon,S.H.Unpublished |
| (Horse and Przewalski's horse) | | mutaions: 355+C | | 3942, 4062, 4536, 4605, 4646, 4669, 5068, 5359, 5527, 5827, 5884, 6004, 6307, 6505, 6529, 6712, 6784, 7001, 7243, 7427, 7612T, 7666, 7898, 7900, 8005, 8076, 8150, 8238, 8358, 8361, 8556T, 8565, 8855, 9053, 9086, 9239, 9696, 9775, 10110, 10173, 10214, 10292, 10376, 10379, 10448, 10631, 10859, 11240, 11378, 11394, 11424, 11492, 11543, 11842, 11879, 11966, 12167, 12230, 12287, 12332, 12404, 12767, 13049, 13304, 13333, 13370, 13463, 13466, 13629, 13696, 13861, 13920, 13933, 14016, 14626, 14803, 14825, 15202, 15342, 15492, 15664, 15700, 15717, 15768, 15774, 15806, 15953, 16035, 16060+C, 16368, 16404d, 16540A, 16560+C, 16626 | | 14016, 14825, 16404d, 16560+C | |
| KF038161 (Horse and Przewalski's horse) | Illumina | Phantom mutaions: 355+C | I2a | 4T, 158, 287, 355+C, 1387d, 1587, 1791, 2614, 2770, 2788, 4062, 4063, 4392, 4646, 4669, 4830, 5061, 5210, 5217d, 5884, 6004, 6175, 6247, 6307, 6784, 7001, 8005, 8358, 8379, 8792, 9239, 9694, 9948, 10083, 10214, 10238, 10376, 11240, 11543, 11827, 11842, 12404, 12443, 12683, 12767, 13049, 13333, 13502, 13761, 14554, 15492, 15535, 15594, 15599, 15647, 15706, 15717, 15768, 15823, 15867, 15960, 15971, 16110, 16368, 16436 | I2a:356, 15582 | I2a:4T, 287, 355+C, 5217d, 15594, 15960 | Yoon,S.H.Unpublished |
| KF038162 (Horse and Przewalski's horse) | Illumina | Phantom mutaions: 355+C, 358+C | D, D1_JN398400 | 158, 287, 355+C, 358+C, 1387d, 1694, 2188, 2217, 2788, 3412, 4062, 4646, 5884, 6004, 6532, 6784, 7001, 7982, 8110, 8867, 9239, 9384, 9402, 10214, 10376, 11240, 11417, 11457, 11543, 11690, 12191, 12218, 12785, 12791, 13049, 13333, 13948, 14200, 15492, 15518, 15599, 15614, 15717, 15734, 15767, 15867, 16108, 16110, 16368, 16391, 16560+C | D:356, 5497, 14825, 15807, 16084; D1_JN398400:356, 3557, 5497, 15807, 16084 | D:287, 355+C, 358+C, 2188, 3412, 7982, 8110, 9384, 11690, 12785, 15614, 16560+C; D1_JN398400:287, 355+C, 358+C, 2188, 3412, 7982, 8110, 9384, 11690, 12785, 15614, 16560+C | Yoon,S.H.Unpublished |
| KF038163 (Horse and Przewalski's horse) | Illumina | Phantom mutaions: 355+C, 358+C | J_JN398418 | 4C, 158, 160, 287, 355+C, 358+C, 384T, 961, 1155, 1243, 1387d, 1427, 2263, 2788, 2863, 3070, 3283, 3942, 3963, 4062, 4188, 4296, 4332, 4536, 4608, 4646, 4669, 4782, 4785, 5527, 5884, 6004, 6283, 6307, 6388, 6472, 6658, 6784, 6877, 7001, 7057, 7201, 7576, 7666, 8005, 8082, 8289, 8358, 8442, 8545, 8565, 9071, 9239, 9248, 9396, 10110, 10214, 10217, 10277, 10376, 11081, 11240, 11543, 11821, 11842, 11879, 11933, 12332, 12767, 13049, 13124, 13229, 13333, 13336, 13696, 13948, 14395, 14551, 14803, 15492, 15529d, 15582, 15599, 15717, 15768, 15804, 15824, 15867, 15868, 16068, 16108, 16110, 16368, 16641 | J_JN398418:356, 1520, 15601 | J_JN398418:4C, 287, 355+C, 358+C, 1427, 6658, 13696 | Yoon,S.H.Unpublished |
| KF038164 (Horse and Przewalski's horse) | Illumina | Phantom mutaions: 355+C, 358+C | D, D1_JN398400 | 158, 287, 355+C, 358+C, 1387d, 1694, 2188, 2217, 2788, 3412, 4062, 4646, 5884, 6004, 6532, 6784, 7001, 7982, 8110, 8867, 9239, 9384, 9402, 10214, 10376, 10631, 11240, 11417, 11457, 11543, 11690, 12191, 12218, 12785, 12791, 13049, 13333, 13948, 14200, 15492, 15518, 15599, 15614, 15717, 15734, 15767, 15867, 16108, 16110, 16368, 16391, 16560+C | D:356, 5497, 14825, 15807, 16084; D1_JN398400:356, 3557, 5497, 15807, 16084 | D:287, 355+C, 358+C, 2188, 3412, 7982, 8110, 9384, 10631, 11690, 12785, 15614, 16560+C; D1_JN398400:287, 355+C, 358+C, 2188, 3412, 7982, 8110, 9384, 10631, 11690, 12785, 15614, 16560+C | Yoon,S.H.Unpublished |
| KF038165 (Horse and Przewalski's horse) | Illumina | Phantom mutaions: 355+C, 358+C | L2a2 | 158, 287, 355+C, 358+C, 961, 1375, 1387d, 2607, 2788, 2899, 3517, 3942, 4062, 4536, 4646, 4669, 5527, 5815, 5884, 6004, 6307, 6784, 7001, 7516, 7666, 7900, 8005, 8058, 8199, 8301, 8319, 8358, 8403, 8565, 9239, 9951, 9999, 10110, 10214, 10292, 10376, 10421, 10613, 11240, 11543, 11693, 11842, 11879, 12119, 12200, 12767, 12896, 12950, 13049, 13333, 13520, 14803, 14995, 15313, 15531, 15582, 15600, 15646, 15717, 15768, 15867, 15868, 15953, 15971, 16065, 16100, 16368, 16403-16404d | L2a2:356, 15491, 15492, 15493, 16626 | L2a2:287, 355+C, 358+C, 9999, 16403-16404d | Yoon,S.H.Unpublished |
| KF038166 (Horse and Przewalski's horse) | Illumina | Phantom mutaions: 355+C, 358+C | H, L2a | 158, 355+C, 358+C, 961, 1375, 1387d, 1546, 1668, 2607, 2788, 3064, 3517, 3942, 4062, 4182, 4536, 4646, 4669, 4830, 5272, 5527, 5815, 5884, 6004, 6076, 6127, 6307, 6565, 6784, 6835, 7001, 7093, 7111, 7258, 7516, 7666, 7900, 8005, 8058, 8301, 8319, 8358, 8403, 8481, 8565, 8741, 8763, 8776, 9209, 9239, 9750, 9856, 9999, 10083, 10110, 10214, 10292, 10376, 10421, 10613, 10914, 11240, 11543, 11693, 11842, 11879, 12119, 12200, 12767, 12896, 12950, 13049, 13333, 13502, 13520, 13761, 13996T, 14335, 14803, 15004, 15313, 15381d, 15523, 15531, 15537, 15582, 15599, 15646, 15717, 15768, 15867, 15868, 15953, 15971, 16065, 16077, 16078, 16100, 16110, 16128, 16368, 16403-16404d, 16626 | H:356, 15492, 15715; L2a:356, 2899, 9951, 14995, 15491, 15492, 15493, 15600 | H:355+C, 358+C, 961, 1375, 2607, 3517, 3942, 4536, 5527, 5815, 7516, 7666, 7900, 8058, 8301, 8319, 8358, 8403, 8565, 9750, 9999, 10110, 10292, 10421, 10613, 11693, 11879, 12119, 12200, 12896, 12950, 13520, 13996T, 14803, 15313, 15531, 15868, 15953, 16065, 16077, 16100, 16128, 16403-16404d, 16626; L2a:355+C, 358+C, 1546, 1668, 3064, 4182, 4830, 5272, 6076, 6127, 6565, 6835, 7093, 7111, 7258, 8481, 8741, 8763, 8776, 9209, 9750, 9856, 9999, 10083, 10914, 13502, 13761, 13996T, 14335, 15004, 15381d, 15523, 15537, 15599, 16077, 16078, 16110, 16128, 16403-16404d | Yoon,S.H.Unpublished |